\documentclass[a4paper,11pt]{article} 
\usepackage[top=1in,left=1in,right=1in,bottom=1in]{geometry}

\RequirePackage{amsthm,amsmath,amsfonts,amssymb}
\RequirePackage{graphicx}
\usepackage{hyperref}
\usepackage[utf8]{inputenc} %
\usepackage{microtype}
\usepackage{mathrsfs}
\usepackage{amscd}     
\usepackage{esint}
\usepackage{enumitem}
\usepackage[dvipsnames]{xcolor}
\usepackage{mathtools}
\usepackage{bbm}
\usepackage[normalem]{ulem} %
\usepackage{bm}
\usepackage{chemarrow}
\usepackage{multirow}
\usepackage{latexsym} 
\usepackage{bigints}
\usepackage{booktabs} 
\usepackage{tabularx} 
\usepackage{threeparttable}
\usepackage{lscape}
\usepackage[font=small,labelfont=bf]{caption}
\usepackage{subcaption}
\usepackage{array} 
\usepackage{makecell}
\usepackage{url}
\usepackage[final]{pdfpages}
\usepackage{afterpage}
\usepackage{lastpage}
\usepackage{fancyhdr}       %
\usepackage[ruled,vlined]{algorithm2e} %
\usepackage{soul}
\usepackage{ragged2e}
\usepackage{import}
\usepackage{orcidlink} 
\usepackage{titletoc}
\usepackage{appendix}

\usepackage{natbib}
\usepackage{hyperref}

\definecolor{MATblue}{HTML}{0072BD}
\definecolor{MATgreen}{HTML}{77AC30}
\definecolor{MATorange}{HTML}{D95319}
\definecolor{MATred}{HTML}{A2142F}
\definecolor{MATyellow}{HTML}{EDB120}
\definecolor{MATcyan}{HTML}{4DBEEE}
\definecolor{MATpurple}{HTML}{7E2F8E}

\hypersetup{
	colorlinks=true,
	citecolor=MATblue,
	linkcolor=MATblue,
	filecolor=MATblue,      
	urlcolor=MATblue,
}

\setulcolor{MATred}
\numberwithin{equation}{section}
\theoremstyle{plain}

\newtheorem{theorem}{Theorem}[section]
\newtheorem{proposition}{Proposition}[section]
\newtheorem{corollary}{Corollary}[section]
\newtheorem{lemma}[theorem]{Lemma}
\theoremstyle{definition}

\newtheorem{assumption}{Assumption}
\newtheorem{remark}{Remark}
\newtheorem{example}{Example}

\DeclareMathOperator*{\argmin}{arg min}

\usepackage[many]{tcolorbox}
\newtcolorbox{colbox}{
    enhanced,
    sharp corners,
    breakable,  %
    borderline west={1pt}{0pt}{gray},
    colback=gray!10,  %
    colframe=gray!10  %
}

\newcommand{\E}{\mathbb{E}}
\newcommand{\de}{\textnormal{d}}
\newcommand{\Prob}{{\mathbb{P}}}

\newcommand{\inv}{^{-1}}

\newcommand{\Ind}{{\mathbbm{1}}}

\newcommand{\toprob}{\overset{p}{\to}}
\newcommand{\todist}{\overset{d}{\to}}
\newcommand{\R}{\mathbb{R}}

\newcommand{\N}{\mathbb{N}}

\newcommand{\abs}[1]{\lvert#1\rvert}
\newcommand{\Abs}[1]{\left\lvert#1\right\rvert}
\newcommand{\norm}[1]{\lVert#1\rVert}
\newcommand{\Norm}[1]{\left\lVert#1\right\rVert}

\newcommand{\vertiii}[1]{{\left\vert\kern-0.25ex\left\vert\kern-0.25ex\left\vert #1 
    \right\vert\kern-0.25ex\right\vert\kern-0.25ex\right\vert}}

\newcommand{\raisedrule}[2][0em]{\leavevmode\leaders\hbox{\rule[#1]{1pt}{#2}}\hfill\kern0pt}

\usepackage{scalerel,stackengine}
\stackMath
\newcommand\reallywidehat[1]{%
\savestack{\tmpbox}{\stretchto{%
  \scaleto{%
    \scalerel*[\widthof{\ensuremath{#1}}]{\kern-.6pt\bigwedge\kern-.6pt}%
    {\rule[-\textheight/2]{1ex}{\textheight}}%
  }{\textheight}%
}{0.5ex}}%
\stackon[1pt]{#1}{\tmpbox}%
}

\newlist{todolist}{itemize}{2}
\setlist[todolist]{label=$\square$}
\usepackage{pifont}

\begin{document}

\title{\vspace{-2em}\normalfont\linespread{1.2}\selectfont%
    Statistical Inference with Ill-Conditioned Random Weight Neural Network Sieves
}

\author{%
    Giovanni Ballarin\,\orcidlink{0000-0003-0312-1169}%
    \thanks{%
    Corresponding author -- University of St.~Gallen, Division of Mathematics and Statistics, Rosenbergstrasse 22, 9000 St.~Gallen, Switzerland. {\texttt{Giovanni.Ballarin@unisg.ch}}}
    \and
    Lukas Gonon\,\orcidlink{0000-0003-3367-2455}%
    \thanks{%
   	University of St.~Gallen, School of Computer Science, Rosenbergstrasse 30, 9000 St.~Gallen, Switzerland. {\texttt{Lukas.Gonon@unisg.ch}}. Honorary Senior Lecturer. Imperial College London, United Kingdom. }
    \and
    Lyudmila Grigoryeva\,\orcidlink{0000-0002-4857-7779}%
    \thanks{%
    University of St.~Gallen, Division of Mathematics and Statistics, Rosenbergstrasse 22, 9000 St.~Gallen, Switzerland. {\texttt{Lyudmila.Grigoryeva@unisg.ch}}. Honorary Associate Professor. University of Warwick, Department of Statistics, Coventry, CV4 7AL, United Kingdom. }
}
\date{\today}

\maketitle

\makeatletter
\let\ofootnote\footnote
\def\footnote{\@ifstar\footnote@star\footnote@nostar}
\def\footnote@star#1{{\let\thefootnote\relax\footnotetext{#1}}}
\def\footnote@nostar{\ofootnote}
\footnote*{%
	\textit{Acknowledgements:} GB is thankful to the Great Minds Postdoctoral Fellowship program of the University of St.~Gallen for financial support.
    We used generative AI to polish the manuscript, proofread, and assist with code development, but not to formulate theoretical models, conceptual frameworks, hypotheses, or mathematical claims. We manually reviewed all AI-assisted work and take full responsibility for the final content.
}%
\makeatother

\vspace{0em}

\begin{abstract}
Random weight neural networks (RWNNs) are shallow networks whose hidden-layer parameters are randomly generated and then kept fixed, while only the output coefficients are estimated. Conditional on the sampled parameters, RWNN regression is a linear sieve problem with random approximation space. We show that such random sieves cannot remain uniformly well-conditioned as their dimension grows. For parametrized feature families satisfying polynomial covering and Lipschitz conditions, the eigenvalues of the population covariance matrix are proven to have polynomial upper bounds determined by the covering geometry of the parameter space, leading to a vanishing smallest eigenvalue. We develop approximation, estimation, and inference theory that explicitly treats the ill-conditioning. Under a $\beta$-H\"older source condition, we obtain both faster-than-Monte-Carlo $L^2$ approximation rates and new $L^\infty$ approximation bounds based on stratified sampling. We then derive improved consistency guarantees for ridge-regularized least squares driven by ridge effective dimension, spectral, and leverage-based stability conditions. For inference, we establish asymptotic linearity and Gaussian limits for linear functional directions while allowing the smallest eigenvalue of the population covariance matrix to vanish. Nondegeneracy is directional: we characterize it with a ridge-normalized Rayleigh quotient and extend it to smooth functionals via Riesz representation. Consistency of ridge-adjusted variance estimators is proved, yielding feasible studentized inference. Numerical experiments illustrate the deterioration of random-feature conditioning and the finite-sample behavior of the resulting estimation and inference procedures.
\end{abstract}


\maketitle

\newpage

\startcontents[mainsections]
\printcontents[mainsections]{l}{1}{\section*{Content}\setcounter{tocdepth}{2}}

\section{Introduction}
\label{section:intro}

Random weight neural networks (RWNNs) are shallow neural networks whose hidden-layer parameters are randomly drawn and subsequently kept fixed, so that only the coefficients of the linear output layer are estimated from the data~\citep{GononGrigoryevaOrtega2023}. Neural networks of this type have been applied
very successfully in a variety of settings; we refer in particular to the seminal works on random feature models \citep{rahimi2007random} and extreme learning machines \citep{Huang2006}. Conditional on the sampled parameters, estimation under squared loss is a linear least-squares problem over the random finite-dimensional space spanned by the $K$ hidden units. When the same sequence of hidden-layer parameters is used as $K$ increases, these spaces are nested and form a random sieve, which we study within the framework of nonparametric sieve regression \citep{grenander1981abstract,gyorfi2006distribution,Chen2007,Tsybakov2009}. However, the population covariance matrix of the randomly generated features need not remain well-conditioned as $K$ increases: additional hidden units may improve approximation while introducing substantial redundancy in the feature representation. Understanding the statistical consequences of such ill-conditioning, particularly for inference on functionals of the regression function, is the central problem addressed in this paper.

Approximation properties of random neural networks have been studied under different assumptions on the target function and the distribution of the hidden-layer parameters \citep{Rahimi2008,GononGrigoryevaOrtega2023,neufeld2023universal2,liu2025integralrepresentationssobolevspaces,deryck2025randomized,mehmood2026random}. In the context of positive definite kernels, \citet{Bach2017} established approximation guarantees for random feature expansions in terms of the spectral properties of the associated integral operator, including results in $L^2$ and $L^\infty$. Related quantitative approximation results for shallow neural networks include those for general activation functions \citep{SiegelXu2020}, for Sobolev classes with ReLU activations \citep{maoApproximationRatesShallow2024}, and for non-polynomial activations on noncompact domains \citep{NeufeldSchmocker2026}.
Statistical convergence rates for random feature estimators have been established in several settings \citep{Rudi2017,Carratino2018,gononRandomFeatureNeural2023,Lanthaler2023}. 
In particular, \citet{BraunKohlerLangerWalk2024} establish an expected squared $L^2(X)$ error bound of order $O((\log n)^4n^{-1/2})$ for a shallow neural network estimator learned by gradient descent, under suitable Fourier-decay assumptions on the regression function. They obtain the same rate for a truncated linear least-squares estimator with randomly sampled hidden-layer parameters. \citet{mei2022generalization} derive precise asymptotics for random features ridge regression. Conditional on the sampled features, a ridge-regularized RWNN estimator is a finite-rank kernel ridge regressor, for which \citet{2023ATheoreticalAnalysisOfTheTestErrorOfFiniteRankKernelRidgeRegression} give sharp non-asymptotic bounds on the test error.

From the kernel point of view---the original setting of the random features approach of \cite{rahimi2007random}---statistical inference for kernel ridge regression has been studied in the recent literature, including uniform confidence bands based on Gaussian and bootstrap approximations \citep{SinghVijaykumar2023} and asymptotic normality for linear functionals, including point evaluations and derivatives \citep{TuoZou2024}. These results concern regularization in a prescribed reproducing kernel Hilbert space, and existing inferential results involving sampled features address particular estimation targets and statistical models; see \cite{ShenBellec2020,Steland2021,GuignardEtAl2021} for recent discussions. In contrast, we develop a theory of functional inference for growing RWNN sieves that accommodates increasingly ill-conditioned or singular population feature covariance matrices.

Randomly generated features present a fundamental and distinctive structural obstacle: a uniform positive lower bound on the population covariance eigenvalues cannot be sustained as the number of features increases. In fact, we show that, for parameterized feature families satisfying polynomial covering and Lipschitz conditions, the eigenvalues of the population feature covariance matrix satisfy polynomial decay upper bounds determined by the covering dimension of the parameter space, uniformly over feature realizations. In particular, its smallest eigenvalue necessarily converges to zero as the number of features increases. Related spectral ill-conditioning has been investigated by \citet{MingYu2026}, who establish rapid singular-value decay for random-feature discretization matrices associated with Fourier and hyperbolic tangent features. Our results instead concern the population regression covariance matrix under general covering and Lipschitz conditions.

Unlike an $L^2(X)$-orthonormal series sieve, whose population Gram matrix is $I_K$, an RWNN sieve does not generally have an orthogonal feature representation. Standard sieve inference typically relies on suitably normalized bases and control of their population Gram matrices \citep{Newey1997,Chen2007,Belloni2015,Chen2015}. Importantly, our spectral deterioration results are not restricted to randomly generated features: they hold uniformly over arbitrary choices of hidden-layer parameters within the prescribed compact feature family and therefore also apply to fully trainable shallow networks satisfying these restrictions. Our subsequent statistical analysis focuses on RWNNs. In this setting, both spectral deterioration and the resulting regularization bias must be accounted for explicitly.

We develop a framework for statistical inference with random sieves and RWNNs based on ridge normalization, in which the relevant quantities are measured relative to the regularized population covariance matrix. Two key results guide our analysis. First, the same covering geometry that induces spectral decay also controls the {\it ridge effective dimension}, which governs the stochastic estimation error and is sublinear in the nominal sieve dimension under suitable ridge scaling. This effective dimension plays a role analogous to that of its counterpart in kernel ridge regression \citep{CaponnettoDeVito2007,Rudi2017}. Second, and more importantly for inference, global spectral nondegeneracy is not required. We formulate a sufficient nondegeneracy condition {\it along the functional direction}, relative to the ridge level, through a ridge-normalized Rayleigh quotient. 
Together with conditions controlling empirical covariance stability, approximation, and ridge bias, we derive asymptotic normality conditional on the sampled hidden-layer parameters, in $P_\theta$-probability, even when the smallest eigenvalue of the population feature covariance matrix converges to zero. Under additional conditions ensuring consistency of the ridge-adjusted variance estimator, we obtain feasible studentized inference. The framework also accommodates irregular sieve functionals, such as point evaluation, whose normalizing variance may vary with the sieve dimension and need not remain bounded.

\subsection{Preview of contributions}

The main contributions of this paper concern four interconnected aspects of growing RWNN sieves: approximation by randomly generated features; spectral properties and $L^2(X)$ estimation; asymptotic inference for linear and smooth functionals; and uniform estimation under prediction-space stability conditions.  Throughout this subsection, $P_\theta$ denotes the law of the sampled hidden-layer parameters, $O_{P_\theta}$ refers to this law, and $O_P$ refers to the joint law of the observations and the hidden-layer parameters (see Subsection~\ref{subsec:notation} below). Table~\ref{tab:comparison_existing_results} summarizes selected approximation benchmarks from the literature and the principal approximation, spectral, estimation, and inference results established in this paper.

First, we establish $L^2$ approximation rates for RWNNs under a $\beta$-H\"older direct source condition. Adapting the control-neighbors Monte Carlo construction of \citet{lelucSpeedingMonteCarlo2024} to the RWNN integral representation yields an approximation error of order $O_{P_\theta}(K^{-1/2-\beta/(d+1)})$, improving upon the classical Monte Carlo order $K^{-1/2}$. Under the additional conditions of the generic sieve theory developed in Appendix~\ref{app:generic_sieve_estimation}, this improvement yields an $o_P(n^{-1/4})$ convergence rate for the constrained unregularized sieve estimator, ensuring that quadratic remainder terms are negligible at the $\sqrt{n}$ scale in smooth-functional inference. We also derive approximation guarantees for integral Sobolev classes (see Subsection~\ref{subsec:approximation_RWNNs_Sobolev}).
Furthermore, we establish sup-norm approximation bounds using stratified sampling of the hidden-layer parameters. Under the $\beta$-H\"older source condition, the expected sup-norm approximation error is bounded by a constant multiple of $(\sqrt{d}+\sqrt{\log K})K^{-1/2-\beta/(d+1)}$. For sign-valued source weights and ReLU$^m$ activations, $m\in\{1,2\}$, the corresponding bound is $(\sqrt{d}+\sqrt{\log K})K^{-1/2-1/(d+1)}$. In the latter case, the construction uses a partition respecting the sign sets of the source function and is therefore oracle unless this partition is known. The proof relies on a single centered empirical-process representation of the approximation error. The construction also provides deterministic $\ell_1$- and $\ell_2$-bounds on the output coefficients, used in the subsequent statistical analysis.

Second, we investigate the spectral properties of the population feature covariance matrix and their implications for statistical estimation. Let $g_K$ denote the vector of $K$ randomly generated features, and define $\Gamma_K:=\E_X[g_K(X)g_K(X)^\top]$ and $\Gamma_{K,\lambda}:=\Gamma_K+\lambda I_K$, where $\E_X$ denotes integration with respect to the distribution of $X$ for fixed hidden-layer parameters. Under polynomial covering and Lipschitz conditions on the feature family, with covering exponent $q_\Pi>0$, we establish spectral-tail bounds implying $\lambda_{\min}(\Gamma_K)\lesssim K^{-2/q_\Pi}$, uniformly over feature realizations. The same bounds control the ridge effective dimension,
\begin{equation*}
    \mathcal{N}_{K,\lambda}
    :=\operatorname{tr}(\Gamma_K\Gamma_{K,\lambda}^{-1})
    \lesssim 1+\left(\frac{K}{\lambda}\right)^{q_\Pi/(q_\Pi+2)}.
\end{equation*}
In particular, $\mathcal{N}_{K_n,\lambda}=o(K_n)$ under the scaling conditions $K_n/\lambda_{K_n,n}\to\infty$ and $\lambda_{K_n,n}K_n^{2/q_\Pi}\to\infty$. Thus, the same covering geometry that rules out uniform lower spectral conditioning also provides bounds on the effective dimension governing ridge estimation.
Building on this spectral analysis, we establish $L^2(X)$ convergence rates for ridge-regularized RWNN least squares estimators. Under the $\beta$-H\"older source condition and suitable ridge-normalized covariance stability, the ridge estimator $\widehat h_{n,\lambda}$ satisfies
\begin{equation*}
    \|\widehat h_{n,\lambda}-h\|_{L^2(X)}
    =
    O_P\left(
        \sqrt{\frac{\mathcal{N}_{K_n,\lambda}}{n}}
        +K_n^{-1/2-\beta/(d+1)}
        +\sqrt{\frac{\lambda_{K_n,n}\log K_n}{K_n}}
    \right).
\end{equation*}
The three terms correspond to stochastic estimation, sieve approximation, and regularization bias. In particular, when $\mathcal{N}_{K_n,\lambda}=o(K_n)$, the stochastic term is of smaller order than the classical $\sqrt{K_n/n}$ sieve term, without requiring a uniform lower bound on the eigenvalues of the population feature covariance matrix.
For comparison, spline and wavelet series estimators attain the minimax-optimal $L^2(X)$ rate $n^{-p/(2p+d)}$ for regression functions with H\"older smoothness $p>0$ under suitable assumptions \citep{stoneOptimalGlobalRates1982,huangLocalAsymptoticsPolynomial2003}. Deep neural network estimators can also achieve this rate up to logarithmic factors on H\"older classes under appropriate architectural restrictions \citep{SchmidtHieber2020}. For shallow neural networks, \citet{BraunKohlerLangerWalk2024} obtain an expected squared $L^2(X)$ error bound of order $(\log n)^4n^{-1/2}$ under Fourier-decay assumptions, including for an estimator with randomly sampled hidden-layer parameters. These benchmarks are not directly comparable with ours, since our $\beta$-H\"older condition concerns the source function in an integral representation of $h$, rather than the smoothness of $h$ itself. The distinguishing feature of our bound is that its stochastic term is governed by the ridge effective dimension $\mathcal{N}_{K_n,\lambda}$, which can be $o(K_n)$, rather than by the nominal sieve dimension $K_n$, even when the population covariance matrix becomes increasingly ill-conditioned.

Third, we establish asymptotic normality for linear functionals of ridge-regularized RWNN estimators. For a prescribed unit direction $\alpha_{K_n}$, measurable with respect to the sampled hidden-layer parameters, we characterize directional nondegeneracy through the ridge-normalized Rayleigh quotient
\begin{equation*}
    \mathcal{R}_{K_n,\lambda}(\alpha_{K_n})
    :=
    \frac{
        \alpha_{K_n}^{\top}
        \Gamma_{K_n,\lambda}^{-1}
        \Gamma_{K_n}
        \Gamma_{K_n,\lambda}^{-1}
        \alpha_{K_n}
    }{
        \alpha_{K_n}^{\top}
        \Gamma_{K_n,\lambda}^{-1}
        \alpha_{K_n}
    }.
\end{equation*}
Under suitable directional nondegeneracy, empirical covariance stability, approximation, and ridge-bias conditions, we obtain asymptotic normality for linear functionals of the estimated coefficients, centered at the corresponding $L^2(X)$ projection coefficients. The convergence holds conditionally on the sampled hidden-layer parameters, in $P_\theta$-probability. Under additional conditions, we establish consistency of a residual-based ridge-adjusted variance estimator, yielding feasible studentized inference. An explicit ReLU example demonstrates that the inferential and variance-consistency conditions can hold even when the population feature covariance matrix is exactly singular.
The directional theory also accommodates sieve-dependent normalizations, including those arising for irregular functionals such as point evaluation. Inference on the corresponding functional of $h$ additionally requires control of the sieve approximation bias. We extend the analysis to smooth nonlinear functionals $\Psi$ whose derivatives admit $L^2(X)$ Riesz representers. For quadratic remainders, an $o_P(n^{-1/4})$ convergence rate in $L^2(X)$ ensures negligibility at the $\sqrt{n}$ scale. Together with suitable functional bias and variance nondegeneracy conditions, this yields asymptotic normality for $\Psi(\widehat h_{n,\lambda})-\Psi(h)$. For the integrated-square functional, we also verify the first-order sieve-bias condition under an explicit choice of sieve dimension.

Fourth, we establish sup-norm error bounds for ridge-regularized RWNN estimators. We first derive bounds under spectral stability conditions and then develop an alternative approach based on prediction-space stability, formulated through ridge smoothing weights, the empirical ridge Lebesgue constant, and leverage-type quantities. Under suitable conditions on these quantities, we obtain uniform convergence bounds without requiring a uniform lower bound on the smallest eigenvalue of the empirical feature covariance matrix. In particular, for fixed $d$ and under the conditions of Corollary~\ref{corollary:uniform_estimation_rate_stable}, we obtain
\begin{equation*}
    \|\widehat h_{n,\lambda}-h\|_\infty
    =
    O_P\left(
        n^{-\frac{1+2\alpha}{4(1+\alpha)}}
        \sqrt{\log n}
    \right),
\end{equation*}
where $\alpha=\beta/(d+1)$ under the H\"older source condition and $\alpha=1/(d+1)$ under the sign source condition. The prediction-space stability quantities entering these bounds are also examined numerically in Section~\ref{section:simulations}.

The remainder of the paper is organized as follows. Section~\ref{sec:rwnn_regression} introduces the regression model and RWNN estimators. Section~\ref{sec:approximation_RWNNs} establishes $L^2$ and sup-norm approximation results. In Section~\ref{sec:Ridge_RWNN_least_squares}, we develop the spectral analysis, estimation bounds, and inferential theory for ridge-regularized RWNNs, including uniform estimation. Section~\ref{section:simulations} presents numerical experiments, and Section~\ref{sec:Conclusion} concludes. Proofs of the main results are provided in Appendices~\ref{app:rwnn_approximation}, \ref{app:ridge_L2_proofs}, \ref{app:inference_proofs}, and \ref{app:supnorm_ridge_proofs}. 
Appendix~\ref{app:generic_sieve_estimation} develops a general sieve framework for estimation and smooth-functional inference.
Additional theoretical and numerical results are collected in Appendices~\ref{app:auxiliary_results} and \ref{appendix:additional_simulations}, respectively.

\begin{table}[t!]
\centering
\begin{threeparttable}
\caption{Selected approximation, estimation, and inference results}
\label{tab:comparison_existing_results}
\scriptsize
\setlength{\tabcolsep}{3pt}
\renewcommand{\arraystretch}{1.06}

\begin{tabularx}{\textwidth}{@{}>{\raggedright\arraybackslash}p{0.225\textwidth}>{\raggedright\arraybackslash}p{0.28\textwidth}>{\centering\arraybackslash}p{0.13\textwidth}>{\raggedright\arraybackslash}X@{}}
\toprule
\multicolumn{4}{@{}l}{\textbf{Panel A: Approximation results}}\\
\midrule
Reference/result & Setting & Error criterion & Rate \\
\midrule
\multicolumn{4}{@{}l}{\textit{Selected existing results}}\\
\addlinespace[2pt]

{}[B93] & Trainable features; Barron class & $L^2$ & $K^{-1/2}$ \\

{}[KB18] & Trainable ReLU features; sign source representation; weighted coefficients & $L^\infty$ & $\sqrt{d+\log(K)}\,K^{-1/2-1/d}$ \\

{}[LPSZ25] & Control-neighbors integration; $\beta$-H\"older integrand on a $(d+1)$-dimensional domain & Root MSE & $K^{-1/2-\beta/(d+1)}$ \\

{}[LMX25] & Random ReLU$^m$ features; integral Sobolev class $\Lambda^q$ & $L^2(\mathcal{X})$ & $(K/\log(K))^{-1/2-(2m+1)/(2d)}$ \\

\addlinespace[3pt]
\multicolumn{4}{@{}l}{\textit{This paper}}\\
\addlinespace[2pt]

Theorem~\ref{theorem:sharp_l2} & I.i.d.\ random features; $\beta$-H\"older source condition & $L^2(X,A,B)$ & $K^{-1/2-\beta/(d+1)}$ \\

Theorem~\ref{theorem:sharp_is_linf_unified} & Stratified random features; $\beta$-H\"older or sign source condition & $\mathbb{E}\|\cdot\|_\infty$ & $a_K K^{-1/2-\alpha}$ \\

\bottomrule
\end{tabularx}

\vspace{6pt}

\begin{tabularx}{\textwidth}{@{}>{\raggedright\arraybackslash}p{0.225\textwidth}>{\raggedright\arraybackslash}p{0.275\textwidth}>{\centering\arraybackslash}p{0.13\textwidth}>{\raggedright\arraybackslash}X@{}}
\multicolumn{4}{@{}l}{\textbf{Panel B: Spectral properties, estimation, and inference}}\\
\midrule
Result & Main conditions & Object & Conclusion \\
\midrule

Proposition~\ref{prop:spectral_tail_parametrized_sieve} & Polynomial covering; Lipschitz features & Spectrum & $\lambda_{\min}(\Gamma_K)\lesssim K^{-2/q_\Pi}$ \\

Corollary~\ref{cor:effective_dimension_spectral} & Spectral decay; suitable ridge scaling & Effective dimension & $\mathcal{N}_{K_n,\lambda}=o(K_n)$ \\

Theorem~\ref{theorem:L2_ridge_sharp_consistency} & I.i.d.\ data; ridge-normalized stability & $L^2(X)$ estimation & $O_P\bigl(\sqrt{\mathcal{N}_{K_n,\lambda}/n}+\Delta_{K_n}+B^{\mathrm{pop}}_{K_n,\lambda}\bigr)$ \\

Corollary~\ref{corollary:RWNN_L2_normality} & Directional nondegeneracy; stability and ridge-bias control & Linear functionals & Gaussian limit allowing $\lambda_{\min}(\Gamma_{K_n})\to0$ \\

Corollaries~\ref{corollary:RWNN_variance_estimator_consistency}, \ref{corollary:RWNN_variance_estimator_consistency_l0} & Directional variance, moment, and growth conditions & Variance estimation & Relative consistency of the ridge-adjusted variance estimator \\

Corollary~\ref{corollary:RWNN_smooth_functional_normality} & Directional CLT; negligible functional bias and remainder & Smooth functionals & Gaussian limit under sieve-dependent normalization \\

Theorem~\ref{theorem:uniform_estimation_bound_stable} & Prediction-space and leverage stability; stratified features & $L^\infty$ estimation & $O_P\bigl(\sqrt{K_n\log(n)/n}+a_{K_n}K_n^{-1/2-\alpha}+\sqrt{\lambda_{K_n,n}}\bigr)$ \\

\bottomrule
\end{tabularx}

\begin{tablenotes}[flushleft]
\footnotesize

\item In Panel A, [B93], [KB18], [LPSZ25], and [LMX25] refer to \citet{barronUniversalApproximationBounds1993}, \citet{klusowskiApproximationCombinationsReLU2018}, \citet{lelucSpeedingMonteCarlo2024}, and \citet{liu2025integralrepresentationssobolevspaces}, respectively. These results concern different function classes, sampling schemes, and error criteria and are not directly comparable.

\item The [KB18] row reports the weighted-coefficient bound for ReLU; sign-restricted coefficients yield $\sqrt{d+\log(K)}\,K^{-1/2-1/(d+2)}$. For squared ReLU, both bounds have prefactor $\sqrt d$. [KB18] uses a $d$-dimensional parameter set, unlike the $(d+1)$-dimensional set $\Pi$ in Theorem~\ref{theorem:sharp_is_linf_unified}; on the $d$-dimensional set $\Pi^{\mathrm{sph}}$, the exponent $1/d$ is recovered; see Remark~\ref{remark:sphere_rates}.

\item Root MSE denotes integration error rather than regression error. The [LMX25] rate holds with high probability for $q=(d+2m+1)/2$, $m\geq1$. In Theorems~\ref{theorem:sharp_is_linf_unified} and~\ref{theorem:uniform_estimation_bound_stable}, $a_K:=\sqrt d+\sqrt{\log(K)}$, and $\alpha=\beta/(d+1)$ under the $\beta$-H\"older source condition, whereas $\alpha=1/(d+1)$ under the sign source condition with a ReLU$^m$ activation, $m\in\{1,2\}$.

\item Panel B lists principal conclusions, subject to the full theorem assumptions. Gaussian limits hold conditionally on $\mathcal F_{K_n}$, in $P_\theta$-probability. A single ridge-adjusted variance estimator is relatively consistent for both directional variances $\sigma_{K_n,\lambda}^{2,(\ell)}(\alpha)$, $\ell\in\{0,1\}$, defined in~\eqref{eq:def_sigma}.

\end{tablenotes}
\end{threeparttable}
\end{table}

\subsection{Notation}
\label{subsec:notation}
We fix a probability space $(\Omega,\mathcal{F},\mathbb{P})$ on which all random elements are defined. We write $\mathbb{N}=\{0,1,2,\ldots\}$ and $\mathbb{N}^{+}=\mathbb{N}\setminus\{0\}$. Unless stated otherwise, $\|\cdot\|$ denotes the Euclidean norm for vectors and the operator $2$-norm for matrices. For symmetric matrices, $A\preceq B$ means that $B-A$ is positive semidefinite. We denote by $\lambda_m$ the Lebesgue measure on $\mathbb{R}^m$ and by $V_m$ the volume of the Euclidean unit ball in $\mathbb{R}^m$. For measurable functions $f:\mathcal{X}\to\mathbb{R}$, we write $\|f\|_{L^2(X)}=(\mathbb{E}_X[|f(X)|^2])^{1/2}$ and $\|f\|_{\infty}=\sup_{x\in\mathcal{X}}|f(x)|$, 
where $P_X$ denotes the marginal distribution of $X$ and $\mathbb{E}_X$ denotes integration with respect to $P_X$, holding the realized hidden-layer parameters fixed. We denote by $P_\theta$ the law of the i.i.d.\ hidden-layer parameter sequence and write $\mathbb{P}_{A,B}$ and $\mathbb{E}_{A,B}$ (or $\mathbb{E}_\theta$) for probability and expectation with respect to this law. For functions $f$ that may depend on the sampled hidden-layer parameters, we define $\|f\|_{L^2(X,A,B)}:=\bigl(\mathbb{E}_{A,B}[\|f\|_{L^2(X)}^2]\bigr)^{1/2}$. The stochastic order symbols $O_P$ and $o_P$ refer to the joint law of the observations and random features, whereas $O_{P_\theta}$ and $o_{P_\theta}$ refer to the law of the random features alone. For positive sequences, $a_n\lesssim b_n$ means $a_n\leq Cb_n$ for a constant $C>0$ that does not depend on $n$, $K$, or other running indices, and $a_n\asymp b_n$ means $a_n\lesssim b_n\lesssim a_n$. Convergence conditional on the hidden-layer parameters is stated explicitly.

\section{Model and random weight neural network sieves}
\label{sec:rwnn_regression}
\subsection{Nonparametric regression model}
\label{subsec:nonparametric_regression}

  Let $\{(X_i, Y_i)\}_{i\geq 1}$ be i.i.d. random variables, where $X_i\in\mathcal{X}\subseteq\mathbb{R}^d$ and $Y_i\in\mathbb{R}$, with $\mathbb{E}[Y_i^2]<\infty$. We work in the standard nonparametric regression setting, cf. \cite{gyorfi2006distribution,Chen2007,Tsybakov2009,Belloni2015,Chen2015}. Consider the regression model
\begin{equation}
\label{eq:regression}
    Y_i = h(X_i) + \epsilon_i, \enspace i\in \N^+,
\end{equation}
where the regression function $h : \mathcal{X} \to \mathbb{R}$ is defined by $h(x):= \E[Y_i \vert X_i = x]$ and the errors $\epsilon_i:=Y_i-h(X_i)$ satisfy $\E[\epsilon_i \vert X_i] = 0$ and $\mathbb{V}[\epsilon_i \vert X_i] = \sigma_\epsilon^2(X_i)<\infty$ a.s. 
Moreover, $\mathbb{E}[Y_i^2]<\infty$ implies $h\in L^2(X)$.

The object of interest is the conditional expectation function (CEF) $h$, which is the minimizer of the population quadratic risk
\begin{equation*}
    h=\arg\min_{f \in L^2(X)}\mathbb{E}\bigl[(Y_i - f(X_i))^2\bigr].
\end{equation*}

\subsection{Shallow neural networks and random weight neural networks}

Let $\sigma:\mathbb{R}\to\mathbb{R}$ be a fixed measurable activation function. For $K \in \mathbb{N}^+$, the class of shallow neural networks of width
$K$ is
\begin{equation}
    \mathcal{H}^{\mathrm{NN}}_K
    :=
    \left\{
        x \mapsto \sum_{k=1}^K W_k \sigma(A_k^\top x + B_k)
        \ \middle| \
        W_k \in \mathbb{R},\;
        A_k \in \mathbb{R}^d,\;
        B_k \in \mathbb{R}
    \right\},
    \label{eq:shallow_nn_class}
\end{equation}
where all parameters $\{(W_k, A_k, B_k)\}_{k=1}^K$ are trainable
\citep{Chen1999, shenAsymptoticPropertiesNeural2023}.

A {\it random weight neural network} (RWNN) is obtained from \eqref{eq:shallow_nn_class} by fixing the inner weights $\{(A_k, B_k)\}_{k=1}^K$ at random draws and retaining only the outer weights $\{W_k\}_{k=1}^K$ as trainable parameters. 
Formally, let $\Pi\subseteq\mathbb{R}^{d+1}$ be a parameter space and let $\pi_\theta$ be a probability measure on $\Pi$. Let $\{(A_k,B_k)\}_{k\ge1}$ be an i.i.d.\ sequence of random variables such that $\theta_k:=(A_k^\top,B_k)^\top\sim\pi_\theta$, independent of the data sequence $\{(X_i,Y_i)\}_{i\geq1}$. All random elements are defined on a common probability space $(\Omega,\mathcal{F},\mathbb{P})$. For each realization $\omega\in\Omega$, the sequence $\{(A_k(\omega),B_k(\omega))\}_{k\ge1}$ defines, for every $K\in \mathbb{N}^+$,  the {\it towering RWNN class}
\begin{equation}
    \mathcal{H}_K(\omega)
    :=
    \left\{
        x \mapsto \sum_{k=1}^K W_k   \sigma(A_k(\omega)^\top x + B_k(\omega))
        \ \middle| \
        W_k \in \mathbb{R}
    \right\}.
    \label{eq:towering_class}
\end{equation}
Since the same weight sequence is reused, the classes are nested: for any $K_1 \leq K_2$ and all $\omega \in \Omega$
\begin{equation*}
    \mathcal{H}_{K_1}(\omega) \subseteq \mathcal{H}_{K_2}(\omega).     
\end{equation*}
This nesting property aligns with the functional sieves framework of \citet{grenander1981abstract}, and allows direct comparison with
nonparametric sieve estimators \citep{Chen2007}. To simplify notation, in the rest of the paper, we suppress the dependence on $\omega$.

\begin{remark}[Aleatory RWNN class]
In practice, RWNNs are typically implemented by drawing a new i.i.d.\ sample $\{(A_k^{(K)}, B_k^{(K)})\}_{k=1}^K $ independently from $\pi_\theta$ for each width $K$. The resulting {\it aleatory} class $\overline{\mathcal{H}}_K$ need not be nested in $K$ and therefore does not fit the functional sieve framework of \citet{grenander1981abstract}. The towering construction provides the nesting required by sieve arguments that explicitly exploit this property. The ridge-regularized results developed in Section~\ref{sec:Ridge_RWNN_least_squares} do not generally require nesting of the feature spaces. For any fixed width $K$, however, $\mathcal{H}_K$ and $\overline{\mathcal{H}}_K$ have the same distribution under $\pi_\theta$. Consequently, all approximation results stated for $ \mathcal{H}_K$ in Subsection~\ref{subsec:approximation_RWNNs_Lipschitz} also hold for the aleatory class $\overline{ \mathcal{H}}_K$ at a fixed $K$.
\end{remark}
We also consider RWNN classes generated by stratified sampling of the inner parameters in Subsections~\ref{subsec:approximation_RWNNs_importance} and \ref{subsec:rwnn_uniform_ridge}. For each $K\in\mathbb{N}^+$, let
$\{\theta_k^{\mathrm{IS},K}\}_{k=1}^K$, with $\theta_k^{\mathrm{IS},K}=((A_k^{\mathrm{IS},K})^\top,B_k^{\mathrm{IS},K})^\top$, be random inner parameters generated by a stratified sampling scheme on $\Pi$. The corresponding {\it stratified RWNN class} is
\begin{equation}
\mathcal{H}^{\mathrm{IS}}_K :=\left\{x\mapsto\sum_{k=1}^{K}W_k
\sigma\bigl((A_k^{\mathrm{IS},K})^\top x+B_k^{\mathrm{IS},K}\bigr) \ \middle| \ W_k\in\mathbb{R}\right\}.
\label{eq:stratified_rwnn_class}
\end{equation}
The superscript $\mathrm{IS}$ is retained to denote this stratified (importance) sampling construction.

Unlike the towering RWNN class $\mathcal{H}_K$, the distribution of the inner parameters in $\mathcal{H}^{\mathrm{IS}}_K$ may depend on $K$, and the resulting sequence of classes need not be nested. The stratified hidden parameters are generated independently of the regression observations.    In case~(ii) of Theorem~\ref{theorem:sharp_is_linf_unified}, the stratification uses the sign partition $\Pi_\pm$ of the source function $f$ in Assumption~\ref{assumption:source_sign}, so the inner parameters depend on the target. The estimation results of Subsection~\ref{subsec:rwnn_uniform_ridge} are therefore oracle results in that case, unless this partition is known.

\subsection{RWNN least squares estimator}
\label{subsec:rwnn_sieve_ls}
Given a sample $\{(Y_i, X_i)\}_{i=1}^n$, we estimate the regression function $h$ by {\it empirical risk minimization} (ERM) over the towering RWNN class $\mathcal{H}_K$ in \eqref{eq:towering_class}, conditional on the realized inner parameters $\theta := \{(A_k, B_k)\}_{k=1}^K$. Let $\ell:\mathbb{R} \times \mathbb{R} \to[0,\infty)$ be a loss function. For a network $h_K^{\theta,W}(x):=\sum_{k=1}^K W_k \sigma(A_k^\top x + B_k)$,  $W=(W_1,\ldots,W_K)^\top\in\mathbb{R}^K$,  the empirical risk is defined as
\begin{equation*}
L_n(h_K^{\theta,W})
=
\frac{1}{n}\sum_{i=1}^n
\ell \left(Y_i,h_K^{\theta,W}(X_i)\right).
\end{equation*}
We focus on the squared loss $\ell(y,t)=(y-t)^2$, while a more general formulation allowing loss functions beyond squared loss is considered in Appendix~\ref{app:generic_sieve_estimation}. The readout weights are obtained as the empirical risk minimizer, which is the linear least squares solution
\begin{equation}
    \widehat{W}_{K,n} \in \arg\min_{W \in \mathbb{R}^K} L_n(h_K^{\theta,W})= \arg\min_{W \in \mathbb{R}^K}
      \frac{1}{n}\sum_{i=1}^n
      \left(Y_i - \sum_{k=1}^K W_k \sigma(A_k^\top X_i + B_k)\right)^2.
    \label{eq:ls_rwnn_problem}
\end{equation}

Following the series (sieve) notation in \cite{Chen2015}, introduce the random feature map
\begin{equation}
    g_K(x)    := \left(\sigma(A_1^\top x + B_1), \ldots,
                  \sigma(A_K^\top x + B_K)\right)^\top \in \mathbb{R}^K,
    \label{eq:feature_map}
\end{equation}
the design matrix and the response vector
\begin{equation}
    G_{K,n}   := \left(g_K(X_1), \ldots, g_K(X_n)\right)^\top
                  \in \mathbb{R}^{n \times K}, \quad Y := (Y_1, \ldots, Y_n)^\top \in \mathbb{R}^n,
    \label{eq:design_matrix}
\end{equation}
respectively.
Then problem~\eqref{eq:ls_rwnn_problem} admits the minimum-Euclidean-norm solution
\begin{equation}
    \widehat{W}_{K,n}
    = (G_{K,n}^\top G_{K,n})^\dagger G_{K,n}^\top {Y},
    \label{eq:sieve_ls_W}
\end{equation}
where $(\cdot)^\dagger$ denotes the Moore-Penrose pseudoinverse.
The {\it RWNN least-squares estimator} $h^{\theta,\widehat{W}}_{K,n}\in \mathcal{H}_K$ of the regression function $h$, corresponding to the minimum-Euclidean-norm solution \eqref{eq:sieve_ls_W}, is defined as
\begin{equation}
    h^{\theta,\widehat{W}}_{K,n}(x)
    := \sum_{k=1}^K  (\widehat{W}_{K,n})_k\sigma(A_k^\top x + B_k)=g_K(x)^\top \widehat{W}_{K,n}
     = g_K(x)^\top (G_{K,n}^\top G_{K,n})^\dagger G_{K,n}^\top {Y},
    \enspace x \in \mathcal{X}.
    \label{eq:sieve_RWNN_estimator}
\end{equation}

The principal object of our statistical analysis is the ridge-regularized RWNN least-squares estimator, which we distinguish from the unregularized estimator $h_{K,n}^{\theta,\widehat W}$ defined in \eqref{eq:sieve_RWNN_estimator}. For the ridge-regularized analysis, we use the coefficient notation $\beta$ in place of $W$, following the conventional notation for linear sieve regression. For deterministic sequences $K_n\to\infty$ and $\lambda_{K_n,n}>0$, define $\widehat h_{n,\lambda}(x):=g_{K_n}(x)^\top\widehat\beta_{K_n,n,\lambda}$, where

\begin{equation}
\label{eq:def_betahat}
\widehat{\beta}_{K_n,n,\lambda}:=\left(n^{-1}G_{K_n,n}^\top G_{K_n,n}+\lambda_{K_n,n}I_{K_n}\right)^{-1}n^{-1}G_{K_n,n}^\top Y.
\end{equation}
The associated population covariance matrix and ridge-normalized quantities are introduced in Section~\ref{subsubsec:ridge_normalization}.

\section{Approximation properties of RWNNs}
\label{sec:approximation_RWNNs}

In this section we consider different approximation regimes for random weight neural networks under related, but distinct, regularity assumptions. 

Subsections~\ref{subsec:approximation_RWNNs_Lipschitz} and~\ref{subsec:approximation_RWNNs_importance} both assume a direct source condition under which the target function admits an integral representation of the form
\begin{equation*}
h(x)=\int_\Pi f(A,B)\sigma(A^\top x+B)\de\pi_\theta(A,B).
\end{equation*}
They differ in terms of how this representation is used. Subsection~\ref{subsec:approximation_RWNNs_Lipschitz} assumes a $\beta$-H\"older product source condition (Assumption~\ref{assumption:direct_source_condition_holder_product}), combining a Lipschitz activation $\sigma$ (Assumption~\ref{assumption:activation_Lipschitz}) with a $\beta$-H\"older weight function $f$ on a compact parameter set $\Pi\subset\mathbb{R}^{d+1}$ equipped with a metric $\rho_\Pi$. It derives $L^2$ approximation rates via the control-neighbors rule of \citet{lelucSpeedingMonteCarlo2024}. Subsection~\ref{subsec:approximation_RWNNs_importance} derives sup-norm approximation rates via a stratified sampling construction, based on partitioning the parameter space and controlling within-cell oscillations.  In that setting two direct source conditions are considered: the same $\beta$-H\"older product source condition as in Subsection~\ref{subsec:approximation_RWNNs_Lipschitz}, and a ReLU$^m$ activation with $m\in\{1,2\}$ (Assumption~\ref{assumption:activation}) combined with a sign weight function $f$ (Assumption~\ref{assumption:source_sign}).

Subsection~\ref{subsec:approximation_RWNNs_Sobolev} takes a different approach, assuming that $h$ belongs to an integral Sobolev class $\Lambda^q$ and working with ReLU$^m$ activations for general $m\geq 1$. 

Proofs of the $L^2$ approximation results under the $\beta$-H\"older source condition, including the control-neighbors weight bounds, are provided in Appendix~\ref{app:L2_approximation_proofs}. The proof of the $L^2(X)$ approximation corollary for integral Sobolev classes is given in Appendix~\ref{app:Sobolev_approximation_proofs}. Proofs of the stratified sup-norm approximation results and the corresponding output-weight bounds are collected in Appendix~\ref{app:Sup-norm_approximation}, while the uniform Monte Carlo approximation bound is established in Appendix~\ref{app:entropy_computation}.

We consider shallow random weight neural networks on a compact domain.
\begin{assumption}
\label{assumption:X_domain}
For some $M > 0$, the regressor space is $\mathcal{X} = [-M, M]^d$.
\end{assumption}
\subsection{$L^2$ approximation under a $\beta$-H\"older source condition}
\label{subsec:approximation_RWNNs_Lipschitz}
Under a $\beta$-H\"older source condition, we show that the RWNN sieve admits an $L^2$ approximation error of order $K^{-1/2-\beta/(d+1)}$.

Whenever the source and regular-sampling assumptions below are invoked jointly, they are understood to hold for the same parameter space $\Pi$, metric $\rho_\Pi$, and sampling measure $\pi_\theta$.

\begin{assumption}[Lipschitz activation]\label{assumption:activation_Lipschitz}
     $\sigma:\mathbb{R}\to\mathbb{R}$ is Lipschitz with constant $L_\sigma>0$.  
\end{assumption}
\begin{assumption}[Product source condition]
\label{assumption:direct_source_condition_holder_product}
    There exist $\beta\in(0,1]$, a compact set  $\Pi\subset\mathbb{R}^{d+1}$ equipped with a metric $\rho_\Pi$ such that $D_\Pi :=\sup_{\theta_1,\theta_2\in\Pi}\rho_\Pi(\theta_1,\theta_2)<\infty$, a probability measure $\pi_\theta$ on $\Pi$, constant $L_f>0$, and a function $f:\Pi\to\mathbb{R}$ satisfying
    \begin{align*}
    	|f(\theta_1)-f(\theta_2)|\leq L_f \rho_\Pi(\theta_1,\theta_2)^\beta, \quad   \theta_1,\theta_2\in\Pi,
    \end{align*}
    such that 
    \begin{equation*}
        h(x)=\int_\Pi\phi_x(\theta) \de\pi_\theta (\theta),\qquad x\in\mathcal{X},
    \end{equation*}
    with $\theta = (A^\top, B)^\top $ and $\phi_x(\theta)=f(A,B)\sigma(A^\top x+B)$. We also define $M_f:=\|f\|_\infty$, $M_\sigma:=\sup_{x\in\mathcal{X}, \theta\in\Pi}|\sigma(A^\top x+B)|$.
\end{assumption}
To establish the approximation result, we use a general integral representation with a H\"older continuous integrand, together with regularity conditions on the sampling distribution. We subsequently verify that the RWNN product source condition implies this abstract representation under a suitable compatibility condition on the metric.
\begin{assumption}[Integral representation]
\label{assumption:integral_representation_phi}
There exist a compact set $\Pi\subset\mathbb{R}^{d+1}$ equipped with a metric $\rho_\Pi$, a probability measure $\pi_\theta$ on $\Pi$, and $\beta\in(0,1]$ such that  for each $x\in\mathcal{X}$ there exists a measurable function $\phi_x:\Pi\rightarrow \mathbb{R}$ such that $\theta\mapsto \phi_x(\theta)$ is $\beta$-H\"older continuous on  $(\Pi,\rho_\Pi)$ with constant $L_{\phi,x}>0$, and
\begin{equation*}
h(x)=\int_\Pi\phi_x(\theta) \de\pi_\theta(\theta), \qquad x\in\mathcal{X}.
\end{equation*}
\end{assumption}

\begin{assumption}[Regular sampling]
\label{assumption:regular_sampling}
Let $\Pi$, $\rho_\Pi$, and $\pi_\theta$ be as in  Assumption~\ref{assumption:integral_representation_phi}. Suppose that $\pi_\theta$ admits a density $\mu_{\pi,\theta}$ with respect to  $\lambda_{d+1}$ satisfying, for constants $c_b,c_u,c_\pi,r_0>0$:
\begin{enumerate}
\item[(i)] $c_b\leq\mu_{\pi,\theta}(\theta)\leq c_u$ for all $\theta\in\Pi$,
\item[(ii)] $\lambda_{d+1}(\Pi\cap B_r^{\|\cdot\|}(\theta))\geq c_\pi\lambda_{d+1}(B_r^{\|\cdot\|}(\theta))$ for all $r\in(0,r_0]$, $\theta\in\Pi$, where $B_r^{\|\cdot\|}(\theta)\subset \mathbb{R}^{d+1}$ denotes the Euclidean ball of radius $r$ centered at $\theta$,
\item[(iii)] There exist constants $0<m_\rho\leq M_\rho<\infty$ such that $m_\rho\|\theta_1-\theta_2\| \leq \rho_\Pi(\theta_1,\theta_2)
\leq M_\rho\|\theta_1-\theta_2\|$
, $\theta_1,\theta_2\in\Pi$.
\end{enumerate}
\end{assumption}

\begin{remark}[Choice of metric $\rho_\Pi$]
\label{remark:metrics}
Two natural choices of $\rho_\Pi$ arise in this paper.
\begin{enumerate}
\item[(i)] {\it Euclidean metric}: $\rho_\Pi(\theta_1,\theta_2)=\|\theta_1-\theta_2\|$. 
\item[(ii)] {\it Weighted $\ell_1$ metric}: $\rho_\Pi(\theta_1,\theta_2)=d_\Pi(\theta_1,\theta_2):=M\|A_1-A_2\|_1+|B_1-B_2|$, with $M$ as in Assumption~\ref{assumption:X_domain}, used in Subsection~\ref{subsec:approximation_RWNNs_importance}.  No regular sampling condition is required there, and, as we show later, this metric is a convenient choice in that setting.
\end{enumerate}
The two metrics are equivalent on  
$\mathbb{R}^{d+1}$ with constants depending 
only on $M$ and $d$, so a $\beta$-H\"older condition formulated with respect to one metric
is equivalent to a $\beta$-H\"older condition with respect to the other,
up to a change in the H\"older constant. Indeed, since $\|A_1-A_2\|\leq\|A_1-A_2\|_1\leq\sqrt{d}
\|A_1-A_2\|$, one has
\begin{equation*}
\begin{aligned}
    \min(M,1)\|\theta_1-\theta_2\|
    \leq d_\Pi(\theta_1,\theta_2) 
    &\leq M\sqrt{d}\|A_1-A_2\| + |B_1-B_2| \\
    &\leq \sqrt{2}\max(M\sqrt{d},1)\|\theta_1-\theta_2\|, \enspace \theta_1, \theta_2 \in \Pi,
\end{aligned}
\end{equation*}
so the H\"older constant transforms by a factor of at 
most $\bigl(\sqrt{2}\,\max(M\sqrt{d},1)/\min(M,1)\bigr)^{\beta}$.
\end{remark}

Notice that, under Assumption~\ref{assumption:direct_source_condition_holder_product}, $M_f<\infty$ by the $\beta$-H\"older continuity of $f$ and the finite $\rho_\Pi$-diameter of $\Pi$, and $D_\Pi<\infty$ by definition.  Moreover, under Assumptions~\ref{assumption:X_domain}, \ref{assumption:activation_Lipschitz}, and
\ref{assumption:direct_source_condition_holder_product}, $M_\sigma<\infty$ by compactness of $\mathcal{X}\times\Pi$ and
continuity of $\sigma$.

\begin{proposition}
\label{prop:product_implies_holder}
    Let Assumptions~\ref{assumption:X_domain},~\ref{assumption:activation_Lipschitz}, and 
    \ref{assumption:direct_source_condition_holder_product} hold, and suppose there exists a constant $L_\rho>0$ such that
    \begin{equation}
    \label{eq:rho_compatibility}
    |(A_1-A_2)^\top x+(B_1-B_2)|
    \leq L_\rho\,\rho_\Pi(\theta_1,\theta_2)\qquad \text{for all } \theta_1,\theta_2\in\Pi,\ x\in\mathcal{X}.
    \end{equation}
    Then Assumption~\ref{assumption:direct_source_condition_holder_product} implies Assumption~\ref{assumption:integral_representation_phi} with H\"older constant that can be chosen uniformly in $x\in\mathcal{X}$ by
    \begin{equation*}
    L_{\phi,x}\leq L_\phi:=M_f L_\sigma L_\rho D_\Pi^{1-\beta}+M_\sigma L_f,
    \end{equation*}
    where $D_\Pi$ and $M_\sigma$ are as in Assumption~\ref{assumption:direct_source_condition_holder_product}.
\end{proposition}

\begin{remark}
    For the two particular choices of $\rho_\Pi$ in Remark~\ref{remark:metrics}, the compatibility condition
    \eqref{eq:rho_compatibility} holds with explicit constants. For the Euclidean metric $\rho_\Pi(\theta_1,\theta_2)=\|\theta_1-\theta_2\|$, one may take $L_\rho=\sup_{x\in\mathcal{X}}\|(x^\top,1)^\top\|=\sqrt{dM^2+1}$, while for the weighted $\ell_1$ metric $\rho_\Pi=d_\Pi$, one may take $L_\rho=1$.
\end{remark}

To construct an approximation based on finitely many sampled features, we build upon the control-neighbors Monte Carlo rule of \citet{lelucSpeedingMonteCarlo2024}. This rule exploits the H\"older regularity of the integrand and the regularity of the sampling distribution to improve upon the standard Monte Carlo rate. The result used and the verification of its sampling conditions are given in Appendix~\ref{app:L2_approximation_proofs}.

The control-neighbors rule assigns weights to the sampled parameters $\theta_1,\dots,\theta_K$ according to
\begin{equation}
\label{eq:control_neigh_def}
\omega_{k,K}:=\frac{1+KV_{K,k}-\widehat{d}_{K,k}}{K},
\end{equation}
where $V_{K,k}:=\pi_\theta(C_{K,k})$ is the probability mass of the Voronoi cell $C_{K,k}$ associated with $\theta_k$, and $\widehat{d}_{K,k}$ is the number of times $\theta_k$ is the nearest neighbor of $\theta_j$, $j\neq k$, with respect to $\rho_\Pi$.

Since the rule is linear in its integrand, applying it to the product source representation yields
\begin{equation}
\label{eq:cnn_rwnn_approximant}
h^{\theta,\omega\texttt{-cnn}}_K(x)
:= \sum_{k=1}^K \omega_{k,K}
\bigl[f(A_k,B_k)\sigma(A_k^\top x+B_k)\bigr] 
= \sum_{k=1}^K W_k\sigma(A_k^\top x+B_k),
\end{equation}
where $W_k:=\omega_{k,K}f(A_k,B_k)$, $k=1,\dots,K$. Hence, conditional on the sampled hidden parameters $\{\theta_k\}_{k=1}^K$, the control-neighbors approximation belongs to the RWNN class.

\begin{theorem}[$L^2$ RWNN approximation]
\label{theorem:sharp_l2}
    Let Assumptions~\ref{assumption:X_domain}, \ref{assumption:activation_Lipschitz}, \ref{assumption:regular_sampling}, and \ref{assumption:direct_source_condition_holder_product} hold, and suppose the compatibility condition \eqref{eq:rho_compatibility} is satisfied. Let $\theta_k=(A_k^\top,B_k)^\top$, $k=1,\dots,K$, be i.i.d.\ from $\pi_\theta$. Then, for $K\geq 4$, the control-neighbors weight vector $W\in\mathbb{R}^K$ defined in \eqref{eq:cnn_rwnn_approximant} is $\sigma(\theta_1,\dots,\theta_K)$-measurable, and the corresponding RWNN approximant $h^{\theta,W}_K=h^{\theta,\omega\texttt{-cnn}}_K$ satisfies
    \begin{equation}
    \label{eq:sharp_l2_rate}
    \|h(X)-h^{\theta,W}_K(X)\|_{L^2(X,A,B)}
    \leq C_q K^{-1/2-\beta/(d+1)},
    \end{equation}
    where $C_q>0$ depends only on $d$, $\beta$, $M_f$, $L_f$, $L_\sigma$, $M_\sigma$, $L_\rho$, $c_b$, $c_u$, $c_\pi$, $r_0$, $m_{\rho}$, $M_{\rho}$, and $D_\Pi$.
\end{theorem}

\begin{remark}[Approximation rates and comparison with Monte Carlo integration]
A standard empirical-process argument based on symmetrization and Dudley's entropy integral applied to the centered process $x\mapsto h(x)-h_K^\theta(x)$, where $h_K^\theta(x)=K^{-1}\sum_{k=1}^K\phi_x(\theta_k)$, yields only the Monte Carlo rate
\begin{equation*}
\mathbb{E}_\theta\left[
\sup_{x\in\mathcal{X}}|h(x)-h_K^\theta(x)|
\right]=\mathcal{O}(K^{-1/2});
\end{equation*}
see Appendix~\ref{app:entropy_computation} for a complete derivation. This rate does not exploit the regularity of $\theta\mapsto\phi_x(\theta)$ on $\Pi$.

The sharper rate in \eqref{eq:sharp_l2_rate}
 arises from the $\beta$-H\"older regularity of $\theta\mapsto\phi_x(\theta)$ on $(\Pi,\rho_\Pi)$,
whose metric small-ball behavior is of order $r^{d+1}$ under Assumption~\ref{assumption:regular_sampling}. The control-neighbors estimator leverages the local spacing of $\theta_1,\dots,\theta_K$ to construct weights $\omega_{k,K}$ that reduce the error by $K^{-\beta/(d+1)}$ relative to the Monte Carlo baseline. 
Since Assumption~\ref{assumption:regular_sampling}~(iii) makes $\rho_\Pi$ bi-Lipschitz equivalent to the Euclidean metric, the corresponding $\beta$-H\"older classes are equivalent up to constants. Hence, for $(d+1)$-dimensional $\beta$-H\"older integrands, the minimax rate under i.i.d.\ sampling is $\mathcal{O}(K^{-1/2-\beta/(d+1)})$ \citep{novakResultsComplexityNumerical2016}, so Theorem~\ref{theorem:sharp_l2} achieves the optimal rate (up to constants) for this class of integrands under i.i.d.\ sampling.
In particular, for $\beta=1$ the rate $K^{-1/2-1/(d+1)}$ is optimal for Lipschitz integrands on a $(d+1)$-dimensional parameter space.
\end{remark}

\begin{theorem}[High-probability $L^2$ approximation]
\label{thm:prob-approx}
    Under the assumptions of Theorem~\ref{theorem:sharp_l2},  let $\theta_k=(A_k^\top,B_k)^\top$, $k=1,\dots,K$, be i.i.d.\ from $\pi_\theta$. Then for every $K\geq 4$, there exists a $\sigma(\theta_1,\dots,\theta_K)$-measurable weight vector $W\in\mathbb{R}^K$ such that for every $\varepsilon\in(0,1)$,
    \begin{equation}
    \label{eq:prob_bound}
    \mathbb{P}_{A,B}
    \left(
    \|h(X)-h^{\theta,W}_K(X)\|_{L^2(X)}
    \leq
    C_q K^{-1/2-\beta/(d+1)}
    \varepsilon^{-1/2}
    \right)
    \geq 1-\varepsilon,
    \end{equation}
    where $C_q>0$ is the constant appearing in Theorem~\ref{theorem:sharp_l2}.
\end{theorem}
\begin{corollary}[RWNN approximation in probability]
\label{cor:rwnn_projection_rate}
    Under the assumptions of Theorem~\ref{thm:prob-approx},
    \begin{equation*}
    \inf_{g\in\mathcal{H}_K}
    \|h-g\|_{L^2(X)}
    =
    O_{P_\theta} \left(K^{-1/2-\beta/(d+1)}\right).
    \end{equation*}
\end{corollary}

The preceding approximation results impose no constraint on the output weights. To establish the same approximation rate for the constrained RWNN sieve (Corollary~\ref{cor:constrained_approx}), which will be used in Section~\ref{sec:Ridge_RWNN_least_squares}, we need to control the Euclidean norm of the control-neighbors weight vector. The following geometric assumption on nearest-neighbor indegrees allows us to derive the required bound.

\begin{assumption}
\label{assumption:bounded_indegree}
There exists a constant	$\psi_\Pi<\infty$ such that for every finite point set $P\subset \Pi$ and every $x\in P$ the number of points in $P$ for which $x$ is a nearest neighbor under $\rho_\Pi$ is at most $\psi_\Pi$.
\end{assumption}
We note that in the case of Euclidean metric $\rho_{\Pi}$, by Lemma~\ref{lemma:KabatianskiZeger}, Assumption~\ref{assumption:bounded_indegree} holds with $\psi_{\Pi}=\psi_{d+1}$, where $\psi_{d+1}$ is the $(d+1)$-dimensional kissing number. 

\begin{lemma}[$\ell_2$-norm bound for control-neighbors weights]
\label{lem:nn_l2_bound}
    Let Assumptions~\ref{assumption:X_domain}, \ref{assumption:regular_sampling}, \ref{assumption:direct_source_condition_holder_product}, and \ref{assumption:bounded_indegree} hold, and let $\theta_k=(A_k^\top,B_k)^\top$, $k=1,\dots,K$, be i.i.d.\ from $\pi_\theta$ with $K \geq 3$.  
Define the weight vector $W\in\mathbb{R}^K$ by
\begin{equation*}
W_k:=\omega_{k,K}f(A_k,B_k), \quad k=1,\dots,K,
\end{equation*}
where $f$ is as in Assumption~\ref{assumption:direct_source_condition_holder_product} and the control-neighbors weights $\omega_{k,K}$ are defined in \eqref{eq:control_neigh_def}.
    Then, for every $\alpha>0$ there exist 
    $C_\alpha,C_\alpha'>0$ such that
    \begin{equation*}
        \mathbb{P}_{A,B}\left(
        \|W\|\leq C_\alpha M_f\sqrt{\frac{\log(K)}{K}}
        \right)\geq 1-C_\alpha' K^{-\alpha},
    \end{equation*}
    where $M_f$ is as in Assumption~\ref{assumption:direct_source_condition_holder_product}. In particular,
    \begin{equation*}
\|W\|=O_{P_\theta}\left(\sqrt{\frac{\log K}{K}}\right).
    \end{equation*}
    Moreover, it holds
    \begin{equation}
    \label{eq:determ_W1_W2}
        \norm{W}_1 \leq M_f(2 + \psi_{\Pi}) 
        \quad\text{and}\quad
        \|W\| \leq M_f \sqrt{ 3(2 + \psi_{\Pi}^2) },
    \end{equation}
    where $\psi_{\Pi} < \infty$ denotes the uniform nearest-neighbor indegree bound corresponding to $\rho_{\Pi}$.
\end{lemma}

\subsection{$L^2$ approximation on integral Sobolev classes}
\label{subsec:approximation_RWNNs_Sobolev}
Recent work by \citet{liu2025integralrepresentationssobolevspaces} derives approximation rates for shallow $\mathrm{ReLU}^m$ networks under additional smoothness assumptions. These results improve on the sampling-based argument used in Subsection~\ref{subsec:approximation_RWNNs_Lipschitz} by working with Barron- and Sobolev-type classes of higher regularity. The underlying idea of relying on suitable~integral representations is similar; such representations are closely related to ridgelet analysis \citep[see, e.g.,][]{sonoda2017neural}. Moreover, \citet{liu2025integralrepresentationssobolevspaces} establish approximation rates for RWNNs with parameters sampled uniformly from the sphere, which further motivates their use in our nonparametric regression setting. Using ridgelet representations, \citet{mehmood2026random} obtain RWNN approximation rates of order $K^{-1/2}$ in time-dependent Sobolev norms for smooth activations such as $\tanh$ and the sigmoid, with hidden-layer parameters drawn from a fixed heavy-tailed distribution; related Sobolev-norm bounds for randomized networks are given in \citet{deryck2025randomized}. The results in this subsection, as well as those in Subsection~\ref{subsec:approximation_RWNNs_importance}, rely on the following assumption on the activation function.
\begin{assumption}\label{assumption:activation}
     For $m\in \mathbb{N}^+$, let $\sigma_m:\mathbb{R}\to\mathbb{R}$ be the ReLU$^m$ activation function $\sigma_m(x) = \max(x,0)^m$. 
\end{assumption}
Subsection~\ref{subsec:approximation_RWNNs_importance} uses this assumption with $m\in\{1,2\}$. %

For $m\in\mathbb{N}^+$, we denote by $\mathcal{H}^{(m)}_K$ the RWNN class \eqref{eq:towering_class} under Assumption~\ref{assumption:activation}, that is with activation $\sigma_m(x)=\max(x,0)^m$. The case $m=1$ coincides with $\mathcal{H}_K$ when the activation in \eqref{eq:towering_class} is chosen to be ReLU.

For $q:=(d+2m+1)/2$, define the class
\begin{equation*}%
    \Lambda^q(\mathcal{X})
    :=
    \left\{ x \mapsto 
        \fint_{\mathbb{S}^{d}} \phi_x^m(\theta) f(\theta) d \theta 
         \ \middle\vert \
        f \in L^2(\mathbb{S}^{d})
    \right\} ,
\end{equation*}
where $\fint_{\mathbb{S}^{d}} := |\mathbb{S}^{d}|^{-1}\int_{\mathbb{S}^{d}}$ 
denotes the surface-normalized integral and $\phi_x^m(\theta) 
:= \sigma_m(\theta_{1:d}^\top x + \theta_0)$ with $\theta = (\theta_{1:d}^\top, \theta_0)^\top \in\mathbb{S}^d$. 
We equip $\Lambda^q(\mathcal{X})$ with the norm \begin{equation*}
 	\|h\|_{\Lambda^q(\mathcal{X})} :=\inf\left\{\|f\|_{L^2(\mathbb{S}^d)} \ \middle\vert \ h(x)=\fint_{\mathbb{S}^d}\phi_x^m(\theta)f(\theta)\,d\theta, \, x\in\mathcal{X}\right\}.
 \end{equation*}

By Theorem~2.3 of \citet{liu2025integralrepresentationssobolevspaces}, $\Lambda^q(\mathcal{X})$ coincides with the classical Sobolev space $H^q(\mathcal{X})$, with equivalent norms. This representation is the key structural property that enables random-feature approximation.

Let $\mu_{\mathbb{S}^d}$ denote the uniform probability measure on $\mathbb{S}^d$, let $\theta_1,\dots,\theta_K\overset{\mathrm{i.i.d.}}{\sim}\mu_{\mathbb{S}^d}$, and write $\theta:=(\theta_1,\dots,\theta_K)$. Define the random feature map $\phi_{\theta,K}^m(x):=\bigl(\phi_x^m(\theta_1),\dots,\phi_x^m(\theta_K)\bigr)^\top\in\mathbb{R}^K$.  Then any function in $\mathcal{H}_K^{(m)}$ can be written in the form $x \mapsto W^\top \phi_{\theta,K}^m(x)$ for some $W \in \mathbb{R}^K$.

The following result, due to 
\citet{liu2025integralrepresentationssobolevspaces}, establishes that $\mathrm{ReLU}^m$ RWNNs with 
parameters sampled uniformly from $\mathbb{S}^{d}$ achieve an $L^2(\mathcal{X})$ approximation rate of $(K/\log(K))^{-1/2-(2m+1)/(2d)}$ with high probability over the parameter draw. Corollary~\ref{corollary:mao_L2_sobol_approx} 
below extends this to the $L^2(X)$ norm required 
for our statistical analysis.

\begin{theorem}[\citealp{liu2025integralrepresentationssobolevspaces}, Theorem 6.2]
\label{theorem:mao_L2_sobol_approx}
    Let Assumption~\ref{assumption:X_domain} hold, let  $h \in \Lambda^q(\mathcal{X})$ with $q = (d + 2m + 1)/2$, and let $\theta_1,\dots,\theta_K\overset{\mathrm{i.i.d.}}{\sim}\mu_{\mathbb{S}^d}$. Then, for any $\delta\in (0,1)$, with probability at least $1-\delta$, there exists a weight vector $W \in \R^{K}$ such that
    \begin{equation*}
        \|W\| \lesssim\left( \frac{K}{\log(K/\delta)} \right)^{-\frac{1}{2}}\|{ h }\|_{ \Lambda^q(\mathcal{X}) }
    \end{equation*}
    and 
    \begin{equation*}
        \|h-W^\top\phi_{\theta,K}^m\|_{L^2(\mathcal{X})}\lesssim\left( \frac{K}{\log(K/\delta)} \right)^{ -\frac{1}{2} - \frac{2m+1}{2d} } \| h \|_{ \Lambda^q(\mathcal{X}) }.
    \end{equation*}
    The implicit constants depend only on $d$ and $m$, and the domain $\mathcal{X}$.
\end{theorem}
\begin{remark}
Theorem~\ref{theorem:mao_L2_sobol_approx} follows from an approximation bound expressed in terms of the covering radius of the sampled parameters on $\mathbb{S}^d$. To make this precise, define the covering radius
\begin{equation*}
    \widetilde{\rho}_{K,d} := \max_{\theta \in \mathbb{S}^{d}} \min_{1 \leq k \leq K} \rho(\theta, \theta_k),
\end{equation*}
where $\rho:\mathbb{S}^{d}\times\mathbb{S}^{d}\to[0,\pi]$ denotes the spherical geodesic metric. The points $\{\theta_k\}_{k=1}^K$ are said to be \textit{well-distributed} on $\mathbb{S}^d$ if $\widetilde{\rho}_{K,d}\leq C_d K^{-1/d}$ for some constant $C_d>0$.
The central approximation result of \citet{liu2025integralrepresentationssobolevspaces}, Theorem~2.2, provides a bound on the approximation error that depends explicitly on  $\widetilde{\rho}_{K,d}$. 
For i.i.d.\ uniformly sampled parameters, the covering radius instead admits high-probability bounds of order $(\log(K/\delta)/K)^{1/d}$, which account for the logarithmic factor in Theorem~\ref{theorem:mao_L2_sobol_approx}.
\end{remark}

Theorem~\ref{theorem:mao_L2_sobol_approx} establishes the approximation rate in $L^2(\mathcal{X})$, that is, with respect to Lebesgue measure on $\mathcal{X}$. Choosing $\delta=K^{-\alpha}$ gives a polynomial failure probability for any $\alpha>0$, and the following corollary transfers the resulting bound to $L^2(X)$ under a bounded density condition on $P_X$.

\begin{corollary}
\label{corollary:mao_L2_sobol_approx}
Under the assumptions of Theorem~\ref{theorem:mao_L2_sobol_approx}, suppose additionally that $P_X$ admits a density $p_X$ with respect to Lebesgue measure on $\mathcal{X}$ satisfying $\|p_X\|_\infty\leq\overline{M}<\infty$. Then, for every $\alpha>0$ and $K\geq 2$, with probability at least $1-K^{-\alpha}$ with respect to the i.i.d.\ uniform sample $\theta_1,\dots,\theta_K$ on $\mathbb{S}^d$, there exists $W\in\mathbb{R}^K$ such that
\begin{equation*}
\|W\|\lesssim\sqrt{\frac{\log (K)}{K}}\|h\|_{\Lambda^q(\mathcal{X})}
\quad\text{and}\quad
\|h-W^\top\phi_{\theta,K}^m\|_{L^2(X)}
\lesssim
\left(\frac{K}{\log(K)}\right)^{-\frac12-\frac{2m+1}{2d}}
\|h\|_{\Lambda^q(\mathcal{X})},
\end{equation*}
where the implicit constants depend only on $\overline{M}$, $\alpha$, $d$, $m$, and $M$.
\end{corollary}

\subsection{Sup-norm approximation via stratified sampling}
\label{subsec:approximation_RWNNs_importance}
In this subsection, we derive sup-norm approximation bounds for RWNNs via a stratified sampling scheme for the hidden-layer parameters, building on \citet{klusowskiApproximationCombinationsReLU2018}. The results complement Theorem~\ref{theorem:sharp_l2}: both subsections assume an integral representation of $h$ of the form $h(x)=\int_\Pi f(A,B)\sigma(A^\top x+B) d\pi_\theta$, but here we obtain $L^\infty$ rather than $L^2$ bounds, under different regularity assumptions on $f$ (H\"older continuity or sign structure) and without requiring the regular sampling condition (Assumption~\ref{assumption:regular_sampling}).

The achieved rate is determined by the regularity of $f$ with respect to $\theta$. In all cases considered below, the approximation error admits a single centered empirical process representation, and the rate follows from bounding the within-cell oscillation of $\phi_x(\theta):=f(A,B)\sigma(A^\top x+B)$.
When $f$ is $\beta$-H\"older continuous (Assumption~\ref{assumption:direct_source_condition_holder_product}) with $\beta\in(0,1]$, the within-cell oscillation scales as $r^\beta$, yielding the rate $K^{-1/2-\beta/(d+1)}$ on $\Pi$. In particular, the Lipschitz case $\beta=1$ achieves the exponent $1/(d+1)$.
When $f$ is sign-valued (Assumption~\ref{assumption:source_sign}), 
the partition is constructed to respect the sign boundary of $f$, so that $f$ is constant on each cell $B_j \subset \Pi_{\pm}$. Within each cell, $\phi_x(\theta)$ is therefore Lipschitz in $\theta$ due to the Lipschitz continuity of $\sigma$, and the same one-term  argument applies. This yields the rate $K^{-1/2-1/(d+1)}$.

Thus, the key structural feature of the analysis is the availability of a single centered empirical process representation of the error, which avoids the need for additional decomposition or balancing arguments.

\begin{assumption}[Parameter space]
\label{assumption:Pi}
For constants $C_A,C_B>0$, define
\begin{equation*}
\Pi:=\{A\in\mathbb{R}^d:\|A\|_1\leq C_A\}\times[-C_B,C_B]
\subset\mathbb{R}^{d+1},
\end{equation*}
and let $\pi_\theta$ be a probability measure on $\Pi$.
\end{assumption}

\begin{assumption}[Direct source condition, sign $f$]
\label{assumption:source_sign}
Let Assumptions~\ref{assumption:X_domain}, \ref{assumption:activation} and \ref{assumption:Pi} hold  with $m\in\{1,2\}$. There exist a constant $M_f>0$ and a measurable function $f:\Pi\to\{-M_f,+M_f\}$ such that
\begin{equation*}
h(x)=\int_\Pi f(A,B) \sigma_m(A^\top x+B) \de\pi_\theta(A,B), \qquad x\in\mathcal{X}.
\end{equation*}
\end{assumption}
Under Assumption~\ref{assumption:source_sign}, the measure $\pi_\theta$ may be chosen as part of the source representation of $h$ and may therefore depend on the target function. Moreover, the stratified construction in case~(ii) of Theorem~\ref{theorem:sharp_is_linf_unified} requires a partition respecting the sign sets $\Pi_\pm:=\{\theta\in\Pi:f(\theta)=\pm M_f\}$. Consequently, this construction is oracle unless the sampling measure and the sign sets are known.

We now describe the stratified sampling construction used in both settings. The idea is to partition the parameter space into strata of controlled diameter and sample independently within each stratum according to the corresponding conditional distribution. Weighting the contribution of each stratum by its probability mass preserves unbiasedness while allowing us to exploit the reduced within-stratum variation of the integrand.

For sufficiently large $K$, set $\widetilde K:=\lfloor K/2\rfloor$ and consider a measurable partition $\{B_j\}_{j=1}^J$ of $\Pi$ into $J\leq\widetilde K$ cells of controlled $d_\Pi$ diameter. The reduced allocation budget allows us to ensure at least one sample per stratum while keeping the total number of sampled parameters at most $K$. Under the sign-valued source condition, the partition is additionally chosen so that each cell is contained in one of the sign sets $\Pi_\pm$.
 Let $P_j:=\pi_\theta(B_j)$ and, for $P_j>0, $ define $\pi_{\theta,j}(E):=\pi_\theta(E\cap B_j)/P_j$. For cells of zero probability mass, the choice of the conditional sampling law is not important.
Let $K_j\in\{\lfloor\widetilde KP_j\rfloor,\lceil\widetilde KP_j\rceil\}$ be independent random allocations satisfying $\mathbb E[K_j]=\widetilde KP_j$, and set $L_j:=K_j+\mathbbm{1}_{\{K_j=0\}}$. Conditional on the allocations, draw $\theta_{j,1},\dots,\theta_{j,L_j}$ independently from $\pi_{\theta,j}$, independently across cells. Writing $\theta_{j,\ell}=(A_{j,\ell}^\top,B_{j,\ell})^\top$ and $\overline K:=\sum_{j=1}^J L_j$, define
\begin{equation}
\label{eq:IS_estimator}
h^{\mathrm{IS}}_{\overline K}(x):=\sum_{j=1}^J\frac{P_j}{L_j}\sum_{\ell=1}^{L_j}f(\theta_{j,\ell})\widetilde{\sigma}(A_{j,\ell}^\top x+B_{j,\ell}),
\end{equation}
where $\widetilde{\sigma}=\sigma$ under the $\beta$-H\"older source condition and $\widetilde{\sigma}=\sigma_m$ under the sign-valued source condition.
 Hence, the corresponding output weights are $W_{j,\ell}:=P_jf(\theta_{j,\ell})/L_j$. The partitions are chosen so that $J\leq\widetilde K$, which gives $\overline K\leq\widetilde K+J\leq K$. Appending neurons with zero output weights yields an element of $\mathcal H_K^{\mathrm{IS}}$. The finitely many smaller values of $K$ are handled separately in the proof.

\begin{theorem}[Sup-norm approximation via stratified sampling]
\label{theorem:sharp_is_linf_unified}
    Let Assumption~\ref{assumption:X_domain} and~\ref{assumption:Pi} hold. Suppose one of the following conditions holds:
    \begin{enumerate}
    \item[(i)] \emph{($\beta$-H\"older continuous $f$, Lipschitz $\sigma$)} Assumptions~\ref{assumption:activation_Lipschitz} and~\ref{assumption:direct_source_condition_holder_product} hold for the same $\Pi$ and $\pi_\theta$ as in Assumption~\ref{assumption:Pi}, with
        \begin{equation*}
    	\rho_\Pi(\theta_1,\theta_2) = d_\Pi(\theta_1,\theta_2) := M\|A_1-A_2\|_1+|B_1-B_2|, \qquad \theta_1,\theta_2\in\Pi,
    \end{equation*}
    where $M$ is the constant from Assumption~\ref{assumption:X_domain}.
    \item[(ii)] \emph{(Sign $f$, $\mathrm{ReLU}^m$ $\sigma$)} 
    Assumption~\ref{assumption:source_sign} holds with $m\in\{1,2\}$.
    \end{enumerate}
    Then, for every  $K\in\mathbb{N}^+$, there exist random inner parameters $\{(A_k,B_k)\}_{k=1}^{ K}$ defining the stratified RWNN class $\mathcal{H}^{\mathrm{IS}}_K$, and  random coefficients  $W=(W_1,\dots,W_{ K})^\top\in\mathbb{R}^{ K}$, measurable with respect to the random allocation variables and the sampled parameters, such that the stratified sampling approximation
    \begin{equation*}
    h^{\mathrm{IS}}_{ K}(x) :=\sum_{k=1}^{ K} W_k \widetilde{\sigma}(A_k^\top x+B_k)\in\mathcal{H}^{\mathrm{IS}}_K,
    \end{equation*}
    where $\widetilde{\sigma}=\sigma$ in case~(i), and $\widetilde{\sigma}=\sigma_m$ with $m\in\{1,2\}$ in case~(ii), satisfies
    \begin{equation}
    \label{eq:sup_norm_unified}
    \E\left[\sup_{x\in\mathcal{X}} \left|h^{\mathrm{IS}}_{ K}(x)-h(x)\right|\right] \le C(\sqrt{d}+\sqrt{\log(K)})K^{-1/2-\alpha},
    \end{equation}
    where
    \begin{equation}
        \label{eq:alpha_case}
    \alpha=\begin{cases} \beta/(d+1), & \text{in case (i)},\\[1mm]
    1/(d+1), & \text{in case (ii)}.
    \end{cases}
    \end{equation}
    The constant $C>0$ in \eqref{eq:sup_norm_unified} in case (i) depends only on $d$, $C_A$, $C_B$, and $M$, as well as on $M_f$, $M_\sigma$, $L_\sigma$, $L_f$, and $\beta$ and is specified in \eqref{eq:C_is_explicit_constant} in the proof.
    In case (ii) $C>0$ depends only on $d$, $C_A$, $C_B$, $M$, $m$, and $M_f$ and is given in \eqref{eq:C_is_explicit_constant_case_ii} in the proof.
\end{theorem}

\begin{remark}[Faster rates on the $\ell_1$-sphere]
\label{remark:sphere_rates}
In both cases of Theorem~\ref{theorem:sharp_is_linf_unified}, if the corresponding source representation and regularity assumptions hold on the $\ell_1$-sphere $\Pi^{\mathrm{sph}}=\{\|A\|_1=C_A\}\times[-C_B,C_B]$ instead of the ball $\Pi=\{\|A\|_1\leq C_A\}\times[-C_B,C_B]$, the effective dimension of the parameter space decreases from $d+1$ to $d$, leading to the following improvements of the rate exponent $\alpha$:
\begin{center}
\begin{tabular}{ccccc}
\toprule
Case & $f$ & $\sigma$& $\alpha$ on ball & $\alpha$ on sphere \\
\midrule
(i) & $\beta$-H\"older, $\beta\in (0,1]$ & Lipschitz & $\beta/(d+1)$ & $\beta/d$ \\
(ii) & Sign & ReLU$^m$, $m\in\{1,2\}$ & $1/(d+1)$ & $1/d$ \\
\bottomrule
\end{tabular}
\end{center}

The improved exponents follow from the covering properties of $\Pi^{\mathrm{sph}}$. Indeed, the $\ell_1$-sphere $\{A\in\mathbb{R}^d:\|A\|_1=C_A\}$ is the union of $2^d$ facets of dimension $d-1$. Including the bias coordinate, each facet of $\Pi^{\mathrm{sph}}$ admits a Lipschitz parametrization by a bounded subset of $\mathbb{R}^d$. Consequently, for sufficiently small $r>0$, $\mathcal{N}(r,\Pi^{\mathrm{sph}},d_\Pi)\leq C_{\mathrm{sph}}r^{-d}$, where $C_{\mathrm{sph}}>0$ depends only on $d$, $M$, $C_A$, and $C_B$. In case~(ii), covering the two sign sets separately preserves this order. Substituting these covering bounds into the stratified approximation argument yields the exponents $\beta/d$ and $1/d$, respectively.

For only bounded $f$ without additional regularity, one should not expect an improvement beyond the classical Monte Carlo rate $K^{-1/2}$ from metric stratification alone: the within-cell oscillation of $\phi_x(\theta)=f(A,B)\sigma(A^\top x+B)$ does not shrink with the cell diameter in the absence of regularity of $f$. Thus the improved exponents in case~(i) rely essentially on the H\"older continuity assumption; if $f$ is only bounded, the natural benchmark is the $K^{-1/2}$ rate both on the ball and on the sphere.

\end{remark}

\begin{remark}[Comparison with Klusowski-Barron Rates]
\label{remark:klusowski_comparison}
\citet{klusowskiApproximationCombinationsReLU2018} consider approximation under a sign-valued representation analogous to Assumption~\ref{assumption:source_sign}, where the inner parameters satisfy $\|A\|_1=1$ and $B\in [-1,0]$. For the $\mathrm{ReLU}^m$ activation with $m\in \{1,2\}$, their Theorem~1 gives (up to a $\sqrt{d + \log(K)}$ factor) the rate $K^{-1/2-1/(d+2)}$ when the coefficients $b_k$ in their finite-network representation are restricted to $b_k\in\{-1,+1\}$, and the rate $K^{-1/2-1/d}$ when $b_k\in[-1,1]$. The latter is the appropriate comparison with our weighted construction and coincides with the exponent $1/d$ obtained here on $\Pi^{\mathrm{sph}}$. For the $b_k\in\{-1,+1\}$ construction, their proof relies on a two-term decomposition of the error into an allocation (or rounding) component and a centered fluctuation term, and the rate follows from balancing these contributions.  When $b_k\in[-1,1]$, the allocation is chosen so that the former term vanishes. The weighted stratified estimator considered here admits an exact centered representation of the approximation error, so that no allocation term arises. As a consequence, the analysis reduces to a single empirical process bound, and the rate is driven by the within-cell oscillation of the integrand. Here, in the sign-valued case, our construction provides a different derivation of the same exponent $1/d$ under the present assumptions. More generally, Theorem~\ref{theorem:sharp_is_linf_unified} also covers the $\beta$-H\"older source condition of case~(i), allows approximation on the full $(d+1)$-dimensional parameter space $\Pi$, where the corresponding exponent is $1/(d+1)$ in the sign-valued case, and records $\ell_1$ and $\ell_2$ bounds on the resulting RWNN output weights used in the subsequent statistical analysis.
\end{remark}

\begin{remark}
For the equal weight Monte Carlo approximant based on i.i.d.\ sampling from $\pi_\theta$, the classical rate $K^{-1/2}$ is obtained under the conditions of Proposition~\ref{prop:entropy_mc_rate}.
 The stratified sampling scheme of Theorem~\ref{theorem:sharp_is_linf_unified} improves this to $K^{-1/2-\beta/(d+1)}$ in case~(i) and $K^{-1/2-1/(d+1)}$ in case~(ii), by allocating neurons proportionally to the cell masses $\pi_\theta(B_j)$, hence ensuring uniform coverage of the parameter space $\Pi$ and reducing  within-cell variability. In case~(ii), the sign structure of $f$ guarantees that it is constant on each cell, yielding Lipschitz behavior within cells regardless of the global regularity of $f$, and thus the full exponent $1/(d+1)$ is achieved.  The gains of $K^{-\beta/(d+1)}$ and $K^{-1/(d+1)}$, respectively, over the equal weight Monte Carlo case reflect the reduction in within-cell variability achieved by stratified sampling, as also exploited in the approximation constructions of \citet{klusowskiApproximationCombinationsReLU2018}.
\end{remark}

\begin{theorem}[High-probability sup-norm approximation]
\label{thm:prob-approx-linf}
    Let the assumptions of Theorem~\ref{theorem:sharp_is_linf_unified} hold. Then for every $K\in\mathbb{N}^+$, there  exist random inner parameters $\{(A_k,B_k)\}_{k=1}^K$ defining the stratified RWNN class $\mathcal{H}^{\mathrm{IS}}_K$ in \eqref{eq:stratified_rwnn_class} and a random vector $W=(W_1,\dots,W_{ K})^\top\in\mathbb{R}^{ K}$, measurable with respect to the random allocation variables and the sampled parameters, such that for every $\varepsilon\in(0,1)$,
    \begin{equation}
    \label{eq:prob_bound_linf}
    \mathbb{P}\left(\sup_{x\in\mathcal{X}} \left|h^{\mathrm{IS}}_{K}(x)-h(x)\right|\leq C(\sqrt{d}+\sqrt{\log(K)})K^{-1/2-\alpha}\varepsilon^{-1}\right)\geq 1-\varepsilon,
    \end{equation}
    where $C>0$ is the constant appearing in Theorem~\ref{theorem:sharp_is_linf_unified}, and $\alpha$ is as in \eqref{eq:alpha_case}.
\end{theorem}

\begin{remark}
Compared to the $L^2$ high-probability bound in Theorem~\ref{thm:prob-approx}, where the probability penalty is $\varepsilon^{-1/2}$, the sup-norm bound 
provides a penalty of $\varepsilon^{-1}$. This reflects the fact that Markov's inequality is applied directly to the sup-norm rather than to its square, and is the standard cost of converting an expectation bound on a non-negative random variable into a high-probability statement without additional second-moment or tail assumptions.
\end{remark}

\begin{corollary}[Stratified RWNN approximation in probability]
\label{cor:stratified_rwnn_projection_rate_linf}
    Under the assumptions of Theorem~\ref{thm:prob-approx-linf},
    \begin{equation*}
    \inf_{g\in\mathcal{H}_{K}^{\mathrm{IS}}}\|h-g\|_{L^\infty(\mathcal{X})}=O_P\left((\sqrt{d}+\sqrt{\log(K)})K^{-1/2-\alpha}\right),
    \end{equation*}
    where $\alpha$ is as in \eqref{eq:alpha_case}.
    The probability is taken with respect to the random allocation variables
    and the sampled parameters in the stratified construction.
\end{corollary}

The following lemma establishes bounds on the output weights of the stratified RWNN approximation. These bounds are subsequently used in the sup-norm analysis of ridge-regularized RWNN estimators in Subsection~\ref{subsec:rwnn_uniform_ridge}.

\begin{lemma}[$\ell_1$- and $\ell_2$-norm bounds for stratified 
sampling weights]
\label{lem:stratified_weight_norms}
    Let the assumptions of Theorem~\ref{theorem:sharp_is_linf_unified} hold, and let
    \begin{equation*}
    h^{\mathrm{IS}}_{ K}(x) = \sum_{k=1}^{ K} W_k \widetilde{\sigma}(A_k^\top x+B_k)\in\mathcal{H}_K^{\mathrm{IS}},
    \end{equation*}
    be the stratified RWNN approximation, where $\widetilde{\sigma}=\sigma$ in case~{(i)} and $\widetilde{\sigma}=\sigma_m$ in case~{(ii)}. Then its output-weight vector $W\in\mathbb{R}^{ K}$ satisfies
    \begin{equation}
    \label{eq:stratified_l1_bound}
    \|W\|_1\leq M_f,
    \end{equation}
    and
    \begin{equation}
    \label{eq:stratified_l2_bound}
    \|W\|\leq \sqrt{6}\,M_f K^{-1/2}.
    \end{equation}
    In particular,
    \begin{equation*}
    \|W\|_1=O(1) \qquad \text{and} \qquad \|W\|=O(K^{-1/2}).
    \end{equation*}
\end{lemma}

\begin{remark}[Comparison with control-neighbors weights]
Lemma~\ref{lem:stratified_weight_norms} and Lemma~\ref{lem:nn_l2_bound} together show that the stratified weights are better controlled than the control-neighbors weights in two respects. First, the stratified construction yields the deterministic $\ell_2$ bound, while the control-neighbors bound holds with high probability. Second, the stratified weights satisfy  $\|W\|\leq\sqrt{6}\,M_fK^{-1/2}$, improving on the order $\|W\|=O_P(M_f\sqrt{\log(K)/K})$ of Lemma~\ref{lem:nn_l2_bound} by a factor of $\sqrt{\log(K)}$. 
\end{remark}

\section{Ridge-regularized RWNN estimation and inference}%
\label{sec:Ridge_RWNN_least_squares}
This section develops the statistical theory of ridge-regularized least-squares estimation with RWNN sieves. We proceed from the spectral properties of random features to statistical estimation and inference. More explicitly, in Section~\ref{subsec:L2_ridge_consistency} we show the spectral deterioration of RWNN sieves with growing sieve dimension, characterize the role played by ridge normalization and the spectrum of the population sieve covariance matrix in determining the effective dimension of the estimation problem by introducing the so-called ridge effective dimension and deriving a bound for it. Subsection~\ref{subsec:L2_estimation} analyzes ridge-normalized covariance stability and uses it to obtain $L^2(X)$ estimation bounds. Subsection~\ref{subsec:linearization_sharper_l2} develops directional inference through linearization and asymptotic normality, allowing the smallest population covariance eigenvalue to vanish. We then establish consistency of a residual-based ridge-adjusted variance estimator, enabling feasible studentized inference, before extending the results to smooth functionals in Subsection~\ref{subsec:inference_smooth_functionals}. Finally, Subsection~\ref{subsec:rwnn_uniform_ridge} addresses uniform estimation, first through spectral stability and then through prediction-space and leverage-based conditions. 

Complementing the ridge-regularized analysis, Appendix~\ref{app:generic_sieve_estimation} develops a general framework for constrained sieve estimation and asymptotic inference on smooth functionals, accommodating loss functions {\it beyond squared-loss regression}. The constrained RWNN approximation results established there are also used in the ridge-regularized estimation and inference analysis of this section.

Proofs of the results on spectral properties and $L^2(X)$ estimation are collected in Appendix~\ref{app:ridge_L2_proofs}, the ones that concern directional and smooth-functional inference and feasible variance estimation are provided in Appendix~\ref{app:inference_proofs}, and, finally, the reader can find the proofs of our results on uniform estimation in Appendix~\ref{app:supnorm_ridge_proofs}.

\subsection{Spectral deterioration and effective dimension}
\label{subsec:L2_ridge_consistency}

We first introduce the ridge-normalized geometry of the population feature covariance matrix. We then establish spectral deterioration for parametrized random-feature sieves and derive corresponding bounds on ridge effective dimension. These results provide the spectral foundation for the estimation and inference theory developed below.

\subsubsection{Ridge normalization and population covariance}
\label{subsubsec:ridge_normalization}

We first introduce the ridge-normalized quantities underlying the $L^2(X)$ analysis. This normalization allows us to formulate the relevant stability and complexity measures without requiring the population sieve covariance matrix to remain uniformly well-conditioned as the sieve dimension $K_n$ increases.

Recall the notation of Section~\ref{subsec:rwnn_sieve_ls}, with $K=K_n$. In particular, the feature map $g_{K_n}$  is defined in \eqref{eq:feature_map}, the design matrix $G_{K_n,n}$, and the response vector $Y$ are defined in \eqref{eq:design_matrix}. We also write $e:=(\epsilon_1,\ldots,\epsilon_n)^\top\in\mathbb{R}^n$. The ridge-regularized RWNN estimator $\widehat{h}_{n,\lambda}$ and its coefficient vector $\widehat{\beta}_{K_n,n,\lambda}$ are defined in \eqref{eq:def_betahat}. We write
\begin{align}
    \label{eq:def_gamma_hat}
    \widehat{\Gamma}_{K_n,n,\lambda} &:= \widehat{\Gamma}_{{K_n},n} + \lambda_{K_n,n} I_{K_n} \enspace \text{and} \enspace \widehat{\Gamma}_{{K_n},n} := n\inv G_{K_n,n}^\top G_{K_n,n}.
\end{align}
Define the population sieve covariance matrix and its ridge-regularized counterpart by %
\begin{equation}
\label{eq:def_Gamma_Gammalambda}
    \Gamma_{{K_n}} := \E_X[ g_{K_n}(X) g_{K_n}(X)^\top ]
    \quad \text{and} \quad
    \Gamma_{{K_n},\lambda} :=  \Gamma_{{K_n}} +  \lambda_{{K_n},n} I_{K_n} ,
\end{equation}
respectively.
The $L^2(X)$-projection coefficient of $h$ onto the linear span of the sieve is given by
\begin{equation}
\label{eq:population_projection}
    \beta_{K_n} \in \argmin_{b \in \R^{K_n}} \, \E_X\big[ ( h(X) - g_{K_n}(X)^\top b )^2 \big].
\end{equation}

If the minimizer is not unique, we choose $\beta_{K_n}$ to be the minimum-Euclidean-norm solution.
We denote the corresponding approximation error by $\Delta_{{K_n}} := \norm{ h - g_{K_n}^\top \beta_{K_n} }_{L^2(X)}$. We also introduce the ridge-normalized feature  map
\begin{equation}
\label{eq:ridge_normalized_feature_map}
    \widetilde{g}_{K_n}(x) := \Gamma_{{K_n},\lambda}^{-1/2} g_{K_n}(x) .
\end{equation}
Additionally, define the uniform norms of the original and ridge-normalized feature maps by
\begin{equation}
\label{eq:def_xi_tildexi}
    \xi_{{K_n}} := \sup_{x \in \mathcal{X}} \norm{ g_{K_n}(x) }
    \quad \text{and} \quad 
    \widetilde{\xi}_{{K_n},\lambda} := \sup_{x \in \mathcal{X}} \norm{ \widetilde{g}_{K_n}(x) } ,
\end{equation}
respectively. Finally, let $s_j^2$, $j=1,\ldots, K_n$, be the eigenvalues of the population sieve covariance matrix $\Gamma_{{K_n}}$ and define
\begin{equation}
\label{eq:def_smallest_population_eigenvalue}
    \overline{s}_{K_n}^2 := \lambda_{\min}( \Gamma_{{K_n}} ).
\end{equation}
Since the ridge regularization is isotropic, $\lambda_{\min}( \Gamma_{{K_n},\lambda} ) = \overline{s}_{K_n}^2 + \lambda_{{K_n},n}$.

\subsubsection{Spectral deterioration of random feature sieves}
\label{subsubsec:spectral_deteriotation}

The smallest population eigenvalue $\overline{s}_{K_n}^2$ defined in \eqref{eq:def_smallest_population_eigenvalue} need not remain bounded away from zero as the sieve dimension $K_n$ increases. More generally, for parameterized sieve families, the covering and Lipschitz conditions below yield explicit control of the spectral tail of the population sieve covariance matrix. The following bound is closely related to classical connections between metric approximation, Kolmogorov widths, and spectral decay (see \citet{Ismagilov1968} and the recent treatment in \citet{Takhanov2026}, in particular the proofs of Theorem~1 and Theorems~3-4). We give the direct finite-dimensional argument below since it applies immediately to the population sieve covariance matrix.

\begin{proposition}
\label{prop:spectral_tail_parametrized_sieve}
    Let $(\Pi,\rho_\Pi)$ be a compact metric space such that, for some constants $C_\Pi, q_\Pi>0$, there exists a constant $\overline{r} > 0$ such that, for all $r < \overline{r}$, there exists an $r/2$-net $\mathcal{N}_{r/2}$ of $\Pi$ under $\rho_\Pi$ satisfying
    \begin{equation}
    \label{eq:parameter_covering}
        |\mathcal{N}_{r/2}|\leq C_\Pi r^{-q_\Pi}.
    \end{equation}
    Let $\{\phi_\theta\mid \theta\in\Pi\}\subset L^2(X)$ satisfy, for some $L_\phi>0$,
    \begin{equation}
    \label{eq:parameter_feature_lipschitz}
        \|\phi_{\theta_1}-\phi_{\theta_2}\|_{L^2(X)}\leq L_\phi\rho_\Pi(\theta_1,\theta_2),\qquad \theta_1,\theta_2\in\Pi.
    \end{equation}
    For arbitrary $\theta_1,\ldots,\theta_K\in\Pi$, define
    \begin{equation*}
        g_K(x):=\big(\phi_{\theta_1}(x),\ldots,\phi_{\theta_K}(x)\big)^\top,\qquad \Gamma_K:=\mathbb{E}_X[g_K(X)g_K(X)^\top],
    \end{equation*}
    and let $s_1^2\geq\cdots\geq s_K^2$ denote the eigenvalues of $\Gamma_K$. 
    
    Then,  for all sufficiently large $K$ and $m<K$,
    \begin{equation}
    \label{eq:spectral_tail_sieve}
        \sum_{j=m+1}^K s_j^2\leq C_{\mathrm{tail}} K m^{-2/q_\Pi},
    \end{equation}
    where $C_{\mathrm{tail}}=L_\phi^2C_\Pi^{2/q_\Pi}$ is independent of $K$ and $\theta_1,\ldots,\theta_K$.
    Moreover, for all sufficiently large $j \leq K$,
    \begin{equation}
    \label{eq:bound_eigenvalue_sieve}
        s_j^2\leq C_{\mathrm{eig}} K j^{-1-2/q_\Pi},
    \end{equation}
    where $C_{\mathrm{eig}}=2\cdot  3^{2/q_\Pi}C_{\mathrm{tail}}$. 
    In particular, for all sufficiently large $K$,
    \begin{equation}
    \label{eq:smallest_eigenvalue_sieve}
        \lambda_{\min}(\Gamma_K)\leq C_{\mathrm{eig}} K^{-2/q_\Pi} .
    \end{equation}
\end{proposition}

The bounds in Proposition~\ref{prop:spectral_tail_parametrized_sieve} are deterministic and hold for arbitrary $\theta_1,\ldots,\theta_K\in\Pi$ whenever $K$ is large. The upper bound on $\lambda_{\min}(\Gamma_K)$ shows that uniform lower spectral conditioning cannot, in general, be maintained for growing parameterized sieves. 
This motivates formulating the subsequent analysis in terms of ridge-normalized quantities that depend on the spectrum beyond its smallest eigenvalue.

\subsubsection{Ridge effective dimension}
\label{subsubsec:ridge_effective_dimension}
Using the ridge-normalized feature map $\widetilde{g}_{K_n}$ defined in \eqref{eq:ridge_normalized_feature_map}, we introduce the {\it ridge effective dimension} of the sieve as
\begin{equation}
\label{eq:def_L2_effective_dimension}
    \mathcal{N}_{K_n,\lambda}:=\E_X\left[\|\widetilde{g}_{K_n}(X)\|^2\right]=\operatorname{tr}\left(\Gamma_{K_n}\Gamma_{K_n,\lambda}^{-1}\right)=\sum_{j=1}^{K_n}
    \frac{s_j^2}{s_j^2+\lambda_{K_n,n}}.
\end{equation}
The spectral-tail bound in Proposition~\ref{prop:spectral_tail_parametrized_sieve} then gives control of the effective dimension.

\begin{corollary}
\label{cor:effective_dimension_spectral}
    Under the conditions of Proposition~\ref{prop:spectral_tail_parametrized_sieve}, let $K_n\to\infty$ and $\lambda_{K_n,n}>0$ satisfy
    \begin{equation*}
        \frac{K_n}{\lambda_{K_n,n}}\to\infty\qquad\text{and}\qquad \lambda_{K_n,n}K_n^{2/q_\Pi}\to\infty.
    \end{equation*}
    Then, for all sufficiently large $n$,
    \begin{equation}
    \label{eq:effective_dimension_spectral_rate}
        \mathcal{N}_{K_n,\lambda}\leq\min\left\{K_n,1+C_{\mathcal{N}}\left(\frac{C_{\mathrm{tail}}K_n}{\lambda_{K_n,n}}\right)^{q_\Pi/(q_\Pi+2)}\right\},
    \end{equation}
    where $C_{\mathcal{N}}=\left(\frac{2}{q_\Pi}\right)^{q_\Pi/(q_\Pi+2)}+\left(\frac{q_\Pi}{2}\right)^{2/(q_\Pi+2)}$. 
    
    Consequently,
    \begin{equation}
    \label{eq:effective_dimension_o_K}
        \mathcal{N}_{K_n,\lambda}=o(K_n).
    \end{equation}
\end{corollary}

\subsection{$L^2$ estimation}
\label{subsec:L2_estimation}
We now study the $L^2(X)$ estimation error of the ridge-regularized sieve estimator $\widehat{h}_{n,\lambda}$. We first establish ridge-normalized covariance stability and derive a general estimation bound in terms of the ridge effective dimension, sieve approximation error, and population ridge bias. We then specialize these results to RWNN sieves, obtaining explicit convergence rates under the approximation conditions developed in Section~\ref{sec:approximation_RWNNs}.

\subsubsection{Ridge-normalized covariance stability}
We start by imposing a condition on the ridge-normalized feature norm $\widetilde{\xi}_{K_n,\lambda}$ defined in \eqref{eq:def_xi_tildexi}, ensuring stability of the empirical covariance matrix in ridge-normalized coordinates.
\begin{assumption}[Ridge-normalized covariance matrix stability]
\label{assumption:L2_ridge_gram_stability}
It holds
    \begin{equation*}
    \frac{\widetilde{\xi}_{K_n,\lambda}^{2}\log(K_n)}{n}=o(1).
    \end{equation*}
\end{assumption}

We notice that, since
\begin{equation*}
\widetilde{\xi}_{K_n,\lambda}\leq\frac{\xi_{K_n}}{\sqrt{\overline s_{K_n}^{\,2}+\lambda_{K_n,n}}},
\end{equation*}
with $\xi_{K_n}$ and $\overline{s}_{K_n}^2$ defined in \eqref{eq:def_xi_tildexi} and \eqref{eq:def_smallest_population_eigenvalue}, respectively, Assumption~\ref{assumption:L2_ridge_gram_stability} is implied by 
\begin{equation}
\frac{\xi_{K_n}^{\,2}\log(K_n)}{n\bigl(\overline s_{K_n}^{\,2}+\lambda_{K_n,n}\bigr)}=o(1).
\label{eq:L2_ridge_stability_rate}
\end{equation}

\begin{remark}
\label{remark:L2_ridge_assumptions_discussion}
    Assumption~\ref{assumption:L2_ridge_gram_stability} is a ridge-normalized analogue of the standard condition ${\xi}_{{K_n}}^2 / {n} \to 0$ and its logarithmic variants commonly imposed in sieve least squares estimation \citep{Newey1997}.
    Widely used well-conditioned sieves---such as splines, wavelets, and other localized series bases---are such that $\overline{s}_{K_n}^2 \geq \underline{c}_{\Gamma} > 0$ and $\xi_{K_n}\lesssim \sqrt{{K_n}}$ (see, for example, \citealp{Chen2007,Chen2015,Belloni2015}). In such cases, \eqref{eq:L2_ridge_stability_rate} is implied by
    \begin{equation*}
        \frac{ \xi_{K_n}^2 \log({K_n}) }{ n } = o(1),
    \end{equation*}
    which, when $\xi_{K_n}\lesssim\sqrt{K_n}$, is in turn implied by $K_n\log(K_n)/n=o(1)$. 
    
    More generally, when $\overline{s}_{K_n}^2$ is allowed to vanish, the sufficient condition~\eqref{eq:L2_ridge_stability_rate} may be used to verify Assumption~\ref{assumption:L2_ridge_gram_stability}. This sufficient condition may be conservative as it controls all feature directions through the smallest eigenvalue of $\Gamma_{K_n}$. To see why Assumption~\ref{assumption:L2_ridge_gram_stability} implies empirical covariance matrix stability, observe that
    \begin{equation*}
        \Gamma_{K_n,\lambda}^{-1/2}\left(\widehat{\Gamma}_{K_n,n}-\Gamma_{K_n}\right)\Gamma_{K_n,\lambda}^{-1/2}=\frac{1}{n}\sum_{i=1}^n\left(\widetilde{g}_{K_n}(X_i)\widetilde{g}_{K_n}(X_i)^\top-\mathbb{E}_X\left[\widetilde{g}_{K_n}(X_i)\widetilde{g}_{K_n}(X_i)^\top\right]\right).
    \end{equation*}
    A matrix-valued law of large numbers, such as Lemma~\ref{lemma:matrix_lln_iid}, together with
    \begin{equation*}
        \left\|\Gamma_{K_n,\lambda}^{-1/2}\Gamma_{K_n}\Gamma_{K_n,\lambda}^{-1/2}\right\|\leq1
    \end{equation*}
    and
    \begin{equation*}
        \left\|\widetilde{g}_{K_n}(X_i)\widetilde{g}_{K_n}(X_i)^\top\right\|=\left\|\widetilde{g}_{K_n}(X_i)\right\|^2\leq\widetilde{\xi}_{K_n,\lambda}^2
    \end{equation*}
    yields
    \begin{equation*}
        \mathbb{E}\left[\left\|\Gamma_{K_n,\lambda}^{-1/2}\left(\widehat{\Gamma}_{K_n,n}-\Gamma_{K_n}\right)\Gamma_{K_n,\lambda}^{-1/2}\right\|\right]\lesssim\frac{\widetilde{\xi}_{K_n,\lambda}^2\log(K_n)}{n}+\sqrt{\frac{\widetilde{\xi}_{K_n,\lambda}^2\log(K_n)}{n}}.
    \end{equation*}
    Therefore, Assumption~\ref{assumption:L2_ridge_gram_stability} and Markov's inequality imply
    \begin{equation}
    \label{eq:nondegeneracy}
        \left\|\Gamma_{K_n,\lambda}^{-1/2}\left(\widehat{\Gamma}_{K_n,n}-\Gamma_{K_n}\right)\Gamma_{K_n,\lambda}^{-1/2}\right\|=O_{P}\left(\sqrt{\frac{\widetilde{\xi}_{K_n,\lambda}^2\log(K_n)}{n}}\right)=o_{P}(1).
    \end{equation}
     This resulting ridge-normalized covariance matrix stability is the key perturbation condition used in Theorem~\ref{theorem:L2_ridge_sharp_consistency}.
\end{remark}

For later use, define the ridge-normalized covariance perturbation matrix and its associated inverse by
\begin{equation}
\label{eq:def_Dn_hat_An}
\begin{aligned}
A_{K_n,n,\lambda}&:=\Gamma_{K_n,\lambda}^{-1/2}(\widehat{\Gamma}_{K_n,n}-\Gamma_{K_n})\Gamma_{K_n,\lambda}^{-1/2},\\
\widehat{D}_{K_n,n,\lambda}&:=\Gamma_{K_n,\lambda}^{1/2}\widehat{\Gamma}_{K_n,n,\lambda}^{-1}\Gamma_{K_n,\lambda}^{1/2}=(I_{K_n}+A_{K_n,n,\lambda})^{-1}.
\end{aligned}
\end{equation}
In particular, \eqref{eq:nondegeneracy} implies $\|A_{K_n,n,\lambda}\|=o_P(1)$ and hence $\|\widehat{D}_{K_n,n,\lambda}\|=O_P(1)$.

\subsubsection{$L^2$ estimation bounds}
\label{subsubsec:L2_error_bound}

We now derive an $L^2(X)$ error bound for the ridge-regularized least-squares sieve estimator. We explicitly separate and analyze the sieve approximation error, the stochastic estimation error, and the effect of ridge regularization. %

\begin{assumption}[Regression errors]
\label{assumption:regression_errors_L2}
    The regression errors satisfy $\mathbb{E}[\epsilon_i\mid X_i]=0$.
    Moreover, there exists a constant $\overline{\sigma}^2<\infty$ such that $\mathbb{E}[\epsilon_i^2\mid X_i]\leq\overline{\sigma}^2$.
\end{assumption}

\begin{theorem}[A refined $L^2(X)$ error bound]
\label{theorem:L2_ridge_sharp_consistency}
    Let $\{Z_i\}_{i\in\mathbb{N}^+}$ be i.i.d.\ observations satisfying Assumption~\ref{assumption:regression_errors_L2}. Let $\{g_{K_n}:\mathcal{X}\to\mathbb R^{K_n}\}_{n\geq1}$ be a sequence of feature maps satisfying
    Assumption~\ref{assumption:L2_ridge_gram_stability}. 
    Define 
    the population ridge bias
    \begin{equation}
    \label{eq:def_L2_population_ridge_bias}
    B^{\mathrm{pop}}_{K_n,\lambda}:=\lambda_{K_n,n}\left\|\Gamma_{K_n}^{1/2}\Gamma_{K_n,\lambda}^{-1}\beta_{K_n}\right\|.
    \end{equation}
    Then
    \begin{equation}
    \label{eq:L2_ridge_refined_bound}
    \left\|\widehat{h}_{n,\lambda}-h\right\|_{L^2(X)}=O_{P}\left(\Delta_{K_n}+\sqrt{
    \frac{\mathcal{N}_{K_n,\lambda}}{n}}+B^{\mathrm{pop}}_{K_n,\lambda}\right).%
    \end{equation}
\end{theorem}

The population ridge bias defined in \eqref{eq:def_L2_population_ridge_bias} admits a direct spectral interpretation. Let $\overline{\beta}_{j,K_n}$ denote the coordinates of $\beta_{K_n}$ with respect to an eigenbasis of $\Gamma_{K_n}$. Then
\begin{equation*}
    \left(B^{\mathrm{pop}}_{K_n,\lambda}\right)^2
    =
    \lambda_{K_n,n}^2 \sum_{j=1}^{K_n}\frac{s_{j}^2}{\left(s_{j}^2+\lambda_{K_n,n}\right)^2} \, \overline{\beta}_{j,K_n}^2.
\end{equation*}
Hence, the population ridge bias is measured in the $L^2(X)$ prediction norm and weights each coefficient direction according to the corresponding eigenvalue of the population covariance matrix. Moreover,
\begin{equation*}
B^{\mathrm{pop}}_{K_n,\lambda}\leq\frac{\lambda_{K_n,n}}{\sqrt{\overline{s}_{K_n}^{\,2}+\lambda_{K_n,n}}}\|\beta_{K_n}\| .
\end{equation*}
Therefore, a na\"ive smallest-eigenvalue bound provides a sufficient, but potentially conservative, upper bound on the population ridge bias.

Since $\mathcal{N}_{K_n,\lambda}\leq K_n$, Theorem~\ref{theorem:L2_ridge_sharp_consistency} implies
\begin{equation*}
\|\widehat{h}_{n,\lambda}-h\|_{L^2(X)}
=O_P\left(\Delta_{K_n}+\sqrt{\frac{K_n}{n}}+B^{\mathrm{pop}}_{K_n,\lambda}\right)
\end{equation*}
and hence if $B^{\mathrm{pop}}_{K_n,\lambda}=O_P\left(\Delta_{K_n}+\sqrt{K_n/n}\right)$, the ridge estimator achieves the classical sieve estimation rate.
Nonetheless, under the conditions of Corollary~\ref{cor:effective_dimension_spectral}, $\mathcal{N}_{K_n,\lambda}=o(K_n)$ and, hence, the stochastic term in Theorem~\ref{theorem:L2_ridge_sharp_consistency},
$\sqrt{\mathcal{N}_{K_n,\lambda}/n}$, is asymptotically strictly smaller than
its classical sieve counterpart, which has order $\sqrt{K_n/n}$.

We conclude this subsection by providing the result, which will be useful later in the analysis of the variance estimator. The following lemma quantifies an empirical prediction bound, which is derived from the same coefficient decomposition used in the proof of Theorem~\ref{theorem:L2_ridge_sharp_consistency}.

\begin{lemma}[Empirical prediction bound for ridge sieves]
\label{lemma:L2_ridge_empirical_prediction}
    Let the assumptions of Theorem~\ref{theorem:L2_ridge_sharp_consistency} hold. Then
    \begin{equation*}
        \left\|\widehat{\Gamma}_{K_n,n}^{1/2}\left(\widehat{\beta}_{K_n,n,\lambda}-\beta_{K_n}\right)\right\|=O_P\left(\Delta_{K_n}+B^{\mathrm{pop}}_{K_n,\lambda}+\sqrt{\frac{\mathcal{N}_{K_n,\lambda}}{n}}\right).
    \end{equation*}
\end{lemma}

\subsubsection{RWNN specialization and primitive rates}
\label{subsubsec:RWWN_specialization_L2}

We now specialize Theorem~\ref{theorem:L2_ridge_sharp_consistency} to the RWNN regression setting. Although the ridge estimator is not constrained to belong to $\mathcal{H}_{K_n}(M_n)$, the constrained approximation constructed in Subsection~\ref{subsec:L2_Chen_Shen} can be used to control both the sieve approximation error $\Delta_{K_n}$ and the population ridge bias $B^{\mathrm{pop}}_{K_n,\lambda}$. Indeed, $\mathcal{H}_{K_n}(M_n)$ is contained in a full linear span $\left\{x\mapsto g_{K_n}(x)^\top\beta:\beta\in\mathbb{R}^{K_n}\right\}$ and hence 
\begin{equation*}
    \Delta_{K_n}
    = \inf_{\beta\in \mathbb{R}^{K_n}}\|h-g_{K_n}^\top\beta\|_{L^2(X)} 
    \:\leq\: \inf_{\{W\in  \mathbb{R}^{K_n}\mid \|W\|\leq M_n\}}\|h-g_{K_n}^\top W\|_{L^2(X)}
\end{equation*}
Consequently, under the source condition of Corollary~\ref{cor:constrained_approx}, $\Delta_{K_n} = O_{P_\theta}\big( K_n^{-1/2-\beta/(d+1)} \big)$. 

The covering properties of the parameter space underlying this approximation result also have a direct implication for the global spectral conditioning of the RWNN sieve. In particular, the polynomial covering bound used in Section~\ref{sec:approximation_RWNNs} implies that, as the number of features grows, the smallest population covariance matrix eigenvalue cannot remain uniformly bounded away from zero.

\begin{corollary}[Smallest population eigenvalue of the RWNN sieve]
\label{cor:RWNN_smallest_eigenvalue}
    Let Assumptions~\ref{assumption:X_domain} and~\ref{assumption:activation_Lipschitz} hold, let $\Pi\subset\mathbb{R}^{d+1}$ be compact, and suppose that~\eqref{eq:rho_compatibility} holds. For  $\theta_k=(A_k^\top,B_k)^\top\in\Pi$, $k=1,\ldots,K$, define $
    g_K(x):=\big(\sigma(A_1^\top x+B_1),\ldots,\sigma(A_K^\top x+B_K)\big)^\top$.
    Then
    \begin{equation*}
    \overline{s}_K^2\lesssim K^{-2/(d+1)}.
    \end{equation*}
    The bound holds for every realization of the hidden-layer parameters and hence, in particular, almost surely when the parameters are randomly sampled.
\end{corollary}

Corollary~\ref{cor:RWNN_smallest_eigenvalue} shows that a uniform positive lower bound on the smallest population covariance eigenvalue cannot be maintained for the compact RWNN parameterizations considered here. This provides additional motivation for the ridge-normalized stability condition in Assumption~\ref{assumption:L2_ridge_gram_stability}, which does not require $\overline{s}_{K_n}^2$ to remain bounded away from zero. 
We next use the constrained RWNN approximant to control the population ridge bias $B^{\mathrm{pop}}_{K_n,\lambda}$ defined in \eqref{eq:def_L2_population_ridge_bias}.

\begin{lemma}[Population ridge bias bound for RWNNs]
\label{lem:RWNN_population_ridge_bias}
    Under the conditions of Corollary~\ref{cor:constrained_approx}, 
    \begin{equation*}
        B^{\mathrm{pop}}_{K_n,\lambda}=O_{P_\theta}\left(K_n^{-1/2-\beta/(d+1)} + \sqrt{\frac{\lambda_{K_n,n}\log (K_n)}{K_n}}\right).
    \end{equation*}
\end{lemma}

Combining the approximation bound for $\Delta_{K_n}$ with Theorem~\ref{theorem:L2_ridge_sharp_consistency} and the preceding result for ridge bias in Lemma~\ref{lem:RWNN_population_ridge_bias}, together with the corresponding RWNN conditions ensuring ridge-normalized covariance matrix stability, yields the following rate for the ridge-regularized RWNN least-squares estimator.

\begin{theorem}
\label{theorem:L2_RWNN_sharp_consistency}
    Let $\{Z_i\}_{i\in\mathbb{N}^+}$ be i.i.d. observations satisfying $Y_i=h(X_i)+\epsilon_i$, $ \E[\epsilon_i\mid X_i]=0$, and $ \E[\epsilon_i^2\mid X_i]\le \bar\sigma^2<\infty.$ Let the conditions of Corollary~\ref{cor:constrained_approx} hold and let $K_n\to\infty$ and $\lambda_{K_n,n}>0$ be deterministic sequences.
    Assume moreover that the defining conditions of Assumption~\ref{assumption:L2_ridge_gram_stability} hold in $P_\theta$-probability along the chosen sequences $K_n$ and $\lambda_{K_n,n}$.
    Then
    \begin{equation}
    \label{eq:L2_RWNN_effective_dimension_rate}
        \norm{\widehat{h}_{n,\lambda}-h}_{L^2(X)}=O_P\left(\sqrt{\frac{\mathcal{N}_{K_n,\lambda}}{n}}+K_n^{-1/2-\beta/(d+1)}+\sqrt{\frac{\lambda_{K_n,n}\log(K_n)}{K_n}}
        \right).
    \end{equation}
    
    Furthermore, let $\beta\in(0,1)$ and suppose that
    \begin{equation}
    \label{eq:RWNN_polynomial_ridge}
        \lambda_{K_n,n}\asymp K_n^{-r},\qquad\frac{2\beta}{d+1}<r<\frac{2}{d+1}.
    \end{equation}
    Then
    \begin{equation}
    \label{eq:RWNN_effective_dimension_polynomial}
        \mathcal{N}_{K_n,\lambda}=O\left(K_n^{\nu_r}\right),\qquad\nu_r:=\frac{(d+1)(1+r)}{d+3}<1,
    \end{equation}
    and
    \begin{equation}
    \label{eq:L2_RWNN_polynomial_effective_dimension_rate}
        \norm{\widehat{h}_{n,\lambda}-h}_{L^2(X)}=O_P\left(\sqrt{\frac{K_n^{\nu_r}}{n}}+K_n^{-1/2-\beta/(d+1)}\right).
    \end{equation}
    
    If
    \begin{equation*}
        K_n\asymp n^{\frac{1}{1+\nu_r+2\beta/(d+1)}},
    \end{equation*}
    then
    \begin{equation}
    \label{eq:L2_RWNN_faster_effective_dimension_rate}
        \norm{\widehat{h}_{n,\lambda}-h}_{L^2(X)}
        =O_P\left(n^{-\frac{1/2+\beta/(d+1)}{1+\nu_r+2\beta/(d+1)}}
        \right).
    \end{equation}
\end{theorem}

\subsection{Directional inference for linear functionals}
\label{subsec:linearization_sharper_l2}
In this subsection, we develop directional inference for ridge-regularized RWNN sieve estimators, including irregular functionals such as point evaluation. We begin by establishing a ridge-normalized linearization for fixed feature maps, separating the leading stochastic term from the remainders arising from empirical covariance perturbations, sieve approximation, and ridge regularization. We then derive directional asymptotic normality under conditions ensuring that these remainders are negligible relative to the corresponding standard deviation. Importantly, we show how this {\it general result can be specialized to RWNN sieves} by conditioning on the sampled hidden-layer parameters and formulating sufficient conditions in terms of RWNN approximation properties, ridge-normalized stability, and directional nondegeneracy characterized through a ridge-normalized Rayleigh quotient.   The resulting Gaussian limits are centered at the corresponding sieve projection coefficients, while inference on functionals of $h$ additionally requires control of the sieve approximation bias.

\subsubsection{Ridge-normalized linearization}

We start by extending the linearization result in \cite{Belloni2015} to the ridge-regularized RWNN least squares estimator. This provides a route to inference for  \textit{linear} sieve functionals, including point evaluation, and is complementary to the smooth-functional CLT developed in  Section~\ref{subsec:rwnn_functional_clt}. In contrast to the \textit{regular} functionals considered there, the linear sieve functionals studied here need not be regular in the sense of \cite{vandervaartDifferentiableFunctionals1991}, that is, with respect to the limiting $L^2(X)$ geometry. %
Point evaluation is a prime example of this case, as its sieve Riesz representer generally depends on $K_n$ and need not be bounded.

The results below are first established for fixed feature maps and subsequently applied conditionally on the sampled hidden-layer parameters.
Throughout this subsection we use the notation from Subsection~\ref{subsec:L2_ridge_consistency}, where $\beta_{K_n}$ denotes the population projection coefficient defined in \eqref{eq:population_projection}. We define  $g_i := g_{K_n}(X_i)$, $\widetilde{g}_i := \widetilde{g}_{K_n}(X_i)=\Gamma_{K_n,\lambda}^{-1/2}g_i$ using the definition of  $\widetilde{g}_{K_n}$ in \eqref{eq:ridge_normalized_feature_map}, and  $\eta_i := \eta_{K_n}(X_i) = h(X_i) - g_{K_n}(X_i)^\top \beta_{K_n}$. 
For any random variable $V_i$, define
\begin{equation*}
    \mathbb{G}_n[ V_i ]
    := \frac{1}{\sqrt{n}} \sum_{i=1}^n \left( V_i - \E[V_i] \right).
\end{equation*}
For a unit vector $\alpha_{K_n}\in \mathbb{S}^{K_n-1}$, we consider the linear  functional $v\mapsto \alpha_{K_n}^\top v$, $v\in \mathbb{R}^{K_n}$. Point evaluation at $x\in \mathcal{X}$ corresponds to $\alpha_{K_n}=g_{K_n}(x)$ and is obtained from the result below by applying it to the normalized direction $g_{K_n}(x)/\|g_{K_n}(x)\|$ and rescaling.
We also define $\widetilde{\alpha}_{K_n,\lambda} := \Gamma_{K_n,\lambda}^{-1/2} \alpha_{K_n}$, which will be used to express the linearization in ridge-normalized coordinates. Whenever there is no possibility of confusion, we further streamline notation and write $\alpha$ and $\widetilde{\alpha}$ in place of $\alpha_{K_n}$ and $\widetilde{\alpha}_{K_n,\lambda}$, respectively.

\begin{lemma}[Linearization]
\label{lemma:linearization}
    Assume that the observations are i.i.d. and satisfy Assumption~\ref{assumption:regression_errors_L2}. Suppose further that 
    \begin{equation}
    \label{eq:linearization_normalized_gram_stability}
        \left\|\Gamma_{K_n,\lambda}^{-1/2}\left(\widehat{\Gamma}_{K_n,n}-\Gamma_{K_n}\right) \Gamma_{K_n,\lambda}^{-1/2}\right\|
        = o_P(1) .
    \end{equation}
    
    Then, for every $\alpha \in \mathbb{S}^{K_n-1}$, the ridge estimator error admits the first-order expansion
    \begin{equation}
    \label{eq:linearization}
        \sqrt{n}\, \alpha^\top \big( \widehat{\beta}_{K_n,n,\lambda} - \beta_{K_n} \big)
        =
        \widetilde{\alpha}^\top \mathbb{G}_n[ \widetilde{g}_i \epsilon_i ] + R_{1,n}(\alpha) + R_{2,n}(\alpha) + R_{\lambda,n}(\alpha) ,
    \end{equation}
    where
    \begin{equation}
    \label{eq:linearization_residuals}
    \begin{aligned}
        R_{1,n}(\alpha) & := \widetilde{\alpha}^\top \big( \widehat{D}_{K_n,n,\lambda} - I_{K_n} \big) \mathbb{G}_n[ \widetilde{g}_i(\epsilon_i + \eta_i) ] , \\
        R_{2,n}(\alpha) & := \widetilde{\alpha}^\top  \mathbb{G}_n[ \widetilde{g}_i \eta_i ] , \\
        R_{\lambda,n}(\alpha) & := - \sqrt{n}\, \lambda_{K_n,n} \widetilde{\alpha}^\top \widehat{D}_{K_n,n,\lambda} {\Gamma}_{K_n,\lambda}^{-1/2} \beta_{K_n} .
    \end{aligned}
    \end{equation}
    Moreover,
    it holds
    \begin{align}
    \label{eq:r1_bound}
        R_{1,n}(\alpha) & = O_P\left(
            \| \widetilde{\alpha} \| \left\| \Gamma_{K_n,\lambda}^{-1/2} \left( \widehat{\Gamma}_{K_n,n}-\Gamma_{K_n} \right) \Gamma_{K_n,\lambda}^{-1/2} \right\| \big( 1 + \widetilde{\xi}_{K_n,\lambda}\Delta_{K_n} \big) \right) , \\
        R_{2,n}(\alpha) & = O_P\left( \| \widetilde{\alpha} \| \widetilde{\xi}_{K_n,\lambda}\Delta_{K_n} \right) ,
        \label{eq:r2_bound}
    \end{align}
    and
    \begin{equation}
    \label{eq:r3_bound}
        R_{\lambda,n}(\alpha) = O_P \left( 
            \sqrt{n} \lambda_{K_n,n}\| \widetilde{\alpha} \|\|{\Gamma}_{K_n,\lambda}^{-1/2} \beta_{K_n}\|
        \right) .
    \end{equation}
\end{lemma}

\subsubsection{Directional asymptotic normality}
\label{subsec:directional_inference}

We can now provide an asymptotic normality result based on the ridge-normalized linearization derived in Lemma~\ref{lemma:linearization}. 
For $\ell\in\{0,1\}$, define
\begin{equation*}
    {\Omega}_{K_n,\lambda}^{(\ell)} := \Gamma_{K_n, \lambda}^{-1} \E\big[ (\epsilon_i + \ell\eta_i)^2 g_i g_i^\top ] \Gamma_{K_n, \lambda}^{-1},
\end{equation*}
and, for $\alpha\in\mathbb R^{K_n}$, define the corresponding directional variance by
\begin{equation}
\label{eq:def_sigma}
    \sigma_{K_n,\lambda}^{2,(\ell)}(\alpha)
    := \alpha^\top\Omega_{K_n,\lambda}^{(\ell)}\alpha
    = \widetilde{\alpha}^\top\E\big[(\epsilon_i+\ell\eta_i)^2\widetilde{g}_i\widetilde{g}_i^\top\big] \widetilde{\alpha} .
\end{equation}
Notice that the specification $\ell=0$ involves the regression errors, while $\ell=1$ also incorporates the sieve approximation error. The latter is particularly relevant for the residual-based variance estimator studied in Subsection~\ref{subsec:feasible_inference}.

\begin{theorem}[Directional asymptotic normality for linear functionals]
\label{theorem:L2_normality}
    Let the assumptions of Lemma~\ref{lemma:linearization} hold, and let $\alpha_{K_n}\in\mathbb{S}^{K_n-1}$ be a prescribed sequence of linear functional directions. 
    Fix $\ell\in\{0,1\}$ and suppose that $\sigma_{K_n,\lambda}^{2,(\ell)}(\alpha_{K_n})>0$ for all sufficiently large $n$ 
    and define
    \begin{equation*}
        \kappa_{K_n,\lambda} := \frac{ \|\widetilde{\alpha}_{K_n,\lambda}\| }{ { \sigma_{K_n,\lambda}^{(\ell)}(\alpha_{K_n}) } } .
    \end{equation*}
    Assume, along this sequence, that:
    \begin{itemize}
        \item[(i)] $ n^{-1/2} \kappa_{K_n,\lambda} \widetilde{\xi}_{K_n,\lambda} = o(1)$.
        \item[(ii)] $\kappa_{K_n,\lambda} \left\|\Gamma_{K_n,\lambda}^{-1/2}\left(\widehat{\Gamma}_{K_n,n}-\Gamma_{K_n}\right)\Gamma_{K_n,\lambda}^{-1/2}\right\|\left(1+\widetilde{\xi}_{K_n,\lambda}\Delta_{K_n}\right)=o_P(1)$. 
        \item[(iii)] $(1-\ell) \kappa_{K_n,\lambda} \widetilde{\xi}_{K_n,\lambda}\Delta_{K_n} = o(1)$.
        \item[(iv)] $\sqrt{n} \kappa_{K_n,\lambda} \lambda_{K_n,n} \|\Gamma_{K_n,\lambda}^{-1/2}\beta_{K_n}\| = o(1)$.
        \item[(v)] $\E[\epsilon_i^2\mid X_i]\geq\underline{\sigma}^2>0$ almost surely.
        \item[(vi)] $\lim_{M\to\infty}\sup_{n\geq1}\sup_{x\in\mathcal{X}}\E\big[ \left(\epsilon_i+\ell\eta_{K_n}(x)\right)^2\Ind\left\{\left|\epsilon_i+\ell\eta_{K_n}(x)\right|>M\right\}\mid X_i=x \big]=0$.
    \end{itemize}
    Then
    \begin{equation*}
        \sqrt{n}\frac{\alpha_{K_n}^\top\left(\widehat{\beta}_{K_n,n,\lambda}-\beta_{K_n}\right)}{\sigma_{K_n,\lambda}^{(\ell)}(\alpha_{K_n})}\todist\mathcal{N}(0,1).
    \end{equation*}
\end{theorem}

\subsubsection{RWNN functional directions}
\label{subsubsec:RWNN_functional_directions}

For the i.i.d.\ RWNN construction, let $\mathcal{F}_{K_n}:=\sigma(\theta_1,\ldots,\theta_{K_n})$ denote the $\sigma$-field generated by the sampled hidden-layer parameters. The prescribed functional directions $\alpha_{K_n}\in\mathbb{S}^{K_n-1}$ considered below are $\mathcal{F}_{K_n}$-measurable.

Define the \textit{directional Rayleigh quotient} of the ridge-normalized population covariance matrix by
\begin{equation}
\label{eq:RWNN_directional_Rayleigh_quotient}
    \mathcal{R}_{K_n,\lambda}(\alpha_{K_n})
    := \frac{\alpha_{K_n}^{\top}\Gamma_{K_n,\lambda}^{-1}\Gamma_{K_n}\Gamma_{K_n,\lambda}^{-1}\alpha_{K_n}}{\alpha_{K_n}^{\top}\Gamma_{K_n,\lambda}^{-1}\alpha_{K_n}}
    = \frac{\widetilde{\alpha}_{K_n}^{\top} \Gamma_{K_n,\lambda}^{-1/2}\Gamma_{K_n}\Gamma_{K_n,\lambda}^{-1/2} \widetilde{\alpha}_{K_n}}{\|\widetilde{\alpha}_{K_n}\|^2}.
\end{equation}
We observe that in \eqref{eq:RWNN_directional_Rayleigh_quotient} we can introduce $\mathfrak{a}_{K_n} := \widetilde{\alpha}_{K_n} / \|\widetilde{\alpha}_{K_n}\|$ with $\mathfrak{a}_{K_n} \in \mathbb{S}^{K_n-1}$, so that $\mathcal{R}_{K_n,\lambda}(\alpha_{K_n}) = \mathfrak{a}_{K_n}^\top \Gamma_{K_n,\lambda}^{-1/2}\Gamma_{K_n}\Gamma_{K_n,\lambda}^{-1/2} \mathfrak{a}_{K_n}$. 
Thus, letting $\overline{\mathfrak{a}}_{j,K_n}$ be the coefficients of $\mathfrak{a}_{K_n}$ in an eigenbasis of $\Gamma_{K_n}$, one has
\begin{equation}
\label{eq:RWNN_Rayleigh_quotient_to_basis}
    \mathcal{R}_{K_n,\lambda}(\alpha_{K_n}) 
    =
    \sum_{j=1}^{K_n} \frac{s_{j}^2}{s_{j}^2+\lambda_{K_n,n}} \, \overline{\mathfrak{a}}_{j,K_n}^2 .
\end{equation}

\begin{corollary}[Asymptotic normality for RWNN linear functionals]
\label{corollary:RWNN_L2_normality}
    Let $\{Z_i\}_{i\in\mathbb N^+}$ be i.i.d.\ observations satisfying Assumption~\ref{assumption:regression_errors_L2} and let the conditions of Corollary~\ref{cor:constrained_approx} hold for some $\beta>0$, so that
    \begin{equation}
    \label{eq:implications_Corollary_approx}
    \Delta_{K_n}=O_{P_\theta}\left(K_n^{-1/2-\beta/(d+1)}\right).%
    \end{equation}
    Let $K_n\to\infty$ and $\lambda_{K_n,n}>0$ be deterministic sequences, and suppose that the defining condition of Assumption~\ref{assumption:L2_ridge_gram_stability} holds in $P_\theta$-probability. Assume moreover that $\E[\epsilon_i^2\mid X_i]\geq\underline{\sigma}^2>0$ almost surely and $\sup_{x\in\mathcal{X}}\E\left[\epsilon_i^2\Ind\{|\epsilon_i|>M\}\mid X_i=x\right]\rightarrow 0$ as $M\to\infty$. Suppose that $\mathcal{R}_{K_n,\lambda}(\alpha_{K_n})>0$ with $P_\theta$-probability tending to one and that
    \begin{equation}
    \label{eq:RWNN_Rayleigh_cond}
        \mathcal{R}_{K_n,\lambda}(\alpha_{K_n})^{-1/2}=O_{P_\theta}(1),\qquad \widetilde{\xi}_{K_n,\lambda}=o_{P_\theta}\left(K_n^{1/2+\beta/(d+1)}\right),
    \end{equation}
    and
    \begin{equation}
    \label{eq:RWNN_ridge_bias}
        \sqrt{n}\lambda_{K_n,n}\left\|\Gamma_{K_n,\lambda}^{-1/2}\beta_{K_n}\right\|=o_{P_\theta}(1). 
    \end{equation}
    Suppose, in addition, that
    \begin{equation}
    \label{eq:RWNN_K_choice}
        K_n^{1+2\beta/(d+1)}=O(n).
    \end{equation}
    
    Then, conditionally on $\mathcal{F}_{K_n}$, in $P_\theta$-probability, for both $\ell\in \{0,1\}$
    \begin{equation}
    \label{eq:RWNN_CLT_directional}
        \sqrt{n}\frac{\alpha_{K_n}^{\top}\left(\widehat{\beta}_{K_n,n,\lambda}-\beta_{K_n}\right)}{\sigma_{K_n,\lambda}^{(\ell)}(\alpha_{K_n})}
        \todist
        \mathcal{N}(0,1).
    \end{equation}
\end{corollary}

\begin{remark}[Admissible choices of sieve dimension]
\label{remark:RWNN_inference_K_choices}
Condition~\eqref{eq:RWNN_K_choice} allows different choices of the sieve dimension. In particular, the choice
\begin{equation*}
    K_n\asymp
n^{\frac{d+1}{2(\beta+d+1)}}
\end{equation*}
satisfies \eqref{eq:RWNN_K_choice} for every $\beta>0$.
Moreover, under the polynomial-ridge regime of Theorem~\ref{theorem:L2_RWNN_sharp_consistency}, with $\beta\in(0,1)$ and $\nu_r=\frac{(d+1)(1+r)}{d+3}<1$, the effective-dimension choice
\begin{equation*}
K_n\asymp n^{\frac{1}{1+\nu_r+2\beta/(d+1)}}
\end{equation*}
also satisfies \eqref{eq:RWNN_K_choice}, since
\begin{equation*}
\frac{ K_n^{1+2\beta/(d+1)}}{n} \asymp n^{-\frac{\nu_r}
{1+\nu_r+2\beta/(d+1)}} \to0.
\end{equation*}
Thus, the sieve dimension associated with the sharper effective-dimension $L^2(X)$ rate is admissible for directional inference as well.
\end{remark}

In view of \eqref{eq:RWNN_Rayleigh_quotient_to_basis}, the first condition in \eqref{eq:RWNN_Rayleigh_cond} requires that the directions $\mathfrak{a}_{K_n}$ are not asymptotically concentrated towards the eigenvectors of $\Gamma_{K_n}$ associated with decreasing eigenvalues. 
To see this, assume, following Proposition~\ref{prop:spectral_tail_parametrized_sieve} and Theorem~\ref{theorem:L2_RWNN_sharp_consistency} for the sake of example, that $\overline{s}_{K_n}^2 \lesssim K_n^{-2/(d+1)}$ and $\lambda_{K_n} \asymp K_n^{-r}$ for $r < 2/(d+1)$,  and that $\overline{\mathfrak{a}}_{j,K_n} = 0$ for all $j <  K_n$ and $\overline{\mathfrak{a}}_{K_n,K_n} = 1$ for all $K_n$ sufficiently large and $P_{\theta}$ almost surely. Then, \eqref{eq:RWNN_Rayleigh_quotient_to_basis} reduces to
\begin{equation*}
    \mathcal{R}_{K_n,\lambda}(\alpha_{K_n}) 
    = \frac{\overline{s}_{K_n}^2}{\overline{s}_{K_n}^2 +\lambda_{K_n,n}}
    \leq \frac{\overline{s}_{K_n}^2}{\lambda_{K_n,n}}
    \lesssim K_n^{r-2/(d+1)} ,
\end{equation*}
meaning $P_{\theta}$ almost surely $\mathcal{R}_{K_n,\lambda}(\alpha_{K_n})^{-1/2} \gtrsim K_n^{1/(d+1)-r/2} \to \infty$ as $K_n$ grows, violating \eqref{eq:RWNN_Rayleigh_cond}. Conversely, if $\mathfrak{a}_{K_n}$ is asymptotically concentrated on eigenspaces whose eigenvalues dominate $\lambda_{K_n,n}$ uniformly, then $\mathcal{R}_{K_n,\lambda}(\alpha_{K_n})\to 1$. Thus, the asymptotic normality of the functional $\sqrt{n} {\alpha_{K_n}^{\top}\big(\widehat{\beta}_{K_n,n,\lambda}-\beta_{K_n}\big)}/{\sigma_{K_n,\lambda}^{(\ell)}(\alpha_{K_n})}$ does not hinge only on the decay of the least eigenvalue, $\overline{s}_{K_n}^2$, but rather on the orientation of $\alpha_{K_n}$ with respect to the \textit{subset of well-conditioned eigenvectors} of $\Gamma_{K_n}$.

The ridge-bias condition in \eqref{eq:RWNN_ridge_bias} is sufficient but not necessary for a prescribed functional direction. 
Indeed, although Theorem~\ref{theorem:L2_normality} imposes the stronger norm-based condition~(iv), its proof only requires the following condition on the directional ridge remainder:
\begin{equation*}
    \frac{\sqrt{n}\lambda_{K_n,n}\left|\widetilde{\alpha}_{K_n,\lambda}^{\top}\widehat{D}_{K_n,n,\lambda}\Gamma_{K_n,\lambda}^{-1/2}\beta_{K_n}\right|}{\sigma_{K_n,\lambda}^{(0)}(\alpha_{K_n})}=o_P(1).
\end{equation*}
Condition~\eqref{eq:RWNN_ridge_bias} provides a convenient direction-independent sufficient condition for this requirement, while sharper conditions may exploit the alignment between the functional direction and the ridge-bias direction.

\subsection{Feasible inference}
\label{subsec:feasible_inference}
We now proceed by moving from asymptotic normality towards feasible inference. More specifically, we  derive consistency of a residual-based ridge-adjusted estimator of the directional asymptotic variance. First, we analyze the effects of approximating the population quantities by their empirical counterparts. Next, we establish relative consistency under general conditions, and, finally, specialize these conditions to RWNNs, showing that the same estimator consistently estimates both directional variance specifications considered in Subsection~\ref{subsec:linearization_sharper_l2}. We conclude with an example demonstrating that feasible inference is possible even when the population feature covariance matrix is singular.
\begin{lemma}
\label{lemma:asymp_var_error_terms}
    Let $\{Z_i\}_{i\in \mathbb{N}^+}$ be i.i.d. and satisfy Assumption~\ref{assumption:regression_errors_L2}, and let $\alpha=\alpha_{K_n}\in\mathbb{S}^{K_n-1}$. Set $\widehat{\epsilon}_i := Y_i - \widehat{h}_{n,\lambda}(X_i)$,  ${r}_i := {\alpha}^\top {\Gamma}_{K_n,\lambda}^{-1} g_i$, and $\widehat{r}_i := {\alpha}^\top \widehat{\Gamma}_{K_n,n,\lambda}^{-1} g_i$. Assume that there exist deterministic positive sequences $\overline{a}_{K_n,n,\lambda}=o(1)$ and $\overline{\mathfrak{M}}_{K_n,n,\lambda}$ such that 
    \begin{equation}
    \label{eq:asymp_var}
        \norm{A_{K_n,n,\lambda}} = O_P\left(\overline{a}_{K_n,n,\lambda}\right) \quad \text{and} \quad \Norm{ \widehat{\Gamma}_{K_n,n}^{1/2} \left(\widehat{\beta}_{K_n,n,\lambda}-\beta_{K_n}\right)}= O_P\left(\overline{\mathfrak{M}}_{K_n,n,\lambda}\right)
    \end{equation}
    Then
    \begin{itemize}
        \item[(i)]  $ \big\lvert n^{-1} \sum_{i=1}^n \big[ \widehat{\epsilon}_i^2 - (\epsilon_i + \eta_i)^2 ] \widehat{r}_i^2 \big\rvert = O_P\left( \norm{ \widetilde{\alpha} }^2 \widetilde{\xi}^2_{K_n,\lambda} \overline{\mathfrak{M}}_{K_n,n,\lambda} \left( \overline{\mathfrak{M}}_{K_n,n,\lambda} + \sqrt{ { \overline{\sigma}^2 + \Delta_{K_n}^2 } } \right) \right) $. 
        \item[(ii)] $ \big\lvert n^{-1} \sum_{i=1}^n (\epsilon_i + \eta_i)^2 \big[ \widehat{r}_i^2 - r_i^2 \big] \big\rvert = O_P\left( \norm{ \widetilde{\alpha} }^2 \widetilde{\xi}_{K_n,\lambda}^2 \overline{a}_{K_n,n,\lambda} \big( \overline{\sigma}^2 + \Delta_{K_n}^2 \big) \right) $.
    \end{itemize}
\end{lemma}

\begin{proposition}[Consistency of the ridge-adjusted variance estimator]
\label{proposition:consistency_ridge_var_estimator}
    Let the assumptions of Lemma~\ref{lemma:asymp_var_error_terms} hold. Fix a possibly $n$-dependent  direction $\alpha=\alpha_{K_n}\in\mathbb S^{K_n-1}$. Let $\sigma_{K_n,\lambda}^{2,(1)}(\alpha)$ be defined as in \eqref{eq:def_sigma}, and assume that $\sigma_{K_n,\lambda}^{2,(1)}(\alpha)>0$ for all sufficiently large $n$. Define the directional Rayleigh quotient and the corresponding factor, respectively, as
    \begin{equation}
    \label{eq:R_L_def}
        \mathcal{R}_{K_n,\lambda}^{(1)}(\alpha):=\frac{\sigma_{K_n,\lambda}^{2,(1)}(\alpha)}{\|\widetilde{\alpha}\|^2} \quad \text{and} \quad \mathcal{L}_{K_n,\lambda}^{(1)}(\alpha):=\frac{\widetilde{\xi}_{K_n,\lambda}^2}{\mathcal{R}_{K_n,\lambda}^{(1)}(\alpha)}.
    \end{equation}
    
    Assume that:
    \begin{itemize}
        \item[(i)] $\mathcal{L}_{K_n,\lambda}^{(1)}(\alpha)\overline{\mathfrak{M}}_{K_n,n,\lambda}\left(\overline{\mathfrak{M}}_{K_n,n,\lambda}+\sqrt{\overline{\sigma}^2+\Delta_{K_n}^2}\right)=o(1)$.
        \item[(ii)] $\mathcal{L}_{K_n,\lambda}^{(1)}(\alpha)\overline{a}_{K_n,n,\lambda}\left(\overline{\sigma}^2+\Delta_{K_n}^2\right)=o(1)$.
        \item[(iii)] $n^{-1/2} \mathcal{L}_{K_n,\lambda}^{(1)}(\alpha) = o(1) $
        \item[(iv)] $\sup_{n \geq 1} \sup_{x \in \mathcal{X}} \E\big[ ( \epsilon_i + \eta_{K_n}(x) )^4 \,\vert\, X_i = x, \mathcal{F}_{K_n} \big] \leq \overline{\sigma}_{\epsilon,\eta}^4 < \infty$.
    \end{itemize}
    Define
    \begin{equation*}
        \widehat{\sigma}_{K_n,n,\lambda}^{2}(\alpha)
        :=
        \alpha^\top \widehat{\Gamma}_{K_n,n,\lambda}^{-1} \left( \frac{1}{n} \sum_{i=1}^n \widehat{\epsilon}^{\,2}_i g_i g_i^\top \right) \widehat{\Gamma}_{K_n,n,\lambda}^{-1} \alpha
    \end{equation*}

    Then,
    \begin{equation*}
        \frac{ \widehat{\sigma}_{K_n,n,\lambda}^{2}(\alpha) }{ \sigma_{K_n,\lambda}^{2,(1)}(\alpha) } 
        \toprob 1.
    \end{equation*}
\end{proposition}

\begin{corollary}[RWNN consistency of the ridge-adjusted variance estimator]
\label{corollary:RWNN_variance_estimator_consistency}
    Let the assumptions of the $\beta$-H\"older RWNN approximation result in Corollary~\ref{cor:constrained_approx} hold, and let $\{Z_i\}_{i\in\mathbb{N}^+}$ be i.i.d. observations satisfying Assumption~\ref{assumption:regression_errors_L2}. Suppose that Assumption~\ref{assumption:L2_ridge_gram_stability} holds and that, for some $\beta\in(0,1)$,
    \begin{equation}
    \label{eq:lambda_r_choice}
        \lambda_{K_n,n}\asymp K_n^{-r}, \qquad \frac{2\beta}{d+1}<r<\frac{2}{d+1}.
    \end{equation}
    Set
    \begin{equation}
    \label{eq:nu_r}
        \nu_r:=\frac{(d+1)(1+r)}{d+3}.
    \end{equation}
    Define
    \begin{equation*}
        \overline{\mathfrak{M}}_{K_n,n,\lambda} 
        :=
        \sqrt{\frac{K_n^{\nu_r}}{n}} + K_n^{-1/2-\beta/(d+1)}\quad \text{and}\quad  \overline{a}_{K_n,n,\lambda}
        :=
        \sqrt{\frac{K_n^{1+r}\log(K_n)}{n}}.
    \end{equation*}
    Fix a possibly $n$-dependent $\mathcal{F}_{K_n}$-measurable direction $\alpha=\alpha_{K_n}\in\mathbb S^{K_n-1}$. Let $\sigma_{K_n,\lambda}^{2,(1)}(\alpha)$ be defined as in \eqref{eq:def_sigma}, and assume that $\sigma_{K_n,\lambda}^{2,(1)}(\alpha)>0$ for all sufficiently large $n$. Assume that, for some $0\leq\rho<\frac{1}{2}+\frac{\beta}{d+1}$,
    \begin{equation}
    \label{eq:primitive_L_growth_l1}
        \mathcal{L}_{K_n,\lambda}^{(1)}(\alpha)=O_{P_\theta}(K_n^\rho),
    \end{equation}
    and that 
    \begin{equation}
    \label{eq:primitive_rate_condition_l1}
        \frac{K_n^{2\rho+1+r}\log(K_n)}{n}=o(1).
    \end{equation}
    Finally, assume that
    \begin{equation}
    \label{eq:primitive_fourth_moment_l1}
        \sup_{n\geq1}\sup_{x\in\mathcal{X}}\E\big[\big(\epsilon_i+\eta_{K_n}(x)\big)^4 \mid X_i=x,  \mathcal{F}_{K_n} \big]\leq \overline{\sigma}_{\epsilon,\eta}^4<\infty.
    \end{equation}
    Then
    \begin{equation*}
        \frac{\widehat{\sigma}_{K_n,n,\lambda}^{2}(\alpha)}{\sigma_{K_n,\lambda}^{2,(1)}(\alpha)}\toprob 1.
    \end{equation*}
\end{corollary}

We next show that the same estimator is also consistent for $\sigma_{K_n,\lambda}^{2,(0)}(\alpha)$, since the contribution of the sieve approximation error to $\sigma_{K_n,\lambda}^{2,(1)}(\alpha)$ is asymptotically negligible.

\begin{corollary}[RWNN consistency for the noise variance]
\label{corollary:RWNN_variance_estimator_consistency_l0}
    Let the assumptions of Corollary~\ref{corollary:RWNN_variance_estimator_consistency} hold. Assume additionally that $\sigma_{K_n,\lambda}^{2,(0)}(\alpha)>0$ for all sufficiently large $n$. Then
    \begin{equation*}
        \frac{\widehat{\sigma}_{K_n,n,\lambda}^{2}(\alpha)}{\sigma_{K_n,\lambda}^{2,(0)}(\alpha)}\toprob 1.
    \end{equation*}
\end{corollary}

The following example demonstrates that singularity of the population covariance matrix is compatible with asymptotically valid inference and consistent variance estimation. 

\begin{example}[Feasible inference with singular population covariance]
\label{example:RWNN_singular_covariance}
Consider $d=1$, $\mathcal X=[-1,1]$, and the ReLU activation $\sigma(z)=z_+$. Let $X_i\sim\mathcal U([-1,1])$ and $Y_i=X_i+\epsilon_i$, where $\epsilon_i\sim\mathcal N(0,\tau^2)$, independently of $X_i$. Independently of the data, draw $A_k\sim\mathcal U([-1,1])$ and $B_k\sim\mathcal U([2,3])$, independently across $k$. Since $Ax+B\geq1$ on the specified domains, the regression function admits the integral representation
\begin{equation*}
    h(x) =\E_{A,B}\big[ 3A(Ax+B)_+ \big] = x, \qquad x\in[-1,1].
\end{equation*}
It is easy to see that the product source condition holds with $f(A,B)=3A$ and $\beta=1/2$, and the regular-sampling assumptions are satisfied.

Let $u_K=(A_1,\ldots,A_K)^\top$ and $v_K=(B_1,\ldots,B_K)^\top$. The corresponding feature map and population covariance matrix satisfy $g_K(x)=xu_K+v_K$ and $ \Gamma_K = \frac13 u_Ku_K^\top+v_Kv_K^\top$, respectively. Consequently, $\operatorname{rank}(\Gamma_K)=2$ almost surely for every $K\geq2$, and hence $\lambda_{\min}(\Gamma_K)=0$ for every $K\geq3$.
Nevertheless, choosing $K_n=\lceil n^{2/5}\rceil$ and $\lambda_{K_n,n}=K_n^{-3/4}$, the conditions of Corollaries~\ref{corollary:RWNN_L2_normality} and~\ref{corollary:RWNN_variance_estimator_consistency} are satisfied along the normalized point evaluation directions and the direction associated with $\Psi_2(h)=\int_{-1}^1h(x)^2\,\de x$. In particular, conditionally on $\mathcal F_{K_n}$, in $P_\theta$-probability, 
\begin{equation*}
    \sqrt n\bigl(\widehat h_{n,\lambda}(x_0)-x_0\bigr)\todist\mathcal N\bigl(0,\tau^2(1+3x_0^2)\bigr),
\end{equation*}
for any fixed $x_0\in[-1,1]$ and
\begin{equation*}
    \sqrt n\left(\Psi_2(\widehat h_{n,\lambda})-\frac23\right)\todist\mathcal N\left(0,\frac{16 \tau^2}{3}\right).
\end{equation*}
Moreover, the corresponding ridge-adjusted directional variance estimators are consistent. We provide detailed verification of this example in Appendix~\ref{appendix:RWNN_singular_example}.
\end{example}

\subsection{Inference for smooth functionals}
\label{subsec:inference_smooth_functionals}
In this subsection, we extend the preceding directional inference results to smooth nonlinear functionals of $h$. The argument uses differentiability and Riesz representation to identify a coefficient direction corresponding to the functional derivative, allowing us to apply the linear-functional theory of Subsection~\ref{subsec:linearization_sharper_l2}. The $L^2(X)$ estimation bounds control the nonlinear remainder, while a separate condition controls the first-order sieve approximation bias. Together, these ingredients yield asymptotic normality centered at the target functional $\Psi(h)$.

\subsubsection{Smooth functionals and Riesz representation}

Let $\Psi:\mathcal{H}\to\mathbb{R}$ be a known functional and assume that the object of interest for inference is $\Psi(h)$. 

\begin{assumption}[Smooth functional and Riesz representer]
\label{assumption:functional_smooth_riesz_ridge}
    Let $\mathcal{V}$ be a closed linear subspace of $L^2(X)$ containing $\mathcal{H}-h$ and the RWNN sieve spaces considered below. 
    There exists a continuous linear functional $\Psi'_h:\mathcal{V}\to\mathbb{R}$ such that, as $\|\widetilde{h}-h\|_{L^2(X)}\to0$,
    \begin{equation}
    \label{eq:func_expansion_L2}
        \Psi(\widetilde{h})-\Psi(h)=\Psi'_h(\widetilde{h}-h)+O(\|\widetilde{h}-h\|_{L^2(X)}^\omega)
    \end{equation}
    for some $\omega>1$. Let $\psi\in\mathcal{V}$ denote the Riesz representer of $\Psi'_h$, so that
    \begin{equation}
    \label{eq:Riesz_representer_L2}
        \Psi'_h(v)=\langle v,\psi\rangle_{L^2(X)},\qquad v\in\mathcal{V}.
    \end{equation}
\end{assumption}

For the RWNN sieve, we use the component-wise notation $
    \Psi'_h(g_{K_n}) := ( \Psi'_h(g_{1,K_n}),\ldots, \allowbreak \Psi'_h(g_{K_n,K_n}) )^\top \in \mathbb{R}^{K_n}$. 
For any $u\in \mathbb{R}^{K_n}$ define $u_g(x):= g_{K_n}(x)^\top u$, $x\in \mathcal{X}$. By the definition of $\Gamma_{K_n}$ in
\eqref{eq:def_Gamma_Gammalambda}, for any $u,v\in\mathbb{R}^{K_n}$,
$\langle u_g,v_g\rangle_{L^2(X)}=u^\top\Gamma_{K_n}v$.
By linearity of $\Psi'_h$, 
\begin{equation}
\label{eq:functional_direction_identity}
    \Psi'_h(u_g)=u^\top \Psi'_h(g_{K_n}).
\end{equation}
Hence, on the RWNN sieve, the derivative of the smooth functional is a linear functional in the coefficient direction $\Psi'_h(g_{K_n})$. When this direction is nonzero, the directional results of Subsection~\ref{subsec:directional_inference} can be applied after normalization, provided their assumptions are satisfied.

We now notice that, by \eqref{eq:Riesz_representer_L2}, $\Psi'_h(g_{K_n})=\E_X[{g_{K_n}(X)\psi(X)}]$.
For the ridge-regularized estimator, define the ridge-adjusted sieve Riesz representer associated with $\Psi'_h$ by 
\begin{equation}
\label{eq:ridge_adjusted_sieve_riesz}
    \psi_{K_n,\lambda}(x):=g_{K_n}(x)^\top \Gamma _{K_n,\lambda}^{-1} \Psi'_h(g_{K_n}), \qquad x\in \mathcal{X}.
\end{equation}
Equivalently, let $b_{\psi,K_n,\lambda}:=\Gamma_{K_n,\lambda}^{-1}\Psi'_h(g_{K_n})$. Then, for every $u\in \mathbb{R}^{K_n}$, one has $u^\top\Gamma_{K_n,\lambda}b_{\psi,K_n,\lambda} = u^\top\Psi'_h(g_{K_n}) = \Psi'_h(u_g)$.

Hence $b_{\psi,K_n,\lambda}$ is the Riesz representer of the restriction of $\Psi'_h$ to the RWNN coefficient space equipped with the $\Gamma_{K_n,\lambda}$ inner product, while $\psi_{K_n,\lambda}$ is the corresponding sieve function.

The regularity of $\Psi'_h$ with respect to the $L^2(X)$ norm controls the associated ridge-normalized coefficient direction. Indeed, for every $u\in \R^{K_n}$, 
\begin{equation*}
    \left|u^\top\Psi'_h(g_{K_n})\right|
    = \left|\langle u_g,\psi\rangle_{L^2(X)}\right| 
    \leq ( u^\top\Gamma_{K_n}u )^{1/2}\|\psi\|_{L^2(X)}
    \leq ( u^\top\Gamma_{K_n,\lambda}u )^{1/2}\|\psi\|_{L^2(X)},
\end{equation*}
and hence $\left\|\Gamma_{K_n,\lambda}^{-1/2}\Psi'_h(g_{K_n})\right\|\le\|\psi\|_{L^2(X)}$. This construction is closely related to methods based on regularized sieves and semiparametric inference via Riesz representation (see, for example,
\citealp{shenMethodsSievesPenalization1997,
chenEfficientEstimationSemiparametric2009,
chenEstimationNonparametricConditional2012,
pouzoNonAsymptoticPropertiesRegularized2016}). 
We apply the empirical counterpart of $\psi_{K_n,\lambda}$ and the associated variance estimator in the simulations of Subsection~\ref{subsec:sims_statistical_inference}.

\subsubsection{Functional linearization and remainder control}
\label{subsubsec:functional_linearization_remainder_control}
We now combine the smooth-functional expansion in Assumption~\ref{assumption:functional_smooth_riesz_ridge} with the $L^2(X)$ convergence results of Subsection~\ref{subsec:L2_estimation}. 
Applying \eqref{eq:func_expansion_L2} to $\widehat h_{n,\lambda}$ yields
\begin{equation}
\label{eq:smooth_functional_expansion}
    \Psi(\widehat{h}_{n,\lambda})-\Psi(h)=\Psi_h'(\widehat{h}_{n,\lambda}-h)+O_P\left(\|\widehat{h}_{n,\lambda}-h\|_{L^2(X)}^\omega\right).
\end{equation}
Hence the functional remainder is negligible at $\sqrt{n}$ scale whenever $\sqrt{n}\, \|\widehat{h}_{n,\lambda}-h\|_{L^2(X)}^\omega=o_P(1)$.

\begin{corollary}[Negligibility of the smooth-functional remainder for RWNNs]
\label{cor:smooth_functional_remainder_ridge}
    Let the conditions of Theorem~\ref{theorem:L2_RWNN_sharp_consistency} and Assumption~\ref{assumption:functional_smooth_riesz_ridge} hold with $\omega\geq 2$.
    Suppose moreover that $\beta\in(0,1)$ and
    \eqref{eq:RWNN_polynomial_ridge} holds, and choose
    \begin{equation}
    \label{eq:smooth_functional_K_choice}
        K_n\asymp n^{\frac{d+1}{2(\beta+d+1)}}.
    \end{equation}
    Then
    \begin{equation}
    \label{eq:smooth_functional_n14_rate}
        \|\widehat h_{n,\lambda}-h\|_{L^2(X)}=o_P(n^{-1/4}),
    \end{equation}
    and consequently
    \begin{equation*}
        \sqrt{n}\big( \Psi(\widehat{h}_{n,\lambda})-\Psi(h)-\Psi_h'(\widehat{h}_{n,\lambda}-h) \big) = o_P(1).
    \end{equation*}
\end{corollary}

We now consider the first-order linear term in \eqref{eq:smooth_functional_expansion}. Define $h_{K_n}(x):=g_{K_n}(x)^\top \beta_{K_n}$ and $a_{K_n}:=\Psi'_h(g_{K_n})$. By linearity of $\Psi'_h$ and \eqref{eq:functional_direction_identity}, 
\begin{equation}
\label{eq:functional_first_order_decomposition}
\begin{aligned}
    \Psi'_h(\widehat{h}_{n,\lambda}-h) &= \Psi'_h(\widehat{h}_{n,\lambda}-h_{K_n})+\Psi'_h(h_{K_n}-h) = a_{K_n}^\top\bigl(\widehat{\beta}_{K_n,n,\lambda}-\beta_{K_n}\bigr)
    +\Psi'_h(h_{K_n}-h).
 \end{aligned}
\end{equation}
Define 
\begin{equation}
\label{eq:functional_sieve_bias_definition}
    B_{\Psi,K_n}:=\Psi'_h(h_{K_n}-h)=\langle h_{K_n}-h,\psi\rangle_{L^2(X)}.
\end{equation} 
Hence, the first-order linear term in \eqref{eq:smooth_functional_expansion} is a linear sieve functional of a type studied in Subsection~\ref{subsec:directional_inference}, with coefficient direction $a_{K_n}$, while $B_{\Psi,K_n}$ is the sieve
approximation bias in the functional direction.

Combining \eqref{eq:smooth_functional_expansion}, \eqref{eq:functional_first_order_decomposition}, and Corollary~\ref{cor:smooth_functional_remainder_ridge}, we obtain
\begin{equation}
\label{eq:smooth_functional_rootn_decomposition}
    \sqrt{n} \big( \Psi(\widehat{h}_{n,\lambda})-\Psi(h) \big) = \sqrt{n}\, a_{K_n}^\top\bigl(\widehat{\beta}_{K_n,n,\lambda}-\beta_{K_n}\bigr) + \sqrt{n}\, B_{\Psi,K_n} + o_P(1).
\end{equation}
\subsubsection{Asymptotic normality for RWNN smooth functionals}
\label{subsubsec:RWNN_smooth_functional_normality}
Suppose that $a_{K_n}\neq 0$ and define the normalized functional direction
\begin{equation}
\label{eq:smooth_functional_direction}
    \mathfrak{a}_{K_n}^{\Psi} := \frac{a_{K_n}}{\|a_{K_n}\|}=\frac{\Psi'_h(g_{K_n})}{\|\Psi'_h(g_{K_n})\|}\in\mathbb{S}^{K_n-1}.
\end{equation}
Since $g_{K_n}$ is determined by the sampled random parameters,
$\mathfrak{a}_{K_n}^{\Psi}$ is $\mathcal{F}_{K_n}$-measurable and hence is an admissible prescribed direction in Corollary~\ref{corollary:RWNN_L2_normality}.
For $\ell\in\{0,1\}$, define the corresponding functional standard deviation by
\begin{equation}
\label{eq:smooth_functional_sigma}
    \sigma_{\Psi,K_n,\lambda}^{(\ell)}:=\|a_{K_n}\|\sigma_{K_n,\lambda}^{(\ell)}(\mathfrak{a}_{K_n}^{\Psi}).
\end{equation}

\begin{corollary}[Asymptotic normality for RWNN smooth functionals]
\label{corollary:RWNN_smooth_functional_normality}
    Let the conditions of Corollary~\ref{cor:smooth_functional_remainder_ridge} hold and suppose that $a_{K_n}\neq 0$ with $P_\theta$-probability tending to one. Assume that the conditions of Corollary~\ref{corollary:RWNN_L2_normality} hold along the direction $\mathfrak{a}_{K_n}^{\Psi}$ defined in \eqref{eq:smooth_functional_direction}. Suppose moreover that
    \begin{equation}
    \label{eq:smooth_functional_variance_nondegeneracy}
        \big( \sigma_{\Psi,K_n,\lambda}^{(0)} \big)^{-1} = O_{P_\theta}(1) ,
    \end{equation}
    and that the sieve approximation bias in the functional direction satisfies
    \begin{equation}
    \label{eq:smooth_functional_sieve_bias}
        { \sqrt{n} \left|\Psi'_h(h_{K_n}-h)\right| }=o_{P_\theta}(1).
    \end{equation}
    
    Then, conditionally on $\mathcal{F}_{K_n}$, in $P_\theta$-probability, for both $\ell\in\{0,1\}$,
    \begin{equation}
    \label{eq:RWNN_smooth_functional_CLT}
        \sqrt{n}\frac{\Psi(\widehat{h}_{n,\lambda})-\Psi(h)}{\sigma_{\Psi,K_n,\lambda}^{(\ell)}}\todist\mathcal{N}(0,1).
    \end{equation}
\end{corollary}

\begin{remark}[Sieve bias for the integrated-square functional]
Consider $\Psi_2(h)=\int_{-1}^1 h(x)^2\,\de x$, with $X\sim\mathcal U([-1,1])$. Since $h_{K_n}$ is the $L^2(X)$-orthogonal projection of $h$ onto the RWNN sieve, the first-order sieve bias satisfies
\begin{equation*}
B_{\Psi_2,K_n}=\Psi'_{2,h}(h_{K_n}-h)=4\langle h,h_{K_n}-h\rangle_{L^2(X)}=-4\|h-h_{K_n}\|_{L^2(X)}^2=-4\Delta_{K_n}^2.
\end{equation*}
Under the assumptions of Corollary~\ref{cor:constrained_approx}, it follows that
\begin{equation*}
\sqrt{n}\,|B_{\Psi_2,K_n}|=O_{P_\theta}\left(\sqrt{n}\,K_n^{-1-2\beta/(d+1)}\right).
\end{equation*}
Consequently, for $K_n$ chosen as in \eqref{eq:smooth_functional_K_choice}, the right-hand side is $O_{P_\theta}\left(n^{-\beta/[2(\beta+d+1)]}\right)=o_{P_\theta}(1)$. Hence the sieve-bias condition \eqref{eq:smooth_functional_sieve_bias} is automatically satisfied for this functional.
\end{remark}

\subsection{Sup-norm analysis}
\label{subsec:rwnn_uniform_ridge}

The sup-norm analysis we develop in this subsection follows two approaches to controlling the stochastic error and ridge-induced bias. First, we derive the bounds using the singular values of the empirical feature matrix. Importantly, these bounds can be conservative when the feature representation is poorly conditioned, which immediately motivates our second approach based on the empirical ridge pseudo-projection. This approach is inspired by the sup-norm stability concepts from sieve estimation \citep{Chen2015} and ridge leverage quantities studied in the kernel regression literature \citep{alaouiFastRandomizedKernel2015,avronRandomFourierFeatures2017,pauwelsRelatingLeverageScores2018}. By controlling the ridge smoothing weights, the empirical ridge Lebesgue constant, and leverage quantities, we obtain uniform estimation bounds under prediction-space stability conditions that do not require a positive lower bound on the smallest empirical covariance eigenvalue.

We start by recalling that $g_{K_n}(x)\in\mathbb{R}^{K_n}$ denotes the vector of random features evaluated at $x\in\mathcal{X}$, and the associated design matrix is
\begin{equation*}
G_{K_n,n}=\bigl(g_{K_n}(X_1),\ldots,g_{K_n}(X_n)\bigr)^\top\in\mathbb{R}^{n\times K_n}.
\end{equation*}

For a deterministic ridge regularization parameter $\lambda_{K_n,n}>0$, the ridge-regularized RWNN least-squares estimator is
\begin{equation}
\label{eq:RWNN_estimator}
\widehat{h}_{n,\lambda}(x)=g_{K_n}(x)^\top\widehat\Gamma_{K_n,n,\lambda}^{-1}\bigl(n^{-1}G_{K_n,n}^\top Y\bigr),
\end{equation}
where
\begin{equation*}
\widehat\Gamma_{K_n,n,\lambda}:=n^{-1}G_{K_n,n}^\top G_{K_n,n}+\lambda_{K_n,n}I_{K_n},\quad \text{and} \quad  Y:=(Y_1,\ldots,Y_n)^\top\in\mathbb{R}^n.
\end{equation*}
To separate the stochastic and conditional bias components of the uniform error, define the ridge pseudo-projection
\begin{equation}
\label{eq:pseudoproj}
\widetilde{h}_{n,\lambda}(x):=g_{K_n}(x)^\top\widehat\Gamma_{K_n,n,\lambda}^{-1}\bigl(n^{-1}G_{K_n,n}^\top H\bigr),
\end{equation}
where
\begin{equation*}
H:=(h(X_1),\ldots,h(X_n))^\top\in\mathbb{R}^n.
\end{equation*}
The uniform estimation error satisfies the following bias-variance decomposition bound
\begin{equation}
\label{eq:sup_error_split}
\|\widehat{h}_{n,\lambda}-h\|_\infty\leq\|\widehat{h}_{n,\lambda}-\widetilde{h}_{n,\lambda}\|_\infty+\|\widetilde{h}_{n,\lambda}-h\|_\infty.
\end{equation}
The first term is the stochastic component of the estimation error (variance), whereas the second term (bias) contains the sieve approximation error and the additional shrinkage bias induced by ridge regularization.

Throughout this subsection, we work under Assumption~\ref{assumption:c_stationarity} on the data, the squared-loss regression model, and Assumption~\ref{assumption:regression_errors_L2} on its error term. Since the uniform stochastic bound proved below is established only for independent observations, we restrict attention to case (i) of Assumption~\ref{assumption:c_stationarity} with $\{Z_i\}_{i\in {\mathbb{N}^+}}$ i.i.d. %
For the uniform stochastic analysis, we impose the following additional moment and sieve-regularity conditions. 

\begin{assumption}
\label{assumption:higher_moment}
    There exists $\delta>0$ such that $\mathbb{E}[|\epsilon_i|^{2+\delta}]<\infty$.
\end{assumption}

\begin{assumption}
\label{assumption:uniform_sieve_regularity}
    There exist deterministic positive sequences $\{\zeta_{K}\}_{K\geq 1}$ and $\{L_{K}\}_{K\geq 1}$ such that, almost surely, 
    \begin{itemize}
        \item [(i)] $\sup_{x\in\mathcal{X}}\|g_{K}(x)\|\leq \zeta_{K}$.
        \item[(ii)] $\|g_{K}(x_1) - g_{K}(x_2)\|\leq L_{K} \|x_1-x_2\|$ for all $x_1,x_2\in \mathcal{X}$.
        \item[(iii)] Along the sequence $K=K_n$, there exist constants $C>0$ and $q>0$ such that 
    \begin{equation}
    \label{eq:pol_growth}
        \frac{L_{K_n}\sqrt{m_n}}{\zeta_{K_n}\sqrt{\log(n)}} \leq C n^q
    \end{equation}
    with $m_n:=\min\{K_n, n\}$, for all sufficiently large $n$. 
    \end{itemize}
\end{assumption}

\begin{remark}
    The rate in Assumption~\ref{assumption:uniform_sieve_regularity}(iii) may alternatively be simplified to 
    \begin{equation*}
       \frac{L_{K_n}}{\zeta_{K_n}} \sqrt{m_n} \:\lesssim\: n^{q'} ,
    \end{equation*}
    for some $q' > 0$, since $\sqrt{\log(n)}\, n^q = o(n^{q'})$ for any $q' > q$.
\end{remark}

\subsubsection{Uniform stochastic error}%
\label{subsubsec:variance_uniform_bound}

We start by bounding the uniform variance term $\|\widehat{h}_{n,\lambda}-\widetilde{h}_{n,\lambda}\|_\infty$ in \eqref{eq:sup_error_split}. Following the strategy of Lemma~2.3 in \cite{Chen2015}, we combine a covering argument on $\mathcal{X}$ with concentration conditional on the regressors. Since ridge regularization does not generate an empirical projection, we use the singular value decomposition of the design matrix. 
The algebraic arguments below do not require $K_n\leq n$. We therefore set $m_n:=\min\{K_n,n\}$ and consider the thin singular value decomposition
\begin{equation}
\label{eq:uniform_design_svd}
    n^{-1/2}G_{K_n,n} = U_{m_n,n} S_{m_n,n} V_{m_n,n}^\top,
\end{equation}
where $U_{m_n,n} \in \R^{n \times m_n}$, $V_{m_n,n} \in \R^{K_n \times m_n}$ and 
\begin{equation*}
    S_{m_n,n}:=\operatorname{diag}(s_{1,n},\ldots,s_{m_n,n}) \in \R^{m_n \times m_n} ,
    \qquad s_{1,n}\geq\cdots\geq s_{m_n,n}\geq0.
\end{equation*}
It holds $V_{m_n,n}^\top V_{m_n,n}=I_{m_n}$ and $U_{m_n,n}^\top U_{m_n,n}=I_{m_n}$. 
The nonzero eigenvalues of matrix $\widehat{\Gamma}_{K_n,n}=n^{-1}G_{K_n,n}^\top G_{K_n,n}$ are then given by the nonzero elements of $\{s_{k,n}^2\}_{k=1}^{m_n}$. When $K_n>n$, this matrix has at least $K_n-n$ additional zero eigenvalues. Accordingly, we extend the sequence to $K_n$ indices by setting $s_{k,n}:=0$, $k=m_n+1,\ldots,K_n$ whenever $K_n>n$; zero singular values arising from rank deficiency are not excluded from $S_{m_n,n}$. Hence, throughout this section, $s_{1,n}^2\geq\cdots\geq s_{K_n,n}^2\geq0$ are the eigenvalues of $\widehat{\Gamma}_{K_n,n}$ counted with multiplicity. In particular, when  $K_n\leq n$, $s_{m_n,n}^2=\lambda_{\text{min}}(\widehat{\Gamma}_{K_n,n})$.

Define the empirical ridge stability factor
\begin{equation}
\label{eq:psi_def}
    \psi_{K_n,n,\lambda}:=\max_{1\leq k\leq m_n}\frac{s_{k,n}}{s_{k,n}^2+\lambda_{K_n,n}}.
\end{equation}
\begin{assumption}
\label{assumption:growth_variance}
    There exists a deterministic positive sequence $\overline{\psi}_{K_n,n}$ such that $\psi_{K_n,n,\lambda}=O_P\bigl(\overline{\psi}_{K_n,n}\bigr)$ and
    \begin{itemize}
        \item[(i)] $\zeta_{K_n}\overline{\psi}_{K_n,n}\sqrt{{\log(n)}/{n}}=o(1)$ ,
        \item[(ii)] $\bigl(\zeta_{K_n}/(\overline{\psi}_{K_n,n} \lambda_{K_n,n})\bigr)^{(2+\delta)/\delta}\lesssim\sqrt{{n}/{\log(n)}}$ ,
        \item[(iii)] $\big( \zeta_{K_n}/(\overline{\psi}_{K_n,n} \lambda_{K_n,n}) \big) \sqrt{{n}/{\log(n)}}\to \infty$ ,
    \end{itemize}
    where $\delta>0$ is the exponent in Assumption~\ref{assumption:higher_moment}.
\end{assumption}

\begin{remark}
    If the sieve is such that $\zeta_{K_n} \asymp \sqrt{K_n}$, Assumption~\ref{assumption:growth_variance}(i) then requires $\overline{\psi}_{K_n,n} \sqrt{K_n \log(n) / n} \allowbreak = o(1)$.
    If $\overline{\psi}_{K_n,n}\asymp 1$, this implies $K_n=o(n/\log(n))$. If $\overline{\psi}_{K_n,n}\to\infty$, the restriction becomes more stringent, requiring
    \begin{equation*}
        K_n=o\left(\frac{n}{\overline{\psi}_{K_n,n}^{\,2}\log(n)}\right).
    \end{equation*}
    Without a lower bound on $\overline{\psi}_{K_n,n}$, condition (i) alone does not imply $K_n/n\to0$. Nevertheless, combining conditions (i) and (ii) yields
    \begin{equation*}
K_n=o\left(\lambda_{K_n,n}\left(\frac{n}{\log(n)}\right)^{\frac{1+\delta}{2+\delta}}\right).
    \end{equation*}
    Consequently, if $\lambda_{K_n,n}=O(1)$, then
    \begin{equation*}
        K_n=o\left(\left(\frac{n}{\log(n)}\right)^{\frac{1+\delta}{2+\delta}}\right).
    \end{equation*}
    In particular, under bounded ridge regularization, Assumption~\ref{assumption:growth_variance} excludes high-dimensional regimes with $K_n/n\to\kappa\in(0,\infty)$ and ultra-high-dimensional regimes with $n=o(K_n)$.
\end{remark}
Notice that since $\sup_{s\geq0}s/(s^2+\lambda_{K_n,n})=(2\sqrt{\lambda_{K_n,n}})^{-1}$, the deterministic choice $\overline{\psi}_{K_n,n}=(2\sqrt{\lambda_{K_n,n}})^{-1}$ always satisfies $\psi_{K_n,n,\lambda}\leq\overline{\psi}_{K_n,n}$. However, conditions~(i)--(iii) of Assumption~\ref{assumption:growth_variance} must still be verified separately for this choice.

Since the domain $\mathcal{X}$ is fixed and compact under Assumption~\ref{assumption:X_domain}, we do not introduce a weighted sup-norm and work instead throughout with the ordinary sup-norm.

\begin{lemma}[Uniform stochastic bound]
\label{lemma:uniform_bound_variance}
    Let Assumptions~\ref{assumption:X_domain}, \ref{assumption:c_stationarity}, \ref{assumption:regression_errors_L2}, \ref{assumption:higher_moment}, \ref{assumption:uniform_sieve_regularity}, and \ref{assumption:growth_variance} hold. Suppose, in addition, that $\{Z_i\}_{i\in\mathbb{N}^+}$ is i.i.d. Then
    \begin{equation*}
    \|\widehat{h}_{n,\lambda}-\widetilde{h}_{n,\lambda}\|_\infty = O_P\left( \zeta_{K_n}\overline{\psi}_{K_n,n} \sqrt{\frac{\log(n)}{n}} \right).
    \end{equation*}
    Consequently,
    \begin{equation*}
    \|\widehat{h}_{n,\lambda}-\widetilde{h}_{n,\lambda}\|_\infty = o_P(1).
    \end{equation*}
\end{lemma}

We now specialize the generic stochastic analysis to the RWNN sieves considered in Section~\ref{sec:approximation_RWNNs}. More specifically, Lemma~\ref{lemma:RWNN_regular} shows that, under either of the two activation and parameter space conditions used in the approximation results, the RWNN feature map satisfies Assumption~\ref{assumption:uniform_sieve_regularity}. Consequently, Corollary~\ref{cor:RWNN_variance} applies Lemma~\ref{lemma:RWNN_regular} to obtain the corresponding variance bound for the ridge-regularized RWNN least squares estimator.

\begin{lemma}
\label{lemma:RWNN_regular}
    Let Assumption~\ref{assumption:X_domain} hold and define 
    \begin{equation*}
        g_K(x) := \left(\widetilde{\sigma}(A_1^\top x+B_1),\ldots, \widetilde{\sigma}(A_{K}^\top x+B_{K})\right)^\top, \quad x\in \mathcal{X}.
    \end{equation*}
    Let one of the following conditions be satisfied:
    \begin{itemize}
        \item[(i)] $\widetilde{\sigma}=\sigma$, Assumption~\ref{assumption:activation_Lipschitz} holds and the parameter space $\Pi\subset\mathbb{R}^{d+1}$ is compact;
        \item[(ii)] $\widetilde{\sigma}=\sigma_m$, $m=1,2$, and Assumption~\ref{assumption:Pi} holds. 
    \end{itemize}
    Then there exist constants $M_g, L_g>0$, independent of $K$ and $n$, such that, almost surely, 
    \begin{equation*}
        \sup_{x\in\mathcal{X}}\|g_K(x)\| \leq M_g \sqrt{K}
    \end{equation*}
    and
    \begin{equation*}
        \|g_K(x_1)-g_K(x_2)\|\leq L_g \sqrt{K} \|x_1-x_2\|, \enspace \text{for}\enspace \text{all} \enspace x_1,x_2\in \mathcal{X}.
    \end{equation*}
    Consequently, Assumption~\ref{assumption:uniform_sieve_regularity} holds with $\zeta_K=M_g \sqrt{K}$ and $L_K=L_g\sqrt{K}$ and, in particular, Assumption~\ref{assumption:uniform_sieve_regularity}~(iii) holds, for example, with $q=1/2$.
\end{lemma}
\begin{corollary}
    \label{cor:RWNN_variance}
    Let the assumptions of Lemma~\ref{lemma:uniform_bound_variance} hold, except that Assumption~\ref{assumption:uniform_sieve_regularity} is replaced by either condition~{(i)} or condition~{(ii)} of Lemma~\ref{lemma:RWNN_regular}. Then
    \begin{equation*}
        \|\widehat{h}_{n,\lambda}-\widetilde{h}_{n,\lambda}\|_\infty = O_P\left( \overline{\psi}_{K_n,n} \sqrt{\frac{K_n \log(n)}{n}} \right).
    \end{equation*}
    If, in addition, $\overline{\psi}_{K_n,n} \sqrt{{K_n \log(n)}/{n}}=o(1)$, then
    \begin{equation*}
        \|\widehat{h}_{n,\lambda}-\widetilde{h}_{n,\lambda}\|_\infty = o_P(1).
    \end{equation*}
\end{corollary}

\subsubsection{Sup-norm bias and the ridge pseudo-projection}%
\label{subsubsec:uniform_bound_bias}

We now derive a bound on the sup-norm bias term, $\|\widetilde{h}_{n,\lambda}-h\|_\infty$, for the ridge-regularized RWNN estimator. 
For an unregularized series least-squares estimator, the standard approach is to exploit the empirical projection structure of the estimator (see, for example, \citealp{Chen2015}). In the present setting, this approach cannot be applied directly for two reasons:
\begin{itemize}
    \item[(a)]
    We do not impose a uniform lower bound on the smallest nonzero singular value of the empirical feature matrix. Consequently, the coefficient bounds involving the inverse empirical covariance may be arbitrarily conservative when the random features are nearly collinear.
    \item[(b)]
    Ridge regularization does not preserve the projection property. More precisely, the corresponding projection operator is not idempotent and does not reproduce every element of the RWNN sieve. An additional ridge-shrinkage term therefore appears in the bias
    decomposition.
\end{itemize}
For any bounded function $f:\mathcal{X}\rightarrow \mathbb{R}$, define the empirical ridge pseudo-projection $\overline{P}_{K_n,n,\lambda}$ by
\begin{equation}
\label{eq:ridge_pseudo_projection}
    \bigl(\overline{P}_{K_n,n,\lambda}f\bigr)(x) := g_{K_n}(x)^\top \widehat{\Gamma}_{K_n,n,\lambda}^{-1} \left( n\inv G_{K_n,n}^\top F \right),
\end{equation}
where $F:=\bigl(f(X_1),\ldots,f(X_n)\bigr)^\top$  and  $\widehat{\Gamma}_{K_n,n,\lambda} := n\inv G_{K_n,n}^\top G_{K_n,n}+\lambda_{K_n,n}I_{K_n}$. In particular, taking $H:=\bigl(h(X_1),\ldots,h(X_n)\bigr)^\top$ gives $\overline{P}_{K_n,n,\lambda}h=\widetilde{h}_{n,\lambda}$.
We view $\overline{P}_{K_n,n,\lambda}$ as an operator on the space of bounded measurable functions on $\mathcal{X}$, equipped with the supremum norm, and denote its induced operator norm by $\|\overline{P}_{K_n,n,\lambda}\|_\infty$.

We first provide a generic coefficient-space bound on this operator norm, which does not exploit any additional structure of the random-feature sieve.

\begin{lemma}
\label{lemma:pseudo-proj_sup_bound}
    Suppose Assumption~\ref{assumption:uniform_sieve_regularity}(i) holds. Then, almost surely, $\|\overline{P}_{K_n,n,\lambda}\|_{\infty}\leq\zeta_{K_n}\psi_{K_n,n,\lambda}$.
\end{lemma}
\begin{remark}
\label{remark:pseudo_projection_bound}
The bound in Lemma~\ref{lemma:pseudo-proj_sup_bound} is generic and may be conservative. It controls the pseudo-projection through the worst amplified direction of the empirical feature covariance and therefore ignores possibly redundant random features.
For localized series bases, such as (tensor products of) splines and wavelets, sharper bounds on the sup-norm of the empirical projection can be obtained by exploiting the intrinsic structure of the basis (see, for example, \cite{huangLocalAsymptoticsPolynomial2003}). For the RWNN sieves considered here, we instead investigate stability directly through the empirical ridge pseudo-projection.

A sharper analysis may instead study directly the condition
\begin{equation*}
    \|\overline{P}_{K_n,n,\lambda}\|_{\infty} = O_P(1) ,
\end{equation*}
and we pursue this direction in more detail below. In particular, this condition appears to be reasonable for some of the settings we explore in the simulations of Section~\ref{section:simulations}.
\end{remark}

We now combine Lemma~\ref{lemma:pseudo-proj_sup_bound} with the sup-norm approximation result of Theorem~\ref{theorem:sharp_is_linf_unified}. We consider the estimator in \eqref{eq:RWNN_estimator} with the stratified random features constructed in Theorem~\ref{theorem:sharp_is_linf_unified}, independently of the regression observations.  In case~(ii) of Theorem~\ref{theorem:sharp_is_linf_unified}, the stratification depends on the sign partition of the source function $f$. Consequently, the corresponding estimation results provide oracle guarantees unless this partition is known.
 Let
\begin{equation*}
    h_{K_n}^{\mathrm{IS}}(x)=g_{K_n}(x)^\top W_{K_n},
\end{equation*}
denote the corresponding stratified RWNN approximant, where $W_{K_n}\in\mathbb{R}^{K_n}$ and $g_{K_n}$ is the same feature map used in \eqref{eq:RWNN_estimator}.

\begin{lemma}
\label{lemma:uniform_bias_bound}
    Let the assumptions of Theorem~\ref{theorem:sharp_is_linf_unified} and Assumption~\ref{assumption:uniform_sieve_regularity}(i) hold.
    Let $h_{K_n}^{\mathrm{IS}}(x) =g_{K_n}(x)^\top W_{K_n}$  be the importance sampling RWNN approximant constructed in  Theorem~\ref{theorem:sharp_is_linf_unified}. Then
    \begin{align}
    \label{eq:uniform_bias_bound}
        \|\widetilde{h}_{n,\lambda}-h\|_\infty 
        & \leq\left(1+\left\|\overline P_{K_n,n,\lambda}\right\|_\infty\right)\left\|h-h_{K_n}^{\mathrm{IS}}\right\|_\infty+R_{K_n,n,\lambda},
    \end{align}
    where
    \begin{equation}
\label{eq:ridge_shrinkage_term}
R_{K_n,n,\lambda}:=\left\|\overline{P}_{K_n,n,\lambda}h_{K_n}^{\mathrm{IS}}
-h_{K_n}^{\mathrm{IS}}
\right\|_\infty.
\end{equation}
    Moreover, 
    \begin{equation}
    \label{eq:ridge_shrinkage_spectral}
        R_{K_n,n,\lambda} \leq \zeta_{K_n}\frac{\lambda_{K_n,n}}{s_{K_n,n}^2+\lambda_{K_n,n}}\|W_{K_n}\|\leq \sqrt{6} M_f \zeta_{K_n}K_n^{-1/2} \frac{\lambda_{K_n,n}} {s_{K_n,n}^2+\lambda_{K_n,n}},
    \end{equation}
    with $M_f$ as in Theorem~\ref{theorem:sharp_is_linf_unified}.
    
    Thus, in particular,
    \begin{align}
        \|\widetilde{h}_{n,\lambda}-h\|_\infty
        &\leq\left(1+\zeta_{K_n}\psi_{K_n,n,\lambda}\right)\left\|h-h_{K_n}^{\mathrm{IS}}\right\|_\infty+\sqrt{6}\,M_f\, \zeta_{K_n}K_n^{-1/2} \frac{\lambda_{K_n,n}} {s_{K_n,n}^2+\lambda_{K_n,n}}.
    \label{eq:uniform_bias_bound_crude}
    \end{align}
\end{lemma}

\subsubsection{Sup-norm estimation bounds}
\label{subsubsec:supnorm_estimation_bounds}
We now combine the stochastic and bias bounds derived in the preceding subsections.

\begin{theorem}[Sup-norm estimation bound]
\label{theorem:uniform_estimation_bound}
    Suppose that the assumptions of Lemmas~\ref{lemma:uniform_bound_variance} and \ref{lemma:uniform_bias_bound} hold. Then
    \begin{align}
    \label{eq:uniform_estimation_bound}
        \|\widehat{h}_{n,\lambda}-h\|_\infty 
        &\leq\|\widehat{h}_{n,\lambda}-\widetilde{h}_{n,\lambda}\|_\infty+\left(1+\zeta_{K_n}\psi_{K_n,n,\lambda}\right)\left\|h-h_{K_n}^{\mathrm{IS}}\right\|_\infty\nonumber\\
        &\quad+\sqrt{6}M_f\zeta_{K_n}K_n^{-1/2}\frac{\lambda_{K_n,n}}{s_{K_n,n}^2+\lambda_{K_n,n}}.
    \end{align}
    
    Moreover, let
    \begin{equation*}
        \alpha :=
        \begin{cases}
            \beta/(d+1),& \text{in case~{(i)} of Theorem~\ref{theorem:sharp_is_linf_unified}},\\[1mm]
            1/(d+1),& \text{in case~{(ii)} of Theorem~\ref{theorem:sharp_is_linf_unified}}.
        \end{cases}
    \end{equation*}
    
    Then
    \begin{align}
    \label{eq:uniform_estimation_rate}
        \|\widehat{h}_{n,\lambda}-h\|_\infty &=O_P\left(\zeta_{K_n}\overline{\psi}_{K_n,n}\sqrt{\frac{\log (n)}{n}}\right)\nonumber\\
        &\quad+O_P\left(\left(1+\zeta_{K_n}\overline{\psi}_{K_n,n}\right)\left(
        \sqrt d+\sqrt{\log (K_n)}\right)K_n^{-1/2-\alpha}\right)\nonumber\\
        &\quad+\sqrt{6}M_f\zeta_{K_n}K_n^{-1/2}\frac{\lambda_{K_n,n}}{s_{K_n,n}^2+\lambda_{K_n,n}}.
    \end{align}
\end{theorem}

\begin{corollary}[Sup-norm rate for RWNN features]
\label{corollary:uniform_estimation_RWNN}
    Suppose that the assumptions of Theorem~\ref{theorem:uniform_estimation_bound} and Lemma~\ref{lemma:RWNN_regular} hold. Then
    \begin{align}
    \left\|\widehat{h}_{n,\lambda}-h\right\|_\infty&=O_P\left(\overline{\psi}_{K_n,n}\sqrt{\frac{K_n\log(n)}{n}}\right)\nonumber\\
    &\quad+O_P\left(\left(1+\sqrt{K_n}\overline{\psi}_{K_n,n}\right)\left(\sqrt{d}+\sqrt{\log(K_n)}\right)K_n^{-1/2-\alpha}\right)\nonumber\\
    &\quad+\sqrt{6} M_f M_g \frac{\lambda_{K_n,n}}{s_{K_n,n}^2+\lambda_{K_n,n}},
    \label{eq:uniform_estimation_RWNN_rate}
    \end{align}
    where $\alpha$ is as in Theorem~\ref{theorem:uniform_estimation_bound}.
    
    In particular, if $\overline{\psi}_{K_n,n}=O(1)$, then
    \begin{align}
    \left\|\widehat{h}_{n,\lambda}-h\right\|_\infty&=O_P\left(\sqrt{\frac{K_n\log(n)}{n}}\right)+O_P\left(\left(\sqrt{d}+\sqrt{\log(K_n)}\right)K_n^{-\alpha}\right)\nonumber\\
    &\quad+\sqrt{6} M_f M_g\frac{\lambda_{K_n,n}}{s_{K_n,n}^2+\lambda_{K_n,n}}.
    \label{eq:uniform_estimation_RWNN_bounded_psi}
    \end{align}
\end{corollary}

\begin{remark}
\label{remark:uniform_estimation_RWNN_interpretation}
The special case in \eqref{eq:uniform_estimation_RWNN_bounded_psi} shows that, when $\overline{\psi}_{K_n,n}=O(1)$, the variance term has the standard sup-norm order. At the same time, notice that the first term of the bias contribution loses a factor $K_n^{-1/2}$ relative to the original RWNN approximation rate. This originates from the spectral estimate $\left\|\overline{P}_{K_n,n,\lambda}\right\|_\infty\leq\zeta_{K_n}\psi_{K_n,n,\lambda}=O_P\left(\sqrt{K_n}\right)$, and a sharper bound $\left\|\overline{P}_{K_n,n,\lambda}\right\|_\infty=O_P(1)$ would retain the original approximation rate.

A separate limitation concerns the ridge-shrinkage term. If $\widehat{\Gamma}_{K_n,n}$ is singular, then $s_{K_n,n}=0$, and the upper bound in \eqref{eq:uniform_estimation_RWNN_rate} does not vanish. Therefore, the smallest singular value argument alone cannot establish that the ridge bias vanishes in the overparametrized regime. A sharper analysis is needed to address this issue.

These limitations motivate the leverage-based analysis developed next, which controls prediction-space stability and ridge shrinkage without relying exclusively on the smallest empirical singular value.
\end{remark}

\subsubsection{Leverage-based sup-norm stability}
\label{subsubsec:supnorm_leverage_stability}

The preceding spectral analysis controls the estimator through the empirical ridge stability factor $\psi_{K_n,n,\lambda}$ and the smallest empirical singular value $s_{K_n,n}$. Such coefficient-space quantities may be conservative for prediction, particularly when the random features are highly redundant. We therefore formulate alternative conditions directly in terms of the empirical ridge pseudo-projection $\overline{P}_{K_n,n,\lambda}$ defined in \eqref{eq:ridge_pseudo_projection}.  This follows the conventional series-estimation perspective in which boundedness of the sup-norm operator norm is referred to as sup-norm stability (see, for example, \citealp{Chen2015}). For the ridge pseudo-projection, we use the empirical ridge Lebesgue constant as a sufficient measure of this stability.

For $x\in\mathcal{X}$ and $i=1,\ldots,n$, define the ridge sieve smoothing weights
\begin{equation}
\label{eq:ridge_smoother_weights}
    p_{i,n,\lambda}(x) := n\inv g_{K_n}(x)^\top \widehat{\Gamma}_{K_n,n,\lambda}^{-1} g_{K_n}(X_i) ,
\end{equation}
and let $p_{n,\lambda}(x):=(p_{1,n,\lambda}(x),\ldots,p_{n,n,\lambda}(x))^\top$.

\begin{lemma}[Properties of the ridge smoothing weights]
\label{lemma:ridge_smoothing_representation}
    For every bounded measurable function $f:\mathcal{X}\rightarrow\mathbb{R}$,
    \begin{equation}
    \label{eq:ridge_smoother_representation}
        \bigl(\overline{P}_{K_n,n,\lambda}f\bigr)(x)=\sum_{i=1}^np_{i,n,\lambda}(x)f(X_i).
    \end{equation}
    Define
    \begin{equation}
    \label{eq:def_Lambda_nu}
        \Lambda_{n,\lambda} := \sup_{x\in\mathcal{X}}\sum_{i=1}^n|p_{i,n,\lambda}(x)| ,
        \quad
        \nu_{n,\lambda} := \sup_{x\in\mathcal{X}}\sum_{i=1}^np_{i,n,\lambda}(x)^2,
    \end{equation}
    as well as
    \begin{equation}
    \label{eq:def_l_m}
        \ell_{n,\lambda} := \sup_{x\in\mathcal{X}}g_{K_n}(x)^\top\widehat{\Gamma}_{K_n,n,\lambda}^{-1}g_{K_n}(x) \quad \text{and} \quad 
        \mathfrak{m}_{n,\lambda} := \sup_{x\in\mathcal{X}}\max_{1\leq i\leq n}|p_{i,n,\lambda}(x)| .
    \end{equation}
    Then
    \begin{equation}
    	\label{eq:ridge_lebesgue_bound}
        \|\overline{P}_{K_n,n,\lambda}\|_\infty\leq\Lambda_{n,\lambda},
    \end{equation}
    and it holds that
    \begin{equation}
    \label{eq:ridge_variance_leverage_bound}
        \nu_{n,\lambda}\leq\frac{\ell_{n,\lambda}}{n}\quad \text{and} \quad \mathfrak{m}_{n,\lambda}\leq\frac{\ell_{n,\lambda}}{n}.
    \end{equation}
\end{lemma}
The quantity $\ell_{n,\lambda}$ introduced in Lemma~\ref{lemma:ridge_smoothing_representation} is studied in random kernel regression literature (see, for example, \citealp{alaouiFastRandomizedKernel2015,avronRandomFourierFeatures2017,pauwelsRelatingLeverageScores2018}). 
In the following, we call $\ell_{n,\lambda}$ the uniform empirical ridge leverage score, and refer to $\Lambda_{n,\lambda}$ as the empirical ridge Lebesgue constant.  The ridge leverage quantities are closely related to the effective dimension introduced in \eqref{eq:def_L2_effective_dimension}. Indeed, the empirical analogue of $\mathcal{N}_{K_n,\lambda}$ satisfies
\begin{equation*}
\frac{1}{n}\sum_{i=1}^n g_{K_n}(X_i)^\top\widehat{\Gamma}_{K_n,n,\lambda}^{-1}g_{K_n}(X_i)=\operatorname{tr}\left(\widehat{\Gamma}_{K_n,n}\widehat{\Gamma}_{K_n,n,\lambda}^{-1}\right).
\end{equation*}
Thus, effective dimension measures average ridge leverage, whereas $\ell_{n,\lambda}$ controls its supremum over $\mathcal{X}$. Moreover, under Assumption~\ref{assumption:L2_ridge_gram_stability}, the ridge-normalized covariance stability in \eqref{eq:nondegeneracy} implies
\begin{equation}
\label{eq:uniform_empirical_leverage_L2}
\ell_{n,\lambda}\leq(1+o_P(1))\widetilde{\xi}_{K_n,\lambda}^2.
\end{equation}
Indeed, by \eqref{eq:def_Dn_hat_An}, on the event $\|A_{K_n,n,\lambda}\|<1$, we have $\widehat{\Gamma}_{K_n,n,\lambda}^{-1}\preceq(1-\|A_{K_n,n,\lambda}\|)^{-1}\Gamma_{K_n,\lambda}^{-1}$. Taking the supremum over $x\in\mathcal{X}$ and using $\|A_{K_n,n,\lambda}\|=o_P(1)$ from \eqref{eq:nondegeneracy} yields \eqref{eq:uniform_empirical_leverage_L2}. Consequently, the additional condition $\widetilde{\xi}_{K_n,\lambda}^2=O_P(K_n)$ is sufficient for $\ell_{n,\lambda}=O_P(K_n)$. This does not, however, imply $\Lambda_{n,\lambda}=O_P(1)$, which requires separate control of the absolute smoothing weights.

 We now note that using the singular value decomposition in \eqref{eq:uniform_design_svd}, the vector of ridge sieve smoothing weights satisfies
\begin{equation*}
    p_{n,\lambda}(x)=\frac{1}{\sqrt{n}}U_{m_n,n}\operatorname{diag}\left(\left(\frac{s_{k,n}}{s_{k,n}^2+\lambda_{K_n,n}}\right)_{k=1}^{m_n}\right)V_{m_n,n}^\top g_{K_n}(x),
\end{equation*}
which, together with Assumption~\ref{assumption:uniform_sieve_regularity}(i), implies $\nu_{n,\lambda}\leq{\zeta_{K_n}^2\psi_{K_n,n,\lambda}^2}/{n}$, while the Cauchy-Schwarz inequality gives $\Lambda_{n,\lambda}\leq\sqrt{n\nu_{n,\lambda}}\leq\zeta_{K_n}\psi_{K_n,n,\lambda}$. Hence, the SVD reproduces the feature-space conditioning bound in Lemma~\ref{lemma:pseudo-proj_sup_bound}, though it does not exploit possible cancellations among the ridge sieve smoothing weights.

\begin{assumption}
\label{assumption:supnorm_leverage_stability}
    The empirical ridge Lebesgue constant and the uniform empirical ridge leverage score satisfy $\Lambda_{n,\lambda}=O_P(1)$ and  $\ell_{n,\lambda}=O_P(K_n)$, respectively.
\end{assumption}

Notice that, by Lemma~\ref{lemma:ridge_smoothing_representation}, Assumption~\ref{assumption:supnorm_leverage_stability} implies $\|\overline{P}_{K_n,n,\lambda}\|_\infty=O_P(1)$, $\nu_{n,\lambda}=O_P\left({K_n}/{n}\right)$, and $\mathfrak{m}_{n,\lambda}=O_P\left({K_n}/{n}\right)$. Unlike a condition of the form $\psi_{K_n,n,\lambda}=O_P(1)$, Assumption~\ref{assumption:supnorm_leverage_stability} controls the fitted values directly and may therefore remain valid when the feature-space problem is poorly conditioned.

\begin{lemma}[Uniform variance bound under sup-norm and leverage stability]
\label{lemma:uniform_bound_variance_stable}
    Let Assumptions~\ref{assumption:X_domain}, \ref{assumption:c_stationarity}, \ref{assumption:regression_errors_L2}, \ref{assumption:higher_moment}, \ref{assumption:uniform_sieve_regularity}~(i)-(ii), and  the condition $\ell_{n,\lambda}=O_P(K_n)$ in Assumption~\ref{assumption:supnorm_leverage_stability} hold. Suppose, in addition, that $\{Z_i\}_{i\in\mathbb{N}^+}$ is i.i.d. and that there exists a deterministic positive sequence $\overline{\psi}_{K_n,n}$ such that $\psi_{K_n,n,\lambda}=O_P(\overline{\psi}_{K_n,n})$. Assume that
    \begin{equation}
    \label{eq:stable_variance_moment_growth}
    \frac{K_n^{1+2/\delta}\log(n)}{n}=O(1),
    \end{equation}
    and that, for some constants $C,q>0$,
    \begin{equation}
    \label{eq:stable_variance_covering_growth}
    \frac{L_{K_n}\overline{\psi}_{K_n,n}\sqrt{m_n}}{\sqrt{K_n\log(n)}}\leq Cn^q
    \end{equation}
    for all sufficiently large $n$. Then
    \begin{equation}
    \label{eq:uniform_variance_stable}
    \|\widehat{h}_{n,\lambda}-\widetilde{h}_{n,\lambda}\|_\infty=O_P\left(\sqrt{\frac{K_n\log(n)}{n}}\right).
    \end{equation}
\end{lemma}

Recall that the ridge remainder in \eqref{eq:ridge_shrinkage_term} is defined by
\begin{equation*}
R_{K_n,n,\lambda}:=\left\|\overline{P}_{K_n,n,\lambda}h_{K_n}^{\mathrm{IS}}-h_{K_n}^{\mathrm{IS}}\right\|_\infty.
\end{equation*}
As shown in \eqref{eq:ridge_shrinkage_spectral}, this remainder admits a bound in terms of the smallest empirical singular value. However, its direct formulation allows us to derive an alternative bound that does not involve this quantity.

\begin{lemma}[Ridge-remainder bound]
\label{lemma:ridge_rremainder_bound}
    It holds that
    \begin{equation}
    \label{eq:ridge_remainder_leverage_bound}
    R_{K_n,n,\lambda}\leq\sqrt{\lambda_{K_n,n}\ell_{n,\lambda}}  \|W_{K_n}\|.
    \end{equation}
    Consequently, for the importance-sampling weights constructed in
    Theorem~\ref{theorem:sharp_is_linf_unified},
    \begin{equation}
    \label{eq:ridge_remainder_IS_bound}
    R_{K_n,n,\lambda}\leq\sqrt{6}M_f\sqrt{\frac{\lambda_{K_n,n}\ell_{n,\lambda}}{K_n}}.
    \end{equation}
    Moreover, under Assumption~\ref{assumption:supnorm_leverage_stability},
    \begin{equation}
    \label{eq:ridge_remainder_rate}
    R_{K_n,n,\lambda}=O_P\left(\sqrt{\lambda_{K_n,n}}\right).
    \end{equation}
\end{lemma}

The bound in Lemma~\ref{lemma:ridge_rremainder_bound} does not involve the smallest empirical singular value and remains valid when $\widehat{\Gamma}_{K_n,n}$ is singular. Under Assumption~\ref{assumption:supnorm_leverage_stability}, it yields $R_{K_n,n,\lambda}=O_P(\sqrt{\lambda_{K_n,n}})$, independently of the smallest empirical singular value.

\begin{theorem}[Sup-norm estimation bound under sup-norm and leverage stability] 
\label{theorem:uniform_estimation_bound_stable} 
    Suppose that the assumptions of Lemmas~\ref{lemma:uniform_bound_variance_stable}, \ref{lemma:uniform_bias_bound}, and \ref{lemma:ridge_rremainder_bound} hold, together with Assumption~\ref{assumption:supnorm_leverage_stability}.
    Then 
    \begin{equation} 
    \label{eq:uniform_estimation_bound_stable} 
        \|\widehat{h}_{n,\lambda}-h\|_\infty\leq\|\widehat{h}_{n,\lambda}-\widetilde{h}_{n,\lambda}\|_\infty+(1+\Lambda_{n,\lambda})\|h-h_{K_n}^{\mathrm{IS}}\|_\infty+\sqrt{6}M_f\sqrt{\frac{\lambda_{K_n,n}\ell_{n,\lambda}}{K_n}}. 
    \end{equation} 
    Let 
    \begin{equation*} 
        \alpha:=
        \begin{cases}
        \beta/(d+1),&\text{in case~(i) of Theorem~\ref{theorem:sharp_is_linf_unified}},\\[1mm]
        1/(d+1),&\text{in case~(ii) of Theorem~\ref{theorem:sharp_is_linf_unified}}.
        \end{cases} 
    \end{equation*} 
    Then 
    \begin{equation} 
    \label{eq:uniform_estimation_rate_stable} 
        \|\widehat{h}_{n,\lambda}-h\|_\infty=O_P\left(\sqrt{\frac{K_n\log(n)}{n}}\right)+O_P\left((\sqrt d+\sqrt{\log(K_n)})K_n^{-1/2-\alpha}\right)+O_P\left(\sqrt{\lambda_{K_n,n}}\right). 
    \end{equation} 
\end{theorem}

\begin{corollary}[Explicit rate under sup-norm and leverage stability] 
\label{corollary:uniform_estimation_rate_stable} 
    Suppose that the assumptions of Theorem~\ref{theorem:uniform_estimation_bound_stable} hold, $d$ is fixed, and 
    \begin{equation} 
    \label{eq:stable_optimal_Kn} K_n\asymp n^{1/(2(1+\alpha))}. 
    \end{equation} 
    Suppose also that 
    \begin{equation} 
    \label{eq:stable_moment_condition} 
    \delta>\frac{2}{1+2\alpha} 
    \end{equation} and 
    \begin{equation} 
    \label{eq:stable_lambda_condition} 
    \lambda_{K_n,n}=O\left(n^{-\frac{1+2\alpha}{2(1+\alpha)}}\log(n)\right). 
    \end{equation} 
    Then 
    \begin{equation} 
    \label{eq:uniform_estimation_rate_stable_explicit} 
    \|\widehat{h}_{n,\lambda}-h\|_\infty=O_P\left(n^{-\frac{1+2\alpha}{4(1+\alpha)}}\sqrt{\log(n)}\right). 
    \end{equation} 
\end{corollary}

\begin{remark}[Comparison with Chen and Christensen] \label{remark:uniform_rate_comparison_Chen} Under sup-norm stability of the empirical projection, \citet{Chen2015} obtain the standard stochastic term $\sqrt{K_n\log(n)/n}$ and show that spline and wavelet series estimators attain the rate $(n/\log(n))^{-p/(2p+d)}$ when the regression function has smoothness $p$. To compare this benchmark with \eqref{eq:uniform_estimation_rate_stable_explicit}, set $p/d=1/2+\alpha$. 
The Chen-Christensen rate then becomes
\begin{equation*}
\left(\frac{n}{\log(n)}\right)^{-p/(2p+d)}=n^{-\frac{1+2\alpha}{4(1+\alpha)}}(\log(n))^{\frac{1+2\alpha}{4(1+\alpha)}}.
\end{equation*}
Hence, the rate in \eqref{eq:uniform_estimation_rate_stable_explicit} has the same polynomial exponent in $n$ and is slower by the logarithmic factor $(\log(n))^{1/(4(1+\alpha))}$. This additional logarithmic factor originates from the $\sqrt{\log(K_n)}$ term in the RWNN sup-norm approximation bound.
 \end{remark}

\section{Numerical experiments}\label{section:simulations}

In this section, we provide simulation evidence to highlight the different structural properties of random weight neural networks empirically.
First, we provide simulation evidence for the sup-norm approximation rates we obtain by means of the importance sampling approximator derived in Section~\ref{subsec:approximation_RWNNs_importance}. When approximating a one-dimensional function with a known integral representation, we find that the IS approach achieves faster approximation rates than na\"ive random sampling.
Then, we consider the effects of \textit{scaling} on the conditioning of RWNNs in terms of the least-squares problem. We find that normalization can substantially improve the conditioning of RWNN sieves, in line with findings from the machine learning literature. Nonetheless, even without normalization, we show that the effective dimension $\mathcal{N}_{K,\lambda}$ is well-behaved and grows moderately with respect to $K$. 
We also study numerically the sup-norm stability conditions introduced in Section~\ref{subsubsec:supnorm_leverage_stability}, and compare the spectral decay of the empirical sieve covariance matrix with stability factors $\Lambda_{n, \lambda}$ and $\ell_{n,\lambda}$.
Finally, we consider inference for both smooth and non-smooth functionals of the regularized random weight neural network least squares estimator, following the theory developed in Section~\ref{sec:Ridge_RWNN_least_squares}. We evaluate confidence intervals obtained both via the asymptotic normality of the estimated functionals, as well as bootstrapping. 

Throughout our simulations---save for the numerical analysis of $L^\infty$ approximation via importance sampling---we consider both compactly-supported RWNN weight sampling distributions (uniform) and laws with unbounded support (Gaussian). Either type of distribution is widely employed in practice, depending on the setting; hence, we wish to provide numerical evidence for both, even though our formal theory only encompasses the former.

Additional numerical results are collected in Appendix~\ref{appendix:additional_simulations}, including comparisons of residual, wild, and score-based bootstrap procedures for pointwise inference, as well as supplementary coverage, bias, and confidence interval results for the inferential experiments.

\subsection{Sup-norm approximation}

We start by providing numerical evidence towards the effectiveness of the stratified sampling approximation scheme that we derived in Section~\ref{subsec:approximation_RWNNs_importance}, cf. Theorem~\ref{theorem:sharp_is_linf_unified}.
Since the stratified sampling approach is adaptive in $h(x)$, we require explicit knowledge of its integral form in order to ensure that we perform stratified sampling correctly. We consider the one-dimensional function 
$ h(x) = \E_{A,B}[(A x + B)_{+}] = 1/4 + x^2/12 $, where $(A, B)^\top \overset{\textnormal{i.i.d.}}{\sim} \mathcal{U}([-1,1]^2)$,
for which it holds $f(A, B) = 1$ by construction. This simple structure allows us to design a straightforward importance sampling scheme: we construct a squared grid over $[-1, 1]^2$ of size $L = \lfloor \sqrt{K} \rfloor$, and we perform uniform sampling within each of the cells. As shown in the proof of Theorem~\ref{theorem:sharp_is_linf_unified}, if $L^2 < K$, we can formally introduce $K - L^2$ additional sieve elements with zero weight to ensure that the RWNN has exactly $K$ hidden units without impacting the approximation error rate. The sup-norm error is computed on a uniform grid over $[-1, 1]$ with 1000 points.

\begin{figure}
    \centering
    \hspace{-2em}
    \includegraphics[width=0.95\textwidth]{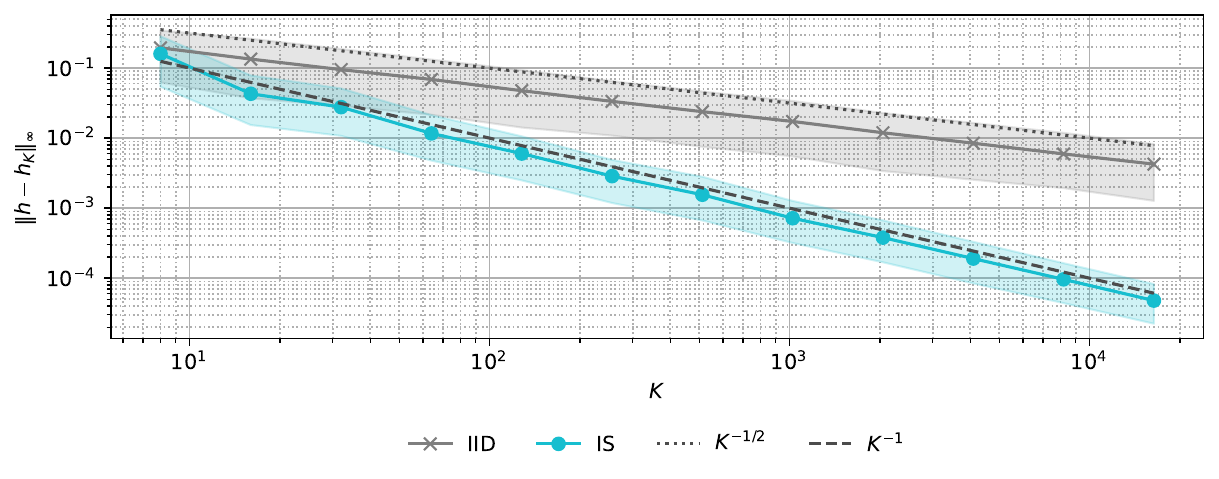}
    \caption{%
    Sup-norm approximation rates in 1D for the importance sampling (IS) approximant (cyan) compared with the na\"ive i.i.d. sampling (IID) approximant (gray). 
    Dashed and dotted lines are used to plot $K^{-1/2}$ and $K^{-1}$ decay rates, respectively.
    For each $K$, we construct $\mathfrak{M} = 1000$ RWNN weight draws. Lines indicate mean values across replications; shaded areas display 90\% quantile bands.
    }
    \label{fig:sup_approx}
\end{figure}

We provide a plot for the approximation performance of $h^{\textnormal{IS}}_K$ versus that of the equivalent i.i.d. sampling approximator $h^{\textnormal{IID}}_K$, with $(A_k, B_k)^\top \overset{\textnormal{i.i.d.}}{\sim} \mathcal{U}([-1,1]^2)$, across $\mathfrak{M} = 1000$ independent draws in Figure~\ref{fig:sup_approx}. Our simulation indicates that indeed the sup-norm approximation rate achieved via stratified sampling is faster than the one that can be obtained through standard Monte Carlo sampling. This improvement, of course, rests on having prior knowledge regarding the regularity of the kernel $f(A,B)$ in the integral representation of $h(x)$. Hence, the design of efficient data-driven sampling schemes that are adaptive to the local smoothness, in the spirit of \citet{2023SWIMSamplingWeightsNeuralNetworks}, appears to be an important avenue for future research on RWNNs.

\subsection{Weight normalization, conditioning, and $L^2$ stability}
\label{subsec:sims_normalization}

Our consistency results from Section~\ref{subsec:L2_ridge_consistency} highlight that the conditioning of the RWNN regression least squares problem is key in determining the quality of $\widehat{h}_{n,\lambda}$, as well as its consistency properties. 
Since the smallest eigenvalue of $\widehat{\Gamma}_{K_n,n}$ depends nonlinearly on both the specific draw of RWNN weights and the data, its decay as $K_n$ grows is hard to characterize theoretically: we focus instead on studying the smallest singular value of the population matrix $\Gamma_K$ for growing $K$, as well as the effective dimension factor $\mathcal{N}_{K, \lambda}$, with $\lambda \in \{10^{-6}, 10^{-4}, 10^{-2}, 1\}$.
For the sake of clarity, we focus on two of the most commonly used activation functions: ReLU and the hyperbolic tangent (tanh). It is well-known that, in practice, weight normalization helps in improving training of neural networks, and we find that this also translates to improved conditioning for the population sieve covariance matrix $\Gamma_K$ of RWNNs. The ReLU activation can be weight-normalized by using the mapping $A_k \mapsto A_k / \norm{A_k}_2 $; for the tanh activation, weight-normalization is achieved via the scaling $A_k \mapsto \sqrt{K} A_k $. 
These simulations allow us to directly investigate the spectral conditioning guarantee in Proposition~\ref{prop:spectral_tail_parametrized_sieve} and the effective dimensionality bound obtained in Corollary~\ref{cor:effective_dimension_spectral}.
We use $n = 10^4$ draws from the uniform distribution over the square $[-1, 1]^2$ to approximate the true population value of $\Gamma_K$, and we sample the random RWNN weights as $A_k \overset{\text{i.i.d.}}{\sim} \mathcal{U}([-1, 1])$ and $B_k \overset{\text{i.i.d.}}{\sim} \mathcal{U}([-1, 1])$. In Appendix~\ref{appendix:additional_figures}, we include additional figures obtained with Gaussian-Gaussian and Gaussian-Uniform samplings of $A_k$ and $B_k$, respectively, and find only minor differences.

\begin{figure}[p!]
    \centering
    \begin{subfigure}{0.97\textwidth}
        \includegraphics[width=\textwidth]{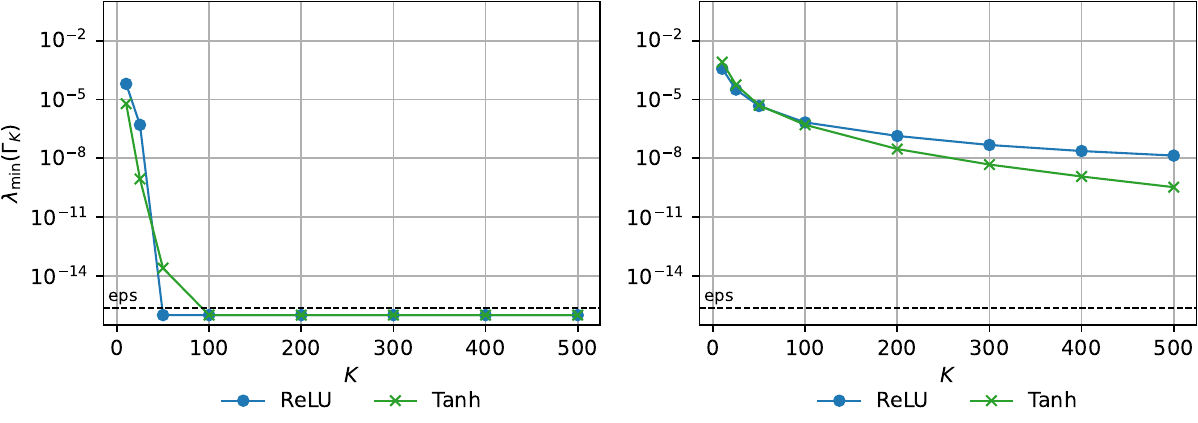}
        \caption{Least singular value of $\Gamma_{K}$ with unnormalized (left) and normalized (right) activations.}
    \end{subfigure}
    \\[1.5em]
    \begin{subfigure}{0.95\textwidth}
        \includegraphics[width=\textwidth]{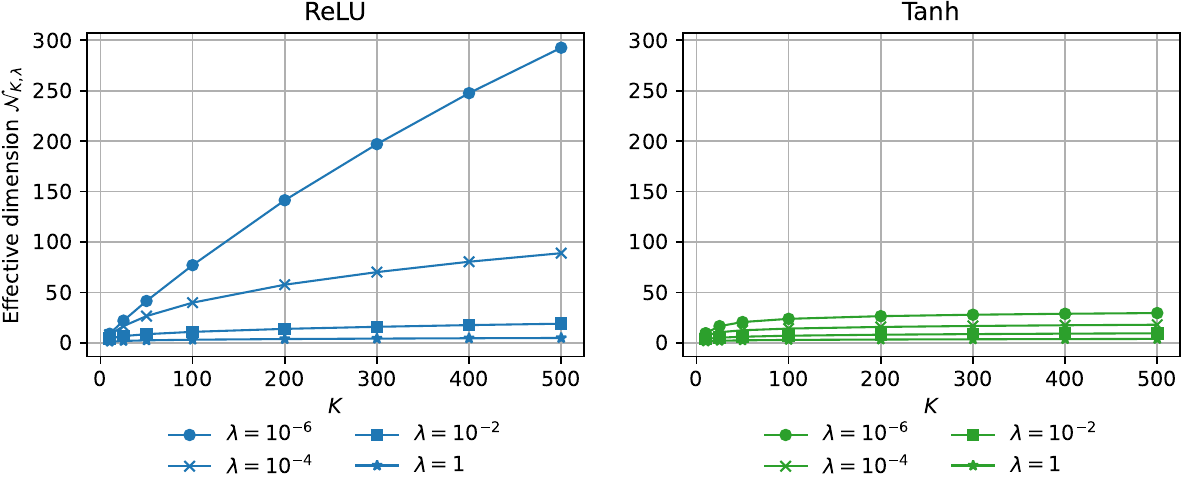}
        \caption{Effective dimension $\mathcal{N}_{K,\lambda}$ computed with ReLU (left) and tanh (right) activations.}
    \end{subfigure}
    \\[1.5em]
    \begin{subfigure}{0.95\textwidth}
        \includegraphics[width=\textwidth]{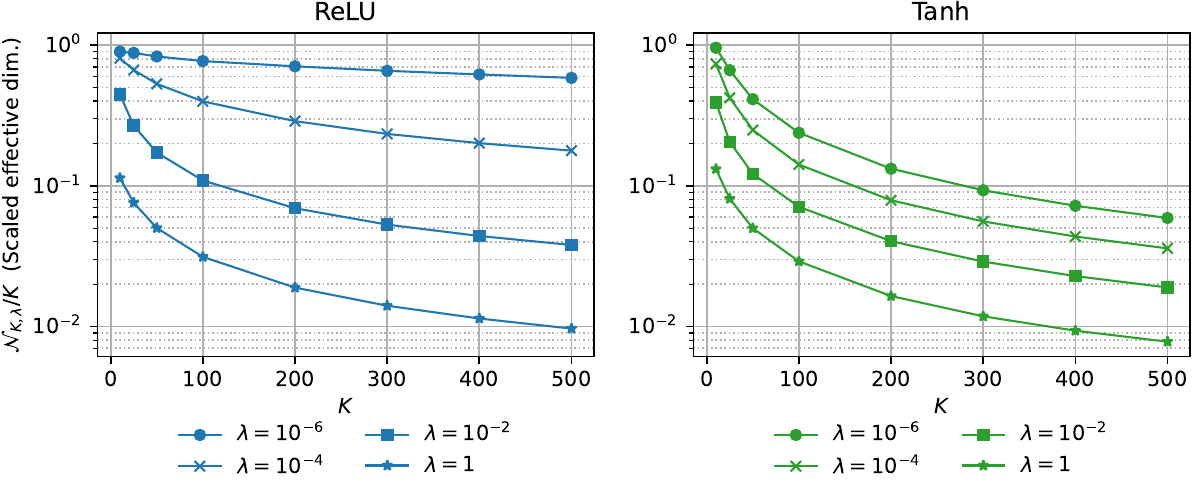}
        \caption{Scaled effective dimension $\mathcal{N}_{K,\lambda} / K$ computed with ReLU (left) and tanh (right) activations.}
    \end{subfigure}
    \caption{%
    Conditioning of RWNN sieves at different sieve dimensions $K$ and $\sigma \in \{\text{ReLU}, \text{tanh}\}$, with $A_k \overset{\text{i.i.d.}}{\sim} \mathcal{U}([-1, 1])$, $B_k \overset{\text{i.i.d.}}{\sim} \mathcal{U}([-1, 1])$. Top panel plots the smallest eigenvalues of $\Gamma_{K}$ computed using $n = 10^4$ i.i.d. draws from $\mathcal{U}([-1,1]^2)$, with and without normalization of activation functions.
    Middle panel shows the effective dimension $\mathcal{N}_{K,\lambda}$ across a range of ridge penalties $\lambda \in \{10^{-6}, 10^{-4}, 10^{-2}, 1\}$ without normalization.
    Bottom panel plots the associated scaled effective dimension $\mathcal{N}_{K,\lambda} / K$.
    For each $K$, we construct $\frak{M} = 100$ RWNN weight draws. Lines indicate mean values across replications.
    }
    \label{fig:plot_conditioning}
\end{figure}

In the top panel of Figure~\ref{fig:plot_conditioning}, we plot the empirical decay curves of $\overline{s}_{K}^2 = \lambda_{\min}(\Gamma_K)$ for each activation function, with both unnormalized and normalized weights, and include the floating-point numerical precision as a reference for numerical ill-conditioning. It is clear that normalization indeed helps substantially in improving the decay rate of $\overline{s}_{K}^2$ compared to the unnormalized baseline. Figure~\ref{fig:plot_conditioning}(b) and (c) nonetheless indicate that the growth of the effective dimension $\mathcal{N}_{K,\lambda}$ is moderate compared to $K$, across different choices of the regularization parameter $\lambda$.

\subsection{Sup-norm stability}
\label{subsec:sims_sup_norm_stability}

As we discuss in detail in Section~\ref{subsec:rwnn_uniform_ridge}, less conservative sup-norm estimation rates can be achieved by introducing additional bounds on sup-norm stability and leverage quantities, as studied in Lemma~\ref{lemma:uniform_bound_variance_stable}.
Since, in this setting, too, studying the empirical conditioning of the RWNN sieve in a theoretical manner is complex, we seek to numerically show that Assumption~\ref{assumption:supnorm_leverage_stability} is realistic.
We draw an i.i.d. sample of $X_i \sim \mathcal{U}([-1,1])$ of size $n$, where $n \in \{100, \allowbreak 200, \allowbreak 400, \allowbreak 800, \allowbreak 1600, \allowbreak 3200, \allowbreak 6400\}$, and for each sample we compute the inner RWNN layer features for $\frak{M} = 100$ draws of $\{ A_k, B_k \}_{k=1}^K$, and these weights are not normalized. 
For our main specification, we let $A_k \overset{\text{i.i.d.}}{\sim} \mathcal{N}(0,1)$ and $B_k \overset{\text{i.i.d.}}{\sim} \mathcal{N}(0,1)$ element-wise, and set $K_n = \lfloor 3 n^{1/3} \rfloor$, $\lambda_{K_n,n} = 10^{-5} n^{-2/3} \log(n)$.
We then compute $\Lambda_{n,\lambda}$ and $\ell_{n,\lambda}$ as defined in \eqref{eq:def_Lambda_nu} and \eqref{eq:def_l_m}, respectively, across $\text{MC}=100$ replications.

\begin{figure}
    \centering
    \vspace{1em}
    \begin{subfigure}{0.33\textwidth}
    \centering
        \includegraphics[width=\textwidth]{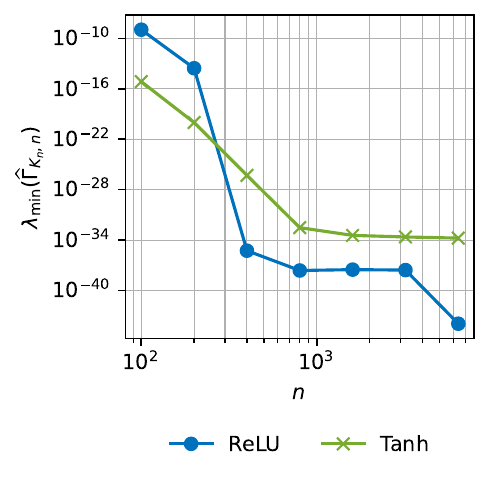}
    \end{subfigure}
    \hfill
    \begin{subfigure}{0.311\textwidth}
    \centering
        \includegraphics[width=\textwidth]{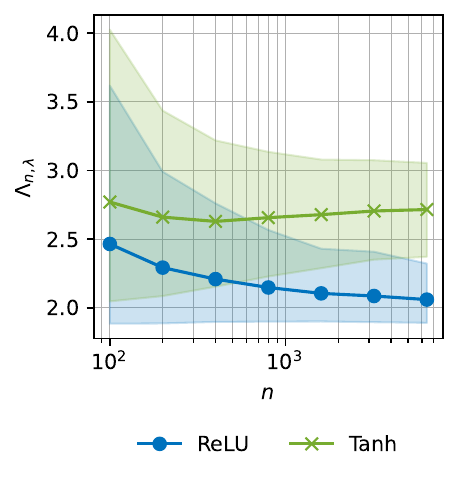}
    \end{subfigure}
    \hfill
    \begin{subfigure}{0.311\textwidth}
    \centering
        \includegraphics[width=\textwidth]{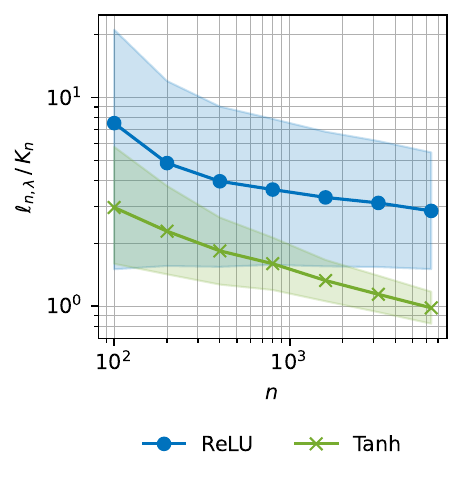}
    \end{subfigure}
    \caption{%
    Sup-norm stability factors computed for $X_i \overset{\text{i.i.d.}}{\sim} \mathcal{U}([-1,1])$ and RWNNs with $A_k \overset{\text{i.i.d.}}{\sim} \mathcal{N}(0,1)$, $B_k \overset{\text{i.i.d.}}{\sim} \mathcal{N}(0,1)$ and $\sigma \in \{\text{ReLU}, \text{tanh}\}$; activation functions are not normalized. For each $n$, we construct $\text{MC} = 100$ Monte Carlo data draws and $\frak{M} = 100$ RWNN weight draws. Lines indicate mean values across replications; shaded areas display 90\% quantile bands.
    }
    \label{fig:sup_stability}
\end{figure}

Figure~\ref{fig:sup_stability} reports our simulation results. We again focus on the ReLU and hyperbolic tangent activation functions. The left panel confirms that the decay of the smallest eigenvalue of the empirical RWNN sieve covariance matrix $\lambda_{\min}(\widehat{\Gamma}_{K_n,n})$ is rapid when we do not explicitly threshold based on the numerical floating-point precision. For both maps, however, from the center and right panels we can see that both $\Lambda_{n,\lambda}$ and the ratio $\ell_{n,\lambda} / K_n$ are well-behaved as the sample size $n$ grows, and sieve width $K_n$ and ridge penalty $\lambda_{K_n,n}$ with it.

\subsection{Asymptotic inference}
\label{subsec:sims_statistical_inference}

The $L^2$ RWNN inference results we derived in Section~\ref{subsec:inference_smooth_functionals} provide direct theoretical justification for inference on smooth functionals. 
Although one could study many different functionals---for instance, partial mean and weighted integral functionals, or consumer surplus in demand analysis (see, for example, \citealp{Newey1997})---we choose the integrated square functional $\Psi_{2}(h) = \int_{\mathcal{X}} h(x)^2 \omega(x) dx$; for simplicity, we employ a uniform weighting function, $\omega(x) = 1$, so that $\Psi_{2}(h) = \int_{X} h(x)^2  dx$. 
We study a one-dimensional regression problem with $n$ i.i.d. draws from $Y_i = g_j(X_i) + \epsilon_i$, where $X_i \sim U([-1,1])$, $\epsilon_i \sim \mathcal{N}(0, \sigma_\epsilon^2)$, $\sigma_\epsilon^2 = 0.04$, and $g_j : \R \to \R$, $j \in \{1,2,3\}$. 
Specifically, we study $g_1(x) := x + 5 \varphi(10 x)$, $g_2(x) := \sin(3 \pi x /2) / [1 + 18 x^2(\text{sgn}(x)+1)]$ and $g_3(x) := \sin(\pi x /2) / [1 + 2 x^2(\text{sgn}(x)+1)]$.\footnote{Here $\varphi$ denotes the standard Gaussian density and $\text{sgn}(x) \in \{-1,0,1\}$ denotes the sign function. These functions, plotted in Appendix~\ref{appendix:additional_figures}, Figure~\ref{fig:hh13_functions}, have been previously used by \cite{Hall2013} and \cite{calonicoEffectBiasEstimation2018}, among others, in Monte Carlo simulations of nonparametric inference procedures.}
For different choices of $\lambda$, we compute $\widehat{h}_{n,\lambda}$ and construct confidence intervals (CIs) for $\Psi(g_j)$ with a nominal coverage probability of $95\%$.

We can perform inference on $\Psi_{2}$ by leveraging the theory from Section~\ref{subsec:inference_smooth_functionals}. Let $\widehat{\epsilon}_i = Y_i - \widehat{h}_{n,\lambda}(X_i)$ be the RWNN regression residuals.
Notice that 
\begin{equation*}
    \Psi_2(\widehat{h}_{n,\lambda})
    = \widehat{\beta}_{K_n,n,\lambda}^\top C_{g,K_n} \widehat{\beta}_{K_n,n,\lambda} ,
    \quad \textnormal{where} \quad
    C_{g,K_n} := \int_{-1}^{1} g_{K_n}(x) g_{K_n}(x)^\top \de x .
\end{equation*}
Since $g_{K_n}$ is a known RWNN sieve basis, $C_{g,K_n}$ can be computed numerically by an appropriate quadrature rule; we term this approach the ``sandwich-type estimator'' (SNW).
Let
\begin{equation*}
    \Psi'_{\widehat{h}_{n,\lambda}}(g_{K_n}) 
    := 
    \big( \Psi'_{\widehat{h}_{n,\lambda}}(g_{1,K_n}),\ldots, \Psi'_{\widehat{h}_{n,\lambda}}(g_{K_n,K_n}) \big)^\top ,
\end{equation*}
so that for $\Psi_2$ we have $\Psi'_{\widehat{h}_{n,\lambda}}(g_{K_n})=2 C_{g,K_n} \widehat{\beta}_{K_n,n,\lambda}$.  The empirical counterpart of the ridge-adjusted sieve Riesz representer defined in \eqref{eq:ridge_adjusted_sieve_riesz} is hence 
\begin{equation*}
\label{eq:def_ridge-adjusted_representer_psi2}
    \widehat{\psi}_{K_n,\lambda}(x)
    \ := \
    g_{K_n}(x)^\top \widehat{\Gamma}_{K_n,n,\lambda}^{-1} \Psi'_{\widehat{h}_{n,\lambda}}(g_{K_n}),
\end{equation*}
or, equivalently, $\widehat{\psi}_{K_n,\lambda}(x) := 2g_{K_n}(x)^\top \widehat{\Gamma}_{K_n,n,\lambda}^{-1} C_{g,K_n} \widehat{\beta}_{K_n,n,\lambda}$.
Let
\begin{equation*}
\label{eq:inference_omega}
    \widehat{\Omega}_{K_n,n,\lambda} 
    := 
    \widehat{\Gamma}_{K_n,n,\lambda}\inv \left( n\inv \sum_{i=1}^n \widehat{\epsilon}_i^{\:2} \, g_{K_n}(X_i) g_{K_n}(X_i)^\top \right) \widehat{\Gamma}_{K_n,n,\lambda}\inv.
\end{equation*}
The variance estimator associated with the empirical ridge-adjusted Riesz representer can be written as 
\begin{align*}
    \widehat{\sigma}^2_{\Psi_2, \text{SNW}} \: &:= \frac{1}{n} \sum_{i=1}^n \widehat{\epsilon}_i^{\,2} \widehat{\psi}_{\Psi_2,\lambda}(X_i)^2  
    =\frac{1}{n} \sum_{i=1}^n \widehat{\epsilon}_i^{\,2} \big( g_{K_n}(X_i)^\top \widehat{\Gamma}_{K_n,\lambda}^{-1} \Psi'_{\widehat{h}_{n,\lambda}}(g_{K_n}) \big)^2\\
    &=\Psi'_{\widehat{h}_{n,\lambda}}(g_{K_n})^\top \widehat{\Omega}_{K_n, n, \lambda}\Psi'_{\widehat{h}_{n,\lambda}}(g_{K_n})
    \: = \: 4 \widehat{\beta}_{K_n,n,\lambda}^\top C_{g,K_n}  \widehat{\Omega}_{K_n, n, \lambda} C_{g,K_n} \widehat{\beta}_{K_n,n,\lambda} ,
\end{align*}
which is then the sandwich-type variance estimator. 

Alternatively, since ${X}\sim U([-1,1])$, the Riesz representer of $\Psi'_h$  with respect to $\norm{\cdot}_{L^2(X)}$ inner product is, for $\Psi\equiv\Psi_2$, directly  
$
    \psi(x) = 4 h(x)
$.
Indeed, for every direction $v\in L^2(X)$, $\Psi'_{h}(v)=2\int_{-1}^{1}h(x)v(x)\de x=4\E[h(X)v(X)]=\langle v, 4h \rangle_{L^2(X)} $. 
Thus, we may consider the direct plug-in functional and variance estimators based on the unregularized $L^2(X)$ Riesz representer $\psi=4h$, given by, respectively,
\begin{equation*}
    \Psi_2(\widehat{h}_{n,\lambda})
    = \int_{-1}^{1} \widehat{h}_{n,\lambda}(x)^2 dx 
    \quad \textnormal{and} \quad
    \widehat{\sigma}^2_{\Psi_2,\mathrm{RSZ}}:=\frac{16}{n}\sum_{i=1}^n\left(\widehat{\epsilon}_i\widehat{h}_{n,\lambda}(X_i)\right)^2 ,
\end{equation*}
where the integral can again be computed by a suitable quadrature rule. We call these the Riesz-type functional and variance estimators (RSZ).

An $100(1-\alpha)\%$ confidence interval for $\Psi_2(h)$ can thus be constructed as
\begin{equation*}
    \text{CI}_{1-\alpha,b}(\Psi_2) 
    = 
    \left[ \Psi_2(\widehat{h}_{n,\lambda}) - \sqrt{\frac{ \widehat{\sigma}^2_{\Psi_2,b} }{n}}\, z_{1-\alpha/2} , \; \Psi_2(\widehat{h}_{n,\lambda}) + \sqrt{ \frac{ \widehat{\sigma}^2_{\Psi_2,b} }{n}}\, z_{1-\alpha/2} \right] ,
\end{equation*}
where $b\in\{\text{SNW},\text{RSZ}\}$ and $z_{\alpha}$ denotes the $\alpha$-quantile of the standard Gaussian distribution. 
Table~\ref{tab:coverage_ridge_lambda_1e-06} summarizes simulation results based on three different RWNN specifications with $\tanh$ activation function. 
We use $\text{MC} = 1000$ replications of $X_i \overset{\textnormal{i.i.d.}}{\sim} \mathcal{U}([-1,1])$ and four combinations of sample size $n$ and RWNN sieve dimension $K_n$. The random weights $\{ A_k, B_k \}_{k=1}^{K_n}$ are resampled at each Monte Carlo replication: 
for the Gaussian-Gaussian specification, $A_k \overset{\text{i.i.d.}}{\sim} \mathcal{N}(0,\sigma_A^2)$ and $B_k \overset{\text{i.i.d.}}{\sim} \mathcal{N}(0,\sigma_B^2)$, with $\sigma_A = 5$ and $\sigma_B = 1$;
for the Gaussian-Uniform, $A_k \overset{\text{i.i.d.}}{\sim} \mathcal{N}(0,\sigma_A^2)$ and $B_k \overset{\text{i.i.d.}}{\sim} \mathcal{U}([-\sigma_B, \sigma_B])$, with $\sigma_A = 5$ and $\sigma_B = 2$; 
for the Uniform-Uniform,
$A_k \overset{\text{i.i.d.}}{\sim} \mathcal{U}([-\sigma_A, \sigma_A])$ and $B_k \overset{\text{i.i.d.}}{\sim} \mathcal{U}([-\sigma_B, \sigma_B])$, with $\sigma_A = 10$ and $\sigma_B = 5$. 
In all settings, we use a fixed ridge regularization factor of $\lambda_{K_n,n} = 10^{-6}$ and a trapezoidal integration scheme to compute both the sandwich- and Riesz-type estimators.\footnote{In either case, we integrate using an equally-spaced grid of 1001 points (so that the zero belongs to the grid) over $[-1, 1]$. In our simulations, we find that this choice is numerically robust and produces no meaningful integration artifacts unless we set $\lambda = 10^{-2}$, a choice that yields substantial bias and poor CI coverage. Even then, the differences in bias are negligible.} Appendix~\ref{appendix:additional_tables} contains results for simulations made with $\lambda \in \{ 10^{-8}, 10^{-4}, 10^{-2} \}$.
For both variance estimators, the coverage probabilities approach the nominal level as $K_n$ and $n$ increase, and interval lengths are extremely similar throughout. In the small-sample regimes where $n=100$ and $K_n=15$ or $n=300$ and $K_n=20$, we note that $\widehat{\sigma}^2_{\Psi_2, \text{SNW}}$ appears to be marginally more robust. At the largest sample-sieve combination, the two procedures produce very similar results, with coverage close to the nominal level for all three functions.

\begin{table}
\caption{Asymptotic inference results for $\Psi_{2}(\widehat{h}_{n, \lambda})$ with $\lambda = 10^{-6}$.}
\label{tab:coverage_ridge_lambda_1e-06}
\centering
\setlength\tabcolsep{4pt}
\setlength\extrarowheight{1pt}
\setlength\extracolsep{0.1pt}
\linespread{1.}\selectfont\centering
\small
\begin{tabular}{llcccccccccc}
\toprule
 &  &  & \multicolumn{3}{c}{Gaussian--Gaussian} & \multicolumn{3}{c}{Gaussian--Uniform} & \multicolumn{3}{c}{Uniform--Uniform} \\
    \cmidrule(lr){4-6} \cmidrule(lr){7-9} \cmidrule(lr){10-12}
 $n$ & $K_n$ & Method & Bias & Coverage & Length & Bias & Coverage & Length & Bias & Coverage & Length \\
\midrule
\multicolumn{12}{l}{$h = g_1$ ~\raisedrule[0.2em]{0.2pt}} \\
\multirow[c]{2}{*}{100} & \multirow[c]{2}{*}{15} & RSZ & 0.089 & 0.891 & 0.242 & 0.086 & 0.896 & 0.244 & -0.022 & 0.877 & 0.269 \\
 &  & SNW & 0.089 & 0.901 & 0.254 & 0.086 & 0.898 & 0.258 & -0.022 & 0.900 & 0.287 \\
\multirow[c]{2}{*}{300} & \multirow[c]{2}{*}{20} & RSZ & 0.004 & 0.930 & 0.146 & 0.002 & 0.929 & 0.146 & -0.040 & 0.928 & 0.151 \\
 &  & SNW & 0.004 & 0.936 & 0.149 & 0.002 & 0.939 & 0.150 & -0.040 & 0.934 & 0.155 \\
\multirow[c]{2}{*}{1000} & \multirow[c]{2}{*}{30} & RSZ & 0.017 & 0.938 & 0.081 & 0.016 & 0.939 & 0.081 & 0.012 & 0.939 & 0.082 \\
 &  & SNW & 0.017 & 0.935 & 0.082 & 0.016 & 0.941 & 0.082 & 0.012 & 0.937 & 0.082 \\
\multirow[c]{2}{*}{3000} & \multirow[c]{2}{*}{40} & RSZ & -0.008 & 0.950 & 0.047 & -0.008 & 0.951 & 0.047 & -0.009 & 0.949 & 0.047 \\
 &  & SNW & -0.008 & 0.951 & 0.047 & -0.008 & 0.954 & 0.047 & -0.009 & 0.951 & 0.047 \\
\multicolumn{12}{l}{$h = g_2$ ~\raisedrule[0.2em]{0.2pt}} \\
\multirow[c]{2}{*}{100} & \multirow[c]{2}{*}{15} & RSZ & 0.095 & 0.891 & 0.152 & 0.104 & 0.894 & 0.152 & 0.094 & 0.876 & 0.152 \\
 &  & SNW & 0.095 & 0.914 & 0.160 & 0.104 & 0.920 & 0.160 & 0.094 & 0.911 & 0.161 \\
\multirow[c]{2}{*}{300} & \multirow[c]{2}{*}{20} & RSZ & 0.039 & 0.935 & 0.091 & 0.041 & 0.937 & 0.091 & 0.043 & 0.933 & 0.091 \\
 &  & SNW & 0.039 & 0.939 & 0.093 & 0.041 & 0.943 & 0.093 & 0.043 & 0.942 & 0.093 \\
\multirow[c]{2}{*}{1000} & \multirow[c]{2}{*}{30} & RSZ & 0.010 & 0.943 & 0.051 & 0.012 & 0.942 & 0.051 & 0.016 & 0.939 & 0.051 \\
 &  & SNW & 0.010 & 0.945 & 0.051 & 0.012 & 0.947 & 0.051 & 0.016 & 0.942 & 0.051 \\
\multirow[c]{2}{*}{3000} & \multirow[c]{2}{*}{40} & RSZ & 0.002 & 0.947 & 0.029 & 0.003 & 0.949 & 0.029 & 0.005 & 0.947 & 0.029 \\
 &  & SNW & 0.002 & 0.948 & 0.030 & 0.003 & 0.947 & 0.030 & 0.005 & 0.951 & 0.029 \\
\multicolumn{12}{l}{$h = g_3$ ~\raisedrule[0.2em]{0.2pt}} \\
\multirow[c]{2}{*}{100} & \multirow[c]{2}{*}{15} & RSZ & 0.085 & 0.913 & 0.159 & 0.094 & 0.907 & 0.159 & 0.105 & 0.891 & 0.157 \\
 &  & SNW & 0.085 & 0.928 & 0.166 & 0.094 & 0.927 & 0.166 & 0.105 & 0.915 & 0.165 \\
\multirow[c]{2}{*}{300} & \multirow[c]{2}{*}{20} & RSZ & 0.034 & 0.923 & 0.095 & 0.034 & 0.922 & 0.095 & 0.039 & 0.923 & 0.095 \\
 &  & SNW & 0.034 & 0.935 & 0.097 & 0.034 & 0.938 & 0.097 & 0.039 & 0.931 & 0.097 \\
\multirow[c]{2}{*}{1000} & \multirow[c]{2}{*}{30} & RSZ & 0.020 & 0.953 & 0.053 & 0.020 & 0.949 & 0.053 & 0.023 & 0.944 & 0.053 \\
 &  & SNW & 0.020 & 0.950 & 0.053 & 0.020 & 0.949 & 0.053 & 0.023 & 0.945 & 0.053 \\
\multirow[c]{2}{*}{3000} & \multirow[c]{2}{*}{40} & RSZ & 0.001 & 0.943 & 0.031 & 0.001 & 0.943 & 0.031 & 0.003 & 0.941 & 0.031 \\
 &  & SNW & 0.001 & 0.941 & 0.031 & 0.001 & 0.942 & 0.031 & 0.003 & 0.943 & 0.031 \\
\bottomrule
\end{tabular}
\\[0.7em]
\justifying
\noindent
{\footnotesize
    \textit{Notes:} 
    Bias, coverage rate, and interval length of asymptotic confidence intervals for $\Psi_2(h)$ estimated using $\text{MC} = 1000$ replications.
    The method column reports the type of functional and variance estimators (sandwich-type or Riesz-type) used to compute asymptotic confidence intervals. 
    The Gaussian-Gaussian, Gaussian-Uniform, and Uniform-Uniform column groupings indicate random sampling laws of $A_k$ and $B_k$, respectively.
    Bias is scaled by a factor of 10 to improve legibility. 
}
\end{table}

\subsection{Bootstrap inference}
\label{subsec:sims_bootstrap}

While our theory applies to both smooth and linear functionals, it generally requires explicit computation of the ridge-adjusted Riesz representer. The bootstrap, by comparison, is an effective and flexible approach that does not generally require explicit computation of functional or directional quantities. It has been widely applied in nonparametric statistics; see, for example, \cite{Horowitz2009,Hall2013} and \cite{cavaliereBootstrapInferencePresence2024} for a recent discussion of bias-correction refinements. 
Although our inferential theory does not encompass bootstrap validity, we believe it is nonetheless interesting to showcase its empirical performance, especially for practitioners.
In this section, we thus focus on a straightforward nonparametric residual bootstrap construction; in Appendix~\ref{appendix:additional_simulations} we provide detailed simulations for the point evaluation functional and apply two additional bootstrapping schemes---a Gaussian wild bootstrap and the score bootstrap of \cite{Chen2018}.
For the sake of space, we limit ourselves to two cases, both involving $g_1$, as defined above: we perform inference on the evaluation functional at $x=0$, $\Psi_0(g_1) = g_1(0)$, and the expected interaction functional $\Psi_1(g_1) = \E[g_1(X) (1 + X)]$.

To implement the residual construct, we generate independent draws of the response variable as $Y_i^* := \widehat{h}_{n,\lambda}(X_i) + \epsilon_i^*$, $i=1,\ldots, n$. 
Let $\Prob_{\widehat{\epsilon},n}$ denote the empirical distribution of the regression residuals.
The innovations are sampled independently from the centered empirical residual distribution, meaning $\epsilon_i^* := \widehat{\epsilon}_{i}^* - \overline{\widehat{\epsilon}}$, with
$\widehat{\epsilon}_i^* \overset{\mathrm{i.i.d.}}{\sim} \Prob_{\widehat{\epsilon},n}$ and $\overline{\widehat{\epsilon}}:= n\inv \sum_{i=1}^n \widehat{\epsilon}_i$. 
Let $\widehat{h}_{n,\lambda}^{*}$ denote the RWNN estimator obtained by refitting the model to the corresponding bootstrap sample (without changing the sieve or the ridge penalty). 
Let $z^{*}_{\alpha}(\Psi)$ denote the conditional $\alpha$-quantile of the bootstrap distribution of $\Psi(\widehat{h}_{n,\lambda}^{*})$, given the original sample. The corresponding residual percentile-bootstrap confidence interval is
\begin{equation*}
    \text{CI}_{1-\alpha}(\Psi) 
    \ := \ 
    \left[ z^{*}_{\alpha/2}, \; z^{*}_{1-\alpha/2} \right], 
    \quad \Psi \in \{ \Psi_0, \Psi_1 \}.
\end{equation*}
Since $\Psi_0$ is irregular with respect to the $L^2(X)$ norm, contrary to $\Psi_1$, we also wish to compare their empirical Rayleigh stability factors to ascertain if they also align with our directional inference theory. Recall from Section~\ref{subsubsec:RWNN_functional_directions} that 
\begin{equation*}
    \mathcal{R}_{K_n,\lambda}(\alpha_{K_n})
    = \frac{\alpha_{K_n}^{\top}\Gamma_{K_n,\lambda}^{-1}\Gamma_{K_n}\Gamma_{K_n,\lambda}^{-1}\alpha_{K_n}}{\alpha_{K_n}^{\top}\Gamma_{K_n,\lambda}^{-1}\alpha_{K_n}} .
\end{equation*}
In our setup, it is simple to show that $\alpha_{K_n}^{(0)} = g_{K_n}(0)$ for $\Psi_0$ and $\alpha_{K_n}^{(1)} = \E[ g_{K_n}(X_i)(1+X_i)]$ for $\Psi_1$. Therefore, the empirical counterpart of $\mathcal{R}_{K_n,\lambda}(\alpha_{K_n})$ can be readily computed by plug-in quantities, allowing us to also quantify the directional conditioning inherent to the inferential problem.
 
\begin{figure}[t!]
    \centering
    \begin{subfigure}{0.98\textwidth}
    \centering
        \includegraphics[width=\textwidth]{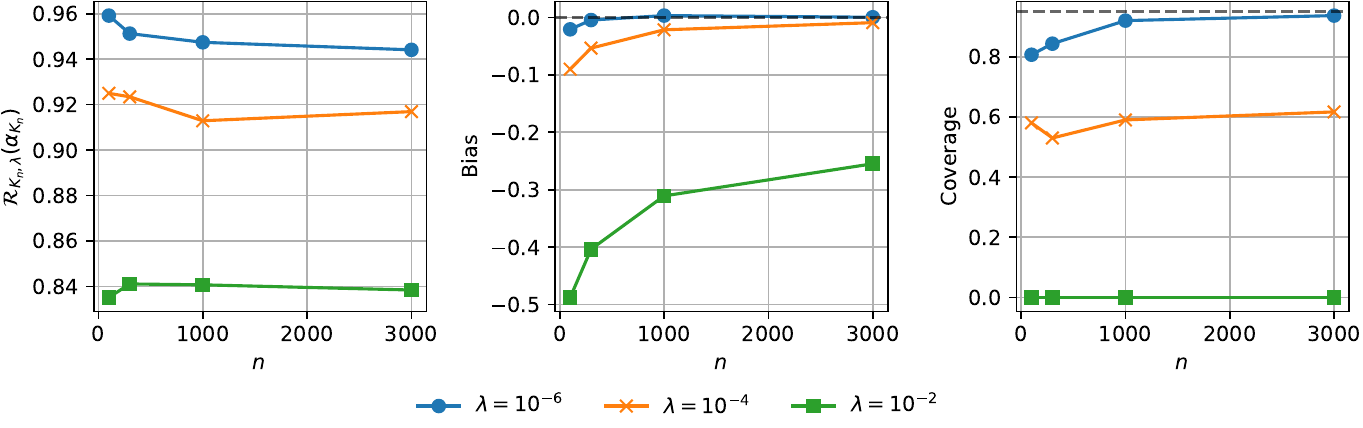}
        \caption{$\Psi_0(g_1) = g_1(0)$}
    \end{subfigure}
    \\[1em]
    \vspace{1em}
    \begin{subfigure}{0.98\textwidth}
    \centering
        \includegraphics[width=\textwidth]{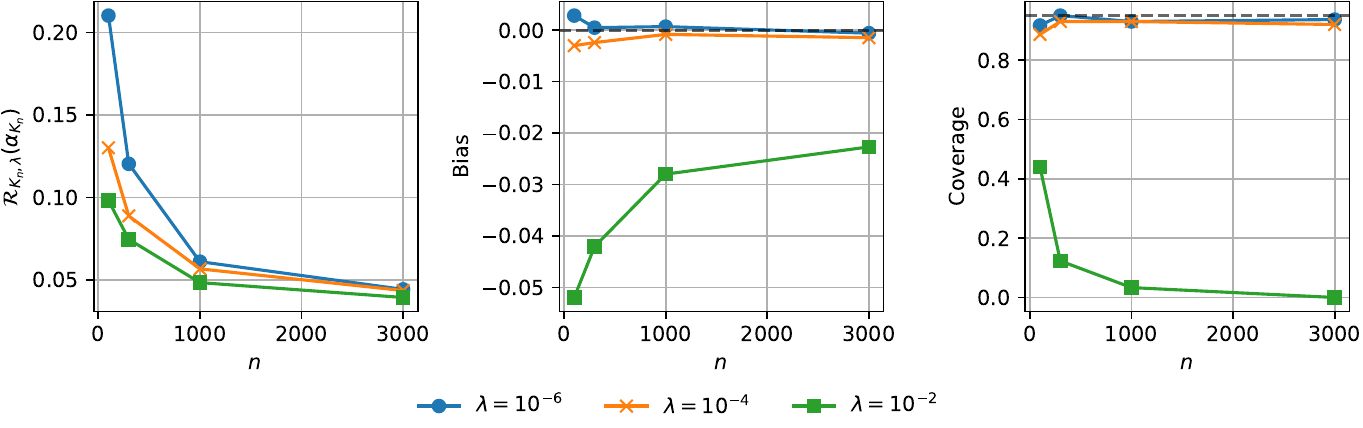}
        \caption{$\Psi_1(g_1) = \E[g_1(X) (1 + X)]$}
    \end{subfigure}
    \\[0.5em]
    \caption{Bootstrap inference results for functionals of $g_1$ with $X_i \overset{\text{i.i.d.}}{\sim} \mathcal{U}([-1,1])$, $\epsilon_i \sim \mathcal{N}(0, \sigma_\epsilon^2)$, $\sigma_\epsilon^2 = 0.04$. RWNNs use $\tanh$ activation, $A_k \overset{\text{i.i.d.}}{\sim} \mathcal{N}(0, \sigma_A^2)$ and $B_k \overset{\text{i.i.d.}}{\sim} \mathcal{N}(0, \sigma_B^2)$, with $\sigma_A = 5$ and $\sigma_B = 1$. Ridge penalty is set as $\lambda \in \{ 10^{-6}, 10^{-4}, 10^{-2} \}$. Lines indicate averages across Monte Carlo replications.}
    \label{fig:inference_rayleigh_bias_cover}
\end{figure}

In the data generating process we use again $X_i \overset{\text{i.i.d.}}{\sim} \mathcal{U}([-1,1])$, $\epsilon_i \sim \mathcal{N}(0, \sigma_\epsilon^2)$, $\sigma_\epsilon^2 = 0.04$. 
The RWNN uses $\tanh$ activation, $A_k \overset{\text{i.i.d.}}{\sim} \mathcal{N}(0, \sigma_A^2)$ and $B_k \overset{\text{i.i.d.}}{\sim} \mathcal{N}(0, \sigma_B^2)$, with $\sigma_A = 5$ and $\sigma_B = 1$, and the weights are not normalized. In Appendix~\ref{appendix:additional_figures}, we provide additional figures for Gaussian-Uniform and Uniform-Uniform weight sampling design.
We set the sample size to $n = \{ 100, 300, 1000, 3000 \}$ and the sieve size to $K_n = \{ 15, 20, 30, 40 \}$, following the simulation designs for the mean-squared functional in Section~\ref{subsec:sims_statistical_inference} above. For both functionals, $\text{MC} = 300$ and $\mathfrak{M} = 300$ Monte Carlo and bootstrap resamplings are used, and the RWNN weights are resampled at each new data draw. 
Similarly to the other simulations, we compare results at different fixed values for the ridge regularization parameter, and use $\lambda \in \{ 10^{-6}, 10^{-4}, {10^-2} \}$.
The results are shown in Figure~\ref{fig:inference_rayleigh_bias_cover}, where we plot values averaged across all Monte Carlo replications.
Panel~(a) shows that the empirical Rayleigh quotients for the evaluation functional $\Psi_0$ are all well-behaved as $n$ grows, and across different scalings of the ridge penalty. 
On the other hand, Panel~(b) indicates that, even though the Rayleigh quotients appear to be decaying as $n$ increases, inference is not adversely affected. Indeed, we can see that bias for $\Psi_1(g_1)$ is substantially smaller than that of the evaluation functional, and CIs retain good coverage across sample sizes with $\lambda = 10^{-6}$.
Setting $\lambda = 10^{-4}$, shows that the point-evaluation functional is much more sensitive to regularization than $\Psi_1$, which is natural given the graph of $g_1$.
Finally, in both panels, we note that the RWNN estimator shows large bias and very poor coverage when we set $\lambda = 10^{-2}$: this is expected, as our asymptotic theory indeed emphasizes that an appropriate choice of $\lambda$ is key in obtaining valid RWNN inference.

\section{Conclusion}
\label{sec:Conclusion}

We have developed approximation, estimation, and inference results for random weight neural network sieves that explicitly accommodate asymptotic ill-conditioning of their population covariance matrices. Under polynomial covering and Lipschitz conditions, we establish spectral tail bounds that imply vanishing smallest eigenvalues and control the ridge effective dimension. We also derive improved $L^2$ approximation rates using control neighbors integration, as well as sup-norm approximation bounds using stratified sampling. For ridge-regularized least squares, our results provide $L^2$ and uniform estimation bounds under appropriate stability conditions, together with conditional asymptotic normality for linear and smooth functionals and consistent ridge-adjusted directional variance estimation. In particular, directional inference can remain valid even when the population feature covariance matrix is singular. Beyond the squared loss setting, we also develop a generic sieve framework for estimation and smooth-functional inference: it accommodates other loss functions, including absolute, Huber, quantile, and logistic losses, provided the corresponding identification, curvature, and empirical process conditions are satisfied.

The results we obtain rely on explicit approximation, regularization bias, and stability conditions, whose verification may be challenging in general random feature designs. Some stratified constructions also require knowledge of the source representation. Developing data-driven sampling schemes and verifying the inferential conditions for broader feature distributions are natural directions for further research. 
As evidenced by our numerical results in Section~\ref{section:simulations}, the choice of the ridge regularization parameter $\lambda \geq 0$ is a fundamental determinant of the inferential properties of the RWNN estimator. In practice, the choice of $\lambda$ is non-trivial, and, even though the use of out-of-sample and cross-validation methods is natural \citep{yangConsistencyCrossValidation2007}, a comprehensive theoretical treatment of model selection procedures for RWNN regression models would be highly desirable as a next step.
Another important extension concerns bias correction procedures for inference, particularly in settings where ridge regularization or random feature approximation induces substantial finite sample bias.
From a practical point of view, we note that our numerical experiments focus on bootstrap procedures without an explicit theory for their validity or an explicit bias correction. Since sieve approximation and ridge regularization can introduce bias whose magnitude also depends on the realized hidden-layer parameters, adapting bootstrap and bias correction methods to RWNN inference is a natural direction for future research \citep{Hall2013,calonicoEffectBiasEstimation2018,cavaliereBootstrapInferencePresence2024,cavaliereImprovedInferenceNonparametric2026}. Another important direction we would like to pursue in the future is to establish Gaussian approximations for the supremum of the normalized estimation error, which could provide a theoretical basis for constructing asymptotically valid uniform confidence bands for RWNN regression functions.

\bibliography{uniform_rwnn.bib,GOLibrary.bib}
\clearpage

\phantomsection
\addcontentsline{toc}{section}{Appendix}

\stopcontents[mainsections]

\appendix

\counterwithin*{table}{section}
\renewcommand{\thetable}{\Alph{section}.\arabic{table}}
\counterwithin*{figure}{section}
\renewcommand{\thefigure}{\Alph{section}.\arabic{figure}}

\begin{center}
    {\LARGE Appendix}\\[1em]
    {\Large%
    	Statistical Inference with Ill-Conditioned Random Weight Neural Network Sieves
    }\\[1em]
    {\large%
        Giovanni Ballarin 
        \hspace{1em}
        Lyudmila Grigoryeva
        \hspace{1em}
        Lukas Gonon
    }\\[3em]
\end{center}

\startcontents[appendices]
\printcontents[appendices]{l}{1}{\section*{Appendices}\setcounter{tocdepth}{2}}

\section{Proofs for RWNN approximation}
\label{app:rwnn_approximation}

\subsection{$L^2$ approximation and control-neighbors weights}
\label{app:L2_approximation_proofs}
In this appendix, we recall the control-neighbors approximation theorem of \citet{lelucSpeedingMonteCarlo2024}, verify the conditions needed for its application, and present the proofs of the approximation results in Subsection~\ref{subsec:approximation_RWNNs_Lipschitz}, including the bounds on the control-neighbors weights.

\begin{theorem}[\citealp{lelucSpeedingMonteCarlo2024}, 
Theorem 1]
\label{theorem:leluc_al_2024_cnn}
    Let Assumptions~\ref{assumption:X_domain},~\ref{assumption:integral_representation_phi} and \ref{assumption:regular_sampling} hold, and let $\theta_k=(A_k^\top,B_k)^\top$, $k=1,\dots,K$, be i.i.d.\ from $\pi_\theta$. Then for $K\geq 4$, the \textit{control-neighbors estimate}
    \begin{equation*}
        {h}^{\theta,\omega\texttt{-cnn}}_{K}(x) := \sum_{k=1}^K \omega_{k,K} \phi_x(\theta_k),
    \end{equation*}
    where $\{\omega_{k,K}\}_{k=1}^K$ are the 
    \textit{nearest-neighbor weights} introduced in Section~2 in 
    \citet{lelucSpeedingMonteCarlo2024}, satisfies
    \begin{equation*}
        \mathbb{E}_{A,B}\big[|h(x)-{h}^{\theta,\omega\texttt{-cnn}}_{K}(x)|^2\big]^{1/2} \leq C_{\texttt{cnn}}(x) K^{-1/2-\beta/(d+1)},
    \end{equation*}
    where
    \begin{equation*}
        C_{\texttt{cnn}}(x):= L_{\phi,x} C_0^{-\beta/(d+1)} \sqrt{\Gamma\left(\frac{2\beta}{d+1} +2 \right)} \left( 2 \sqrt{\frac{\beta}{d+1} +1} + 2^{\beta/(d+1)}\sqrt{10 \frac{C_1}{C_0}} \right),
    \end{equation*}
    where $C_0,C_1>0$ are constants such that $C_0 r^{d+1}\leq \pi_\theta\bigl(B_r^{\rho_\Pi}(\theta)\bigr)\leq C_1 r^{d+1}$ for any $\theta\in\Pi$ and $r\in(0,D_\Pi]$, with $D_\Pi=\sup_{\theta_1,\theta_2\in\Pi}\rho_\Pi(\theta_1,\theta_2)<\infty$.
\end{theorem}

We note that, by Assumption~\ref{assumption:regular_sampling}~(iii), $B_{r/M_\rho}^{\|\cdot\|}(\theta)\subseteq B_r^{\rho_\Pi}(\theta) \subseteq B_{r/m_\rho}^{\|\cdot\|}(\theta)$.  Hence Assumption~\ref{assumption:regular_sampling}~(i)-(ii) implies that, for all sufficiently small $r>0$, $c_b c_\pi V_{d+1}M_\rho^{-(d+1)}r^{d+1}\leq\pi_\theta\bigl(B_r^{\rho_\Pi}(\theta)\bigr)\leq c_uV_{d+1}m_\rho^{-(d+1)}r^{d+1}$,  uniformly in $\theta\in\Pi$. Since $\Pi$ is bounded under $\rho_\Pi$, these local bounds imply, after adjusting the constants, the metric small-ball condition required in Theorem~\ref{theorem:leluc_al_2024_cnn}.

\begin{proof}[Proof of Proposition~\ref{prop:product_implies_holder}]
For $\theta_1,\theta_2\in\Pi$ and $x\in\mathcal{X}$,
\begin{align*}
|\phi_x(\theta_1)-\phi_x(\theta_2)|
&\leq M_f|\sigma(A_1^\top x+B_1)-\sigma(A_2^\top x+B_2)|+M_\sigma|f(\theta_1)-f(\theta_2)|\\
&\leq M_f L_\sigma|(A_1-A_2)^\top x+(B_1-B_2)|
+M_\sigma L_f\rho_\Pi(\theta_1,\theta_2)^\beta\\
&\leq M_f L_\sigma L_\rho\rho_\Pi(\theta_1,\theta_2)
+M_\sigma L_f\rho_\Pi(\theta_1,\theta_2)^\beta\\
&\leq\bigl(M_f L_\sigma L_\rho D_\Pi^{1-\beta}
+M_\sigma L_f\bigr)\rho_\Pi(\theta_1,\theta_2)^\beta
=L_\phi\,\rho_\Pi(\theta_1,\theta_2)^\beta,
\end{align*}
where the first inequality uses the triangle inequality together with Assumption~\ref{assumption:direct_source_condition_holder_product}, the second uses the same assumption and Assumption~\ref{assumption:activation_Lipschitz}, the third uses \eqref{eq:rho_compatibility}, and the fourth uses $\rho_\Pi(\theta_1,\theta_2)\leq D_\Pi$ to bound $\rho_\Pi(\theta_1,\theta_2)\le D_\Pi^{1-\beta}\rho_\Pi(\theta_1,\theta_2)^\beta$. This proves the claim.
\end{proof}
\begin{proof}[Proof of Theorem~\ref{theorem:sharp_l2}]
By the hypotheses of the theorem, Proposition~\ref{prop:product_implies_holder} yields that 
Assumption~\ref{assumption:direct_source_condition_holder_product} implies Assumption~\ref{assumption:integral_representation_phi} with the integrand $\phi_x(\theta):=f(A,B)\sigma(A^\top x+B)$ being $\beta$-H\"older in $\theta$ under $\rho_\Pi$, uniformly in $x\in\mathcal{X}$, with constant
\begin{equation*}
L_{\phi,x}\leq L_\phi
:=M_f L_\sigma L_{\rho} D_\Pi^{1-\beta}+M_\sigma L_f,
\end{equation*}
where $D_\Pi$ and 
$M_\sigma$ are as in Assumption~\ref{assumption:direct_source_condition_holder_product}.

Assumption~\ref{assumption:regular_sampling} provides the measure conditions of Theorem~\ref{theorem:leluc_al_2024_cnn}. Applying Theorem~\ref{theorem:leluc_al_2024_cnn} and replacing $L_{\phi,x}$ by the uniform bound $L_\phi$ yields, for each $x\in\mathcal{X}$,
\begin{equation}
\label{eq:cnn_pointwise}
    \mathbb{E}_{A,B}\big[|h(x)-
    h^{\theta,\omega\texttt{-cnn}}_K(x)|^2\big]
    \leq C_{\texttt{cnn}}^2 K^{-1-2\beta/(d+1)},
\end{equation}
where $C_{\texttt{cnn}}>0$ depends only on $d$, $\beta$, $L_\phi$, $c_b$, $c_u$, $c_\pi$, and 
hence only on $d$, $\beta$, $M_f$, $L_f$, $L_\sigma$, $L_\rho$, $M_\sigma$, $c_b$, $c_u$, $c_\pi$, $r_0$, $m_{\rho}$, $M_{\rho}$, $D_\Pi$.

Since the right-hand side of \eqref{eq:cnn_pointwise} does not depend on $x$, Fubini's theorem yields
\begin{equation*}
    \|h(X)-h^{\theta,\omega\texttt{-cnn}}_K(X)
    \|^2_{L^2(X,A,B)}
    =\mathbb{E}_{A,B}\left[
    \mathbb{E}_X\left[|h(X)-
    h^{\theta,\omega\texttt{-cnn}}_K(X)|^2\right]
    \right]
    \leq C_{\texttt{cnn}}^2 K^{-1-2\beta/(d+1)}.
\end{equation*}
Taking a square root yields the stated bound with 
$C_q=C_{\texttt{cnn}}$.

Finally, by \eqref{eq:cnn_rwnn_approximant}, $h^{\theta,\omega\texttt{-cnn}}_K \in \mathcal{H}_K$, with output coefficients $W_k:=\omega_{k,K}f(A_k,B_k)$, $k=1,\dots,K$. The weights $\omega_{k,K}$ are measurable functions of $\{\theta_j\}_{j=1}^K$ by construction \citep{lelucSpeedingMonteCarlo2024}, and $f$ is measurable. Hence $W$ is measurable with respect to $\sigma(\theta_1,\dots,\theta_K)$, completing the proof.
\end{proof}
\begin{proof}[Proof of Theorem~\ref{thm:prob-approx}]
By Theorem~\ref{theorem:sharp_l2},
\begin{equation*}
\|h(X)-h^{\theta,W}_K(X)\|^2_{L^2(X,A,B)}
\leq C_q^2 K^{-1-2\beta/(d+1)}.
\end{equation*}
Since $\|\cdot\|^2_{L^2(X,A,B)}=\mathbb{E}_{A,B}[\|\cdot\|^2_{L^2(X)}]$, this gives
\begin{equation*}
\mathbb{E}_{A,B}
\big[
\|h(X)-h^{\theta,W}_K(X)\|_{L^2(X)}^2
\big]
\leq C_q^2 K^{-1-2\beta/(d+1)}.
\end{equation*}
Applying Markov's inequality yields
\begin{equation*}
\mathbb{P}_{A,B}
\left(
\|h(X)-h^{\theta,W}_K(X)\|_{L^2(X)}^2
>
C_q^2 K^{-1-2\beta/(d+1)}\varepsilon^{-1}
\right)
\leq\varepsilon.
\end{equation*}
Taking a square root on both sides of the inequality inside the probability proves \eqref{eq:prob_bound}.
\end{proof}
\begin{proof}[Proof of Corollary~\ref{cor:rwnn_projection_rate}]
Fix $\varepsilon\in(0,1)$. By Theorem~\ref{thm:prob-approx}, there exists a $\sigma(A,B)$-measurable weight vector $W$ such that
\begin{equation*}
\mathbb{P}_{A,B} \left(\mathcal{E}_{K,\varepsilon}
\right)\geq 1-\varepsilon,
\end{equation*}
where
\begin{equation*}
\mathcal{E}_{K,\varepsilon}
:=\left\{
\|h-h_K^{\theta,W}\|_{L^2(X)}
\leq
C_q K^{-1/2-\beta/(d+1)}\varepsilon^{-1/2}
\right\},
\end{equation*}
and $C_q>0$ is the constant of Theorem~\ref{theorem:sharp_l2}. On 
$\mathcal{E}_{K,\varepsilon}$,
\begin{equation*}
\inf_{g\in\mathcal{H}_K}
\|h-g\|_{L^2(X)}
\leq
\|h-h_K^{\theta,W}\|_{L^2(X)},
\end{equation*}
and hence
\begin{equation*}
\mathbb{P}_{A,B}
\left(
\inf_{g\in\mathcal{H}_K}
\|h-g\|_{L^2(X)}
>
C_q K^{-1/2-\beta/(d+1)}\varepsilon^{-1/2}
\right)
\leq\varepsilon,
\end{equation*}
which proves the $O_P$ rate.
\end{proof}

\begin{proof}[Proof of Lemma~\ref{lem:nn_l2_bound}]
Let $C_{K,k}$ denote the Voronoi cell of $\theta_k$ with $V_{K,k}:=\pi_\theta(C_{K,k})$.
Notice that since $\{C_{K,k}\}_{k=1}^K$ forms a partition of $\Pi$, one has $V_{K,k} \ge 0$ and $\sum_{k=1}^K V_{K,k} = 1$. Hence
\begin{equation}
\label{eq:Vkbound}	
\sum_{k=1}^K V_{K,k}^2
\le
\max_{1 \le k \le K} V_{K,k}.
\end{equation}

By Assumption~\ref{assumption:direct_source_condition_holder_product}, it holds that
\begin{equation*}
    \|W\|_1 = \sum_{k=1}^K \abs{\omega_{k,K}} \abs{f(A_k,B_k)} \le M_f \sum_{k=1}^K \abs{ \omega_{k,K} } .
\end{equation*}
The definition of control-neighbor weights \eqref{eq:control_neigh_def} yields
\begin{align*}
    \sum_{k=1}^K \abs{ \omega_{k,K} } 
    & = \frac{1}{K} \sum_{k=1}^K \Abs{ 1 + K V_{K,k} - \widehat{d}_{K,k} } \leq 1 + \sum_{k=1}^K V_{K,k} + \frac{1}{K} \sum_{k=1}^K \widehat{d}_{K,k} \leq 2 + \psi_{\Pi},
\end{align*}
where the last inequality follows from Assumption~\ref{assumption:bounded_indegree}, which gives $\widehat{d}_{K,k}\leq\psi_\Pi$ for all $k$. This proves the first bound in \eqref{eq:determ_W1_W2}.

Then, again by Assumption~\ref{assumption:direct_source_condition_holder_product}, 
\begin{equation*}
    \|W\|^2 = \sum_{k=1}^K \omega_{k,K}^2 f(A_k,B_k)^2 \le M_f^2 \sum_{k=1}^K \omega_{k,K}^2.
\end{equation*}
Using the weight definition \eqref{eq:control_neigh_def} and $(a+b+c)^2 \le 3(a^2+b^2+c^2)$, one obtains that
\begin{align}
	\sum_{k=1}^K \omega_{k,K}^2 &= \frac{1}{K^2} \sum_{k=1}^K (1 + K V_{K,k} - \widehat{d}_{K,k})^2 \le \frac{3}{K^2} \sum_{k=1}^K \left(1 + K^2 V_{K,k}^2 + \widehat{d}_{K,k}^2\right)\nonumber \\
	&\leq \frac{3}{K} + 3 \sum_{k=1}^K V_{K,k}^2 + \frac{3}{K^2} \sum_{k=1}^K \widehat{d}_{K,k}^2.
\label{eq:omega_split}
\end{align}
Since $\sum_{k=1}^K V_{K,k}^2\leq1$ and $\widehat{d}_{K,k}\leq\psi_\Pi$, it follows from \eqref{eq:omega_split} that 
\begin{equation*}
   \sum_{k=1}^K \omega_{k,K}^2 \leq \frac{3}{K} + 3 + 3\frac{\psi_{\Pi}^2}{K} \leq 3(2+\psi_{\Pi}^2),
\end{equation*}
for $K\geq 1$. Hence
\begin{equation*}
    \|W\|^2_2 \leq 3M_f^2(2+\psi_\Pi^2),
\end{equation*}
which proves the second bound in \eqref{eq:determ_W1_W2}.

Let $\rho_K:=\sup_{\theta\in\Pi}\min_{1\leq k\leq K}\rho_\Pi(\theta,\theta_k)$ be the covering radius of $\{\theta_1,\ldots,\theta_K\}$ in $(\Pi,\rho_\Pi)$. By Assumption~\ref{assumption:regular_sampling}~(iii),  $B_{r/M_\rho}^{\|\cdot\|}(\theta)\subseteq B_{r}^{\rho_\Pi}(\theta)\subseteq B_{r/m_\rho}^{\|\cdot\|}(\theta)$, $\theta\in\Pi$, $r>0$.
Since $C_{K,k}\subseteq B_{\rho_K}^{\rho_\Pi}(\theta_k)$ by definition of the covering radius, Assumption~\ref{assumption:regular_sampling} gives
\begin{equation}
\label{eq:Vsq}	
    V_{K,k} = \pi_\theta(C_{K,k})
    \leq c_u \lambda_{d+1}(C_{K,k})
    \leq c_u V_{d+1} m_\rho^{-(d+1)} 
    \rho_K^{d+1}.
\end{equation}
By a standard covering argument, for every $\alpha>0$ there exist constants $\bar C_\alpha,\bar C'_\alpha>0$ such that
\begin{equation}
\label{event}
    \mathbb{P}_{A,B} \left(\rho_K^{d+1}\le \bar C_\alpha \frac{\log(K)}{K}\right)
    \ge 1-\bar C'_\alpha K^{-\alpha}.
\end{equation}

Indeed, by Assumption~\ref{assumption:regular_sampling}~(iii), an $r/(2 M_\rho)$-net of $\Pi$ under the Euclidean metric is an $r/2$-net under $\rho_\Pi$. Moreover, since $D_\Pi$ is as in
Assumption~\ref{assumption:direct_source_condition_holder_product}, $\sup_{\theta_1,\theta_2\in\Pi}
\|\theta_1-\theta_2\|
\leq {D_\Pi}/{m_\rho}$.  Hence, by a standard volumetric bound (see, for example, Corollary 4.2.11 in~\cite{vershynin2026high}), there exists an $r/2$-net $\mathcal{N}_{r/2}$ of $\Pi$ under $\rho_\Pi$ such that, for $r\leq 2M_{\rho}r_0$, 
\begin{equation*}
    |\mathcal{N}_{r/2}| 
    \leq \left( 1 + \frac{4 M_{\rho} D_\Pi}{m_\rho r} \right)^{d+1} 
    = r^{-(d+1)}\left( r+ \frac{4 M_{\rho} D_\Pi}{m_\rho} \right)^{d+1} 
    \leq  C_\Pi r^{-(d+1)} ,
\end{equation*}
where $C_{\Pi}:=\left(2M_{\rho}r_0+({4M_\rho D_\Pi})/{m_\rho}\right)^{d+1}$. 
If $\rho_K>r$, then there exists $z\in\mathcal{N}_{r/2}$ such that $B_{r/2}^{\rho_\Pi}(z)\cap\Pi$ contains no sample point.

For $r\leq 2 M_{\rho} r_0$, Assumption~\ref{assumption:regular_sampling}~(i)-(iii) gives
\begin{equation*}
    \pi_\theta\bigl(B_{r/2}^{\rho_\Pi}(z)\bigr) \geq  \pi_\theta\big( B_{r/(2M_{\rho})}^{\|\cdot\|}(z)\cap\Pi \big)
    \geq c_b c_\pi V_{d+1}\big( r/(2M_{\rho}) \big)^{d+1},
\end{equation*}
and hence
\begin{equation*}
    \mathbb{P}_{A,B}(\rho_K>r)
    \leq C_\Pi r^{-(d+1)}
    \exp \left(-Kc_b c_\pi V_{d+1}\big( r/{(2M_{\rho})} \big)^{d+1}\right).
\end{equation*}
Set $r^{d+1}=\bar C_\alpha\log(K)/K$ and choose $\bar C_\alpha=(2M_{\rho})^{d+1}(\alpha+1)/(c_b c_\pi V_{d+1})$. 
Since $r=\left(\bar C_\alpha{\log(K)}/{K}\right)^{1/(d+1)} \allowbreak \rightarrow 0$, there exists $K_0\geq 3$, depending only on $\bar C_\alpha$, $M_\rho$, and $r_0$, such that $r\leq 2M_\rho r_0$ for all $K\geq K_0$. Hence, for $K\geq K_0$, using $1/\log(K)\leq 1$, $\mathbb{P}_{A,B}(\rho_K>r)\leq ({C_\Pi}/{\bar C_\alpha}) K^{-\alpha}$. By choosing $\bar C'_\alpha \geq \max\{C_\Pi/\bar C_{\alpha}, K_0^\alpha\}$ the same bound with right-hand side ${\bar C_\alpha}' K^{-\alpha}$ holds also for finitely many values  $3\leq K<K_0$, and this yields that \eqref{event} holds for all $K\geq 3$.

On the event \eqref{event}, substituting into \eqref{eq:Vsq}, yields
\begin{equation}\label{eq:maxVbound}
    \max_{1\leq k\leq K}V_{K,k}
    \leq c_u V_{d+1} m_\rho^{-(d+1)} \bar{C}_\alpha \frac{\log(K)}{K}
    =: D_\alpha \frac{\log(K)}{K}.
\end{equation}
Again, we note that $\widehat{d}_{K,k} \le \psi_{\Pi}$ and hence
\begin{equation}
\label{eq:degree_bound}	
    \sum_{k=1}^K \widehat{d}_{K,k}^2
    \le
    K{\psi_{\Pi}^2} .
\end{equation}
Thus, substituting \eqref{eq:Vkbound}, \eqref{eq:maxVbound} and \eqref{eq:degree_bound} into \eqref{eq:omega_split}, and once more using $1/K\leq\log(K)/K$ for $K\geq e$,
\begin{equation*}
    \sum_{k=1}^K\omega_{k,K}^2
    \leq \frac{3D_\alpha\log(K)}{K}
    +\frac{3\log(K)}{K}
    +\frac{3 \psi_{\Pi}^2\log(K)}{K}
    = \widetilde{C}_\alpha \frac{\log(K)}{K},
\end{equation*}
with $\widetilde{C}_\alpha:=3 D_\alpha+3+3 \psi_{\Pi}^2$ on an event of probability at least $1-\bar{C}_\alpha'K^{-\alpha}$. Consequently,
\begin{equation*}
    \|W\|^2\leq M_f^2 \widetilde{C}_\alpha \frac{\log(K)}{K},
\end{equation*}
which yields the result with $C_\alpha:=\sqrt{\widetilde{C}_\alpha}$ and $C_\alpha':=\bar{C}_\alpha'$.
\end{proof}

\subsection{$L^2$ approximation on integral Sobolev classes}
\label{app:Sobolev_approximation_proofs}

\begin{proof}[Proof of Corollary~\ref{corollary:mao_L2_sobol_approx}]
Apply Theorem~\ref{theorem:mao_L2_sobol_approx} with $\delta=K^{-\alpha}$. Since $\log(K/\delta)=(1+\alpha)\log K$, the theorem gives the asserted coefficient bound and the corresponding $L^2(\mathcal{X})$ approximation bound with probability at least $1-K^{-\alpha}$. The conclusion follows from
\begin{equation*}
\|h-W^\top\phi_{\theta,K}^m\|_{L^2(X)}
\leq \sqrt{\overline{M}}\,
\|h-W^\top\phi_{\theta,K}^m\|_{L^2(\mathcal{X})}.
\end{equation*}
\end{proof}

\subsection{Sup-norm approximation via stratified sampling}
\label{app:Sup-norm_approximation}
\begin{proof}[Proof of Theorem~\ref{theorem:sharp_is_linf_unified}]

\smallskip

\noindent\textbf{Proof of case (i).}

For $K\geq 2$, set $\widetilde{K}:=\lfloor K/2\rfloor$. We first carry out the stratified construction with budget $\widetilde{K}$.
The proof proceeds in four steps. First, we construct a stratified sampling scheme on $\Pi$ with allocations proportional to the cell masses, partitioning it into $J$ cells and allocating neurons approximately proportionally to the cell masses $\pi_\theta(B_j)$ while ensuring that each cell receives at least one sample. Second, we express the error as a centered empirical process indexed by $x\in\mathcal{X}$ and symmetrize it. Third, we bound the expected supremum of the symmetrized process using Dudley's entropy integral. Finally, we choose the partition scale so that the resulting number $ \overline{K}$ of active neurons satisfies $\overline{K} \leq 2 \widetilde{K}\leq K$, augment the construction with $K-\overline{K}$ additional inner parameters, and assign to the remaining neurons zero output weights. We then  obtain the rate $K^{-1/2-\beta/(d+1)}$ up to logarithmic factors.

\smallskip

\noindent\textsc{Step 1.} Fix $\epsilon>0$. For convenience, in this proof write $\phi(\theta)(x):=\phi_x(\theta)=f(A,B)\sigma(A^\top x+B)$ for $\theta=(A^\top,B)^\top\in\Pi$ and $x\in\mathcal{X}$.

Let $\mathcal{U}:=\{A^\top x+B:x\in\mathcal{X},\ (A,B)\in\Pi\}$, which is compact as the continuous image of $\mathcal{X}\times\Pi$, and recall  $M_\sigma=\sup_{u\in\mathcal{U}}|\sigma(u)|<\infty$ from Assumption~\ref{assumption:direct_source_condition_holder_product}. Since $\sigma$ is Lipschitz with constant $L_\sigma$ and 
$\|A\|_1\le C_A$, for fixed $\theta\in\Pi$ and 
$x,\widetilde{x}\in\mathcal{X}$,
\begin{equation*}
|\sigma(A^\top x+B)-\sigma(A^\top\widetilde{x}+B)|
\le L_\sigma\|A\|_1\|x-\widetilde{x}\|_\infty\le L_\sigma C_A\|x-\widetilde{x}\|_\infty,
\end{equation*}
and for $\theta,\widetilde{\theta}\in\Pi$ and $x\in\mathcal{X}$,
\begin{equation}
\label{eq:sigma_theta_lip}
|\sigma(A^\top x+B)-\sigma(\widetilde{A}^\top x+\widetilde{B})|
\le L_\sigma\bigl|(A-\widetilde{A})^\top x+(B-\widetilde{B})\bigr|
\le L_\sigma d_\Pi(\theta,\widetilde{\theta}),
\end{equation}
by Assumption~\ref{assumption:activation_Lipschitz} and the definition of $d_\Pi$, using $\|x\|_\infty\le M$.

For $\theta=(A^\top,B)^\top,\widetilde{\theta}=(\widetilde{A}^\top,\widetilde{B})^\top\in\Pi$ and  $x\in\mathcal{X}$, it holds that
\begin{align*}
|\phi(\theta)(x)-\phi(\widetilde{\theta})(x)|
&\le |f(A,B)||\sigma(A^\top x+B)
-\sigma(\widetilde{A}^\top x+\widetilde{B})|\\
&\qquad+|\sigma(\widetilde{A}^\top x+\widetilde{B})|
|f(A,B)-f(\widetilde{A},\widetilde{B})|\\
&\le M_f L_\sigma d_\Pi(\theta,\widetilde{\theta})
+M_\sigma L_f d_\Pi(\theta,\widetilde{\theta})^\beta\\
&\le M_f L_\sigma (2MC_A+2C_B)^{1-\beta}d_\Pi(\theta,\widetilde{\theta})^\beta
+M_\sigma L_f d_\Pi(\theta,\widetilde{\theta})^\beta,
\end{align*}
where in the second inequality we used $|f(A,B)|\le M_f$ and\eqref{eq:sigma_theta_lip} for the first term, and $|\sigma(\widetilde{A}^\top x+\widetilde{B})|\le M_\sigma$ together with Assumption~\ref{assumption:direct_source_condition_holder_product} for the second one; in the last inequality we used $d_\Pi(\theta,\widetilde{\theta})\le 2MC_A+2C_B$ for $\theta,\widetilde{\theta}\in\Pi$.

Set $L_u:=M_f L_\sigma (2MC_A+2C_B)^{1-\beta}+M_\sigma L_f$. Then
\begin{equation}
\label{eq:u_lip}
\sup_{x\in\mathcal{X}}|\phi(\theta)(x)-\phi(\widetilde{\theta})(x)|\le L_u  d_\Pi(\theta,\widetilde{\theta})^{\beta}.
\end{equation}

Since $\Pi=\{A\in\mathbb{R}^d:\|A\|_1\le C_A\}\times[-C_B,C_B]$ and
$d_\Pi(\theta,\widetilde{\theta})=M\|A-\widetilde{A}\|_1+|B-\widetilde{B}|$ is additive across the two components, the covering number factorizes. Covering the $\ell_1$-ball $\{A:\|A\|_1\le C_A\}$ under the metric $M\|\cdot\|_1$ with radius $r/2$ is equivalent to covering it under $\|\cdot\|_1$ with radius $r/(2M)$, which yields at most $(1+4MC_A/r)^d$ by a standard volumetric argument (see, for example, Corollary 4.2.11 in~\cite{vershynin2026high}). Similarly, covering $[-C_B,C_B]$ under $|\cdot|$ with radius $r/2$ yields at most $(1+4C_B/r)$ points. Taking the product covering gives
\begin{equation}
\label{eq:Pi_covering}
\mathcal{N}(r,\Pi,d_\Pi)\le \left(1+\frac{4MC_A}{r}\right)^d\left(1+\frac{4C_B}{r}\right), \qquad r>0.
\end{equation}
Let $r_\epsilon:=\left(\epsilon/L_u\right)^{1/\beta}$. Let $\{\bar\theta_j\}_{j=1}^J$ be an $r_\epsilon$-net of $\Pi$ under $d_\Pi$ of minimal cardinality, so that $J= \mathcal{N}(r_\epsilon,\Pi,d_\Pi)$. Let $\{B_j\}_{j=1}^J$ be a measurable partition of $\Pi$ such that 
$B_j\subset\{\theta\in\Pi:d_\Pi(\theta,\bar\theta_j)\leq r_\epsilon\}$ for each $j=1,\ldots,J$. Such a partition 
exists since $\{\bar\theta_j\}$ is an $r_\epsilon$-net. Therefore, by \eqref{eq:u_lip},
\begin{equation}
\label{eq:u_cell_bound}
\sup_{\theta\in B_j}\sup_{x\in\mathcal{X}}|\phi(\theta)(x)-\phi(\bar\theta_j)(x)|\le \epsilon.
\end{equation}
If $r_\epsilon\le \min(2MC_A,2C_B)$, then
\begin{align*}
1+\frac{4MC_A}{r_\epsilon}\le \frac{{ 6}MC_A}{r_\epsilon},\qquad 1+\frac{{ 4}C_B}{r_\epsilon}\le \frac{6 C_B}{r_\epsilon}.
\end{align*}
Then, by \eqref{eq:Pi_covering}, 
\begin{align*}
J { =} \mathcal{N}(r_\epsilon, \Pi, d_\Pi)
&\le \left(1 + \frac{{ 4}MC_A}{r_\epsilon}\right)^d \left(1 + \frac{{ 4}C_B}{r_\epsilon}\right).
\end{align*}
Combining these bounds yields
\begin{equation}
\label{eq:J_bound}
J\le ({ 6}MC_A)^d({ 6}C_B)  r_\epsilon^{-(d+1)}= C_\Pi r_\epsilon^{-(d+1)}, \enspace \text{with} \enspace C_\Pi:={ 6}C_B({ 6}MC_A)^d.
\end{equation}
For each $j=1,\ldots,J$, define $P_j:=\pi_\theta(B_j)$.
If $P_j>0$, define
\begin{align*}
\pi_{\theta,j}(E):=\frac{\pi_\theta(E\cap B_j)}{P_j},
\qquad E\subset\Pi\ \text{measurable},
\end{align*}
and if $P_j=0$, let $\pi_{\theta,j}$ be any probability measure on $B_j$. { Let $\{K_j\}_{j=1}^J$ be independent random variables satisfying $K_j\in\{\lfloor \widetilde{K}P_j\rfloor,\lceil \widetilde{K}P_j\rceil\}$, $\E[K_j]=\widetilde{K}P_j$,}
and define
\begin{align*}
L_j:=K_j+\Ind_{\{K_j=0\}}, \qquad \underline{K}:=(K_1,\ldots,K_J)^\top.
\end{align*}
Conditional on $\underline{K}$, draw $\theta_{j,1},\ldots,\theta_{j,L_j}\stackrel{\mathrm{i.i.d.}}{\sim}\pi_{\theta,j}$, 
independently across $j$. Define
\begin{align*}
\overline{K}:=\sum_{j=1}^J L_j.
\end{align*}
Then
\begin{align}
\label{eq:Kbar_bound}
\overline{K}
&=\sum_{j=1}^J K_j\Ind_{\{K_j>0\}}+\sum_{j=1}^J \Ind_{\{K_j=0\}}\le\sum_{j=1}^J ({ \widetilde{K}}P_j+1)\Ind_{\{K_j>0\}}+\sum_{j=1}^J \Ind_{\{K_j=0\}}\nonumber\\
&={ \widetilde{K}}\sum_{j=1}^J P_j\Ind_{\{K_j>0\}}+J\le { \widetilde{K}}+J,
\end{align}
where in the first inequality we used that $K_j \le \lceil { \widetilde{K}}P_j\rceil \le { \widetilde{K}}P_j+1$.

Recall the stratified approximant $h^{\mathrm{IS}}_{\overline{K}}$ defined in \eqref{eq:IS_estimator}, where $\phi(\theta)(x)=f(\theta)\widetilde{\sigma}(A^\top x+B)$. Since $\theta_{j,\ell}$ are i.i.d.\ from $\pi_{\theta,j}$,
\begin{align}
\label{eq:hx_expectation}	
\E[h^{\mathrm{IS}}_{ \overline{K}}(x)]=\sum_{j=1}^J P_j  \E_{\pi_{\theta,j}}[\phi(\theta)(x)] =h(x).
\end{align}

\smallskip
\noindent\textsc{Step 2.} Define the centered summands
\begin{align*}
  \widetilde{\phi}_{j,\ell}(x):=\phi(\theta_{j,\ell})(x)-\E_{\pi_{\theta,j}}[\phi(\theta)(x)].
\end{align*}
Since $\theta_{j,\ell}\sim \pi_{\theta,j}$, $\E_{\pi_{\theta,j}}[  \widetilde{\phi}_{j,\ell}(x)]=0$.

Let $\mu_j(x):=\phi(\bar\theta_j)(x)$. Since $\theta_{j,\ell}\in B_j$ $\pi_{\theta,j}$-almost surely, \eqref{eq:u_cell_bound} yields
\begin{align*}
|\phi(\theta_{j,\ell})(x)-\mu_j(x)|\le \epsilon,
\qquad x\in\mathcal{X}.
\end{align*}
Taking expectations and using Jensen's inequality,
\begin{align*}
\bigl|\E_{\pi_{\theta,j}}[\phi(\theta)(x)]-\mu_j(x)\bigr|=
\bigl|\E_{\pi_{\theta,j}}[\phi(\theta)(x)-\mu_j(x)]\bigr|
\le\E_{\pi_{\theta,j}}\bigl[|\phi(\theta)(x)-\mu_j(x)|\bigr]\le \epsilon.
\end{align*}
Combining these bounds gives
\begin{equation}
\label{eq:utilde_bound}
\sup_{x\in\mathcal{X}}|  \widetilde{\phi}_{j,\ell}(x)|\le\sup_{x\in\mathcal{X}}|\phi(\theta_{j,\ell})(x)-\mu_j(x)|+
\sup_{x\in\mathcal{X}}\bigl|\mu_j(x)-\E_{\pi_{\theta,j}}[\phi(\theta)(x)]\bigr|\le 2\epsilon.
\end{equation}
Moreover, by \eqref{eq:IS_estimator} and \eqref{eq:hx_expectation}
\begin{equation}
\label{eq:error_centered}
h^{\mathrm{IS}}_{ \overline{K}}(x)-h(x)=\sum_{j=1}^J \frac{P_j}{L_j} \sum_{\ell=1}^{L_j}\left(\phi(\theta_{j, \ell})(x)-\mathbb{E}_{\pi_{\theta, j}}[\phi(\theta)(x)]\right)=\sum_{j=1}^J \frac{P_j}{L_j}\sum_{\ell=1}^{L_j}  \widetilde{\phi}_{j,\ell}(x).
\end{equation}
Conditional on $\underline{K}$, the family $\{  \widetilde{\phi}_{j,\ell}\}_{j,\ell}$ is independent and centered. By the Rademacher symmetrization inequality~\citep[Lemma~2.3.6]{vandervaartWeakConvergenceEmpirical2023},
\begin{equation}
\label{eq:symmetrization_main}
    \E \left[\sup_{x\in\mathcal{X}}|h^{\mathrm{IS}}_{ \overline{K}}(x)-h(x)| \ \middle| \ \underline{K}\right] \le 2 \E \left[\sup_{x\in\mathcal{X}} \left|\sum_{j,\ell}\varepsilon_{j,\ell}\frac{P_j}{L_j}  \widetilde{\phi}_{j,\ell}(x)\right| \ \middle| \ \underline{K}\right],
\end{equation}
where $\{\varepsilon_{j,\ell}\}$ are i.i.d.\ Rademacher variables independent of $ \underline{\theta}$ and $\underline{K}$.

\smallskip
\noindent\textsc{Step 3.}
For $x\in\mathcal{X}$, define the Rademacher process
\begin{equation*}
Z(x):=\sum_{j=1}^J\frac{P_j}{L_j}\sum_{\ell=1}^{L_j}\varepsilon_{j,\ell}  \widetilde{\phi}_{j,\ell}(x),
\end{equation*}
and write $ \underline{\theta}:=(\theta_{j,\ell})_{j,\ell}$. Conditional on $( \underline{\theta},\underline{K})$, the variables $\varepsilon_{j,\ell}$ are i.i.d.\ Rademacher and the coefficients $  \widetilde{\phi}_{j,\ell}(x)$ are deterministic. Define the pseudometric
\begin{align}
\label{eq:rho_metric}
\rho_{ \underline{\theta},\underline{K}}(x,\widetilde{x})^2 &:= \sum_{j=1}^J\frac{P_j^2}{L_j^2} \sum_{\ell=1}^{L_j} \bigl[  \widetilde{\phi}_{j,\ell}(x)-  \widetilde{\phi}_{j,\ell}(\widetilde{x})\bigr]^2, \qquad x,\widetilde{x}\in\mathcal{X}.
\end{align}
By Hoeffding's lemma, $Z$ is sub-Gaussian with respect to $\rho_{ \underline{\theta}, \underline{K}}$. { Fix $x_0\in\mathcal{X}$. Setting
\begin{equation*}
A_{\underline{K}}^2:=\sum_{j=1}^J\frac{P_j^2}{L_j},
\end{equation*}
by Cauchy-Schwarz and \eqref{eq:utilde_bound}, $\E_\varepsilon\left[|Z(x_0)|\mid\underline\theta,\underline K\right]\leq 2\epsilon A_{\underline K}$.} Hence Dudley's entropy integral \citep[Corollary~2.2.9]{vandervaartWeakConvergenceEmpirical2023} { applied to $Z(x)-Z(x_0)$ } yields, for each fixed realization of $( \underline{\theta},\underline{K})$,
\begin{equation}
\label{eq:dudley}
    \E_\varepsilon \left[\sup_{x\in\mathcal{X}}|Z(x)| \ \middle| \ \underline{\theta},\underline{K}\right] \le { 2\epsilon A_{\underline K}+}C_1\int_0^{\Delta_{ \underline{\theta},\underline{K}}}\sqrt{\log\mathcal{N}(s,\mathcal{X},\rho_{ \underline{\theta},\underline{K}})} \de s,
\end{equation}
where $C_1>0$ is a universal constant and $\Delta_{ \underline{\theta},\underline{K}}:=\sup_{x,\widetilde{x}\in\mathcal{X}}\rho_{ \underline{\theta},\underline{K}}(x,\widetilde{x})$.

Since $|  \widetilde{\phi}_{j,\ell}(x)|\le 2\epsilon$ by \eqref{eq:utilde_bound}, the triangle
inequality gives $|  \widetilde{\phi}_{j,\ell}(x)-  \widetilde{\phi}_{j,\ell}(\widetilde{x})|\le 4\epsilon$,
and hence $\rho_{ \underline{\theta},\underline{K}}(x,\widetilde{x})^2\le 16\epsilon^2 A_{\underline{K}}^2$,
so that
\begin{equation}
\label{eq:Delta_bound}
\Delta_{ \underline{\theta},\underline{K}}\le 4\epsilon A_{\underline{K}}.
\end{equation}

For fixed $\theta=(A^\top,B)^\top\in\Pi$ and $x, \widetilde{x}\in\mathcal{X}$,
\begin{align*}
|\phi(\theta)(x)-\phi(\theta)(\widetilde{x})|
&= |f(A,B)| |\sigma(A^\top x+B)-\sigma(A^\top \widetilde{x}+B)| \le M_f L_\sigma |A^\top(x-\widetilde{x})| \\
&\le M_f L_\sigma \|A\|_1 \|x-\widetilde{x}\|_\infty \le
M_f L_\sigma C_A \|x-\widetilde{x}\|_\infty.
\end{align*}
Hence, by the triangle inequality and Jensen's inequality applied to 
$\E_{\pi_{\theta,j}}$,
\begin{align*}
|\widetilde{\phi}_{j,\ell}(x)-\widetilde{\phi}_{j,\ell}(\widetilde{x})|
&\le |\phi(\theta_{j,\ell})(x)-\phi(\theta_{j,\ell})(\widetilde{x})|
+\bigl|\E_{\pi_{\theta,j}}\bigl[\phi(\theta)(x)-\phi(\theta)(\widetilde{x})\bigr]\bigr|\\
&\le 2M_f L_\sigma C_A\|x-\widetilde{x}\|_\infty.
\end{align*}
Substituting into the definition of $\rho_{ \underline{\theta},\underline{K}}$ in \eqref{eq:rho_metric} yields
\begin{equation*}
\rho_{ \underline{\theta},\underline{K}}(x,\widetilde{x}) \le 2M_f L_\sigma C_A A_{\underline{K}} \|x-\widetilde{x}\|_\infty.
\end{equation*}
Therefore, since $\mathcal{X}=[-M,M]^d$, a standard volumetric argument (see, for example, Corollary 4.2.11 in~\cite{vershynin2026high}) gives
\begin{equation}
\label{eq:covering_bound}
\mathcal{N}(s,\mathcal{X},\rho_{ \underline{\theta},\underline{K}})\le \left(1+\frac{4MM_f L_\sigma C_A A_{\underline{K}}}{s}\right)^d.
\end{equation}
{  Set {$C_2 :=MM_f L_\sigma C_A$}. Next, combining \eqref{eq:dudley}, \eqref{eq:Delta_bound}, and \eqref{eq:covering_bound}, and using the nonnegativity of the integrand, we obtain
\begin{align}
\label{eq:Dudley_full_temp}
    \E_\varepsilon \left[\sup_{x\in\mathcal{X}}|Z(x)| \ \middle| \ \underline{\theta},\underline{K}\right] 
    &\le 2\epsilon A_{\underline K}+C_1 \int_0^{4\epsilon A_{\underline K}} \sqrt{ d\log\left(1+\frac{4C_2A_{\underline K}}{s}\right)}\de s\nonumber \\
    &\leq  2\epsilon A_{\underline K}+4C_1\epsilon A_{\underline K}\sqrt{d} \int_0^1 \sqrt{\log\left(1+\frac{C_2}{\epsilon v}\right)} \de v \nonumber\\
    &\le 2\epsilon A_{\underline K}+ 4C_1\epsilon A_{\underline K}\sqrt{d} \int_0^1 \left[
\sqrt{\log\left(1+\frac{C_2}{\epsilon}\right)} + \sqrt{\log\left(\frac{1}{v}\right)} \right]\de v.
\end{align}
Here, in the second line we applied the change of variables $v:=s/(4\epsilon A_{\underline{K}})$ so that $s=4\epsilon A_{\underline{K}}v$ and $\de s=4\epsilon A_{\underline{K}}\de v$.  In the third line  we used that setting  $a:=C_2/\epsilon$, for every $v\in(0,1]$ it holds that $\log(1+a/v)\le \log(1+a)+\log(1/v)$ and that $\sqrt{u+w}\le \sqrt{u}+\sqrt{w}$ for $u,w\geq0$. We proceed by considering the second summand in the integrand of the last expression. More explicitly, using the change of variables $u:=\log(1/v)$, so that $v=e^{-u}$ and $\de v=-e^{-u}\de u$, we get that 
\begin{equation*}
\int_0^1\sqrt{\log(1/v)} \de v=\int_0^\infty u^{1/2}e^{-u}\de u=\Gamma(3/2)=\sqrt\pi/2<1.
\end{equation*}
We apply this inequality to \eqref{eq:Dudley_full_temp}, noting that $\sqrt{\log \left(1+C_2 / \epsilon\right)}+1 \geq 1$ for $\sqrt{d} \geq 1$, yielding
}
\begin{equation}
\label{eq:dudley_bound}
    \E_\varepsilon \left[\sup_{x\in\mathcal{X}}|Z(x)| \ \middle| \ \underline{\theta},\underline{K}\right] 
    \le { (4C_1+2) } \epsilon A_{\underline{K}}\sqrt{d} \left(\sqrt{\log \left( 1+\dfrac{{ C_2}}{\epsilon} \right)}+1\right) =:F(\underline{K}),
\end{equation}
where 
the right-hand side depends on $( \underline{\theta},\underline{K})$ only through $\underline{K}$.

If ${ \widetilde{K}}P_j\le 2$, then $P_j^2/L_j\le 2P_j/{ \widetilde{K}}$. If ${ \widetilde{K}}P_j>2$, then $L_j\ge K_j\ge { \widetilde{K}}P_j/2$, so again $P_j^2/L_j\le 2P_j/{ \widetilde{K}}$. Summing over $j$ and using $\sum_jP_j=1$ gives
\begin{equation}
\label{eq:AK_bound}
A_{\underline{K}}^2\le \frac{2}{{ \widetilde{K}}}.
\end{equation}
Taking conditional expectation with respect to $ \underline{\theta}$ given $ \underline{K}$ in \eqref{eq:dudley_bound}, we obtain
\begin{equation*}
    \E \left[\E_\varepsilon \left[\sup_{x\in\mathcal{X}}|Z(x)| \ \middle| \ \underline{\theta},\underline{K}\right] \ \middle| \ \underline{K}\right]\le F(\underline{K}).
\end{equation*}
Substituting \eqref{eq:AK_bound} into $F(\underline{K})$, combining with \eqref{eq:symmetrization_main}, and taking expectation over $\underline{K}$ yields
\begin{equation}
\label{eq:step3_bound}
    \E \left[\sup_{x\in\mathcal{X}}|h^{\mathrm{IS}}_{ \overline{K}}(x)-h(x)|\right] 
    \le 
    { C_3 } \frac{\epsilon}{\sqrt{{ \widetilde{K}}}}\sqrt{d}\left(\sqrt{\log \left(1+\frac{{ C_2}}{\epsilon}\right)}+1\right),
\end{equation}
where { $C_3 := \sqrt{8}(4C_1+2) $ } and is { a universal constant}.

\noindent\textsc{Step 4.} 
{
Set
\begin{equation*}
    \epsilon := L_u r_\epsilon^\beta = c_0^\beta L_u { \widetilde{K}}^{-\beta/(d+1)}.
\end{equation*}
and} $r_\epsilon:=c_0 { \widetilde{K}}^{-1/(d+1)}$, where $c_0:=C_\Pi^{1/(d+1)}$. {  Since $r_\epsilon$ tends to zero as ${ \widetilde{K}}\to \infty$, there exists $K_0\in\mathbb{N}^+$, depending only on $d$, $C_A$, $C_B$, and $M$, such that $r_\epsilon\leq\min\{2MC_A,2C_B\}$ for all ${ \widetilde{K}}\geq K_0$. Hence, for ${ \widetilde{K}}\geq K_0$, \eqref{eq:J_bound} applies and gives} $J\leq C_\Pi r_\epsilon^{-(d+1)}=C_\Pi c_0^{-(d+1)}{ \widetilde{K}}={ \widetilde{K}}$. Then \eqref{eq:Kbar_bound} implies $\overline{K}\le 2{ \widetilde{K}}\leq K$. { Augment the $\overline K$ active inner parameters by $K-\overline K$ additional inner parameters in $\Pi$ and assign zero output weights to the corresponding neurons. Denoting the resulting network by $h_K^{\mathrm{IS}}\in\mathcal{H}_K^{\mathrm{IS}}$, we have $h_K^{\mathrm{IS}}(x)=h_{\overline K}^{\mathrm{IS}}(x)$, $x\in \mathcal{X}$.  }

Substituting this choice into \eqref{eq:step3_bound} and using
$h_K^{\mathrm{IS}}=h_{\overline K}^{\mathrm{IS}}$, we obtain
\begin{align*}
    \E \left[\sup_{x\in\mathcal{X}}|h^{\mathrm{IS}}_{{K}}(x)-h(x)|\right] 
    & \le { C_4} { \widetilde{K}}^{-1/2-\beta/(d+1)}\sqrt{d} \left(\sqrt{\log \left(1+ { C_5} { \widetilde{K}}^{\beta/(d+1)}\right)}+1\right),
\end{align*}
{ where ${ C_4} := { C_3} c_0^\beta L_u$ and ${ C_5 } := { C_2}/{(c_0^\beta L_u)}$}. Since {  $\log(1+ { C_5} { \widetilde{K}}^{\beta/(d+1)})\leq \log(1+ { C_5})+\frac{\beta}{d+1}\log { \widetilde{K}}$, it holds that  
\begin{align*}
    \sqrt{d}\left(\sqrt{\log\left(1+ { C_5} { \widetilde{K}}^{\beta/(d+1)}\right)}+1\right)
    &\leq\left(1+\sqrt{\log(1+ { C_5})}\right)\sqrt{d}+\sqrt{\frac{d\beta}{d+1}}\sqrt{\log({ \widetilde{K}})}\\
    &\qquad\leq\left(1+\sqrt{\log(1+ { C_5})}\right)\left(\sqrt{d}+\sqrt{\log ({ \widetilde{K}})}\right),
\end{align*}
where the last inequality uses $\beta\leq1$. Hence for $K\geq K_*:=\max\{2K_0,2\}$, since  $\widetilde{K}=\lfloor K/2\rfloor\geq K/3$ for every $K\geq2$
and $\log(\widetilde{K})\leq\log(K)$,} 
\begin{align*}
    \E \left[\sup_{x\in\mathcal{X}}|h^{\mathrm{IS}}_{{K}}(x) -h(x)|\right] 
    \le 
    C_{\mathrm{IS}}^{(0)} \big( \sqrt{d}+\sqrt{\log(K)} \big) K^{-1/2-\beta/(d+1)},
\end{align*}
where { $C_{\mathrm{IS}}^{(0)}:=3^{1/2+\beta/(d+1)} { C_4} \left(1+\sqrt{\log(1+{ C_5})}\right)$}. 

{  It remains to consider the finitely many $1\leq K<K_*$. For these values of $K$, take the trivial partition $J=1$ with $B_1=\Pi$, choose any $\bar{\theta}_1\in\Pi$, and set $r_\epsilon:=D_\Pi$, which is finite by Assumption~\ref{assumption:direct_source_condition_holder_product}. Since $\rho_\Pi=d_\Pi$ in case~{(i)}, the definition of $D_\Pi$ gives $B_1=\Pi\subseteq\left\{\theta\in\Pi\mid d_\Pi(\theta,\bar\theta_1)\leq r_\epsilon\right\}$, 
so this is an admissible one-cell instance of the preceding construction, and taking $\epsilon:=L_u D_\Pi^\beta$ ensures that \eqref{eq:u_cell_bound} holds. Moreover, $P_1=\pi_\theta(B_1)=1$, and hence $K_1\in\left\{\lfloor KP_1\rfloor,\lceil KP_1\rceil\right\}=\{K\}$, so that $K_1=K$. Since $K\geq 1$, the definition $L_1=K_1+\Ind_{\{K_1=0\}}$ gives $L_1=K$, and consequently $\overline{K}=L_1=K\leq 2K$. The corresponding estimator is $h^{\mathrm{IS}}_{{K}}(x)={1}/{K}\sum_{\ell=1}^{K}\phi(\theta_{1,\ell})(x)$,  which is the one-cell instance of the same stratified sampling construction. In particular, its unbiasedness follows from \eqref{eq:hx_expectation}. Next, by the integral representation of $h$ and the bound $|\phi(\theta)(x)|\leq M_f M_\sigma$,
\begin{equation*}
    |h(x)|\leq\int_\Pi|\phi(\theta)(x)|\de\pi_\theta(\theta)\leq M_f M_\sigma
    \quad\text{and}\quad
    \left|h^{\mathrm{IS}}_{{K}}(x)\right|\leq\frac{1}{K}\sum_{\ell=1}^{K}|\phi(\theta_{1,\ell})(x)|\leq M_f M_\sigma,
\end{equation*}
so that for  $1\leq K<K_*$, we have $\E \left[\sup_{x\in\mathcal{X}}|h^{\mathrm{IS}}_{{K}}(x)-h(x)|\right]\leq 2M_f M_\sigma$.

 Since $K^{1/2+\beta/(d+1)}\leq K_*^{1/2+\beta/(d+1)}$ and $\sqrt{d}+\sqrt{\log K}\geq 1$, it follows that \eqref{eq:sup_norm_unified} holds for $1\leq K<K_*$ with $C_{\mathrm{IS}}^{(1)}:=2M_fM_\sigma K_*^{1/2+\beta/(d+1)} $. Consequently, \eqref{eq:sup_norm_unified} holds for all $K\in\mathbb{N}^+$ with $C_{\mathrm{IS}}:=\max\left\{
C_{\mathrm{IS}}^{(0)}, C_{\mathrm{IS}}^{(1)} \right\}$. Equivalently,
\begin{equation}
\label{eq:C_is_explicit_constant}
    C_{\mathrm{IS}}
    :=
    \max\left\{ 
        3^{1/2+\beta/(d+1)} { C_3} c_0^\beta L_u \left(1+\sqrt{\log\left(1+\frac{{ C_2}}{c_0^\beta L_u}\right)}\right),
        2M_fM_\sigma K_*^{1/2+\beta/(d+1)}
    \right\},
\end{equation}
where we recall that ${ C_3} >0$ is a universal constant, $C_\Pi={ 6}C_B({ 6}MC_A)^d$, $c_0=C_\Pi^{1/(d+1)}$, ${ C_2}=M M_fL_\sigma C_A$, $L_u=M_fL_\sigma(2MC_A+2C_B)^{1-\beta}+M_\sigma L_f$, 
\begin{equation*} 
    K_0=\left\lceil \left( {c_0}/{\min\{2MC_A,2C_B\}}\right)^{d+1}\right\rceil ,
\end{equation*} 
and $K_*=\max\{2K_0,2\}$.

For $K\geq K_*$, the active stratified construction uses $\overline K\leq K$ neurons and the zero-appending described above yields an element of $\mathcal{H}_K^{\mathrm{IS}}$ without changing the approximating function. For $1\leq K<K_*$, the one-cell construction already uses exactly $K$ neurons. This completes the proof.
}

\smallskip

\noindent\textbf{Proof of case (ii).}

{ For $K\geq 2$, set $\widetilde{K}:=\lfloor K/2\rfloor$, as in the proof of case (i), and carry out the stratified construction with budget $\widetilde{K}$. The proof follows that of case~(i), with the main modification in Step~1, where the H\"older continuity of $f$ is replaced by its sign structure. Steps~2--4 are analogous, with the corresponding changes in the constants.}

\smallskip
\noindent\textsc{Step 1.}
Define $\phi(\theta)(x):=\phi_x(\theta)=f(A,B)\sigma_m(A^\top x+B)$ for $\theta=(A^\top,B)^\top\in\Pi$ and, as in case~(i), equip $\Pi$ with the metric $d_\Pi$ defined in Remark~\ref{remark:metrics}(ii).

For $\theta,\widetilde\theta\in\Pi$ and $x\in\mathcal{X}$, set $s:=A^\top x+B$ and $\widetilde{s}:=\widetilde{A}^\top x+\widetilde{B}$, so that $|s|,|\widetilde{s}|\le C_AM+C_B=:R$. For $m=1$, 
non-expansiveness of the ReLU activation function gives 
$|\sigma_1(s)-\sigma_1(\widetilde{s})|\le|s-\widetilde{s}|$. For $m=2$, 
the mean value theorem applied to $\sigma_2(t)=\max(t,0)^2$, which 
has derivative $2\max(t,0)$, gives 
$|\sigma_2(s)-\sigma_2(\widetilde{s})|\le 2R|s-\widetilde{s}|$. In both 
cases,
\begin{equation}
\label{eq:relu_m_lip_app}
|\sigma_m(s)-\sigma_m(\widetilde{s})|
\le mR^{m-1}|s-\widetilde{s}|
\le L_{\sigma,m} d_\Pi(\theta,\widetilde\theta),
\end{equation}
where $L_{\sigma,m}:=m(C_AM+C_B)^{m-1}$ and the second inequality 
uses
\begin{align*}
|s-\widetilde{s}|
&=|(A-\widetilde{A})^\top x+(B-\widetilde{B})|\le M\|A-\widetilde{A}\|_1+|B-\widetilde{B}|=d_\Pi(\theta,\widetilde\theta),
\end{align*}
by H\"older's inequality and $\|x\|_\infty\le M$ by Assumption~\ref{assumption:X_domain}.

Since $f:\Pi\to\{-M_f,+M_f\}$, partition $\Pi$ into $\Pi_\pm:=\{(A,B)\in\Pi:f(A,B)=\pm M_f\}$, 
so that $f$ is constant on each part.

Set $L_u^*:=M_f L_{\sigma,m}$. Fix $\epsilon>0$ and 
$r_\epsilon:=\epsilon/L_u^*$. Cover $\Pi_+$ and $\Pi_-$ separately with $r_\epsilon$-nets under $d_\Pi$, and choose measurable partitions subordinate to the corresponding $d_\Pi$-balls, yielding a measurable 
partition $\{B_j\}_{j=1}^J$ of $\Pi$ with $B_j\subset\Pi_\pm$ and representatives $\bar\theta_j\in B_j$ satisfying $\sup_{\theta\in B_j}d_\Pi(\theta,\bar\theta_j)\le r_\epsilon$. Since $f$ is constant on each $B_j$, for $\theta\in B_j$,
\begin{align*}
|\phi(\theta)(x)-\phi(\bar\theta_j)(x)|
&=M_f|\sigma_m(A^\top x+B)
-\sigma_m(\bar A_j^\top x+\bar B_j)|\\
&\le M_f L_{\sigma,m} d_\Pi(\theta,\bar\theta_j)
\le L_u^* r_\epsilon
=\epsilon,
\end{align*}
using \eqref{eq:relu_m_lip_app}. Hence
\begin{equation}
\label{eq:u_cell_bound_sign_app}
\sup_{\theta\in B_j}\sup_{x\in\mathcal{X}}
|\phi(\theta)(x)-\phi(\bar\theta_j)(x)|\le\epsilon.
\end{equation}

Covering $\Pi_+$ and $\Pi_-$ separately  by a standard volumetric argument (see, for example, Corollary 4.2.11 in~\cite{vershynin2026high}) gives, for $r_\epsilon\le\min(2MC_A,2C_B)$, the following relation
\begin{equation}
\label{eq:J_bound_sign_app}
    J\le { 2\left(1+\frac{2(MC_A+C_B)}{r_\epsilon}\right)^{d+1}}\leq C_\Pi^* r_\epsilon^{-(d+1)},
    \quad \text{with} \enspace 
    C_\Pi^* := { 2\bigl(4(MC_A+C_B)\bigr)^{d+1}}.
\end{equation}
The remainder of Step~1 follows precisely the proof of case~(i), { with $\widetilde{K}$ as the sampling budget, in terms of defining $P_j$, $\pi_{\theta,j}$, $K_j$, $L_j$, $\underline{K}$, the conditional draws, the estimator $h^{\mathrm{IS}}_{\overline{K}}$ in \eqref{eq:IS_estimator}, and its unbiasedness in \eqref{eq:hx_expectation}. In particular, \eqref{eq:Kbar_bound} gives  $\overline{K}\le \widetilde{K}+J$.}

\smallskip
\noindent\textsc{Step 2.}
Given \eqref{eq:u_cell_bound_sign_app} in place of 
\eqref{eq:u_cell_bound}, this step follows as Step~2 of 
case~(i), yielding $\sup_x|\widetilde{\phi}_{j,\ell}(x)|\le 2\epsilon$, the error representation \eqref{eq:error_centered}, and the symmetrization inequality \eqref{eq:symmetrization_main}.

\smallskip
\noindent\textsc{Step 3.}
The Rademacher process $Z(x)$, pseudometric 
$\rho_{ \underline{\theta},\underline{K}}$, sub-Gaussianity argument, Dudley bound \eqref{eq:dudley}, $\Delta$ bound \eqref{eq:Delta_bound}, and tower property argument follow  as in Step~3 of case~(i). The only difference is the Lipschitz constant of $x\mapsto \phi_x(\theta)$. For fixed $\theta\in\Pi$ and $x,\widetilde{x}\in\mathcal{X}$, we write 
\begin{align*}
|\phi(\theta)(x)-\phi(\theta)(\widetilde{x})|
&=M_f|\sigma_m(A^\top x+B)-\sigma_m(A^\top\widetilde{x}+B)|\le M_f L_{\sigma,m}|A^\top(x-\widetilde{x})|\\
&\le M_f L_{\sigma,m}\|A\|_1\|x-\widetilde{x}\|_\infty\le L_u^* C_A\|x-\widetilde{x}\|_\infty.
\end{align*} 
Setting ${ C_2^*} := M L_u^* C_A$ and applying the same integral 
evaluation as in Step~3 of case~(i) gives
\begin{equation}
\label{eq:step3_bound_sign_app}
    \E\left[\sup_{x\in\mathcal{X}}|h^{\mathrm{IS}}_{\overline{K}}(x)-h(x)|\right]\le { C_3} \frac{\epsilon}{\sqrt{{ \widetilde{K}}}}\sqrt{d}\left(\sqrt{\log\left(1+\frac{{ C_2^*}}{\epsilon}\right)}+1\right),
\end{equation}
where ${ C_3:=\sqrt{8}(4C_1+2)}>0$ { is the universal constant} as in case (i).

\smallskip
\noindent\textsc{Step 4.}
{ Set $\epsilon:= c_0^*L_u^*{ \widetilde{K}}^{-1/(d+1)}$ and} $r_\epsilon:=c_0^* { \widetilde{K}}^{-1/(d+1)}$, with $c_0^*:=(C_\Pi^*)^{1/(d+1)}$. 
The condition
$r_\epsilon\le \min(2MC_A,2C_B)$ holds for all ${ \widetilde{K}}\ge K_0$,  with { \begin{equation*}
    K_0:=\left\lceil\left(\frac{c_0^*}{\min\{2MC_A,2C_B\}}\right)^{d+1}\right\rceil .
    \end{equation*}} 
For such ${ \widetilde{K}}$, \eqref{eq:J_bound_sign_app} applies and gives
\begin{align*}
J\le C_\Pi^* r_\epsilon^{-(d+1)}= C_\Pi^*(c_0^*)^{-(d+1)}{ \widetilde{K}}
= { \widetilde{K}}.
\end{align*}
{ Hence, by \eqref{eq:Kbar_bound}, $\overline K\le { \widetilde{K}}+J\le 2{ \widetilde{K}} \leq K$. Augment the $\overline K$ active inner parameters by $K-\overline K$ additional inner parameters in $\Pi$ and assign zero output weights to the corresponding neurons. Denoting the resulting network by $h_K^{\mathrm{IS}}\in\mathcal{H}_K^{\mathrm{IS}}$, we have $h_K^{\mathrm{IS}}(x)=h_{\overline K}^{\mathrm{IS}}(x)$, $x\in \mathcal{X}$.} Substituting into \eqref{eq:step3_bound_sign_app} yields
\begin{align*}
    \E\left[\sup_{x\in\mathcal{X}}
    |h^{\mathrm{IS}}_{ K}(x)-h(x)|\right]
    \le 
    { C_4^*} { \widetilde{K}}^{-1/2-1/(d+1)}\sqrt{d} \left(\sqrt{\log\left(1+ { C_5^*} { \widetilde{K}}^{1/(d+1)}\right)}+1
    \right),
\end{align*}
where { ${ C_4^*} := { C_3} c_0^*L_u^* $ and} ${ C_5^*} := { C_2^*}/({c_0^*L_u^*}) = MC_A/c_0^*$.

{ Following the same argument as in Step~4 of case~{(i)}, for $K\geq K_*:=\max\{2K_0,2\}$ using
$\widetilde{K}=\lfloor K/2\rfloor\geq K/3$ and
$\log(\widetilde{K})\leq\log{K}$, we obtain
\begin{equation*}
\E\left[\sup_{x\in\mathcal{X}}
|h^{\mathrm{IS}}_{ K}(x)-h(x)|\right]\leq C_{\mathrm{IS}}^{*,(0)}\left(\sqrt d+\sqrt{\log K}\right) K^{-1/2-1/(d+1)},
\end{equation*}
where $C_{\mathrm{IS}}^{*,(0)}:=3^{1/2+1/(d+1)} { C_4^*} \left(1+\sqrt{\log(1+ { C_5^*} )}\right)$}.

{ It remains to consider the finitely many $1\leq K<K_*$. For these values of $K$, take the trivial one-cell stratification $B_1=\Pi$, sample $K$ inner parameters from $\pi_\theta$, and set all output weights equal to zero. Then $h_K^{\mathrm{IS}}\in\mathcal{H}_K^{\mathrm{IS}}$ and $h_K^{\mathrm{IS}}\equiv0$. Recalling that $R:=MC_A+C_B$, we have $|h(x)|\leq \int_\Pi |f(\theta)\sigma_m(A^\top x+B)|\de\pi_\theta(\theta) \leq M_fR^m$ and hence $\E\left[\sup_{x\in\mathcal{X}} |h_K^{\mathrm{IS}}(x)-h(x)| \right] \leq M_fR^m$.

We proceed  analogously to the proof of case (i). Since $K^{1/2+1/(d+1)}\leq K_*^{1/2+1/(d+1)}$ and $\sqrt{d}+\sqrt{\log(K)}\geq 1$, it follows that \eqref{eq:sup_norm_unified} holds for $1\leq K<K_*$ with $C_{\mathrm{IS}}^{*,(1)}:=M_fR^m K_*^{1/2+1/(d+1)} $. Consequently, \eqref{eq:sup_norm_unified} holds for all $K\in\mathbb{N}^+$ with $C_{\mathrm{IS}}^*:=\max\left\{
C_{\mathrm{IS}}^{*,(0)}, C_{\mathrm{IS}}^{*,(1)} \right\}$. Equivalently,
\begin{equation}
\label{eq:C_is_explicit_constant_case_ii}
    C_{\mathrm{IS}}^*
    :=
    \max\left\{
        3^{1/2+1/(d+1)} { C_3} c_0^* L_u^* \left(1+\sqrt{\log\left(1+ { C_5^*}\right)}\right),
        M_fR^m K_*^{1/2+1/(d+1)}
    \right\},
\end{equation}
where we recall that ${ C_3} >0$ is a universal constant, ${ C_5^*} =MC_A/c_0^*$, $c_0^*=(C_\Pi^*)^{1/(d+1)}$, $C_\Pi^*=2\bigl(4(MC_A+C_B)\bigr)^{d+1}$, $L_u^*=M_fL_{\sigma,m}$, $R=MC_A+C_B$,  and $K_*=\max\{2K_0,2\}$, with $K_0=\left\lceil\left({c_0^*}/{\min\{2MC_A,2C_B\}}\right)^{d+1}\right\rceil$.

Combining the large- and small-$K$ bounds as in the final paragraph of the proof of case (i) completes the proof.}
\end{proof}
\begin{proof}[Proof of Theorem~\ref{thm:prob-approx-linf}]
By Theorem~\ref{theorem:sharp_is_linf_unified},
\begin{equation*}
\E\left[\sup_{x\in\mathcal{X}} \left|h^{\mathrm{IS}}_{ K}(x)-h(x)\right|\right]\leq C(\sqrt{d}+\sqrt{\log(K)})K^{-1/2-\alpha}.
\end{equation*}
Applying Markov's inequality to $\sup_{x\in\mathcal{X}} \big|h^{\mathrm{IS}}_{ K}(x)-h(x)\big|$ yields
\begin{equation*}
\mathbb{P}\left(\sup_{x\in\mathcal{X}} \left|h^{\mathrm{IS}}_{ K}(x)-h(x)\right|>C(\sqrt{d}+\sqrt{\log(K)})K^{-1/2-\alpha}\varepsilon^{-1}\right)\leq \varepsilon,
\end{equation*}
which proves \eqref{eq:prob_bound_linf} as required.
\end{proof}
\begin{proof}[Proof of Corollary~\ref{cor:stratified_rwnn_projection_rate_linf}]
Fix $\varepsilon\in(0,1)$. By Theorem~\ref{thm:prob-approx-linf}, there exists $h_K^{\mathrm{IS}}\in\mathcal{H}_K^{\mathrm{IS}}$ such that
\begin{equation*}
\mathbb{P}\left(\mathcal{E}_{K,\varepsilon}^{\infty}\right)\geq 1-\varepsilon,
\end{equation*}
where
\begin{equation*}
\mathcal{E}_{K,\varepsilon}^{\infty}:=\left\{\|h-h^{\mathrm{IS}}_{ K}\|_{L^\infty(\mathcal{X})}\leq C(\sqrt{d}+\sqrt{\log(K)})K^{-1/2-\alpha} \varepsilon^{-1} \right\}.
\end{equation*}
On $\mathcal{E}_{K,\varepsilon}^{\infty}$, since $h_K^{\mathrm{IS}}\in\mathcal{H}_K^{\mathrm{IS}}$, one has
\begin{equation*}
\inf_{g\in\mathcal{H}_K^{\mathrm{IS}}} \|h-g\|_{L^\infty(\mathcal{X})} \leq \|h-h^{\mathrm{IS}}_{ K}\|_{L^\infty(\mathcal{X})}.
\end{equation*}
Hence
\begin{equation*}
\mathbb{P} \left( \inf_{g\in\mathcal{H}_K^{\mathrm{IS}}} \|h-g\|_{L^\infty(\mathcal{X})} > C(\sqrt{d}+\sqrt{\log(K)})K^{-1/2-\alpha} \varepsilon^{-1}
\right) \leq \varepsilon,
\end{equation*}
which proves the stated $O_P$ rate.
\end{proof}
\begin{proof}[Proof of Lemma~\ref{lem:stratified_weight_norms}]
For the large $K$ construction in the proof of Theorem~\ref{theorem:sharp_is_linf_unified}, by \eqref{eq:IS_estimator}, after identifying $k$ with $(j,\ell)$,
\begin{equation*}
W_k=W_{j,\ell}:=\frac{P_j}{L_j}f(\theta_{j,\ell}).
\end{equation*}
The remaining $(K-\overline{K})$ output weights introduced by appending zeros are equal to zero.

In case~{(i)}, Assumption~\ref{assumption:direct_source_condition_holder_product} gives $|f|\leq M_f$, while in case~{(ii)},  Assumption~\ref{assumption:source_sign} gives  $f\in\{-M_f,+M_f\}$. In both cases it holds that  $|f|\leq M_f$ and one obtains
\begin{equation*}
\|W\|_1\leq M_f\sum_{j=1}^J\sum_{\ell=1}^{L_j}\frac{P_j}{L_j} = M_f\sum_{j=1}^J P_j = M_f,
\end{equation*}
which yields \eqref{eq:stratified_l1_bound}. Additionally, we write that 
\begin{equation*}
\|W\|^2 \leq M_f^2\sum_{j=1}^J\frac{P_j^2}{L_j}.
\end{equation*}
We now show that $P_j^2/L_j\leq 2P_j/\widetilde{K}$ for $j=1\ldots, J$. Recall that $K_j\in\{\lfloor \widetilde{K}P_j\rfloor,\lceil \widetilde{K}P_j\rceil\}$ and $L_j=K_j+\Ind_{\{K_j=0\}}$. If $\widetilde{K}P_j\leq 2$, then $L_j\geq 1$, so
\begin{equation*}
\frac{P_j^2}{L_j}\leq P_j^2\leq \frac{2P_j}{\widetilde{K}}.
\end{equation*}
If $\widetilde{K}P_j>2$, then $K_j\geq \lfloor \widetilde{K}P_j\rfloor\geq \widetilde{K}P_j/2$, hence $L_j=K_j$ and ${P_j^2}/{L_j}\leq {2P_j}/{\widetilde{K}}$. Therefore,
\begin{equation*}
\sum_{j=1}^J\frac{P_j^2}{L_j} \leq \frac{2}{\widetilde{K}}\sum_{j=1}^J P_j = \frac{2}{\widetilde{K}},
\end{equation*}
{ which implies that $\|W\|\leq \sqrt{2}M_f \widetilde{K}^{-1/2}\leq \sqrt{6}M_f K^{-1/2}$, where the last inequality uses $\widetilde{K}=\lfloor K/2\rfloor\geq K/3$ for $K\geq 2$. Hence \eqref{eq:stratified_l1_bound} and \eqref{eq:stratified_l2_bound} hold for the large $K$ construction. 

For the values $1\leq K<K_*$ considered separately at the end of Step~4 of the proof of Theorem~\ref{theorem:sharp_is_linf_unified}, case~(i) uses the one-cell construction with 
\begin{equation*}
W_k=\frac{1}{K}f(\theta_k), \qquad k=1,\dots, K,
\end{equation*}
and hence 
\begin{equation*}
    \|W\|_1\leq M_f, \qquad  \|W\|\leq M_f {K}^{-1/2}\leq \sqrt{6}M_f K^{-1/2}. 
\end{equation*}
In case~(ii), all output weights are zero. 

We conclude that \eqref{eq:stratified_l1_bound} and \eqref{eq:stratified_l2_bound} hold for all $K\in\mathbb{N}^+$.
}
\end{proof}

\subsection{Uniform Monte Carlo approximation via metric entropy}
\label{app:entropy_computation}
We establish that the standard Monte Carlo estimator
\begin{equation*}
    h_K(x) := \frac{1}{K}\sum_{k=1}^K \phi_x(\theta_k), \qquad \theta_k \overset{\mathrm{iid}}{\sim} \pi_\theta,
\end{equation*}
satisfies a uniform Monte Carlo bound over the compact covariate domain.

\begin{proposition}\label{prop:entropy_mc_rate}
Let Assumptions~\ref{assumption:X_domain}, \ref{assumption:activation_Lipschitz} and \ref{assumption:direct_source_condition_holder_product} hold. Define $M_f := \sup_{\theta\in\Pi}|f(\theta)|$.
Then there exists a constant $C_{\mathrm{ent}}>0$ such that
\begin{equation*}
    \mathbb{E}_{\theta} \left[\sup_{x\in\mathcal{X}}\left|h(x) - h_K(x)\right|\right]\le C_{\mathrm{ent}} K^{-1/2}.
\end{equation*}
The constant $C_{\mathrm{ent}}$ depends only on $d$, $M$, $M_f$, $M_\sigma$, $L_{\sigma}$, and the support $\Pi$.
\end{proposition}

\begin{proof}
By Assumption~\ref{assumption:direct_source_condition_holder_product}, $h(x)=\mathbb{E}_{\theta}[\phi_x(\theta)]$, and one can write that 
\begin{equation*}
h(x)-h_K(x)=\mathbb{E}_{\theta}[\phi_x(\theta)]-\frac{1}{K}\sum_{k=1}^K \phi_x(\theta_k).
\end{equation*}
Hence, it suffices to bound
\begin{equation*}
\mathbb{E}_{\theta}\left[\sup_{x\in\mathcal{X}}\left|
\frac{1}{K}\sum_{k=1}^K \phi_x(\theta_k)-\mathbb{E}_{\theta}[\phi_x(\theta)]\right|\right].
\end{equation*}
Next, using the symmetrization inequality (\citealp{Boucheron2013}, Lemma 11.4) one obtains
\begin{equation}
\label{eq:symmetrization}
	\mathbb{E}_{\theta} \left[\sup_{x\in\mathcal{X}}\left|
    \frac{1}{K}\sum_{k=1}^K \phi_x(\theta_k)-
    \mathbb{E}_{\theta}[\phi_x(\theta)]\right|\right]\le2 
    \mathbb{E}_{\theta,\varepsilon}\left[\sup_{x\in\mathcal{X}}\left|\frac{1}{K}\sum_{k=1}^K \varepsilon_k \phi_x(\theta_k)\right|\right],
\end{equation}
where $\varepsilon_1,\dots,\varepsilon_K$ are i.i.d.\ Rademacher random variables independent of $\theta_1,\dots,\theta_K$.

Conditional on $\theta_1,\dots,\theta_K$, we define
\begin{equation}
\label{eq:def_Wx}
W_x :=\frac{1}{K}\sum_{k=1}^K \varepsilon_k \phi_x(\theta_k), \qquad x\in\mathcal{X}.
\end{equation}
Since $\Pi$ is compact, define $M_A := \sup_{(A^\top,B)^\top\in\Pi}\|A\|$ and recall that $M_f := \sup_{\theta\in\Pi}|f(\theta)|$.

Under Assumption~\ref{assumption:activation_Lipschitz} and~\ref{assumption:direct_source_condition_holder_product}, for all $x_1,x_2\in\mathcal{X}$ it holds that
\begin{eqnarray}
\label{eq:phi_Lipschitz}
	|\phi_{x_1}(\theta)-\phi_{x_2}(\theta)|& =|f(A,B)| |\sigma(A^\top x_1+B)-\sigma(A^\top x_2+B)|\nonumber \\&\le M_f L_{\sigma}|A^\top(x_1-x_2)| \le M_f L_{\sigma}M_A \|x_1-x_2\|.
\end{eqnarray}
We proceed by noticing that for $x_1,x_2\in\mathcal{X}$
\begin{equation*}
	W_{x_1}-W_{x_2}=\sum_{k=1}^K \varepsilon_k a_k(x_1,x_2), \enspace \text{with} \enspace a_k(x_1,x_2) :=
\frac{1}{K}\left(\phi_{x_1}(\theta_k)-\phi_{x_2}(\theta_k)\right)
\end{equation*}
and hence for every $\lambda\in\mathbb{R}$,
\begin{align*}
	\mathbb{E}_\varepsilon \left[\exp\left(\lambda(W_{x_1}-W_{x_2})\right) \right]&=\prod_{k=1}^K\mathbb{E}_\varepsilon \left[\exp\left(\lambda \varepsilon_k a_k(x_1,x_2)\right) \right]=\prod_{k=1}^K\operatorname{cosh}\left(\lambda a_k(x_1,x_2)\right)\\
    &\le \exp\left( \frac{\lambda^2}{2} \sum_{k=1}^K a_k(x_1,x_2)^2 \right),
\end{align*}
where we used the definition of the moment generating function of Rademacher random variables and where the last step follows from $\operatorname{cosh}(t)\leq \exp({t^2/2})$ for all $t\in \mathbb{R}$.

Hence $(W_x)_{x\in\mathcal{X}}$ has conditionally sub-Gaussian increments with respect to the pseudometric defined as 
\begin{equation*}
	d_K(x_1,x_2):=\left(\sum_{k=1}^K a_k(x_1,x_2)^2\right)^{1/2}=\left(\frac{1}{K^2}\sum_{k=1}^K|\phi_{x_1}(\theta_k)-\phi_{x_2}(\theta_k)|^2\right)^{1/2}, \quad x_1,x_2\in \mathcal{X},
\end{equation*}
and for which \eqref{eq:phi_Lipschitz} implies
\begin{equation}
\label{eq:metric_bound}
	d_K(x_1,x_2) \le \frac{M_f L_{\sigma} M_A}{\sqrt{K}}\|x_1-x_2\| .
\end{equation}
We now apply Dudley's entropy bound (\citealp{vandervaartWeakConvergenceEmpirical2023}, Corollary 2.2.9). For every $x_0\in \mathcal{X}$, it holds 
\begin{align}
\label{eq:expectationW}
	\mathbb{E}_\varepsilon \left[ \sup_{x\in\mathcal{X}} | W_x |
    \right] &\le \mathbb{E}_\varepsilon \left[  | W_{x_0} |
    \right]+ C \int_0^\infty \sqrt{ \log \mathcal{N}(r,\mathcal{X},d_K) } dr \nonumber\\
    &= \mathbb{E}_\varepsilon \left[  | W_{x_0} |
    \right]+ C \int_0^{\mathrm{diam}(\mathcal{X},d_K)}\sqrt{ \log \mathcal{N}(r,\mathcal{X},d_K) } dr,
\end{align} 
where $\mathcal{N}(r,\mathcal{X},d_K)$ is the covering number of $\mathcal{X}$ with respect to the pseudometric $d_K$, $C>0$ is a universal constant which absorbs factor $2$ from the standard relation between covering and packing numbers, $\mathcal{N}(r,\mathcal{X},d_K)\leq \mathcal{D}(r,\mathcal{X},d_K)\leq \mathcal{N}(r/2,\mathcal{X},d_K)$, with $\mathcal{D}$ denoting the packing number  (see Definition 2.2.3 in  \citealp{vandervaartWeakConvergenceEmpirical2023}). The equality in~\eqref{eq:expectationW} is obtained by noticing that $\mathcal{N}(r,\mathcal{X},d_K)=1$ for $r>\mathrm{diam}(\mathcal{X},d_K)$. By \eqref{eq:metric_bound}, one has
\begin{equation}
\label{eq:N_volume}
	\mathcal{N}(r,\mathcal{X},d_K) \le \mathcal{N} \left( \frac{r \sqrt{K}}{M_f L_{\sigma} M_A}, \mathcal{X}, \|\cdot\| \right).
\end{equation}
By Assumption~\ref{assumption:X_domain}, $\mathcal{X}=[-M,M]^d$, and hence the volumetric covering bound (see, for example, Corollary 4.2.11 in~\cite{vershynin2026high}) gives for any $s>0$,
\begin{equation}
\label{eq:logN}
	\log \mathcal{N}(s,\mathcal{X},\|\cdot\|) \le d\log \left(1+\frac{2M\sqrt{d}}{s}\right).
\end{equation}
Notice that \eqref{eq:metric_bound} implies that 
$
\mathrm{diam}(\mathcal{X},d_K)
\le
\frac{M_f L_{\sigma} M_A}{\sqrt{K}} \mathrm{diam}(\mathcal{X},\|\cdot\|)
=
\frac{2M\sqrt{d} M_f L_{\sigma} M_A}{\sqrt{K}}$. Additionally, by Cauchy-Schwarz,   definition~\eqref{eq:def_Wx}, and Assumption~\ref{assumption:direct_source_condition_holder_product},
\begin{equation}
\label{eq:W_0_term}
    \mathbb{E}_\varepsilon \left[  | W_{x_0} |\right]\leq \left(\mathbb{E}_\varepsilon \left[  W_{x_0} ^2\right]\right)^{1/2}=\dfrac{1}{K}\left(\sum_{k=1}^K|\phi_{x_0}(\theta_{k})|^{2}\right)^{1/2}\leq \dfrac{1}{\sqrt{K}}M_fM_{\sigma}.
\end{equation}

Taking $s=r\sqrt{K}/(M_f L_{\sigma} M_A)$ in \eqref{eq:logN} and using \eqref{eq:N_volume} and \eqref{eq:W_0_term} in \eqref{eq:expectationW} yields
\begin{equation*}
	\mathbb{E}_\varepsilon \left[ \sup_{x\in\mathcal{X}} | W_x |
\right] \le \frac{1}{\sqrt{K}} \left(M_f M_{\sigma}+C M_f L_{\sigma} M_A \int_0^{2M\sqrt{d}}
\sqrt{ d  \log \left(1+\frac{2M\sqrt{d}}{s}\right) } ds\right),
\end{equation*}
which is finite.
 
Combining this expression with \eqref{eq:symmetrization} yields the required result with 
\begin{equation*}
    C_{\mathrm{ent}} := 2\left(M_f M_{\sigma}+C M_f L_{\sigma} M_A \int_0^{2M\sqrt{d}}
\sqrt{ d \log \left(1+\frac{2M\sqrt{d}}{s}\right) } ds\right).
\end{equation*}
\end{proof}

\clearpage

\section{Proofs for spectral properties and $L^2$ ridge estimation}
\label{app:ridge_L2_proofs}

\subsection{Spectral deterioration and ridge effective dimension}
\label{app:spectral_proofs}

\begin{proof}[Proof of Proposition~\ref{prop:spectral_tail_parametrized_sieve}]
Fix $m<K$ and set $r_m:=(C_\Pi/m)^{1/q_\Pi}$, with both $m$ and $K$ sufficiently large so that $r_m < \overline{r}$. 
By \eqref{eq:parameter_covering}, there exists an  $r_m/2$-net $\{\vartheta_1,\ldots,\vartheta_{M_m}\}$ with $M_m\leq m$. For each $\theta_k$, choose $\vartheta_{c(k)}$ such that $\rho_\Pi(\theta_k,\vartheta_{c(k)})\leq r_m/2$, and set $\overline{g}_K(x):=\bigl(\phi_{\vartheta_{c(1)}}(x),\ldots,\phi_{\vartheta_{c(K)}}(x)\bigr)^\top$. Since $\overline{g}_K(x)$ lies in a subspace of $\R^K$ of dimension at most $m$, Theorem~1 of \citet{Fan1949} and \eqref{eq:parameter_feature_lipschitz} yield
\begin{equation*}
    \sum_{j=m+1}^K s_j^2\leq\E_X\|g_K(X)-\overline{g}_K(X)\|^2=\sum_{k=1}^K\|\phi_{\theta_k}-\phi_{\vartheta_{c(k)}}\|_{L^2(X)}^2\leq K L_\phi^2 r_m^2=K L_\phi^2 C_\Pi^{2/q_\Pi} m^{-2/q_\Pi},
\end{equation*}
which with $C_{\mathrm{tail}}:=L_\phi^2C_\Pi^{2/q_\Pi}$ proves \eqref{eq:spectral_tail_sieve}.  
For all sufficiently large $j$, taking $m=\lfloor j/2\rfloor$ and assuming $r_m < \overline{r}$ gives
\begin{equation*}
    (j-m)s_j^2 \leq \sum_{\ell=m+1}^K s_\ell^2 \leq C_{\mathrm{tail}}Km^{-2/q_\Pi}.
\end{equation*}
Since $j-m\geq j/2$ and $m\geq j/3$, taking $C_{\mathrm{eig}}=2\cdot 3^{2/q_\Pi}C_{\mathrm{tail}}$ yields \eqref{eq:bound_eigenvalue_sieve}. Finally, taking $j=K$ proves \eqref{eq:smallest_eigenvalue_sieve}, as required.
\end{proof}
\begin{proof}[Proof of Corollary~\ref{cor:effective_dimension_spectral}]
We start by noticing that, since $0\leq\frac{s_j^2}{s_j^2+\lambda_{K_n,n}}\leq 1$, the definition of $\mathcal{N}_{K_n,\lambda}$ in \eqref{eq:def_L2_effective_dimension} immediately gives the first term in the minimum in \eqref{eq:effective_dimension_spectral_rate}.

Next, again, using \eqref{eq:def_L2_effective_dimension} and Proposition~\ref{prop:spectral_tail_parametrized_sieve}, we obtain
\begin{equation}
\label{eq:N_cal_bound_proof}
\mathcal{N}_{K_n,\lambda}\leq m+\frac{1}{\lambda_{K_n,n}}\sum_{j=m+1}^{K_n}s_j^2\leq m+\frac{C_{\mathrm{tail}}K_n}{\lambda_{K_n,n}}m^{-2/q_\Pi}.
\end{equation}
Minimizing the two terms on the right-hand side with respect to $m>0$ gives
\begin{equation}
\label{eq:mn_def}
	m_n^*:=\left(\frac{2C_{\mathrm{tail}}K_n}{q_\Pi\lambda_{K_n,n}}\right)^{q_\Pi/(q_\Pi+2)}\qquad \text{and} \qquad m_n:=\lceil m_n^*\rceil .
\end{equation}
Since $K_n/\lambda_{K_n,n}\to\infty$, we have $m_n\to\infty$, while $\lambda_{K_n,n}K_n^{2/q_\Pi}\to\infty$ implies $m_n/K_n\to0$. Hence $m_n$ is admissible for all sufficiently large $n$. Substituting $m_n$ into \eqref{eq:N_cal_bound_proof} and using that, by \eqref{eq:mn_def}, $m_n\leq m_n^*+1$ and $m_n^{-2/q_\Pi}\leq(m_n^*)^{-2/q_\Pi}$ gives
\begin{align*}
\mathcal{N}_{K_n,\lambda}&\leq 1+m_n^*+\frac{C_{\mathrm{tail}}K_n}{\lambda_{K_n,n}}(m_n^*)^{-2/q_\Pi}\\
&=1+\left(\frac{2C_{\mathrm{tail}}K_n}{q_\Pi\lambda_{K_n,n}}\right)^{q_\Pi/(q_\Pi+2)}+\frac{C_{\mathrm{tail}}K_n}{\lambda_{K_n,n}}\left(\frac{2C_{\mathrm{tail}}K_n}{q_\Pi\lambda_{K_n,n}}\right)^{-2/(q_\Pi+2)}\\
&=1+\left[\left(\frac{2}{q_\Pi}\right)^{q_\Pi/(q_\Pi+2)}+\left(\frac{q_\Pi}{2}\right)^{2/(q_\Pi+2)}\right]\left(\frac{C_{\mathrm{tail}}K_n}{\lambda_{K_n,n}}\right)^{q_\Pi/(q_\Pi+2)},
\end{align*}
where defining $C_{\mathcal{N}}$ as a constant in front of the second summand gives the second argument in  \eqref{eq:effective_dimension_spectral_rate}.

Finally, dividing both sides of the inequality above by $K_n$ gives
\begin{equation*}
    \frac{\mathcal{N}_{K_n,\lambda}}{K_n}\leq \frac{1}{K_n} + C_{\mathcal{N}}C_{\mathrm{tail}}^{q_\Pi/(q_\Pi+2)}\left(\lambda_{K_n,n}K_n^{2/q_\Pi}\right)^{-q_\Pi/(q_\Pi+2)}=o(1),
\end{equation*}
which proves \eqref{eq:effective_dimension_o_K}, as required.
\end{proof}

\subsection{Ridge-normalized covariance stability and $L^2$ estimation}
\label{app:L2_ridge_proofs}
\begin{proof}[Proof of Theorem~\ref{theorem:L2_ridge_sharp_consistency}]
We follow the proof of Theorem~4.1 in \cite{Belloni2015}, keeping track of the ridge normalization. First, by the triangle inequality,
\begin{align}\label{eq:l2_ridge_initial_bound}
    \norm{ \widehat{h}_{n,\lambda} - h }_{L^2(X)}
    &\leq \norm{ {g}_{K_n}^\top (\widehat{\beta}_{{K_n},n,\lambda} - \beta_{K_n}) }_{L^2(X)} + \norm{ {g}_{K_n}^\top \beta_{K_n} - h }_{L^2(X)} \nonumber\\
    &= \norm{ \Gamma_{K_n}^{1/2} (\widehat{\beta}_{{K_n},n,\lambda} - \beta_{K_n}) } + \Delta_{K_n},
\end{align}
where the equality follows from $\|g_{K_n}^\top b\|_{L^2(X)}^2=b^\top\Gamma_{K_n}b=\|\Gamma_{K_n}^{1/2}b\|^2$ for every $b\in\mathbb{R}^{K_n}$.

By \eqref{eq:def_Dn_hat_An} and \eqref{eq:nondegeneracy}, $\|\widehat{D}_{K_n,n,\lambda}\|=O_P(1)$.

Define the approximation residuals $\eta_i := h(X_i) - g_{{K_n}}(X_i)^\top \beta_{K_n}$, and $\eta:=(\eta_1,\ldots, \eta_n)^\top$. Then $Y=G_{K_n,n}\beta_{K_n}+\eta+e$. Using this together with the definition of $ \widehat{{\beta}}_{K_n,n,\lambda}$ in \eqref{eq:def_betahat}, we obtain 
\begin{align*} 
    \Gamma_{K_n}^{1/2}\left( 
    \widehat{\beta}_{K_n,n,\lambda} - \beta_{K_n} \right)
    &= \Gamma_{K_n}^{1/2} \widehat\Gamma_{K_n,n,\lambda}^{-1} \left(n^{-1}G_{K_n,n}^\top Y\right) - \Gamma_{K_n}^{1/2}\beta_{K_n} \nonumber\\ 
    &= \Gamma_{K_n}^{1/2} \widehat\Gamma_{K_n,n,\lambda}^{-1} \left(n^{-1}G_{K_n,n}^\top e\right) + \Gamma_{K_n}^{1/2} \widehat\Gamma_{K_n,n,\lambda}^{-1} \left(n^{-1}G_{K_n,n}^\top \eta\right) \nonumber\\ 
    &\quad\quad+ \Gamma_{K_n}^{1/2} \widehat\Gamma_{K_n,n,\lambda}^{-1} \left(n^{-1}G_{K_n,n}^\top G_{K_n,n}\beta_{K_n}\right) - \Gamma_{K_n}^{1/2}\beta_{K_n}\nonumber\\ 
    &= \Gamma_{K_n}^{1/2} \widehat\Gamma_{K_n,n,\lambda}^{-1} \left(n^{-1}G_{K_n,n}^\top e\right) + \Gamma_{K_n}^{1/2} \widehat\Gamma_{K_n,n,\lambda}^{-1} \left(n^{-1}G_{K_n,n}^\top \eta\right) \nonumber\\ 
    &\quad\quad+ \Gamma_{K_n}^{1/2} \widehat\Gamma_{K_n,n,\lambda}^{-1} \left( \widehat\Gamma_{K_n,n,\lambda} - \lambda_{K_n,n}I_{K_n} \right)\beta_{K_n} - \Gamma_{K_n}^{1/2}\beta_{K_n}\nonumber\\ 
    &= \Gamma_{K_n}^{1/2} \widehat\Gamma_{K_n,n,\lambda}^{-1} \left(n^{-1}G_{K_n,n}^\top e\right) + \Gamma_{K_n}^{1/2} \widehat\Gamma_{K_n,n,\lambda}^{-1} \left(n^{-1}G_{K_n,n}^\top \eta\right) \nonumber\\ 
    &\quad\quad - \lambda_{K_n,n} \Gamma_{K_n}^{1/2} \widehat\Gamma_{K_n,n,\lambda}^{-1} \beta_{K_n}, 
\end{align*}
where in the third equality we used that $n^{-1}G_{K_n,n}^\top G_{K_n,n} = \widehat\Gamma_{K_n,n,\lambda} - \lambda_{K_n,n}I_{K_n}$.
Consequently, for the first term in the right-hand side of \eqref{eq:l2_ridge_initial_bound} it holds that
\begin{align}
\label{eq:l2_ridge_beta_bound}
\norm{\Gamma_{K_n}^{1/2}\left( 
    \widehat{\beta}_{K_n,n,\lambda} - \beta_{K_n} \right)} \leq & \: \norm{\Gamma_{K_n}^{1/2} \widehat\Gamma_{K_n,n,\lambda}^{-1} \left(n^{-1}G_{K_n,n}^\top e\right)} +\norm{\Gamma_{K_n}^{1/2} \widehat\Gamma_{K_n,n,\lambda}^{-1} \left(n^{-1}G_{K_n,n}^\top \eta\right)} \nonumber \\
    &\quad +\lambda_{K_n,n}\norm{\Gamma_{K_n}^{1/2} \widehat\Gamma_{K_n,n,\lambda}^{-1} \beta_{K_n}}.
\end{align}
We now proceed in steps and study each term in the right-hand side of \eqref{eq:l2_ridge_beta_bound} separately.

\smallskip
\noindent\textsc{Step 1.}
Using
\begin{equation*}
	\widehat{\Gamma}_{K_n,n,\lambda}^{-1}=\Gamma_{K_n,\lambda}^{-1/2}\widehat{D}_{K_n,n,\lambda}\Gamma_{K_n,\lambda}^{-1/2},
\end{equation*}
we obtain
\begin{align}
\norm{\Gamma_{K_n}^{1/2} \widehat\Gamma_{K_n,n,\lambda}^{-1} \left(n^{-1}G_{K_n,n}^\top e\right)}&=\left\|\Gamma_{K_n}^{1/2}\Gamma_{K_n,\lambda}^{-1/2}\widehat{D}_{K_n,n,\lambda}\Gamma_{K_n,\lambda}^{-1/2}\left(n^{-1}G_{K_n,n}^{\top}e\right)\right\|\nonumber\\
&\le\left\|\Gamma_{K_n}^{1/2}\Gamma_{K_n,\lambda}^{-1/2}\right\|\left\|\widehat{D}_{K_n,n,\lambda}\right\| \left\|\Gamma_{K_n,\lambda}^{-1/2}\left(n^{-1}G_{K_n,n}^{\top}e\right)\right\|.
\label{eq:term_first}
\end{align}
We recall that we have already established that $\|\widehat{D}_{{K_n},n,\lambda}\|=O_P(1)$. Since $\Gamma_{K_n}\preceq\Gamma_{K_n,\lambda}$, it follows that $\left\|\Gamma_{K_n}^{1/2}\Gamma_{K_n,\lambda}^{-1/2}\right\| \le 1$ and hence it is enough to bound the last factor in \eqref{eq:term_first}, for which we write
\begin{align*}
    \E\big[ \norm{\Gamma_{K_n,\lambda}^{-1/2}\left(n^{-1}G_{K_n,n}^{\top}e\right)}^2 \big]
    & = \E\bigg[  \bigg\| n^{-1}\sum_{i=1}^n \Gamma_{K_n,\lambda}^{-1/2}{g}_{K_n}(X_i)\epsilon_i \bigg\|^2 \bigg]\nonumber \\
    &=n^{-1}\E\big[  \epsilon_i^2 {g}_{K_n}(X_i)^\top \Gamma_{K_n,\lambda}^{-1} {g}_{K_n}(X_i) \big]\nonumber\\
    & \leq n^{-1}\bar{\sigma}^2  \operatorname{tr}\left(  \Gamma_{K_n} \Gamma_{K_n,\lambda}^{-1}  \right),
\end{align*}
where we used that, under the hypotheses of the theorem, $\E[\epsilon_i|X_i]=0$, the summands are centered, and the data are i.i.d. 

Hence, by Markov's inequality, with \eqref{eq:def_L2_effective_dimension}, one has
\begin{align}
\label{eq:term_first_Op}
\norm{\Gamma_{K_n,\lambda}^{-1/2}\left(n^{-1}G_{K_n,n}^{\top}e\right)} & = O_P\left(\sqrt{\frac{ \mathcal{N}_{K_n, \lambda}}{n}}\right),
\end{align}
which gives the same bound for \eqref{eq:term_first}.

\smallskip
\noindent\textsc{Step 2.}
We handle the second term in \eqref{eq:l2_ridge_beta_bound} in a similar way to the first. Using \eqref{eq:def_Dn_hat_An}, we first notice that $\widehat{\Gamma}_{K_n,n,\lambda}^{-1}=\Gamma_{K_n,\lambda}^{-1/2} \widehat{D}_{K_n,n,\lambda}\Gamma_{K_n,\lambda}^{-1/2}$ and write 
\begin{equation*}
    \norm{\Gamma_{K_n}^{1/2} \widehat\Gamma_{K_n,n,\lambda}^{-1} \left(n^{-1}G_{K_n,n}^\top \eta\right)}
    \leq \norm{\Gamma_{K_n}^{1/2}\Gamma_{K_n,\lambda}^{-1/2}} \norm{ \widehat{D}_{{K_n},n,\lambda} } \norm{ \Gamma_{K_n,\lambda}^{-1/2} \left(n\inv {G}_{{K_n},n}^\top \eta \right)} .
\end{equation*}
Since $\|\widehat{D}_{{K_n},n,\lambda}\| = O_P(1)$ and $\norm{\Gamma_{K_n}^{1/2}\Gamma_{K_n,\lambda}^{-1/2}}\leq 1$, it remains to analyze the last factor. 

Using the definition of $\eta_i$ and the fact that $\beta_{K_n}$ is an $L^2(X)$ projection coefficient and $\E\left[g_{K_n}(X_i)\eta_i\right]=0$, we immediately have that
\begin{equation*}
    \E[ g_{K_n}(X_i) \eta_i ] = \E_X[ g_{K_n}(X_i) ( h(X_i) - g_{K_n}(X_i)^\top \beta_{K_n} ) ] = 0 ,
\end{equation*}
which further implies $\E_X [\Gamma_{K_n,\lambda}^{-1/2}g_{K_n}(X)\eta(X)]=0$. We can now write 
\begin{align}
\label{eq:expect_gamma}
    \E\big[ \norm{ \Gamma_{K_n,\lambda}^{-1/2}\left(n\inv {G}_{{K_n},n}^\top \eta\right) }^2 \big]
    & = n^{-2}\E\bigg[ \bigg\| \sum_{i=1}^n \widetilde{g}_{K_n}(X_i)\eta_i \bigg\|^2 \bigg] = n^{-1} \E\big[ \eta_i^2 \|\widetilde{g}_{K_n}(X_i)\|^2 \big] \nonumber\\
    & \leq \frac{ \widetilde{\xi}_{K_n,\lambda}^2 }{ n } \, \E[\eta_i^2] 
        = \frac{ \widetilde{\xi}_{K_n,\lambda}^2 }{ n } \, \Delta_{K_n}^2 ,
\end{align}
where in the first equality we applied again the fact that cross terms are zero by independence; in the inequality, we used $\norm{ \widetilde{g}_{K_n}(X_i) } \leq \sup_{x \in \mathcal{X}} \norm{ \widetilde{g}_{K_n}(x) } \leq \widetilde{\xi}_{{K_n},\lambda}$, while the last equality follows from the definition $\E[\eta_i^2] = \|h-g_{K_n}^\top\beta_{K_n}\|_{L^2(X)}^2 = \Delta_{K_n}^2$.

By \eqref{eq:expect_gamma}, Markov's inequality implies $\norm{ \Gamma_{K_n,\lambda}^{-1/2}\left(n\inv {G}_{{K_n},n}^\top \eta\right) } = O_P( n^{-1/2}  \widetilde{\xi}_{K_n,\lambda} \Delta_{K_n}  )$, while Assumption~\ref{assumption:L2_ridge_gram_stability} yields $n^{-1/2}\widetilde{\xi}_{K_n,\lambda}=o(1)$ (see again Remark~\ref{remark:L2_ridge_assumptions_discussion}), and hence one gets 
\begin{align}
\label{eq:term_second_Op}
\norm{\Gamma_{K_n}^{1/2} \widehat\Gamma_{K_n,n,\lambda}^{-1} \left(n^{-1}G_{K_n,n}^\top \eta\right)} & = O_P\left(\frac{\widetilde{\xi}_{K_n,\lambda}}{\sqrt{n}}\Delta_{K_n}\right)=O_P\left(\Delta_{K_n}\right).
\end{align}

\smallskip
\noindent\textsc{Step 3.}
For the third and last term in \eqref{eq:l2_ridge_beta_bound}, adding and subtracting $\lambda_{K_n,n}\Gamma_{K_n}^{1/2}\Gamma_{K_n,\lambda}^{-1}\beta_{K_n}$ yields
\begin{align}
\label{eq:term_III}
    \lambda_{{K_n},n}\norm{  \Gamma_{{K_n}}^{1/2} \widehat{\Gamma}_{{K_n},n,\lambda}\inv \beta_{K_n} }
    &\leq \lambda_{{K_n},n}\norm{  \Gamma_{{K_n}}^{1/2} {\Gamma}_{{K_n},\lambda}\inv \beta_{K_n} }+\lambda_{{K_n},n}
    \norm{  \Gamma_{K_n}^{1/2}\left(\widehat{\Gamma}_{K_n,n,\lambda}^{-1}-\Gamma_{K_n,\lambda}^{-1}\right)\beta_{K_n}},
\end{align}
where the first term on the right-hand side is the population ridge bias measured in the $L^2(X)$ prediction norm, $B^{\mathrm{pop}}_{K_n,\lambda}$. For the second term, we use that $\widehat{\Gamma}_{K_n,n,\lambda}^{-1}-\Gamma_{K_n,\lambda}^{-1}= - \widehat{\Gamma}_{K_n,n,\lambda}^{-1} \left(\widehat{\Gamma}_{K_n,n}-{\Gamma}_{K_n}\right){\Gamma}_{K_n,\lambda}^{-1}$ and apply once more the identity for $\widehat{\Gamma}_{K_n,n,\lambda}^{-1}$ in the beginning of Step 1 to get
\begin{align*}
    & \lambda_{{K_n},n}
    \norm{  \Gamma_{K_n}^{1/2}\left(\widehat{\Gamma}_{K_n,n,\lambda}^{-1}-\Gamma_{K_n,\lambda}^{-1}\right)\beta_{K_n}} \\
    & \qquad\quad \leq \lambda_{{K_n},n}\norm{\Gamma_{K_n}^{1/2}\Gamma_{K_n,\lambda}^{-1/2}}\norm{ \widehat{D}_{{K_n},n,\lambda} } \norm{\Gamma_{K_n,\lambda}^{-1/2}\left(\widehat{\Gamma}_{K_n,n}-\Gamma_{K_n}\right)\Gamma_{K_n,\lambda}^{-1}\beta_{K_n}}\\
    & \qquad\quad \leq \lambda_{{K_n},n}\norm{ \widehat{D}_{{K_n},n,\lambda} }\norm{\Gamma_{K_n,\lambda}^{-1/2}\left(\widehat{\Gamma}_{K_n,n}-\Gamma_{K_n}\right)\Gamma_{K_n,\lambda}^{-1}\beta_{K_n}}.
\end{align*}
Set  $v_{K_n,\lambda}:=\Gamma_{K_n,\lambda}^{-1}\beta_{K_n}$ and define 
\begin{equation*}
    \zeta_{i,K_n,\lambda}
    :=
    \Gamma_{K_n,\lambda}^{-1/2} \left(g_{K_n}(X_i)g_{K_n}(X_i)^\top-\Gamma_{K_n}\right) v_{K_n,\lambda} , 
    \quad i=1,\ldots,n.
\end{equation*}
Then $\E[\zeta_{i,K_n,\lambda}]=0$, since $\E\left[g_{K_n}(X_i)g_{K_n}(X_i)^\top\right]=\Gamma_{K_n}$. Additionally, it holds that $\Gamma_{K_n,\lambda}^{-1/2} \allowbreak (\widehat{\Gamma}_{K_n,n} \allowbreak -\Gamma_{K_n} ) v_{K_n,\lambda}=n\inv \sum_{i=1}^n \zeta_{i,K_n,\lambda}$. Moreover,
\begin{align*}
    \E\left[\|\zeta_{i,K_n,\lambda}\|^2\right]
    &= \E_X\left[ \big\| \widetilde{g}_{K_n}(X) g_{K_n}(X)^\top v_{K_n,\lambda} - \Gamma_{K_n,\lambda}^{-1/2} \Gamma_{K_n} v_{K_n,\lambda} \big\|^2 \right] \\
    &= \E_X\left[ \|\widetilde{g}_{K_n}(X)\|^2\left(g_{K_n}(X)^\top v_{K_n,\lambda}\right)^2 \right] + v_{K_n,\lambda}^\top \Gamma_{K_n} \Gamma_{K_n,\lambda}^{-1} \Gamma_{K_n} v_{K_n,\lambda} \\
    &\qquad\quad - 2\, v_{K_n,\lambda}^\top \E_X\left[ g_{K_n}(X) \widetilde{g}_{K_n}(X)^\top \right] \Gamma_{K_n,\lambda}^{-1/2} \Gamma_{K_n} v_{K_n,\lambda} \\
    &\leq \E_X\left[ \|\widetilde{g}_{K_n}(X)\|^2\left(g_{K_n}(X)^\top v_{K_n,\lambda}\right)^2 \right]\\
    &\leq \widetilde{\xi}_{K_n,\lambda}^2 \left( v_{K_n,\lambda}^\top\Gamma_{K_n}v_{K_n,\lambda} \right) = \widetilde{\xi}_{K_n,\lambda}^2\left(\frac{B^{\mathrm{pop}}_{K_n,\lambda}}{\lambda_{K_n,n}}\right)^2 ,
\end{align*}
where the first inequality follows from the fact that $\E_X\left[ g_{K_n}(X_i) \widetilde{g}_{K_n}(X)^\top \right] = \Gamma_{K_n} \Gamma_{K_n,\lambda}^{-1/2}$.
Hence, by independence and Markov's inequality,
\begin{equation*}
	\left\|\Gamma_{K_n,\lambda}^{-1/2}\left(\widehat{\Gamma}_{K_n,n}-\Gamma_{K_n}\right)v_{K_n,\lambda}\right\|=O_P\left(\frac{\widetilde{\xi}_{K_n,\lambda}}{\sqrt{n}}\frac{B^{\mathrm{pop}}_{K_n,\lambda}}{\lambda_{K_n,n}}\right).
\end{equation*}
Since $\|\widehat D_{K_n,n,\lambda}\|=O_P(1)$, it follows that
\begin{equation*}
\lambda_{K_n,n}\left\|\Gamma_{K_n}^{1/2}\left(\widehat{\Gamma}_{K_n,n,\lambda}^{-1}-\Gamma_{K_n,\lambda}^{-1}\right)\beta_{K_n}\right\|=O_P\left(B^{\mathrm{pop}}_{K_n,\lambda}\frac{\widetilde{\xi}_{K_n,\lambda}}{\sqrt n}\right).
\end{equation*}
Assumption~\ref{assumption:L2_ridge_gram_stability} implies $\widetilde{\xi}_{K_n,\lambda}/\sqrt n=o(1)$, and therefore the third term in \eqref{eq:l2_ridge_beta_bound} is $O_P(B^{\mathrm{pop}}_{K_n,\lambda})$.

\medskip
Combining the initial error bound \eqref{eq:l2_ridge_initial_bound} with our derivations for the three terms in \eqref{eq:l2_ridge_beta_bound}, namely, \eqref{eq:term_first_Op}, \eqref{eq:term_second_Op}, and the third one immediately above, we obtain the required bound in \eqref{eq:L2_ridge_refined_bound}, completing the proof.
\end{proof}
\begin{proof}[Proof of Lemma~\ref{lemma:L2_ridge_empirical_prediction}]
Similarly to the  decomposition \eqref{eq:l2_ridge_beta_bound} in the proof of Theorem~\ref{theorem:L2_ridge_sharp_consistency}, we write 
\begin{align}
\label{eq:empirical_prediction_beta_decomposition}
    \widehat{\Gamma}_{K_n,n}^{1/2}\left(\widehat{\beta}_{K_n,n,\lambda}-\beta_{K_n}\right)
    &=\widehat{\Gamma}_{K_n,n}^{1/2}\widehat{\Gamma}_{K_n,n,\lambda}^{-1}\left( n^{-1} G_{K_n,n}^{\top}e\right)+\widehat{\Gamma}_{K_n,n}^{1/2}\widehat{\Gamma}_{K_n,n,\lambda}^{-1}\left( n^{-1} G_{K_n,n}^{\top}\eta\right)\nonumber\\
    &\qquad -\lambda_{K_n,n}\widehat{\Gamma}_{K_n,n}^{1/2}\widehat{\Gamma}_{K_n,n,\lambda}^{-1}\beta_{K_n}.
\end{align}
Next, since $\widehat{\Gamma}_{K_n,n}\preceq \widehat{\Gamma}_{K_n,n,\lambda}$, for every $v\in\mathbb{R}^{K_n}$ it holds that
\begin{align}
\label{eq:empirical_prediction_operator_bound}
    \left\|\widehat{\Gamma}_{K_n,n}^{1/2}
    \widehat{\Gamma}_{K_n,n,\lambda}^{-1}v
    \right\|^2&\leq v^\top\widehat{\Gamma}_{K_n,n,\lambda}^{-1}v=
    \left(\Gamma_{K_n,\lambda}^{-1/2}v\right)^\top \widehat{D}_{K_n,n,\lambda}\left(\Gamma_{K_n,\lambda}^{-1/2}v\right)\nonumber\\
    &\leq\|\widehat{D}_{K_n,n,\lambda}\|\left\|\Gamma_{K_n,\lambda}^{-1/2}v\right\|^2.
\end{align}
Under Assumption~\ref{assumption:L2_ridge_gram_stability}, $\|\widehat D_{K_n,n,\lambda}\|=O_P(1)$, and applying  \eqref{eq:empirical_prediction_operator_bound} to the first term in \eqref{eq:empirical_prediction_beta_decomposition}, together with the corresponding bound \eqref{eq:term_first_Op} derived in Step~1 of the proof of Theorem~\ref{theorem:L2_ridge_sharp_consistency}, gives
\begin{equation}
\label{eq:term_first_Op_empirical}
    \left\|\widehat{\Gamma}_{K_n,n}^{1/2}\widehat{\Gamma}_{K_n,n,\lambda}^{-1}\left(\frac{1}{n}G_{K_n,n}^{\top}e\right)\right\|=O_P\left(\sqrt{\frac{\mathcal{N}_{K_n,\lambda}}{n}}\right).
\end{equation}
Similarly, for the second term in \eqref{eq:empirical_prediction_beta_decomposition}, we use the corresponding bound \eqref{eq:term_second_Op} derived in Step~2 of the proof of Theorem~\ref{theorem:L2_ridge_sharp_consistency} and obtain 
\begin{equation}
\label{eq:term_second_Op_empirical}
    \left\|\widehat{\Gamma}_{K_n,n}^{1/2}\widehat{\Gamma}_{K_n,n,\lambda}^{-1}\left(\frac{1}{n}G_{K_n,n}^{\top}\eta\right)\right\|=O_P\left(\frac{\widetilde{\xi}_{K_n,\lambda}\Delta_{K_n}}{\sqrt n}\right)=O_P\left(\Delta_{K_n}\right),
\end{equation}
where in the last step we used that, by Assumption~\ref{assumption:L2_ridge_gram_stability}, ${\widetilde{\xi}_{K_n,\lambda}}/{\sqrt n}=O_P(1)$.

For the last term, again, similarly to \eqref{eq:term_III}, we write 
\begin{align}
\label{eq:third_term_empirical}
    \lambda_{K_n,n}\widehat{\Gamma}_{K_n,n}^{1/2}\widehat{\Gamma}_{K_n,n,\lambda}^{-1}\beta_{K_n}&=\lambda_{K_n,n}\widehat{\Gamma}_{K_n,n}^{1/2}\Gamma_{K_n,\lambda}^{-1}\beta_{K_n}\nonumber\\
    &\qquad+\lambda_{K_n,n}\widehat{\Gamma}_{K_n,n}^{1/2}\left(\widehat{\Gamma}_{K_n,n,\lambda}^{-1}-\Gamma_{K_n,\lambda}^{-1}\right)\beta_{K_n}.
\end{align}
For the first term on the right-hand side of \eqref{eq:third_term_empirical}, we write
\begin{align*}
    \E\left[\left\|\lambda_{K_n,n}\widehat{\Gamma}_{K_n,n}^{1/2}
    \Gamma_{K_n,\lambda}^{-1}\beta_{K_n}\right\|^2\right]
    &=\left\|\lambda_{K_n,n} \Gamma_{K_n}^{1/2}
    \Gamma_{K_n,\lambda}^{-1}\beta_{K_n} \right\|^2=\left(B^{\mathrm{pop}}_{K_n,\lambda}\right)^2,
\end{align*}
where in the first equality we used that $\E\big[ \widehat{\Gamma}_{K_n,n} \big] = \Gamma_{K_n}$.
Consequently, by Markov's inequality,
\begin{equation}
\label{eq:term_third_Op_empirical}
    \left\|\lambda_{K_n,n}\widehat{\Gamma}_{K_n,n}^{1/2}
    \Gamma_{K_n,\lambda}^{-1}\beta_{K_n}\right\|=O_P\left(B^{\mathrm{pop}}_{K_n,\lambda}\right).
\end{equation}
For the second term in \eqref{eq:third_term_empirical}, again similarly to Step~3 of the proof of  Theorem~\ref{theorem:L2_ridge_sharp_consistency}, we obtain 
\begin{equation}
\begin{aligned}
\label{eq:term_forth_Op_empirical}
    & \lambda_{K_n,n}\left\|\widehat{\Gamma}_{K_n,n}^{1/2}\left(\widehat{\Gamma}_{K_n,n,\lambda}^{-1}-\Gamma_{K_n,\lambda}^{-1}\right)\beta_{K_n}\right\| \\ 
    &\qquad\qquad \leq \lambda_{K_n,n}\|\widehat{D}_{K_n,n,\lambda}\|^{1/2}\left\|\Gamma_{K_n,\lambda}^{-1/2}\left(\widehat{\Gamma}_{K_n,n}-\Gamma_{K_n}\right)\Gamma_{K_n,\lambda}^{-1}\beta_{K_n}\right\| \\
    &\qquad\qquad = O_P\left(\frac{\widetilde{\xi}_{K_n,\lambda}B^{\mathrm{pop}}_{K_n,\lambda}}{\sqrt n}\right)=O_P\left(B^{\mathrm{pop}}_{K_n,\lambda}\right),
\end{aligned}
\end{equation}
where the last step follows from noting that, under Assumption~\ref{assumption:L2_ridge_gram_stability}, ${\widetilde{\xi}_{K_n,\lambda}}/{\sqrt n}=O_P(1)$.

Combining \eqref{eq:term_first_Op_empirical}, \eqref{eq:term_second_Op_empirical}, \eqref{eq:term_third_Op_empirical}, and \eqref{eq:term_forth_Op_empirical} in \eqref{eq:empirical_prediction_beta_decomposition} proves the claim, as required.
\end{proof}
\subsection{RWNN specialization and primitive rates}
\label{app:RWNN_L2_proofs}

\begin{proof}[Proof of Corollary~\ref{cor:RWNN_smallest_eigenvalue}]
As in the proof of Lemma~\ref{lem:nn_l2_bound}, compactness of $\Pi\subset\mathbb R^{d+1}$ and the equivalence of $\rho_\Pi$ with the Euclidean metric established in Remark~\ref{remark:metrics} gives
$|\mathcal{N}_{r/2}|\lesssim r^{-(d+1)}$. Moreover, Assumption~\ref{assumption:activation_Lipschitz} and \eqref{eq:rho_compatibility} imply $\|\phi_{\theta_1}-\phi_{\theta_2}\|_{L^2(X)} \leq L_\sigma L_\rho\rho_\Pi(\theta_1,\theta_2)$. The claim follows from Proposition~\ref{prop:spectral_tail_parametrized_sieve} with $q_\Pi=d+1$.
\end{proof}
\begin{proof}[Proof of Lemma~\ref{lem:RWNN_population_ridge_bias}]
Let $W_{K_n}^*$ be the control-neighbors coefficient vector used in the proof of Corollary~\ref{cor:constrained_approx}, and set $\mathfrak{D}^*_{K_n}:=\|h-g_{K_n}^\top W_{K_n}^*\|_{L^2(X)}$. Define $\beta^{\mathrm{pop}}_{K_n,\lambda}:=\Gamma_{K_n,\lambda}^{-1}\Gamma_{K_n}\beta_{K_n}$. Then $\beta_{K_n}-\beta^{\mathrm{pop}}_{K_n,\lambda} =\lambda_{K_n,n}\Gamma_{K_n,\lambda}^{-1}\beta_{K_n}$, and hence
$\|g_{K_n}^\top(\beta_{K_n}-\beta^{\mathrm{pop}}_{K_n,\lambda})\|_{L^2(X)}
=B^{\mathrm{pop}}_{K_n,\lambda}$. Since $\beta_{K_n}$ is the $L^2(X)$ projection coefficient,
\begin{equation*}
    \|h-g_{K_n}^\top\beta^{\mathrm{pop}}_{K_n,\lambda}\|_{L^2(X)}^2=\Delta_{K_n}^2+\big(B^{\mathrm{pop}}_{K_n,\lambda}\big)^2.
\end{equation*}
Moreover, by optimality of $\beta^{\mathrm{pop}}_{K_n,\lambda}$ for the population ridge criterion,
\begin{equation*}
    \|h-g_{K_n}^\top \beta^{\mathrm{pop}}_{K_n,\lambda}\|_{L^2(X)}^2 + \lambda_{K_n,n} \norm{ \beta^{\mathrm{pop}}_{K_n,\lambda} }^2 
    \leq
    \|h-g_{K_n}^\top b\|_{L^2(X)}^2 + \lambda_{K_n,n} \norm{ b }^2  
    \quad  \forall \ b \in \R^{K_n} ,
\end{equation*}
and thus
\begin{equation*}
    \Delta_{K_n}^2+\big(B^{\mathrm{pop}}_{K_n,\lambda}\big)^2+\lambda_{K_n,n}\|\beta^{\mathrm{pop}}_{K_n,\lambda}\|^2\leq (\mathfrak{D}^*_{K_n})^2+\lambda_{K_n,n}\|W_{K_n}^*\|^2.
\end{equation*}
As $\Delta_{K_n} \leq \mathfrak{D}^*_{K_n}$, it follows, as required, that 
\begin{equation*}
    B^{\mathrm{pop}}_{K_n,\lambda}
    \leq \sqrt{2} \left( \mathfrak{D}^*_{K_n}+\sqrt{\lambda_{K_n,n}}\|W_{K_n}^*\| \right)
    =O_{P_\theta}\left(K_n^{-1/2-\beta/(d+1)}+\sqrt{\frac{\lambda_{K_n,n}\log(K_n)}{K_n}}\right),
\end{equation*}
where in the last equality we used the bounds for $W_{K_n}^*$ established in the proof of Corollary~\ref{cor:constrained_approx}, namely $\mathfrak{D}^*_{K_n}=O_{P_\theta}(K_n^{-1/2-\beta/(d+1)})$ and  $\|W_{K_n}^*\|=O_{P_\theta}(\sqrt{\log(K_n)/K_n})$, where the latter follows by Lemma~\ref{lem:nn_l2_bound}.
\end{proof}
\begin{proof}[Proof of Theorem~\ref{theorem:L2_RWNN_sharp_consistency}]
By Theorem~\ref{theorem:L2_ridge_sharp_consistency}, together with Corollary~\ref{cor:constrained_approx},  Lemma~\ref{lem:RWNN_population_ridge_bias}, and the assumed $P_\theta$-probability conditions on the random feature maps, \eqref{eq:L2_RWNN_effective_dimension_rate} immediately follows.

Suppose now that \eqref{eq:RWNN_polynomial_ridge} holds. By using the effective-dimension bound in \eqref{eq:effective_dimension_spectral_rate} with $q_\Pi=d+1$, \eqref{eq:RWNN_effective_dimension_polynomial} holds. 
Next, since $r>2\beta/(d+1)$ by \eqref{eq:RWNN_polynomial_ridge}
\begin{equation*}
        \sqrt{\frac{\lambda_{K_n,n}\log(K_n)}{K_n}}=o\left(K_n^{-1/2-\beta/(d+1)}\right).
    \end{equation*}
    Hence,
\begin{equation}
\label{eq:L2_RWNN_nu_intermediate_rate2}
\norm{\widehat{h}_{n,\lambda}-h}_{L^2(X)}=O_P\left(\sqrt{\frac{K_n^{\nu_r}}{n}}+K_n^{-1/2-\beta/(d+1)}\right).
\end{equation}
    We choose $K_n$ by balancing the stochastic and approximation terms, so that $n^{-1/2}K_n^{\nu_r/2}\asymp K_n^{-1/2-\beta/(d+1)}$
   or, equivalently, 
    \begin{equation*}
        K_n\asymp n^{\frac{1}{1+\nu_r+2\beta/(d+1)}}.
    \end{equation*}
    Substituting this choice into either term in \eqref{eq:L2_RWNN_nu_intermediate_rate2} yields
    \begin{align*}
        K_n^{-1/2-\beta/(d+1)}&\asymp\left(n^{\frac{1}{1+\nu_r+2\beta/(d+1)}}\right)^{-1/2-\beta/(d+1)}=n^{-\frac{1/2+\beta/(d+1)}{1+\nu_r+2\beta/(d+1)}}.
    \end{align*}
Noticing that, by construction, the stochastic term has the same order, this gives \eqref{eq:L2_RWNN_faster_effective_dimension_rate}, completing the proof.
\end{proof}

\clearpage

\section{Proofs for directional and smooth functional inference}
\label{app:inference_proofs}

\subsection{Ridge-normalized linearization and directional asymptotic normality}
\label{app:directional_inference_proofs}

\begin{proof}[Proof of Lemma~\ref{lemma:linearization}]
We start by writing
\begin{align*}
     \widehat{\beta}_{K_n,n,\lambda} - \beta_{K_n} 
    & =  \widehat{\Gamma}_{K_n,n,\lambda}^{-1} (n^{-1} G_{K_n,n}^\top Y ) - \beta_{K_n}  \\
    & =  \widehat{\Gamma}_{K_n,n,\lambda}^{-1} \big[ n^{-1} G_{K_n,n}^\top G_{K_n,n}\beta_{K_n} + n\inv G_{K_n,n}^\top (\eta+e) \big] - \beta_{K_n} \\
    & =  \widehat{\Gamma}_{K_n,n,\lambda}^{-1} \big[\, \widehat{\Gamma}_{K_n,n} \beta_{K_n} + n\inv G_{K_n,n}^\top (\eta+e) \big] - \beta_{K_n} \\
    & =  \widehat{\Gamma}_{K_n,n,\lambda}^{-1} \big[ (\widehat\Gamma_{K_n,n,\lambda}-\lambda_{K_n,n}I_{K_n}) \beta_{K_n}+ n\inv G_{K_n,n}^\top (\eta+e) \big] - \beta_{K_n} \\
    & =  \widehat{\Gamma}_{K_n,n,\lambda}^{-1} \big( n\inv G_{K_n,n}^\top (\eta+e) \big) - \lambda_{K_n,n}\widehat{\Gamma}_{K_n,n,\lambda}^{-1}\beta_{K_n} ,
\end{align*}
where in the third and fourth equalities we used \eqref{eq:def_gamma_hat}. Multiplying by $\sqrt{n}\alpha^\top$ gives
\begin{align*}
     \sqrt{n}\alpha^\top\left(\widehat{\beta}_{K_n,n,\lambda} - \beta_{K_n} \right)
    & =  \alpha^\top \widehat{\Gamma}_{K_n,n,\lambda}^{-1} \mathbb{G}_{n}\left[ g_i(\eta_i+\epsilon_i)\right]  - \sqrt{n}\lambda_{K_n,n}\alpha^\top\widehat{\Gamma}_{K_n,n,\lambda}^{-1}\beta_{K_n} ,
\end{align*}
where 
\begin{equation*}
    \mathbb{G}_n[ g_i (\epsilon_i + \eta_i) ] 
    =
    \frac{1}{\sqrt{n}} \sum_{i=1}^n \big( g_i (\epsilon_i + \eta_i) - \E[ g_i (\epsilon_i + \eta_i) ] \big) 
    = 
    \frac{1}{\sqrt{n}} \sum_{i=1}^n  g_i (\epsilon_i + \eta_i),
\end{equation*}
since $\E[ g_i \epsilon_i ] = \E\big[ g_i \, \E[\epsilon_i \mid X_i] \big] = 0$ by the law of iterated expectations and by Assumption~\ref{assumption:regression_errors_L2}, and $\E[g_i \eta_i]=0$ by the $L^2(X)$ projection property of $\beta_{K_n}$. 

Next, note that 
\begin{align*}
    \alpha^\top \widehat{\Gamma}_{K_n,n,\lambda}^{-1}
    = \alpha^\top  \Gamma_{{K_n},\lambda}^{-1/2}\Gamma_{{K_n},\lambda}^{1/2} \widehat{\Gamma}_{{K_n},n,\lambda}\inv \Gamma_{{K_n},\lambda}^{1/2} \Gamma_{{K_n},\lambda}^{-1/2}
    = \widetilde{\alpha}^\top \widehat{D}_{K_n,n,\lambda}\Gamma_{{K_n},\lambda}^{-1/2}.
\end{align*}
Similarly to the proof of Theorem~\ref{theorem:L2_ridge_sharp_consistency}, %
recall from \eqref{eq:def_Dn_hat_An} that $\widehat{D}_{K_n,n,\lambda}=(I_{K_n}+A_{K_n,n,\lambda})^{-1}$.

Hence,
\begin{align*}
     \sqrt{n}\alpha^\top\left(\widehat{\beta}_{K_n,n,\lambda} - \beta_{K_n} \right)
    & = \widetilde{\alpha}^\top \widehat{D}_{K_n,n,\lambda} \left(\Gamma_{{K_n},\lambda}^{-1/2} \mathbb{G}_{n}\left[ g_i(\eta_i+\epsilon_i)\right]- \sqrt{n}\lambda_{K_n,n}\Gamma_{{K_n},\lambda}^{-1/2}\beta_{K_n}\right) \\
    &= \widetilde{\alpha}^\top \widehat{D}_{K_n,n,\lambda} \left(\mathbb{G}_{n}\left[ \widetilde{g}_i(\eta_i+\epsilon_i)\right] -\sqrt{n}\lambda_{K_n,n}\Gamma_{{K_n},\lambda}^{-1/2}\beta_{K_n}\right)\\
    &= \widetilde{\alpha}^\top \widehat{D}_{K_n,n,\lambda} \mathbb{G}_{n}\left[ \widetilde{g}_i(\eta_i+\epsilon_i)\right]+ \widetilde{\alpha}^\top \mathbb{G}_{n}\left[ \widetilde{g}_i\epsilon_i\right]\\
    &\quad - \widetilde{\alpha}^\top \mathbb{G}_{n}\left[ \widetilde{g}_i\epsilon_i\right] - \sqrt{n}\lambda_{K_n,n} \widetilde{\alpha}^\top \widehat{D}_{K_n,n,\lambda}\Gamma_{{K_n},\lambda}^{-1/2}\beta_{K_n}\\
    &= \widetilde{\alpha}^\top \mathbb{G}_{n}\left[ \widetilde{g}_i\epsilon_i\right] + \widetilde{\alpha}^\top \left( \widehat{D}_{K_n,n,\lambda}-I_{K_n}\right)\mathbb{G}_{n}\left[ \widetilde{g}_i(\eta_i+\epsilon_i)\right] \\
    &\quad + \widetilde{\alpha}^\top \mathbb{G}_{n}\left[ \widetilde{g}_i\eta_i\right] - \sqrt{n}\lambda_{K_n,n} \widetilde{\alpha}^\top  \widehat{D}_{K_n,n,\lambda}\Gamma_{{K_n},\lambda}^{-1/2}\beta_{K_n},
\end{align*}
where we used the linearity of $\mathbb{G}_n$ and the identity
$\Gamma_{K_n,\lambda}^{-1/2}g_i=\widetilde{g}_i$.
The last equality then gives the expansion \eqref{eq:linearization}, with the last three terms of the right-hand side corresponding to $R_{1,n}(\alpha)$, $R_{2,n}(\alpha)$, and $R_{\lambda,n}(\alpha)$ in \eqref{eq:linearization_residuals}, respectively.
We now bound each residual term separately.

\smallskip
\noindent\textsc{Step 1.} 
We first bound 
\begin{equation*}
    R_{1,n}(\alpha)
    = \widetilde{\alpha}^\top \left( \widehat{D}_{K_n,n,\lambda}-I_{K_n}\right)\mathbb{G}_{n}\left[ \widetilde{g}_i(\eta_i+\epsilon_i)\right] .
\end{equation*}
By \eqref{eq:linearization_normalized_gram_stability}, $\|A_{K_n,n,\lambda}\|=o_P(1)$, $\|\widehat{D}_{K_n,n,\lambda}\|=O_P(1)$ and $\|\widehat{D}_{K_n,n,\lambda}-I_{K_n}\|=O_P\left(\|A_{K_n,n,\lambda}\|\right)$.
We now split the term $R_{1,n}(\alpha)$ as
\begin{align*}
    R_{1,n}(\alpha)=R'_{1,n}(\alpha)+R''_{1,n}(\alpha),
\end{align*}
where
\begin{align*}
    R'_{1,n}(\alpha)
    & := \widetilde{\alpha}^\top \left(\widehat{D}_{K_n,n,\lambda}-I_{K_n}\right)\mathbb{G}_n[\widetilde{g}_i\epsilon_i], \\
    R''_{1,n}(\alpha)
    & := \widetilde{\alpha}^\top \left(\widehat{D}_{K_n,n,\lambda}-I_{K_n}\right)\mathbb{G}_n[\widetilde{g}_i\eta_i].
\end{align*}
Note that, conditional on the sample $X_{1:n}$,
\begin{align*}
    \E\left[ R'_{1,n}(\alpha) \,\middle\vert\, X_{1:n} \right] 
    & = \widetilde{\alpha}^\top \left(\widehat{D}_{K_n,n,\lambda}-I_{K_n}\right) \mathbb{G}_n\big[ \widetilde{g}_i \, \E[\epsilon_i \mid X_i ] \big] = 0 .
\end{align*}
Moreover, exactly as in the proof of Theorem~\ref{theorem:L2_ridge_sharp_consistency},
\begin{align*}
    & \Gamma_{K_n,\lambda}^{-1/2}\widehat{\Gamma}_{K_n,n}\Gamma_{K_n,\lambda}^{-1/2}=\Gamma_{K_n,\lambda}^{-1/2}\Gamma_{K_n}\Gamma_{K_n,\lambda}^{-1/2}+A_{K_n,n,\lambda},
\end{align*}
so that 
\begin{equation*}
    \Norm{ \Gamma_{{K_n},\lambda}^{-1/2} \widehat{\Gamma}_{K_n,n} \Gamma_{{K_n},\lambda}^{-1/2} }
    \leq  \Norm{ \Gamma_{{K_n},\lambda}^{-1/2} {\Gamma}_{K_n} \Gamma_{{K_n},\lambda}^{-1/2} } + o_P(1)
    = O_P(1).
\end{equation*}
We write 
\begin{align*}
    \E\left[ \mathbb{G}_n[ \widetilde{g}_i \epsilon_i ] \mathbb{G}_n[ \epsilon_i \widetilde{g}_i^\top ]  \,\middle\vert\, X_{1:n} \right]
    & = \frac{1}{n} \E\left[ \sum_{i=1}^n \epsilon_i^2 \widetilde{g}_i \widetilde{g}_i^\top  \,\middle\vert\, X_{1:n} \right] 
    \leq \overline{\sigma}^2  \Gamma_{{K_n},\lambda}^{-1/2} \widehat{\Gamma}_{K_n,n} \Gamma_{{K_n},\lambda}^{-1/2} .
\end{align*}
and we thus obtain
\begin{align*}
     \text{Var}( R'_{1,n}(\alpha) \,\big\vert\, X_{1:n})
    & \leq \overline{\sigma}^2 \, \widetilde{\alpha}^\top \left(\widehat{D}_{K_n,n,\lambda}-I_{K_n}\right) \Gamma_{{K_n},\lambda}^{-1/2} \widehat{\Gamma}_{K_n,n} \Gamma_{{K_n},\lambda}^{-1/2} \left(\widehat{D}_{K_n,n,\lambda}-I_{K_n}\right) \widetilde{\alpha} \\
    & \leq \overline{\sigma}^2 \norm{ \widetilde{\alpha} }^2 \, \norm{ \widehat{D}_{K_n,n,\lambda}-I_{K_n} }^2 \, \Norm{ \Gamma_{{K_n},\lambda}^{-1/2} \widehat{\Gamma}_{K_n,n} \Gamma_{{K_n},\lambda}^{-1/2} } .
\end{align*}
Hence, conditional Chebyshev's inequality, together with $\|\widehat{D}_{K_n,n,\lambda}-I_{K_n}\|=O_P(\|A_{K_n,n,\lambda}\|)$, yields 
\begin{equation}
\label{eq:R1_first_term}
    R'_{1,n}(\alpha)
    = O_P\left( \norm{ \widetilde{\alpha} } \, \norm{ A_{K_n,n,\lambda} } \right) .
\end{equation}
Exactly as in the proof of Theorem~\ref{theorem:L2_ridge_sharp_consistency},
\begin{equation*}
    \left\|\mathbb{G}_n[\widetilde g_i\eta_i]
    \right\|=O_P\left(\widetilde{\xi}_{K_n,\lambda}\Delta_{K_n}\right).
\end{equation*}
Consequently, 
\begin{align}
\label{eq:R1_second_term}
    R''_{1,n}(\alpha)=O_P\left(\| \widetilde{\alpha} \|\|A_{K_n,n,\lambda}\|\widetilde{\xi}_{K_n,\lambda}\Delta_{K_n}\right).
\end{align}
Combining \eqref{eq:R1_first_term} and \eqref{eq:R1_second_term} gives
\begin{align*}
    R_{1,n}(\alpha)
    =
    O_P\left(\| \widetilde{\alpha} \|\|A_{K_n,n,\lambda}\|\left(1+\widetilde{\xi}_{K_n,\lambda}\Delta_{K_n}\right)\right),
\end{align*}
which proves the bound in \eqref{eq:r1_bound}.

\smallskip
\noindent\textsc{Step 2.} For the term $R_{2,n}(\alpha) = \widetilde{\alpha}^\top \mathbb{G}_{n}\left[ \widetilde{g}_i\eta_i\right]$, by the Cauchy-Schwarz inequality, it holds that $|R_{2,n}(\alpha)|\leq\| \widetilde{\alpha} \|\left\|\mathbb{G}_n[\widetilde{g}_i\eta_i]\right\|$. 
Hence, using $\left\|\mathbb{G}_n[\widetilde{g}_i\eta_i]\right\|=O_P\left(\widetilde{\xi}_{K_n,\lambda}\Delta_{K_n}\right)$ derived in Step~1, gives \eqref{eq:r2_bound}, as required.

\smallskip
\noindent\textsc{Step 3.} 
Finally, we apply the Cauchy-Schwarz inequality to 
\begin{equation*}
    R_{\lambda,n}(\alpha) = -\sqrt{n}\lambda_{K_n,n} \widetilde{\alpha}^\top \allowbreak \widehat{D}_{K_n,n,\lambda} \Gamma_{{K_n},\lambda}^{-1/2}\beta_{K_n} 
\end{equation*}
and obtain 
\begin{align*}
    |R_{\lambda,n}(\alpha)|&\leq\sqrt{n}\,\lambda_{K_n,n}\| \widetilde{\alpha} \|\|\widehat{D}_{K_n,n,\lambda}\|\left\|\Gamma_{K_n,\lambda}^{-1/2}\beta_{K_n}\right\|.
\end{align*}
Noticing that  $\|\widehat{D}_{K_n,n,\lambda}\|=O_P(1)$ yields the bound in \eqref{eq:r3_bound}, as required.
\end{proof}
\begin{proof}[Proof of Theorem~\ref{theorem:L2_normality}]
\label{theorem:L2_linear_functionals_normality}

Fix $\ell\in\{0,1\}$ and, for the sake of convenience, define
\begin{equation*}
    \zeta_{n,i}^{(\ell)} := \epsilon_i+\ell\eta_i
    \qquad\text{and}\qquad
    \sigma_{n,\ell} := \sqrt{ \sigma_{K_n,\lambda}^{2,(\ell)}(\alpha_{K_n}) }.
\end{equation*}
Since $\E[\widetilde g_i\epsilon_i]=0$ and $\E[\widetilde g_i\eta_i]=0$, we have $\E\left[\widetilde g_i\zeta_{n,i}^{(\ell)}\right]=0$.
By Lemma~\ref{lemma:linearization}, and since
\begin{equation*}
    \widetilde{\alpha}_{K_n,\lambda}^{\top}\mathbb G_n[\widetilde g_i\epsilon_i]+R_{2,n}(\alpha_{K_n})=\widetilde{\alpha}_{K_n,\lambda}^{\top}\mathbb G_n[\widetilde g_i\zeta_{n,i}^{(\ell)}]+(1-\ell)R_{2,n}(\alpha_{K_n}),
\end{equation*}
we obtain
\begin{align}
\label{eq:directional_normality_linearization_general}
    \sqrt n\frac{\alpha_{K_n}^{\top}(\widehat\beta_{K_n,n,\lambda}-\beta_{K_n})}{\sigma_{n,\ell}}
    =&
    \frac{\widetilde\alpha_{K_n,\lambda}^{\top}}{\sigma_{n,\ell}}\mathbb G_n[\widetilde g_i\zeta_{n,i}^{(\ell)}]
    +\frac{R_{1,n}(\alpha_{K_n})}{\sigma_{n,\ell}}\nonumber \\
    &+\frac{(1-\ell)R_{2,n}(\alpha_{K_n})}{\sigma_{n,\ell}}
    +\frac{R_{\lambda,n}(\alpha_{K_n})}{\sigma_{n,\ell}}.
\end{align}
Using the bounds in Lemma~\ref{lemma:linearization}, condition~(ii) yields
\begin{equation*}
    \frac{R_{1,n}(\alpha_{K_n})}{\sigma_{n,\ell}}=O_P\left(\frac{\|\widetilde\alpha_{K_n,\lambda}\|}{\sigma_{n,\ell}}\left\|\Gamma_{K_n,\lambda}^{-1/2}(\widehat\Gamma_{K_n,n}-\Gamma_{K_n})\Gamma_{K_n,\lambda}^{-1/2}\right\|\left(1+\widetilde{\xi}_{K_n,\lambda}\Delta_{K_n}\right)\right)=o_P(1).
\end{equation*}
Similarly, condition~(iii) gives
\begin{equation*}
    \frac{(1-\ell)R_{2,n}(\alpha_{K_n})}{\sigma_{n,\ell}}=O_P\left(\frac{(1-\ell)\|\widetilde\alpha_{K_n,\lambda}\|\widetilde{\xi}_{K_n,\lambda}\Delta_{K_n}}{\sigma_{n,\ell}}\right)=o_P(1).
\end{equation*}
Notice that this condition is vacuous when $\ell=1$. Finally, condition~(iv) yields
\begin{equation*}
    \frac{R_{\lambda,n}(\alpha_{K_n})}{\sigma_{n,\ell}}=O_P\left(\frac{\sqrt n\lambda_{K_n,n}\|\widetilde\alpha_{K_n,\lambda}\|\|\Gamma_{K_n,\lambda}^{-1/2}\beta_{K_n}\|}{\sigma_{n,\ell}}\right)=o_P(1).
\end{equation*}
Consequently,
\begin{equation}
\label{eq:directional_normality_leading_term_general}
    \sqrt n\frac{\alpha_{K_n}^{\top}(\widehat\beta_{K_n,n,\lambda}-\beta_{K_n})}{\sigma_{n,\ell}}=\frac{\widetilde\alpha_{K_n,\lambda}^{\top}}{\sigma_{n,\ell}}\mathbb G_n[\widetilde g_i\zeta_{n,i}^{(\ell)}]+o_P(1).
\end{equation}
Define
\begin{equation}
\label{eq:def_directional_weights_general}
    \omega_{n,i}^{(\ell)}:=\frac{\widetilde\alpha_{K_n,\lambda}^{\top}\widetilde g_i}{\sqrt n\,\sigma_{n,\ell}}.
\end{equation}
Then
\begin{equation}
\label{eq:directional_weighted_sum_general}
    \frac{\widetilde\alpha_{K_n,\lambda}^{\top}}{\sigma_{n,\ell}}\mathbb G_n[\widetilde g_i\zeta_{n,i}^{(\ell)}]=\sum_{i=1}^n\omega_{n,i}^{(\ell)}\zeta_{n,i}^{(\ell)}.
\end{equation}
The summands have mean zero. Moreover, by the definition of $\sigma_{n,\ell}$,
\begin{equation*}
\text{Var}\left(\sum_{i=1}^n\omega_{n,i}^{(\ell)}\zeta_{n,i}^{(\ell)}\right)=\sum_{i=1}^n\E\left[\omega_{n,i}^{(\ell)2}\zeta_{n,i}^{(\ell)2}\right]=1.
\end{equation*}
Condition~(i) implies
\begin{equation}
\label{eq:directional_weight_bound_general}
    \max_{1\leq i\leq n}|\omega_{n,i}^{(\ell)}|\leq\frac{\|\widetilde\alpha_{K_n,\lambda}\|\widetilde{\xi}_{K_n,\lambda}}{\sqrt n\,\sigma_{n,\ell}}=:a_{n,\ell}=o(1).
\end{equation}
We now verify the Lindeberg condition. For any $\delta>0$,
\begin{align*}
    \sum_{i=1}^n\E\left[\omega_{n,i}^{(\ell)2}\zeta_{n,i}^{(\ell)2}\Ind\{|\omega_{n,i}^{(\ell)}\zeta_{n,i}^{(\ell)}|>\delta\}\right]
    &\leq
    \sum_{i=1}^n\E\left[\omega_{n,i}^{(\ell)2}\zeta_{n,i}^{(\ell)2}\Ind\{|\zeta_{n,i}^{(\ell)}|>\delta/a_{n,\ell}\}\right] \\
    &\leq
    \left(\sup_{x\in\mathcal{X}}\E\left[\zeta_{n,i}^{(\ell)2}\Ind\{|\zeta_{n,i}^{(\ell)}|>\delta/a_{n,\ell}\}\mid X_i=x\right]\right)
    \sum_{i=1}^n\E\left[\omega_{n,i}^{(\ell)2}\right].
\end{align*}
Since $\E[\epsilon_i\mid X_i]=0$ and $\ell\in\{0,1\}$,
\begin{equation*}
    \E\left[\zeta_{n,i}^{(\ell)2}\mid X_i\right]=\E[\epsilon_i^2\mid X_i]+\ell\eta_i^2\geq\underline\sigma^2.
\end{equation*}
It follows that
\begin{equation*}
    \sum_{i=1}^n\E\left[\omega_{n,i}^{(\ell)2}\right]=\frac{\E[(\widetilde\alpha_{K_n,\lambda}^{\top}\widetilde g_i)^2]}{\sigma_{n,\ell}^2}\leq\frac{1}{\underline\sigma^2}.
\end{equation*}
Therefore,
\begin{equation*}
    \sum_{i=1}^n\E\left[\omega_{n,i}^{(\ell)2}\zeta_{n,i}^{(\ell)2}\Ind\{|\omega_{n,i}^{(\ell)}\zeta_{n,i}^{(\ell)}|>\delta\}\right]\leq\frac{1}{\underline\sigma^2}\sup_{x\in\mathcal{X}}\E\left[\zeta_{n,i}^{(\ell)2}\Ind\{|\zeta_{n,i}^{(\ell)}|>\delta/a_{n,\ell}\}\mid X_i=x\right].
\end{equation*}
Since $a_{n,\ell}\to0$, condition~(vi) implies that the right-hand side converges to zero. Hence, the Lindeberg condition holds, and the Lindeberg-Feller central limit theorem gives
\begin{equation*}
    \sum_{i=1}^n\omega_{n,i}^{(\ell)}\zeta_{n,i}^{(\ell)}\todist\mathcal{N}(0,1).
\end{equation*}
Combining this with \eqref{eq:directional_normality_leading_term_general} and applying Slutsky's theorem yields
\begin{equation*}
    \sqrt n\frac{\alpha_{K_n}^{\top}(\widehat\beta_{K_n,n,\lambda}-\beta_{K_n})}{\sigma_{K_n,\lambda}^{(\ell)}(\alpha_{K_n})}\todist\mathcal{N}(0,1),
\end{equation*}
as required.
\end{proof}
\begin{proof}[Proof of Corollary~\ref{corollary:RWNN_L2_normality}]
We first verify conditions~(i)-(vi) of Theorem~\ref{theorem:L2_normality} for $\ell=0$ along the prescribed sequence $\alpha_{K_n}$.  We write
\begin{align}
\label{eq:dir_sigma_bar_bound}
\sigma_{K_n,\lambda}^{2,(0)}(\alpha_{K_n})&=\widetilde{\alpha}_{K_n,\lambda}^\top\E\left[\epsilon_i^2\widetilde{g}_i\widetilde{g}_i^\top\right]\widetilde{\alpha}_{K_n,\lambda}\geq\underline{\sigma}^2\widetilde{\alpha}_{K_n,\lambda}^\top\Gamma_{K_n,\lambda}^{-1/2}\Gamma_{K_n}\Gamma_{K_n,\lambda}^{-1/2}\widetilde{\alpha}_{K_n,\lambda},
\end{align}
where in the inequality we used that, by hypotheses, $\E[\epsilon_i^2\mid X_i]\geq\underline{\sigma}^2>0$ almost surely. Using the definition of $\mathcal{R}_{K_n,\lambda}(\alpha_{K_n})$ in \eqref{eq:RWNN_directional_Rayleigh_quotient}, we obtain
\begin{equation}
\label{eq:alpha_sigma_ratio}
\frac{\|\widetilde{\alpha}_{K_n,\lambda}\|}{\sigma_{K_n,\lambda}^{(0)}(\alpha_{K_n})}\leq\underline{\sigma}^{-1}\mathcal{R}_{K_n,\lambda}(\alpha_{K_n})^{-1/2}=O_{P_\theta}(1).
\end{equation}
By \eqref{eq:RWNN_K_choice}, ${K_n^{1/2+\beta/(d+1)}}/{\sqrt n}=O(1)$. Hence, the second condition in \eqref{eq:RWNN_Rayleigh_cond} implies
\begin{equation*}
	\frac{\widetilde{\xi}_{K_n,\lambda}}{\sqrt n}=o_{P_\theta}(1).
\end{equation*}
Combining this with \eqref{eq:alpha_sigma_ratio}, we obtain
\begin{equation*}
\frac{\norm{\widetilde{\alpha}_{K_n,\lambda}}\widetilde{\xi}_{K_n,\lambda}}{\sqrt n\,\sigma_{K_n,\lambda}^{(0)}(\alpha_{K_n})}=o_{P_\theta}(1),
\end{equation*}
which verifies condition~(i) in $P_\theta$-probability.

Moreover, the second condition in \eqref{eq:RWNN_Rayleigh_cond} and \eqref{eq:implications_Corollary_approx} imply
\begin{equation*}
\widetilde{\xi}_{K_n,\lambda}\Delta_{K_n}=o_{P_\theta}(1).
\end{equation*}
Together with \eqref{eq:alpha_sigma_ratio} and the ridge-normalized empirical covariance matrix stability implied by Assumption~\ref{assumption:L2_ridge_gram_stability}, this verifies condition~(ii) in $P_\theta$-probability. Furthermore,
\begin{equation*}
\frac{\|\widetilde{\alpha}_{K_n,\lambda}\|\widetilde{\xi}_{K_n,\lambda}\Delta_{K_n}}{\sigma_{K_n,\lambda}^{(0)}(\alpha_{K_n})}=o_{P_\theta}(1),
\end{equation*}
so condition~(iii) is also satisfied. Finally, \eqref{eq:alpha_sigma_ratio} together with \eqref{eq:RWNN_ridge_bias} verifies condition~(iv). Conditions~(v) and~(vi) for $\ell=0$ are imposed directly in the hypotheses of the corollary.

Thus, the conditions of Theorem~\ref{theorem:L2_normality} hold in $P_\theta$-probability for $\ell=0$ and hence Theorem~\ref{theorem:L2_normality} yields \eqref{eq:RWNN_CLT_directional} for $\ell=0$. It remains only to show that replacing $\sigma_{K_n,\lambda}^{(0)}(\alpha_{K_n})$ by $\sigma_{K_n,\lambda}^{(1)}(\alpha_{K_n})$ does not affect the limit. Since $\E[\epsilon_i\mid X_i]=0$,
\begin{equation*}
\sigma_{K_n,\lambda}^{2,(1)}(\alpha_{K_n})-\sigma_{K_n,\lambda}^{2,(0)}(\alpha_{K_n})=\E\left[\eta_i^2\left(\widetilde{\alpha}_{K_n,\lambda}^\top\widetilde g_i\right)^2\right].
\end{equation*}
Hence
\begin{equation*}
0\leq\frac{\sigma_{K_n,\lambda}^{2,(1)}(\alpha_{K_n})-\sigma_{K_n,\lambda}^{2,(0)}(\alpha_{K_n})}{
\sigma_{K_n,\lambda}^{2,(0)}(\alpha_{K_n})}\leq\underline{\sigma}^{-2}\mathcal{R}_{K_n,\lambda}(\alpha_{K_n})^{-1}\widetilde{\xi}_{K_n,\lambda}^2\Delta_{K_n}^2=o_{P_\theta}(1),
\end{equation*}
which implies that
\begin{equation}
\label{eq:sigma_ratio_o1}
\frac{\sigma_{K_n,\lambda}^{(1)}(\alpha_{K_n})}{\sigma_{K_n,\lambda}^{(0)}(\alpha_{K_n})}=1+o_{P_\theta}(1).
\end{equation}
Combining this with \eqref{eq:RWNN_CLT_directional} for $\ell=0$ and Slutsky's theorem gives the same conditional limit with $\sigma_{K_n,\lambda}^{(1)}(\alpha_{K_n})$ in the denominator. Hence \eqref{eq:RWNN_CLT_directional} holds for both $\ell\in\{0,1\}$.
\end{proof}

\subsection{Ridge-adjusted variance estimation and studentization}
\label{app:variance_estimation_proofs}

\begin{proof}[Proof of Lemma~\ref{lemma:asymp_var_error_terms}]
To show (i) holds, we develop the sum as
\begin{equation*}
    \frac{1}{n} \sum_{i=1}^n \big[ \widehat{\epsilon}_i^2 - (\epsilon_i + \eta_i)^2 ] \widehat{r}_i^2
    = \frac{1}{n} \sum_{i=1}^n \big[ g_i^\top (\widehat{\beta}_{K_n,n,\lambda} -  \beta_{K_n}) \big]^2 \widehat{r}_i^2 - \frac{2}{n} \sum_{i=1}^n (\epsilon_i + \eta_i) g_i^\top (\widehat{\beta}_{K_n,n,\lambda} -  \beta_{K_n}) \widehat{r}_i^2 ,
\end{equation*}
where we have used that $\widehat{\epsilon}_i = (\eta_i + \epsilon_i) - g_i^\top (\widehat{\beta}_{K_n,n,\lambda} -  \beta_{K_n})$.
We have
\begin{equation}
\label{eq:bound_ri_hat2}
\begin{aligned}
    \sup_{1\leq i\leq n}\widehat r_i^2
    &= \sup_{1\leq i\leq n}\big\lvert \alpha^\top \widehat{\Gamma}_{K_n,n,\lambda}^{-1} g_i \big\rvert^2 \\
    &= \sup_{1\leq i\leq n}\big\lVert \alpha^\top  \Gamma_{K_n,\lambda}^{-1/2} \big( \Gamma_{K_n,\lambda}^{1/2} \widehat{\Gamma}_{K_n,n,\lambda}^{-1} \Gamma_{K_n,\lambda}^{1/2} \big) \Gamma_{K_n,\lambda}^{-1/2} g_i \big\rVert^2 \\
    &\leq \norm{ \widetilde{\alpha} }^2 \norm{ \widehat{D}_{K_n,n,\lambda} }^2 \sup_{1\leq i\leq n}\norm{ \widetilde{g}_i }^2 \leq \norm{ \widetilde{\alpha} }^2 \norm{ \widehat{D}_{K_n,n,\lambda} }^2 \widetilde{\xi}^2_{K_n,\lambda} ,
\end{aligned}
\end{equation}
with $\widehat{D}_{K_n,n,\lambda}$ defined as in \eqref{eq:def_Dn_hat_An}.
By the second condition in \eqref{eq:asymp_var},
\begin{equation}
\label{eq:bound_gi_beta_betahat}
\begin{aligned}
    \frac{1}{n} \sum_{i=1}^n \big[ g_i^\top (\widehat{\beta}_{K_n,n,\lambda} -  \beta_{K_n}) \big]^2 
    &=O_P\left(
    \overline{\mathfrak M}_{K_n,n,\lambda}^2
    \right),
\end{aligned}
\end{equation}
Moreover, by the first condition in \eqref{eq:asymp_var} and $\overline a_{K_n,n,\lambda}=o(1)$, $\|A_{K_n,n,\lambda}\|=o_P(1)$. Since $ \widehat{D}_{K_n,n,\lambda} =  \left(I_{K_n} + A_{{K_n},n,\lambda}\right)^{-1}$ it follows that  $\norm{ \widehat{D}_{K_n,n,\lambda} } = O_P(1)$ (see also the proof of Theorem~\ref{theorem:L2_ridge_sharp_consistency}).
Thus, combining bounds \eqref{eq:bound_ri_hat2} and \eqref{eq:bound_gi_beta_betahat}, we obtain
\begin{align*}
    \frac{1}{n} \sum_{i=1}^n \big[ g_i^\top (\widehat{\beta}_{K_n,n,\lambda} -  \beta_{K_n}) \big]^2 \widehat{r}_i^2
    & \leq \left[  \frac{1}{n} \sum_{i=1}^n \big[ g_i^\top (\widehat{\beta}_{K_n,n,\lambda} -  \beta_{K_n}) \big]^2 \right] \sup_{1\leq i\leq n}\widehat r_i^2 \\
    & = O_P\left( \norm{ \widetilde{\alpha} }^2 \widetilde{\xi}^2_{K_n,\lambda} \overline{\mathfrak M}_{K_n,n,\lambda}^2 \right)
\end{align*}
For the cross term, we use Cauchy-Schwarz:
\begin{align*}
    &\left \vert\frac{1}{n}\sum_{i=1}^n(\epsilon_i+\eta_i)g_i^\top
    (\widehat{\beta}_{K_n,n,\lambda}-\beta_{K_n})\widehat r_i^2\right\vert \\
    &\qquad\leq \left[ \frac1n\sum_{i=1}^n(\epsilon_i+\eta_i)^2 \right]^{1/2} \left[ \frac1n\sum_{i=1}^n\left[g_i^\top(\widehat{\beta}_{K_n,n,\lambda}-\beta_{K_n})\right]^2 \right]^{1/2} \sup_{1\leq i\leq n}\widehat r_i^2 .
\end{align*}
We now use that 
\begin{equation*}
    \E\big[ (\epsilon_i+\eta_i)^2 \big ]=\E[\epsilon_i^2]+\E[\eta_i^2]
    \leq
    \overline\sigma^2+\Delta_{K_n}^2 ,
\end{equation*}
since
$ \E[\epsilon_i\eta_i]=\E\big[\eta_i \ \E[\epsilon_i \,\vert\, X_i ] \big] = 0 $, with Markov's inequality then yielding
\begin{equation}
\label{eq:bound_avg_ei2_etai2}
    \frac{1}{n} \sum_{i=1}^n (\epsilon_i + \eta_i)^2 
    = O_P\left( \overline{\sigma}^2 + \Delta_{K_n}^2 \right) .
\end{equation}
Hence, \eqref{eq:bound_ri_hat2}, \eqref{eq:bound_gi_beta_betahat}, and  \eqref{eq:bound_avg_ei2_etai2} give
\begin{equation*}
    \left\vert \frac{1}{n}\sum_{i=1}^n(\epsilon_i+\eta_i)g_i^\top
    (\widehat{\beta}_{K_n,n,\lambda}-\beta_{K_n})\widehat r_i^2 \right\vert
    = O_P\left( \norm{ \widetilde{\alpha} }^2  \widetilde{\xi}^2_{K_n,\lambda} \overline{\mathfrak M}_{K_n,n,\lambda} \sqrt{ \overline\sigma^2+\Delta_{K_n}^2 } \, \right) .
\end{equation*}
By merging bounds, we then find 
\begin{equation*}
    \left\lvert \frac{1}{n} \sum_{i=1}^n \big[ \widehat{\epsilon}_i^2 - (\epsilon_i + \eta_i)^2 ] \widehat{r}_i^2 \right\rvert
    = 
    O_P\left( \norm{ \widetilde{\alpha} }^2 \widetilde{\xi}^2_{K_n,\lambda} \overline{\mathfrak M}_{K_n,n,\lambda} \left( \overline{\mathfrak M}_{K_n,n,\lambda} + \sqrt{ { \overline{\sigma}^2 + \Delta_{K_n}^2 } } \, \right) \right) ,
\end{equation*}
which establishes (i).

To prove (ii), we first write
\begin{equation*}
    \left\lvert \frac{1}{n} \sum_{i=1}^n (\epsilon_i + \eta_i)^2 \big[ \widehat{r}_i^2 - r_i^2 \big] \right\rvert 
    \leq \frac{1}{n} \sum_{i=1}^n (\epsilon_i + \eta_i)^2 \big\lvert \widehat{r}_i^2 - r_i^2 \big\rvert ,
\end{equation*}
and note
\begin{align*}
    \widehat{r}_i^2 
    &= \big( \widetilde{\alpha}^\top \widehat{D}_{K_n,n,\lambda} \widetilde{g}_i \big)^2 \\
    &= \big( \widetilde{\alpha}^\top \widetilde{g}_i \big)^2 + \big( \widetilde{\alpha}^\top \big( \widehat{D}_{K_n,n,\lambda} - I_{K_n} \big) \widetilde{g}_i \big)^2 + 2 \widetilde{\alpha}^\top \big( \widehat{D}_{K_n,n,\lambda} - I_{K_n} \big) \widetilde{g}_i \, \widetilde{\alpha}^\top \widetilde{g}_i .
\end{align*}
Since $r_i^2 = \big( \widetilde{\alpha}^\top \widetilde{g}_i \big)^2$, it readily follows 
\begin{equation}
\label{eq:bound_rihat2_ri2}
\begin{aligned}
    \big\lvert \widehat{r}_i^2 - r_i^2 \big\rvert
    & \leq \norm{ \widetilde{\alpha} }^2 \norm{ \widetilde{g}_i }^2 \norm{ \widehat{D}_{K_n,n,\lambda} - I_{K_n} }^2 + 2 \norm{ \widetilde{\alpha} }^2 \norm{ \widetilde{g}_i }^2 \norm{ \widehat{D}_{K_n,n,\lambda} - I_{K_n} } \\
    & \leq \norm{ \widetilde{\alpha} }^2 \widetilde{\xi}_{K_n,\lambda}^2 \big( \norm{ \widehat{D}_{K_n,n,\lambda} - I_{K_n} }^2 + 2 \norm{ \widehat{D}_{K_n,n,\lambda} - I_{K_n} } \big) 
\end{aligned}
\end{equation}
Combining \eqref{eq:bound_rihat2_ri2} and \eqref{eq:bound_avg_ei2_etai2} we find
\begin{equation*}
    \left\lvert \frac{1}{n} \sum_{i=1}^n (\epsilon_i + \eta_i)^2 \big[ \widehat{r}_i^2 - r_i^2 \big] \right\rvert
    = O_P\left( \norm{ \widetilde{\alpha} }^2 \widetilde{\xi}_{K_n,\lambda}^2 \bar{a}_{K_n, n, \lambda} \big( \overline{\sigma}^2 + \Delta_{K_n}^2 \big) \right).
\end{equation*}
In the last expression we used that, since $\left\|A_{K_n, n, \lambda}\right\|=O_P\left(\bar{a}_{K_n, n, \lambda}\right)$ by \eqref{eq:asymp_var}, $\bar{a}_{K_n, n, \lambda}=o(1)$ by the hypotheses of the lemma, and $\widehat{D}_{K_n, n, \lambda}=\left(I_{K_n}+A_{K_n, n, \lambda}\right)^{-1}$, we further have $\|\widehat{D}_{K_n, n, \lambda}-I_{K_n}\| \leq \|\widehat{D}_{K_n, n, \lambda}\| \|A_{K_n, n, \lambda}\|=O_P\left(\bar{a}_{K_n, n, \lambda}\right)$ and hence $\|\widehat{D}_{K_n, n, \lambda}-I_{K_n}\|^2 + 2\|\widehat{D}_{K_n, n, \lambda}-I_{K_n}\|=O_P\left(\bar{a}_{K_n, n, \lambda}\right)$.
\end{proof}

\begin{proof}[Proof of Proposition~\ref{proposition:consistency_ridge_var_estimator}]
To streamline notation, we drop some subscripts and write $\sigma_\lambda^2 := \sigma_{K_n,\lambda}^{2,(1)}(\alpha)$. By~\eqref{eq:def_sigma} and by the definition of $r_i$ in Lemma~\ref{lemma:asymp_var_error_terms}, $\sigma_\lambda^2 = \E\left[(\epsilon_i+\eta_i)^2 r_i^2\right]$. Accordingly, by the definition of $\widehat{r}_i$ in Lemma~\ref{lemma:asymp_var_error_terms}, $\widehat{\sigma}^2_{\lambda} = \widehat\sigma_{K_n,n,\lambda}^2(\alpha) = \frac{1}{n}\sum_{i=1}^n \widehat{\epsilon}_i^2\widehat r_i^2$.
Dividing by $\sigma^2_{\lambda}$ and then adding and subtracting, we can then find
\begin{equation}
\label{eq:var_ratio}
\begin{aligned}
    \frac{\widehat{\sigma}^2_{\lambda}}{\sigma^2_{\lambda}}
    &= \frac{1}{n \sigma^2_{\lambda}} \sum_{i=1}^n \big[ \widehat{\epsilon}_i^2 - (\epsilon_i + \eta_i)^2  \big] \widehat{r}_i^2 + \frac{1}{n \sigma^2_{\lambda}} \sum_{i=1}^n (\epsilon_i + \eta_i)^2 \big( \widehat{r}_i^2 - r_i^2 \big) \\
    &\qquad + \frac{1}{n \sigma^2_{\lambda}} \sum_{i=1}^n (\epsilon_i + \eta_i)^2 r_i^2 \\
    &=: E_{1,n}(\alpha) + E_{2,n}(\alpha) + E_{3,n}(\alpha) .
\end{aligned}
\end{equation}
We now proceed to show that both $E_{1,n}(\alpha)$ and $E_{2,n}(\alpha)$ are asymptotically negligible, and that $E_{3,n}(\alpha) = 1 + o_P(1)$.

We first consider $E_{1,n}(\alpha)$.  Using the definitions in \eqref{eq:R_L_def}, Lemma~\ref{lemma:asymp_var_error_terms}~(i), and hypothesis~(i),
\begin{equation}
\label{eq:E1_bound}
    \left|E_{1,n}(\alpha)\right| = O_P\left(\mathcal{L}_{K_n,\lambda}^{(1)}(\alpha)\overline{\mathfrak{M}}_{K_n,n,\lambda}\left(\overline{\mathfrak{M}}_{K_n,n,\lambda}+\sqrt{\overline{\sigma}^2+\Delta_{K_n}^2}\right)\right)=o_P(1),
\end{equation}
by the first negligibility condition.

We next control $E_{2,n}(\alpha)$. Again, by the definitions in \eqref{eq:R_L_def}, Lemma~\ref{lemma:asymp_var_error_terms}~(ii), and hypothesis~(ii),
\begin{equation}
\label{eq:E2_bound}
    \left|E_{2,n}(\alpha)\right| = O_P\left(\mathcal{L}_{K_n,\lambda}^{(1)}(\alpha)\overline{a}_{K_n,n,\lambda}\left(\overline{\sigma}^2+\Delta_{K_n}^2\right)\right)= o_P(1).
\end{equation}
 It remains to control $E_{3,n}(\alpha)$. Define $q_i := (\epsilon_i+\eta_i)^2 r_i^2$. By the definition of $\sigma_\lambda^2$, $\E[q_i]=\sigma_\lambda^2$.
Therefore,
\begin{equation}
    E_{3,n}(\alpha)-1 = \frac{1}{n\sigma_\lambda^2}\sum_{i=1}^n\left(q_i-\E[q_i]\right).
\end{equation}
By independence,
\begin{align*}
    \operatorname{Var}\left(E_{3,n}(\alpha)-1  \right) &= \frac{1}{n\sigma_\lambda^4}\operatorname{Var}(q_i ) \leq \frac{1}{n\sigma_\lambda^4}\E\left[(\epsilon_i+\eta_i)^4 r_i^4\right]\\
    &\leq \frac{1}{n\sigma_\lambda^4}\overline{\sigma}_{\epsilon,\eta}^4\|\widetilde{\alpha}\|^4\widetilde{\xi}_{K_n,\lambda}^4=\frac{\overline\sigma_{\epsilon,\eta}^4}{n}\left(\mathcal{L}_{K_n,\lambda}^{(1)}(\alpha)\right)^2,
\end{align*}
where in the last inequality we used hypothesis~(iv) and in the last equality we applied the definition in \eqref{eq:R_L_def}.

By Chebyshev's inequality and hypothesis~(iii),
\begin{equation}
\label{eq:E3_bound}
    E_{3,n}(\alpha)-1 = O_P\left(\frac{\mathcal{L}_{K_n,\lambda}^{(1)}(\alpha)}{\sqrt{n}}\right)=o_P(1),
\end{equation}
Combining the bounds \eqref{eq:E1_bound}, \eqref{eq:E2_bound}, and \eqref{eq:E3_bound} in \eqref{eq:var_ratio} concludes the proof.
\end{proof}

\begin{proof}[Proof of Corollary~\ref{corollary:RWNN_variance_estimator_consistency}]
We verify the conditions of Proposition~\ref{proposition:consistency_ridge_var_estimator}.

By Corollary~\ref{cor:constrained_approx},
\begin{equation}
\label{eq:var_cons_Delta_rate}
    \Delta_{K_n}=O_{P_\theta}\left(K_n^{-1/2-\beta/(d+1)}\right).
\end{equation}
Moreover, under \eqref{eq:lambda_r_choice}, Theorem~\ref{theorem:L2_RWNN_sharp_consistency} gives
\begin{equation}
\label{eq:var_cons_effective_dimension}
    \mathcal{N}_{K_n,\lambda}=O_{P_\theta}\left(K_n^{\nu_r}\right),
    \qquad
    \nu_r=\frac{(d+1)(1+r)}{d+3}.
\end{equation}
Since $r<2/(d+1)$, $1-\nu_r=\frac{2-(d+1)r}{d+3}>0$, and hence $\nu_r<1$.

Additionally, Lemma~\ref{lem:RWNN_population_ridge_bias} yields
\begin{equation*}
    B^{\mathrm{pop}}_{K_n,\lambda}=O_{P_\theta}\left(K_n^{-1/2-\beta/(d+1)}+\sqrt{\frac{\lambda_{K_n,n}\log(K_n)}{K_n}}\right).
\end{equation*}
Again, with the choice of $\lambda_{K_n,n}$ in  \eqref{eq:lambda_r_choice},
\begin{equation*}
    \frac{\sqrt{\lambda_{K_n,n}\log(K_n)/K_n}}{K_n^{-1/2-\beta/(d+1)}}\asymp K_n^{\beta/(d+1)-r/2}\sqrt{\log(K_n)}\rightarrow 0,
\end{equation*}
using the lower bound $r>2\beta/(d+1)$ in \eqref{eq:lambda_r_choice}.
Consequently,
\begin{equation}
\label{eq:var_cons_ridge_bias_rate}
    B^{\mathrm{pop}}_{K_n,\lambda}=O_{P_\theta}\left(K_n^{-1/2-\beta/(d+1)}\right).
\end{equation}

By Lemma~\ref{lemma:L2_ridge_empirical_prediction},
\begin{equation*}
    \left\|\widehat{\Gamma}_{K_n,n}^{1/2}\left(\widehat{\beta}_{K_n,n,\lambda}-\beta_{K_n}\right)\right\|=O_P\left(\Delta_{K_n}+B^{\mathrm{pop}}_{K_n,\lambda}+\sqrt{\frac{\mathcal{N}_{K_n,\lambda}}{n}}\right)=O_P\left(\overline{\mathfrak{M}}_{K_n,n,\lambda}\right),
\end{equation*}
where in the second equality we used  \eqref{eq:var_cons_Delta_rate},
\eqref{eq:var_cons_effective_dimension}, and \eqref{eq:var_cons_ridge_bias_rate}. This renders the second condition in \eqref{eq:asymp_var} in Lemma~\ref{lemma:asymp_var_error_terms} satisfied.

Next, by \eqref{eq:nondegeneracy} in Remark~\ref{remark:L2_ridge_assumptions_discussion}, Assumption~\ref{assumption:L2_ridge_gram_stability} gives
\begin{equation*}
    \|A_{K_n,n,\lambda}\|=O_P\left(\sqrt{\frac{\widetilde{\xi}_{K_n,\lambda}^{\,2}\log(K_n)}{n}}\right).
\end{equation*}
Since
\begin{equation*}
    \widetilde{\xi}_{K_n,\lambda}^{\,2}\leq\frac{\xi_{K_n}^2}{\overline s_{K_n}^{\,2}+\lambda_{K_n,n}}\leq\frac{\xi_{K_n}^2}{\lambda_{K_n,n}},
\end{equation*}
with  $\xi_{K_n}^2=O_{P_\theta}(K_n)$, and $\lambda_{K_n,n}\asymp K_n^{-r}$ by \eqref{eq:lambda_r_choice}, we obtain
\begin{equation*}
    \widetilde{\xi}_{K_n,\lambda}^{2}=O_{P_\theta}(K_n^{1+r}).
\end{equation*}
Hence
\begin{equation}
\label{eq:var_cons_A_rate}
    \|A_{K_n,n,\lambda}\|=O_P\left(\sqrt{\frac{K_n^{1+r}\log(K_n)}{n}}\right)=O_P\left(\overline{a}_{K_n,n,\lambda}\right).
\end{equation}
Moreover, since $\rho\geq0$ by the hypotheses of the corollary,
\eqref{eq:primitive_rate_condition_l1} implies $\overline{a}_{K_n,n,\lambda}=o(1)$. Hence, the first condition in \eqref{eq:asymp_var}  in Lemma~\ref{lemma:asymp_var_error_terms} is also satisfied.

It remains to verify conditions (i)-(iii) of Proposition~\ref{proposition:consistency_ridge_var_estimator}.
By \eqref{eq:primitive_L_growth_l1}, it holds that 
\begin{equation*}
    \mathcal{L}_{K_n,\lambda}^{(1)}(\alpha)\overline{a}_{K_n,n,\lambda}=O_{P_\theta}\left(\sqrt{\frac{K_n^{2\rho+1+r}\log(K_n)}{n}}\right)=o_{P_\theta}(1)
\end{equation*}
where the last equality follows from  \eqref{eq:primitive_rate_condition_l1}.
Next,
\begin{equation}
\label{eq:var_cons_LM_negligible}
    \mathcal{L}_{K_n,\lambda}^{(1)}(\alpha)\overline{\mathfrak{M}}_{K_n,n,\lambda}=O_{P_\theta}\left(\sqrt{\frac{K_n^{2\rho+\nu_r}}{n}}+K_n^{\rho-1/2-\beta/(d+1)}\right)=o_{P_\theta}(1),
\end{equation}
where to get the last equality we used that since $\nu_r<1+r$ by \eqref{eq:nu_r}, \eqref{eq:primitive_rate_condition_l1} implies
${K_n^{2\rho+\nu_r}}/{n} = o(1)$, while
$\rho < 1/2 + {\beta}/{(d+1)}$ implies
$K_n^{\rho-1/2-\beta/(d+1)}\rightarrow 0$.
Also, \eqref{eq:primitive_rate_condition_l1} implies $K_n^{2\rho}/n=o(1)$, and hence
\begin{equation}
\label{eq:var_cons_L_sqrtn_negligible}
    \frac{\mathcal{L}_{K_n,\lambda}^{(1)}(\alpha)}{\sqrt n}=o_{P_\theta}(1).
\end{equation}
Finally, \eqref{eq:var_cons_Delta_rate} implies $\Delta_{K_n}=o_{P_\theta}(1)$, and therefore $\sqrt{\overline{\sigma}^2+\Delta_{K_n}^2}=O_{P_\theta}(1)$ which yields  $\overline{\sigma}^2+\Delta_{K_n}^2=O_{P_\theta}(1)$.

Together with \eqref{eq:var_cons_LM_negligible}, this verifies
condition~(i) of Proposition~\ref{proposition:consistency_ridge_var_estimator}. At the same time, \eqref{eq:var_cons_A_rate} and \eqref{eq:primitive_rate_condition_l1} verify condition~(ii), while \eqref{eq:var_cons_L_sqrtn_negligible} verifies condition~(iii). Condition~(iv) is exactly \eqref{eq:primitive_fourth_moment_l1}. Hence all the conditions of Proposition~\ref{proposition:consistency_ridge_var_estimator} are satisfied, which proves the claim, as required.
\end{proof}

\begin{proof}[Proof of Corollary~\ref{corollary:RWNN_variance_estimator_consistency_l0}]
Since $\E[\epsilon_i\mid X_i]=0$, $\sigma_{K_n,\lambda}^{2,(1)}(\alpha)-\sigma_{K_n,\lambda}^{2,(0)}(\alpha)=
\E\left[\eta_i^2r_i^2\right]$. 
Hence
\begin{align*}
    0&\le \frac{\sigma_{K_n,\lambda}^{2,(1)}(\alpha)-\sigma_{K_n,\lambda}^{2,(0)}(\alpha)}{\sigma_{K_n,\lambda}^{2,(1)}(\alpha)}\le\frac{\|\widetilde{\alpha}\|^2\widetilde{\xi}_{K_n,\lambda}^2\Delta_{K_n}^2}{\sigma_{K_n,\lambda}^{2,(1)}(\alpha)}=
    \mathcal{L}_{K_n,\lambda}^{(1)}(\alpha)\Delta_{K_n}^2.
\end{align*}
By Corollary~\ref{cor:constrained_approx} and \eqref{eq:primitive_L_growth_l1}, $\mathcal{L}_{K_n,\lambda}^{(1)}(\alpha)\Delta_{K_n}^2=O_{P_\theta}\left(K_n^{\rho-1-2\beta/(d+1)}\right)=o_{P_\theta}(1)$.
Hence, ${\sigma_{K_n,\lambda}^{2,(1)}(\alpha)} /\allowbreak {\sigma_{K_n,\lambda}^{2,(0)}(\alpha)}\toprob 1$.
Combining this with Corollary~\ref{corollary:RWNN_variance_estimator_consistency} and Slutsky's theorem proves the claim.
\end{proof}

\subsection{Smooth-functional inference}
\label{app:smooth_functional_proofs}

\begin{proof}[Proof of Corollary~\ref{cor:smooth_functional_remainder_ridge}]
By \eqref{eq:L2_RWNN_polynomial_effective_dimension_rate}, one has $\norm{\widehat{h}_{n,\lambda}-h}_{L^2(X)} = O_P\big( \sqrt{{K_n^{\nu_r}}/{n}}+K_n^{-1/2-\beta/(d+1)} \big)$, where $\nu_r=\frac{(d+1)(1+r)}{d+3}<1$. Let $b:=\beta/(d+1)>0$. Under \eqref{eq:smooth_functional_K_choice}, $K_n\asymp n^{1/(2(1+b))}$, and therefore
\begin{align*}
    \sqrt{\frac{K_n^{\nu_r}}{n}} & \asymp n^{-\frac{1}{2}+\frac{\nu_r}{4(1+b)}},\\
    K_n^{-\frac{1}{2}-b} & \asymp n^{-\frac{1+2b}{4(1+b)}}.
\end{align*}
Since $\nu_r<1$ and $b>0$, $\frac{1}{2}-\frac{\nu_r}{4(1+b)}>\frac{1}{4}$, $\frac{1+2b}{4(1+b)}>\frac{1}{4}$. Hence both terms are $o(n^{-1/4})$, which proves \eqref{eq:smooth_functional_n14_rate}.
Since $\omega\geq2$, it thus holds $\sqrt{n}\|\widehat h_{n,\lambda}-h\|_{L^2(X)}^\omega=o_P(1)$,  and the conclusion follows from
\eqref{eq:smooth_functional_expansion}.
\end{proof}

\begin{proof}[Proof of Corollary~\ref{corollary:RWNN_smooth_functional_normality}]
By \eqref{eq:sigma_ratio_o1} in the proof of Corollary~\ref{corollary:RWNN_L2_normality} and by  \eqref{eq:smooth_functional_sigma}, it follows that
\begin{equation*}
\frac{\sigma_{\Psi,K_n,\lambda}^{(1)}}{\sigma_{\Psi,K_n,\lambda}^{(0)}}=1+o_{P_\theta}(1).
\end{equation*}
Therefore, \eqref{eq:smooth_functional_variance_nondegeneracy} implies 
\begin{equation}
\label{eq:smooth_functional_variance_nondegeneracy_both}
    \big( \sigma_{\Psi,K_n,\lambda}^{(\ell)} \big)^{-1}=O_{P_\theta}(1),\qquad \ell\in\{0,1\}.
\end{equation}
By \eqref{eq:smooth_functional_rootn_decomposition},
\eqref{eq:functional_sieve_bias_definition},
\eqref{eq:smooth_functional_sieve_bias}, and
\eqref{eq:smooth_functional_variance_nondegeneracy_both},
\begin{equation}
\label{eq:smooth_functional_reduction_to_direction}
    \sqrt{n}\frac{\Psi(\widehat{h}_{n,\lambda})-\Psi(h)}{
    \sigma_{\Psi,K_n,\lambda}^{(\ell)}}=\sqrt{n}\frac{a_{K_n}^{\top}
    \bigl(\widehat{\beta}_{K_n,n,\lambda}-\beta_{K_n}\bigr)}{\sigma_{\Psi,K_n,\lambda}^{(\ell)}}+o_P(1),\qquad \ell\in\{0,1\}.
\end{equation}
Then, \eqref{eq:smooth_functional_sigma} and \eqref{eq:smooth_functional_direction} allow us to write
\begin{equation*}
    \sqrt{n}\frac{a_{K_n}^{\top}\bigl(\widehat\beta_{K_n,n,\lambda}-\beta_{K_n}\bigr)}{\sigma_{\Psi,K_n,\lambda}^{(\ell)}}=
    \sqrt{n}\frac{ (\mathfrak{a}_{K_n}^{\Psi})^{\top}\bigl(\widehat\beta_{K_n,n,\lambda}-\beta_{K_n}\bigr)}{\sigma_{K_n,\lambda}^{(\ell)}(\mathfrak{a}_{K_n}^{\Psi})}.
\end{equation*}
Hence, applying Corollary~\ref{corollary:RWNN_L2_normality}, we find that the right-hand side of the last display
converges in distribution to $\mathcal{N}(0,1)$, conditionally on
$\mathcal{F}_{K_n}$, in $P_\theta$-probability. Combining this with
\eqref{eq:smooth_functional_reduction_to_direction} and Slutsky's
theorem proves \eqref{eq:RWNN_smooth_functional_CLT}.
\end{proof}

\subsection{Proof of Example~\ref{example:RWNN_singular_covariance}}
\label{appendix:RWNN_singular_example}
\begin{proof}
We first verify the approximation assumptions. Since $Ax+B\geq1$ for every $(A,B,x)\in[-1,1]\times[2,3]\times[-1,1]$, the ReLU activation satisfies $\sigma(Ax+B)=Ax+B$ throughout the relevant domains. Using independence of $A$ and $B$, $\E[A]=0$, and $\E[A^2]=1/3$, we obtain
\begin{equation*}
\E_{A,B}[3A(Ax+B)_+]=3x\E[A^2]+3\E[A]\E[B]=x=h(x).
\end{equation*}
The function $f(A,B)=3A$ is Lipschitz and hence $\beta$-H\"older with $\beta=1/2$ on the compact parameter space $\Pi=[-1,1]\times[2,3]$. Moreover, the uniform distribution on this rectangle satisfies the regular sampling assumptions for the Euclidean metric, and the nearest-neighbor indegree is uniformly bounded. Hence the assumptions of Corollary~\ref{cor:constrained_approx} are satisfied.

Write $u_K=(A_1,\ldots,A_K)^\top$, $v_K=(B_1,\ldots,B_K)^\top$, $S_K=[u_K\ \ v_K]$, and $Q=\operatorname{diag}(1/3,1)$. Then $g_K(x)=S_K(x,1)^\top$ and $\Gamma_K=S_KQS_K^\top$. By the strong law of large numbers,
\begin{equation*}
\frac{1}{K}S_K^\top S_K\longrightarrow\begin{pmatrix}1/3&0\\0&19/3\end{pmatrix},\qquad P_\theta\text{-a.s.}
\end{equation*}
Consequently, $\operatorname{rank}(S_K)=2$ a.s. for every $K\geq2$, and $\Gamma_K$ has exactly two positive eigenvalues. The nonzero eigenvalues of $\Gamma_K$ coincide with those of $Q^{1/2}S_K^\top S_KQ^{1/2}$, so both are of order $K$, $P_\theta$-almost surely. Therefore $\operatorname{rank}(\Gamma_K)=2$ and $\lambda_{\min}(\Gamma_K)=0$ for every $K\geq3$, almost surely.

For $K\geq2$, the regression function $h(x)=x$ belongs to the realized RWNN sieve. Its minimum-Euclidean-norm projection coefficient is
\begin{equation*}
    \beta_K=S_K(S_K^\top S_K)^{-1} (1, 0)^\top,
\end{equation*}
and hence $\Delta_K=0$ and $\eta_K(x)=0$ for all $x\in\mathcal X$. Since $\|S_K\|=O_{P_\theta}(\sqrt K)$ and $\|(S_K^\top S_K)^{-1}\|=O_{P_\theta}(K^{-1})$, it follows that $\|\beta_K\|=O_{P_\theta}(K^{-1/2})$. Moreover, $\beta_K\in\operatorname{Range}(\Gamma_K)$, and the positive eigenvalues of $\Gamma_K$ are of order $K$. Thus, uniformly for $\lambda>0$,
\begin{equation*}
\|\Gamma_{K,\lambda}^{-1/2}\beta_K\|=O_{P_\theta}(K^{-1}).
\end{equation*}
The same spectral bounds yield
\begin{equation*}
\widetilde{\xi}_{K,\lambda}=\sup_{x\in[-1,1]}\|\Gamma_{K,\lambda}^{-1/2}g_K(x)\|=O_{P_\theta}(1),
\end{equation*}
since $g_K(x)\in\operatorname{Range}(\Gamma_K)$ and $\sup_{x\in[-1,1]}\|g_K(x)\|=O_{P_\theta}(\sqrt K)$.

For every nonzero $\alpha_K\in\operatorname{Range}(\Gamma_K)$, the spectral representation \eqref{eq:RWNN_Rayleigh_quotient_to_basis} implies
\begin{equation*}
    \frac{\lambda_{\min}^+(\Gamma_K)}{\lambda_{\min}^+(\Gamma_K)+\lambda}\leq\mathcal R_{K,\lambda}(\alpha_K)\leq1,
\end{equation*}
where $\lambda_{\min}^+(\Gamma_K)$ denotes the smallest strictly positive eigenvalue of $\Gamma_K$. Since $\lambda_{\min}^+(\Gamma_K)\asymp K$, the Rayleigh quotient converges to one whenever $\lambda/K\to0$. In particular, this applies to the normalized point evaluation direction $\alpha_K=g_K(x_0)/\|g_K(x_0)\|$ for every fixed $x_0\in[-1,1]$. It also applies to the normalized derivative direction associated with $\Psi_2$, since
\begin{equation*}
    \Psi'_{2,h}(g_K)=2\int_{-1}^1xg_K(x)\,\de x=\frac43u_K \ \in \ \operatorname{Range}(\Gamma_K).
\end{equation*}

Now choose $K_n=\lceil n^{2/5}\rceil$ and $\lambda_{K_n,n}=K_n^{-3/4}$. With $d=1$, $\beta=1/2$, and $r=3/4$, the polynomial-ridge restriction \eqref{eq:lambda_r_choice} is satisfied since $1/2<3/4<1$. Furthermore,
\begin{equation*}
    \frac{K_n^{1+2\beta/(d+1)}}{n}=\frac{K_n^{3/2}}{n}=O(n^{-2/5})=o(1)
    \quad \text{and} \quad
    \sqrt n\,\lambda_{K_n,n}\|\Gamma_{K_n,\lambda}^{-1/2}\beta_{K_n}\|=O_{P_\theta}(n^{-1/5})=o_{P_\theta}(1).
\end{equation*}
The normalized feature condition in \eqref{eq:RWNN_Rayleigh_cond} follows from $\widetilde{\xi}_{K_n,\lambda}=O_{P_\theta}(1)$, while the directional Rayleigh condition follows from the preceding spectral argument. Moreover,
\begin{equation*}
\frac{\widetilde{\xi}_{K_n,\lambda}^{\,2}\log K_n}{n}=O_{P_\theta}\left(\frac{\log K_n}{n}\right)=o_{P_\theta}(1).
\end{equation*}
Thus Assumption~\ref{assumption:L2_ridge_gram_stability} holds in $P_\theta$-probability. The regression errors satisfy the required second moment and uniform integrability conditions as they are independent Gaussian random variables with variance $\tau^2>0$. This verifies the conditions of Corollary~\ref{corollary:RWNN_L2_normality} along both prescribed directions.

We next verify the conditions for feasible inference. Since $\eta_{K_n}=0$ for all sufficiently large $n$ and $\E[\epsilon_i^2\mid X_i]=\tau^2$, for every nonzero direction $\alpha_K\in\operatorname{Range}(\Gamma_K)$ we have
\begin{equation*}
    \mathcal R_{K,\lambda}^{(1)}(\alpha_K)=\tau^2\mathcal R_{K,\lambda}(\alpha_K)
    \quad \text{and} \quad
    \mathcal L_{K,\lambda}^{(1)}(\alpha_K)=\frac{\widetilde{\xi}_{K,\lambda}^{\,2}}{\tau^2\mathcal R_{K,\lambda}(\alpha_K)}=O_{P_\theta}(1).
\end{equation*}
Hence \eqref{eq:primitive_L_growth_l1} holds with $\rho=0$. The corresponding rate condition is
\begin{equation*}
\frac{K_n^{2\rho+1+r}\log K_n}{n}=\frac{K_n^{7/4}\log K_n}{n}=O(n^{-3/10}\log n)=o(1).
\end{equation*}
The conditional fourth moment condition \eqref{eq:primitive_fourth_moment_l1} follows from Gaussianity and $\eta_{K_n}=0$ for $K_n\geq2$; the finitely many initial values are uniformly bounded since the feature parameters lie in the compact rectangle and the population projection is bounded. Consequently, Corollaries~\ref{corollary:RWNN_variance_estimator_consistency} and~\ref{corollary:RWNN_variance_estimator_consistency_l0} apply to both normalized directions.

It remains to identify the limiting distributions. Let $q(x)=(x,1)^\top$, $\vartheta_0=(1,0)^\top$, $T_K=S_K^\top S_K$, and $\widehat Q_n=n^{-1}\sum_{i=1}^nq(X_i)q(X_i)^\top$. Since every function in the realized sieve has the form $x\mapsto q(x)^\top\vartheta$, the minimum-Euclidean-norm coefficient vector representing $\vartheta$ is $S_KT_K^{-1}\vartheta$. Therefore the original ridge regression problem is equivalent to minimizing
\begin{equation*}
\frac1n\sum_{i=1}^n\bigl(Y_i-q(X_i)^\top\vartheta\bigr)^2+\lambda_{K_n,n}\vartheta^\top T_{K_n}^{-1}\vartheta
\end{equation*}
over $\vartheta\in\mathbb R^2$. Its minimizer satisfies
\begin{equation*}
\widehat\vartheta_n=(\widehat Q_n+\lambda_{K_n,n}T_{K_n}^{-1})^{-1}\left(\frac1n\sum_{i=1}^nq(X_i)Y_i\right),
\end{equation*}
and $\widehat h_{n,\lambda}(x)=q(x)^\top\widehat\vartheta_n$. Since $\|T_{K_n}^{-1}\|=O_{P_\theta}(K_n^{-1})$, we have
\begin{equation*}
\sqrt n\,\|\lambda_{K_n,n}T_{K_n}^{-1}\|=O_{P_\theta}(n^{-1/5})=o_{P_\theta}(1).
\end{equation*}
Using $Y_i=q(X_i)^\top\vartheta_0+\epsilon_i$, we obtain the exact decomposition
\begin{equation*}
\sqrt n(\widehat\vartheta_n-\vartheta_0)=(\widehat Q_n+\lambda_{K_n,n}T_{K_n}^{-1})^{-1}\frac1{\sqrt n}\sum_{i=1}^nq(X_i)\epsilon_i-\sqrt n(\widehat Q_n+\lambda_{K_n,n}T_{K_n}^{-1})^{-1}\lambda_{K_n,n}T_{K_n}^{-1}\vartheta_0.
\end{equation*}
By the law of large numbers, $\widehat Q_n\toprob Q=\operatorname{diag}(1/3,1)$. The second term is $o_P(1)$ conditionally on the features, while the multivariate central limit theorem gives
\begin{equation*}
\frac1{\sqrt n}\sum_{i=1}^nq(X_i)\epsilon_i\todist\mathcal N(0,\tau^2Q).
\end{equation*}
Consequently, conditionally on $\mathcal F_{K_n}$, in $P_\theta$-probability,
\begin{equation*}
\sqrt n(\widehat\vartheta_n-\vartheta_0)\todist\mathcal N(0,\tau^2Q^{-1}),\qquad Q^{-1}=\begin{pmatrix}3&0\\0&1\end{pmatrix}.
\end{equation*}
Since $\widehat h_{n,\lambda}(x_0)=q(x_0)^\top\widehat\vartheta_n$, it follows that
\begin{equation*}
\sqrt n\bigl(\widehat h_{n,\lambda}(x_0)-x_0\bigr)\todist\mathcal N\bigl(0,\tau^2(1+3x_0^2)\bigr).
\end{equation*}

Finally, writing $H=\operatorname{diag}(2/3,2)$, we have $\Psi_2(\widehat h_{n,\lambda})=\widehat\vartheta_n^\top H\widehat\vartheta_n$ and $\Psi_2(h)=\vartheta_0^\top H\vartheta_0=2/3$. The derivative at $\vartheta_0$ is $2H\vartheta_0=(4/3,0)^\top$. The delta method therefore gives
\begin{equation*}
\sqrt n\left(\Psi_2(\widehat h_{n,\lambda})-\frac23\right)\todist\mathcal N\left(0,(2H\vartheta_0)^\top\tau^2Q^{-1}(2H\vartheta_0)\right)=\mathcal N\left(0,\frac{16\tau^2}{3}\right).
\end{equation*}
The same conclusion follows from Corollary~\ref{corollary:RWNN_smooth_functional_normality}: the quadratic functional satisfies Assumption~\ref{assumption:functional_smooth_riesz_ridge} with $\omega=2$, its sieve approximation bias vanishes, its nonlinear remainder is negligible since $\|\widehat h_{n,\lambda}-h\|_{L^2(X)}=O_P(n^{-1/2})$, and its asymptotic variance converges to $16\tau^2/3>0$.

For completeness, we note that the feasible variance estimators are also consistent for the unnormalized point evaluation and smooth functional limits. Indeed, let $\widehat\epsilon_i=Y_i-q(X_i)^\top\widehat\vartheta_n$ and $D_n=\lambda_{K_n,n}T_{K_n}^{-1}$. For either $c_n=q(x_0)$ or $c_n=2H\widehat\vartheta_n$, the corresponding ridge-adjusted sandwich estimator can be written as
\begin{equation*}
\widehat V_n(c_n)=\frac1n\sum_{i=1}^n\widehat\epsilon_i^2\left[c_n^\top(\widehat Q_n+D_n)^{-1}q(X_i)\right]^2.
\end{equation*}
Since $\widehat\vartheta_n\toprob\vartheta_0$, $D_n=o_{P_\theta}(1)$, and $n^{-1}\sum_{i=1}^n\widehat\epsilon_i^2q(X_i)q(X_i)^\top\toprob\tau^2Q$, it follows that
\begin{equation*}
\widehat V_n(q(x_0))\toprob\tau^2(1+3x_0^2),\qquad \widehat V_n(2H\widehat\vartheta_n)\toprob\frac{16\tau^2}{3}.
\end{equation*}
All convergence statements above hold conditionally on $\mathcal F_{K_n}$, in $P_\theta$-probability, as claimed.
\end{proof}

\clearpage

\section{Proofs for sup-norm ridge estimation}
\label{app:supnorm_ridge_proofs}

\subsection{Uniform stochastic error}
\label{app:supnorm_stochastic_proofs}

\begin{proof}[Proof of Lemma~\ref{lemma:uniform_bound_variance}]
\smallskip
\noindent\textsc{Step 1.} %
First, we define $\check{h}_{n,\lambda}:=\widehat{h}_{n,\lambda}-\widetilde{h}_{n,\lambda}$ and observe
\begin{equation}
\label{eq:h_check}
\begin{aligned}
    \check{h}_{n,\lambda}(x)
    & =g_{K_n}(x)^\top\widehat{\Gamma}_{K_n,n,\lambda}^{-1}\left(n^{-1}G_{K_n,n}^\top e\right) \\
    & =g_{K_n}(x)^\top\widehat{\Gamma}_{K_n,n,\lambda}^{-1}\left(n^{-1/2} V_{m_n,n} S_{m_n,n} U_{m_n,n}^\top e\right) 
    \\ & = n^{-1/2} g_{K_n}(x)^\top V_{m_n,n} \operatorname{diag}\left(\left(\frac{s_{k,n}}{s_{k,n}^2+\lambda_{K_n,n}}\right)_{k=1}^{m_n}\right) U_{m_n,n}^\top e,
\end{aligned}
\end{equation}
where in the first equality we used the definitions of $\widehat{h}_{n,\lambda}$ and $\widetilde{h}_{n,\lambda}$ in \eqref{eq:RWNN_estimator} and \eqref{eq:pseudoproj}, respectively, together with $Y-H=e$, while in the second one we applied the singular value decomposition in \eqref{eq:uniform_design_svd}. 
From the second equality in \eqref{eq:h_check}, we see that
\begin{equation}
\label{eq:norm_gamma_G}
\left\|\widehat{\Gamma}_{K_n,n,\lambda}^{-1}\left(n^{-1}G_{K_n,n}^\top e\right)\right\|\leq \psi_{K_n,n,\lambda}\left\|n^{-1/2}U_{m_n,n}^\top e\right\| = \psi_{K_n,n,\lambda} \widetilde{U}_n ,
\end{equation}
where $\widetilde{U}_n := \left\|n^{-1/2}U_{m_n,n}^\top e\right\|$. 
Notice that, $U_{m_n,n}$ is measurable with respect to $X_{1:n}:=(X_1,\ldots,X_n)$ and $\theta_{1:K_n}=(\theta_1,\ldots,\theta_{K_n})$. Since the random features
are independent of the observations, Assumption~\ref{assumption:regression_errors_L2} and the i.i.d. hypothesis imply
\begin{align}
    \E\left[\widetilde{U}_n^2 \middle| X_{1:n},\theta_{1:K_n} \right]&=\frac{1}{n}\operatorname{tr}\left(U_{m_n,n}^\top \E\left[ee^\top\mid X_{1:n}\right] U_{m_n,n} \right)\nonumber \\
    &\leq \frac{\overline\sigma^2}{n} \operatorname{tr}\left( U_{m_n,n}^\top U_{m_n,n} \right)= \overline\sigma^2\frac{m_n}{n},
\end{align}
which, by conditional Markov inequality applied to random variable $\widetilde{U}_n^2$, followed by the tower property, together with using the fact that the columns of $U_{m_n,n}$ are orthonormal vectors, implies 
$\widetilde{U}_n = O_P\left(\sqrt{{m_n}/{n}}\right)$.

Further, for any $x_1,x_2\in\mathcal{X}$, Assumption~\ref{assumption:uniform_sieve_regularity}~(i)-(ii), together with \eqref{eq:norm_gamma_G}, give
\begin{align}
\label{eq:mean_value_h_check}
    |\check{h}_{n,\lambda}(x_1)-\check{h}_{n,\lambda}(x_2)|&\leq \|g_{K_n}(x_1)-g_{K_n}(x_2)\| \left\|
    \widehat{\Gamma}_{K_n,n,\lambda}^{-1} \left(n^{-1}G_{K_n,n}^\top e\right) \right\|\nonumber \\ 
    &\leq L_{K_n} \|x_1-x_2\| \psi_{K_n,n,\lambda} \widetilde{U}_n.
\end{align}
This implies that the sample paths of  $\check{h}_{n,\lambda}$ are Lipschitz on $\mathcal{X}$, with a random Lipschitz coefficient controlled by $\psi_{K_n,n,\lambda}\widetilde U_n$.

\smallskip
\noindent\textsc{Step 2.}
We now define
\begin{equation}
\label{eq:def_tn}
    t_n:=\zeta_{K_n}\overline{\psi}_{K_n,n}\sqrt{\frac{\log(n)}{n}}.
\end{equation}
Since $\psi_{K_n,n,\lambda}=O_P\bigl(\overline{\psi}_{K_n,n}\bigr)$ by Assumption~\ref{assumption:growth_variance} and $\widetilde{U}_n=O_P\left(\sqrt{{m_n}/{n}}\right)$, for every $\eta>0$ there exists a constant $M_\eta>0$ such that, for all sufficiently large $n$,
\begin{equation*}
    \mathbb{P}\left(\psi_{K_n,n,\lambda}>M_\eta\overline{\psi}_{K_n,n}\right)\leq\frac{\eta}{2} \quad \text{and} \quad \mathbb{P}\left(\widetilde U_n>M_\eta\sqrt{\frac{m_n}{n}}\right)\leq\frac{\eta}{2}.
\end{equation*}
Next, define $\mathcal{A}_{n,\eta}:= \left\{\psi_{K_n,n,\lambda}\leq M_\eta\overline{\psi}_{K_n,n}\right\}$ and $\mathcal{B}_{n,\eta}:= \mathcal{A}_{n,\eta} \cap \left\{\widetilde U_n\leq M_\eta\sqrt{{m_n}/{n}}\right\}$. Then one gets that $\mathbb{P}\left(\mathcal{B}_{n,\eta}^{\complement}\right)\leq\eta$ for all sufficiently large $n$. Moreover, on the event $\mathcal{B}_{n,\eta}$, \eqref{eq:mean_value_h_check} implies the following bound:
\begin{equation}
\label{eq:localized_lipschitz_bound}
    |\check{h}_{n,\lambda}(x_1)-\check{h}_{n,\lambda}(x_2)|
    \leq 
    M_\eta^2 L_{K_n} \overline{\psi}_{K_n,n} \sqrt{\frac{m_n}{n}} \, \|x_1-x_2\|.
\end{equation}

Let $\mathcal{C}_n=\{x_{1,n},\ldots, x_{N_n,n}\}$ be a minimal $\rho_n$-net of $\mathcal{X}$ with respect to the Euclidean norm, with $N_n:=|\mathcal{C}_n|$ and covering radius $\rho_n > 0$.
We choose $\rho_n$ such that the right-hand side of \eqref{eq:localized_lipschitz_bound} is at most $t_n/4$ whenever $\|x_1-x_2\|\leq\rho_n$, which gives
\begin{equation*} 
    \rho_n := \frac{ \zeta_{K_n}\sqrt{\log(n)} }{ 4M_\eta^2 L_{K_n}\sqrt{m_n} }. 
\end{equation*} 
Indeed, one can verify that $\rho_n   M_\eta^2 L_{K_n} \overline{\psi}_{K_n,n} \sqrt{{m_n}/{n}}  = {t_n}/{4}$ and hence, on $\mathcal{B}_{n,\eta}$, whenever $\|x_1-x_2\|\leq\rho_n$ one has from \eqref{eq:localized_lipschitz_bound} that $|\check{h}_{n,\lambda}(x_1)-\check{h}_{n,\lambda}(x_2)| \leq {t_n}/{4}$. 
Therefore, for every $x\in\mathcal{X}$ there exists an index $j(x)\in\{1,\ldots, N_n\}$ such that $ \|x-x_{j(x),n}\|\leq\rho_n$. 
Since $\mathcal{X}=[-M,M]^d$ by Assumption~\ref{assumption:X_domain}, the same standard volumetric argument (see, for example, Corollary 4.2.11 in~\cite{vershynin2026high}) used in \eqref{eq:covering_bound} gives
\begin{equation} 
\label{eq:uniform_covering_number} 
    N_n = \mathcal{N}(\rho_n,\mathcal{X},\|\cdot\|) \leq \left(1+\frac{2M\sqrt{d}}{\rho_n}\right)^d,
\end{equation} 
and hence 
\begin{align*} 
    \log(N_n) &\leq d\log\left(1+\frac{2M\sqrt d}{\rho_n}\right)=d\log\left(1+8M M_{\eta}^2 \sqrt{d}\frac{L_{K_n}\sqrt{m_n}}{\zeta_{K_n}\sqrt{\log(n)}}\right)\nonumber\\
    &\leq d\log\left(1+8CM M_{\eta}^2 \sqrt{d}n^q\right)\leq d\log\left(1+C_\eta n^q\right)\leq d \log\big( (1 + C_\eta) n^q \big) \lesssim \log(n)
\end{align*} 
where in the second inequality we used  Assumption~\ref{assumption:uniform_sieve_regularity}~(iii), while in the third one defined $C_\eta:=8CM M_{\eta}^2 \sqrt{d}$.
Therefore, by the construction of $\rho_n$, on $\mathcal{B}_{n,\eta}$, it holds 
\begin{equation*} 
    \sup_{x\in\mathcal{X}} |\check{h}_{n,\lambda}(x)-\check{h}_{n,\lambda}(x_{j(x),n})| \leq \frac{t_n}{4}. 
\end{equation*}
Hence, on $\mathcal{B}_{n,\eta}$, 
\begin{equation} 
\label{eq:h_check_inf}
    \|\check{h}_{n,\lambda}\|_\infty \leq \max_{x_{j,n}\in\mathcal{C}_n} |\check{h}_{n,\lambda}(x_{j,n})| + \frac{t_n}{4}. 
\end{equation} 

\smallskip
\noindent\textsc{Step 3.} For each $x\in\mathcal{X}$ and $1\leq i\leq n$, define $ a_{i,n}(x):=g_{K_n}(x)^\top\widehat{\Gamma}_{K_n,n,\lambda}^{-1} g_{K_n}(X_i)$. Then, by the first equality in \eqref{eq:h_check}, we have that
\begin{equation*}
    \check{h}_{n,\lambda}(x) = \frac{1}{n}\sum_{i=1}^n a_{i,n}(x)\epsilon_i.
\end{equation*}
Let $\{\tau_n\}_{n\geq1}$ be a deterministic positive sequence satisfying $\tau_n\to\infty$, a precise choice of which we will specify below. Define
\begin{equation}
\label{eq:centered_eps_trunc}
\epsilon_{i,n}^{(1)} := \epsilon_i\Ind\{|\epsilon_i|\leq\tau_n\} - \E\left[ \epsilon_i\Ind\{|\epsilon_i|\leq\tau_n\} \mid X_i \right]
\end{equation}
and
\begin{equation}
\label{eq:centered_eps_remainder}
\epsilon_{i,n}^{(2)} := \epsilon_i-\epsilon_{i,n}^{(1)}.
\end{equation}
We may hence write $\check{h}_{n,\lambda}(x) = \check h_{n,\lambda}^{(1)}(x) + \check{h}_{n,\lambda}^{(2)}(x)$, where, for $\ell\in\{1,2\}$,
\begin{equation}
\label{eq:h_tilde_l}	
\check{h}_{n,\lambda}^{(\ell)}(x):=\frac{1}{n}\sum_{i=1}^n a_{i,n}(x)\epsilon_{i,n}^{(\ell)}.
\end{equation}
We first analyze the random coefficients $a_{i,n}(x)$ in \eqref{eq:h_tilde_l}. Let $u_{i,n}^\top$ denote the $i$th row of $U_{m_n,n}$, $i\in \{1, \ldots, n\}$. Since the singular value decomposition in \eqref{eq:uniform_design_svd} gives
\begin{equation*}
g_{K_n}(X_i)=\sqrt{n} V_{m_n,n}S_{m_n,n}u_{i,n},
\end{equation*}
we have
\begin{align*}
\frac{1}{n}|a_{i,n}(x)|&=\frac{1}{\sqrt{n}}\left|g_{K_n}(x)^\top V_{m_n,n} \operatorname{diag}\left(\left(\frac{s_{k,n}}{s_{k,n}^2+\lambda_{K_n,n}} \right)_{k=1}^{m_n}\right) u_{i,n} \right|\\
&\leq \frac{\|g_{K_n}(x)\|}{\sqrt{n}} \psi_{K_n,n,\lambda}\|u_{i,n}\| \leq \frac{\zeta_{K_n}\psi_{K_n,n,\lambda}}{\sqrt{n}},
\end{align*}
where the first inequality follows from the Cauchy-Schwarz inequality and the definition of $\psi_{K_n,n,\lambda}$ in \eqref{eq:psi_def}, while the second one uses $\|g_{K_n}(x)\|\leq\zeta_{K_n}$ for any $x\in \mathcal{X}$ due to Assumption~\ref{assumption:uniform_sieve_regularity} and that $\|u_{i,n}\|\leq1$ since $U_{m_n,n}$ has orthonormal columns.

Moreover, we have immediately that
\begin{align}
\label{eq:conditional_variance_coefficients}
\frac{1}{n^2}\sum_{i=1}^n a_{i,n}(x)^2&=\frac{1}{n^2}\sum_{i=1}^n g_{K_n}(x)^\top \widehat{\Gamma}_{K_n,n,\lambda}^{-1} g_{K_n}(X_i) g_{K_n}(X_i)^\top \widehat{\Gamma}_{K_n,n,\lambda}^{-1} g_{K_n}(x)\nonumber\\
&=\frac{1}{n } g_{K_n}(x)^\top \widehat{\Gamma}_{K_n,n,\lambda}^{-1} \widehat{\Gamma}_{K_n,n} \widehat{\Gamma}_{K_n,n,\lambda}^{-1} g_{K_n}(x)\nonumber\\
&\leq\frac{\zeta_{K_n}^2\psi_{K_n,n,\lambda}^2}{n},
\end{align}
where we used  that
\begin{equation*}
\left\|\widehat{\Gamma}_{K_n,n,\lambda}^{-1}\widehat{\Gamma}_{K_n,n}\widehat{\Gamma}_{K_n,n,\lambda}^{-1} \right\|=\max_{1\leq k\leq m_n}\frac{s_{k,n}^2}{(s_{k,n}^2+\lambda_{K_n,n})^2}=\psi_{K_n,n,\lambda}^2
\end{equation*}
and that $\|g_{K_n}(x)\|\leq\zeta_{K_n}$ by Assumption~\ref{assumption:uniform_sieve_regularity}.

We will use two different coefficient bounds below. The sharper bound in \eqref{eq:conditional_variance_coefficients} controls the conditional variance in the application of the Bernstein inequality. For the maximal size of the remainder term $\epsilon_{i,n}^{(2)}$ in \eqref{eq:centered_eps_remainder}, we will derive a uniform bound. Since $\widehat{\Gamma}_{K_n,n}=n^{-1}G_{K_n,n}^\top G_{K_n,n}$ is positive semidefinite and $\lambda_{K_n,n}>0$, we have $ \lambda_{\min}(\widehat{\Gamma}_{K_n,n,\lambda})=\lambda_{\min}(\widehat{\Gamma}_{K_n,n}+\lambda_{K_n,n}I_{K_n})\geq\lambda_{K_n,n}$ and hence, $ \left\|\widehat{\Gamma}_{K_n,n,\lambda}^{-1}\right\|=\lambda_{\min}(\widehat{\Gamma}_{K_n,n,\lambda})^{-1}\leq\lambda_{K_n,n}^{-1}$. Recalling that by the definition $ a_{i,n}(x)=g_{K_n}(x)^\top\widehat{\Gamma}_{K_n,n,\lambda}^{-1}g_{K_n}(X_i)$ in the beginning of Step~3,
the Cauchy-Schwarz inequality gives, for all $x\in \mathcal{X}$ and $1\leq i\leq n$,
\begin{align} 
\label{eq:a_abs}
|a_{i,n}(x)|&\leq\|g_{K_n}(x)\|\left\|\widehat{\Gamma}_{K_n,n,\lambda}^{-1}\right\|\|g_{K_n}(X_i)\| \leq \zeta_{K_n}\lambda_{K_n,n}^{-1}\zeta_{K_n}=\frac{\zeta_{K_n}^2}{\lambda_{K_n,n}}, 
\end{align} 
where we used $\|g_{K_n}(x)\|\leq\zeta_{K_n}$ and, since $X_i\in\mathcal{X}$, also $\|g_{K_n}(X_i)\|\leq\zeta_{K_n}$.

\smallskip
\noindent\textsc{Step 4.} 
We first analyze random variables $\epsilon_{i,n}^{(1)}$ in \eqref{eq:centered_eps_trunc} which by construction are centered conditionally on $X_i$, that is $\E[\epsilon_{i,n}^{(1)}\mid X_i]=0$. 
Moreover, by the conditional triangle inequality, 
\begin{align} 
\label{eq:eps_1_abs}
|\epsilon_{i,n}^{(1)}|&\leq|\epsilon_i|\Ind\{|\epsilon_i|\leq\tau_n\}+\left|\E\left[\epsilon_i\Ind\{|\epsilon_i|\leq\tau_n\}\mid X_i\right]\right|\nonumber \\ &\leq\tau_n+\E\left[|\epsilon_i|\Ind\{|\epsilon_i|\leq\tau_n\}\mid X_i\right]\leq2\tau_n. 
\end{align} 
Since the observations are i.i.d.\ and the random features are independent of the data, conditionally on $X_{1:n}$ and $\theta_{1:K_n}$, the variables $\epsilon_{1,n}^{(1)},\ldots,\epsilon_{n,n}^{(1)}$ are independent. In addition, for every covering point $x_{j,n}\in \mathcal{C}_n$, the coefficients $a_{i,n}(x_{j,n})$ are measurable with respect to the same conditioning variables. Consequently, the random variables $n^{-1}a_{i,n}(x_{j,n})\epsilon_{i,n}^{(1)}$, $i=1,\ldots,n$, are conditionally independent and centered, since 
\begin{equation*} 
\E\left[\left.\frac{1}{n}a_{i,n}(x_{j,n})\epsilon_{i,n}^{(1)}\right|X_{1:n},\theta_{1:K_n}\right]=\frac{1}{n}a_{i,n}(x_{j,n})\E[\epsilon_{i,n}^{(1)}\mid X_i]=0. 
\end{equation*}
Combining the uniform bound in \eqref{eq:a_abs} with \eqref{eq:eps_1_abs} yields, uniformly over $x\in\mathcal{X}$ and $1\leq i\leq n$, 
\begin{equation} 
\label{eq:uniform_max_summand} 
    \frac{1}{n}|a_{i,n}(x)\epsilon_{i,n}^{(1)}|\leq\frac{2\zeta_{K_n}^2\tau_n}{n\lambda_{K_n,n}}.
\end{equation}

We now bound the conditional variance of $\epsilon_{i,n}^{(1)}$. Since $\epsilon_{i,n}^{(1)}$ is obtained by conditionally centering the truncated error, it holds that
\begin{align*}
    \operatorname{Var}\big(\epsilon_{i,n}^{(1)}\mid X_i\big)
    &=\E\big[(\epsilon_{i,n}^{(1)})^2\mid X_i\big]\leq \E\left[\epsilon_i^2\Ind\{|\epsilon_i|\leq\tau_n\}\mid X_i\right]
    \leq \E[\epsilon_i^2\mid X_i]
    \leq\overline{\sigma}^2.
\end{align*}
where the last inequality follows from Assumption~\ref{assumption:regression_errors_L2}. 
Since $a_{i,n}(x)$ is measurable with respect to $\sigma(X_{1:n},\theta_{1:K_n})$, it follows from \eqref{eq:conditional_variance_coefficients} that
\begin{align}
\label{eq:uniform_conditional_variance}
    \sum_{i=1}^n\E\left[ \left(\frac{1}{n}a_{i,n}(x)\epsilon_{i,n}^{(1)}\right)^2 \,\middle|\, X_{1:n},\theta_{1:K_n}\right]&= \frac{1}{n^2} \sum_{i=1}^n a_{i,n}(x)^2 \E\left[ (\epsilon_{i,n}^{(1)})^2 \,\middle|\, X_{1:n},\theta_{1:K_n} \right]\nonumber\\
    &\leq\frac{\overline{\sigma}^2}{n^2}\sum_{i=1}^n a_{i,n}(x)^2\leq\frac{\overline{\sigma}^2\zeta_{K_n}^2 \psi_{K_n,n,\lambda}^2}{n},
\end{align}
where in the final inequality we used the bound from \eqref{eq:conditional_variance_coefficients}.
Recall now that the event $\mathcal{A}_{n,\eta}:=\{\psi_{K_n,n,\lambda}\leq M_\eta\overline\psi_{K_n,n}\}$ is measurable with respect to $X_{1:n}$ and $\theta_{1:K_n}$. Therefore, on $\mathcal{A}_{n,\eta}$, the conditional variance bound in \eqref{eq:uniform_conditional_variance} becomes
\begin{equation}
\label{eq:uniform_conditional_variance_Aneta}
    \sum_{i=1}^n \E\left[  \left( \frac{1}{n}a_{i,n}(x)\epsilon_{i,n}^{(1)} \right)^2 \middle| X_{1:n}, \theta_{1:K_n}\right] \leq \frac{\overline{\sigma}^2M_\eta^2\zeta_{K_n}^2 \overline{\psi}_{K_n,n}^2}{n}.
\end{equation}

We can now apply the conditional Bernstein inequality in Lemma~\ref{lemma:conditional_bernstein}. Indeed, conditionally on $X_{1:n}$ and $\theta_{1:K_n}$, the random variables $n^{-1}a_{i,n}(x_{j,n})\epsilon_{i,n}^{(1)}$, $i=1,\ldots,n$, are independent and centered, their absolute values are bounded by \eqref{eq:uniform_max_summand}, and their conditional variance sum is bounded by \eqref{eq:uniform_conditional_variance_Aneta}. Hence, for every $1\leq j\leq N_n$ and every $c>0$,
\begin{align}
\label{eq:uniform_bernstein}
    &\Prob\left(\left|\check{h}_{n,\lambda}^{(1)}(x_{j,n})\right|>ct_n \,\middle|\, X_{1:n},\theta_{1:K_n}\right)\Ind_{\mathcal{A}_{n,\eta}} \nonumber\\
    &\qquad\leq2\exp\left(-\frac{c^2t_n^2}{2\left(\overline{\sigma}^2M_\eta^2\zeta_{K_n}^2\overline\psi_{K_n,n}^2/n+2c\zeta_{K_n}^2\tau_nt_n/(3n\lambda_{K_n,n})\right)}\right) \nonumber\\
    &\qquad = 2\exp\left(-\frac{c^2\log(n)}{2\left(\overline{\sigma}^2M_\eta^2 +\frac{2}{3}c\frac{\zeta_{K_n}\tau_n}{\lambda_{K_n,n}\overline{\psi}_{K_n,n}}\sqrt{\frac{\log(n)}{n}}\right)}\right)
\end{align}
where in the last equality we used the definition of $t_n$ in \eqref{eq:def_tn}. 

We choose the truncation $\tau_n$ as    
\begin{equation}
\label{eq:def_tau}
    \tau_n^{1+\delta}:=\frac{\zeta_{K_n}}{\lambda_{K_n,n}\overline{\psi}_{K_n,n}}\sqrt{\frac{n}{\log(n)}},
\end{equation}
where $\delta>0$ is the exponent in Assumption~\ref{assumption:higher_moment}, and recall that, by Assumption~\ref{assumption:growth_variance}, $\tau_n\to \infty$. Notice that with this choice it holds that  
\begin{align} 
\label{eq:cutoff_choice} \frac{\zeta_{K_n}\tau_n} {\lambda_{K_n,n}\overline{\psi}_{K_n,n}} \sqrt{\frac{\log(n)}{n}} &= \left[ \frac{ \left( \zeta_{K_n}/ (\lambda_{K_n,n}\overline{\psi}_{K_n,n}) \right)^{(2+\delta)/\delta} }{ \sqrt{n/\log(n)} } \right]^{\delta/(1+\delta)} = O(1), 
\end{align} where the final equality follows from Assumption~\ref{assumption:growth_variance}. It follows from \eqref{eq:uniform_bernstein} and \eqref{eq:cutoff_choice}, for every fixed $c>0$, there exists a constant $C_{\eta,c}>0$ such that 
\begin{equation*}
\Prob\left(\left|\check{h}_{n,\lambda}^{(1)}(x_{j,n})\right|>ct_n\middle|X_{1:n}\right)\Ind_{\mathcal{A}_{n,\eta}}\leq  2n^{-C_{\eta, c}},
\end{equation*}
and $C_{\eta, c}\to \infty$ as $c\to \infty$. Taking expectations and applying the union bound over the covering points in $\mathcal{C}_n$, we now get that 
\begin{equation*}
\Prob\left(\left\{\max_{1\leq j\leq N_n}|\check h_{n,\lambda}^{(1)}(x_{j,n})|>ct_n\right\}\cap  \mathcal{A}_{n,\eta} \right) \leq  2N_n n^{-C_{\eta, c}}.
\end{equation*}
Using the fact that if $\rho_n^{-1}$ grows at most polynomially in $n$, then $\log(N_n) \lesssim \log(n)$, as we noticed below \eqref{eq:uniform_covering_number}, results in the right-hand side converging to zero for $c>0$ sufficiently large, namely 
\begin{equation}
\label{eq:first_summand}
\Prob\left(\left\{\max_{1\leq j\leq N_n}|\check h_{n,\lambda}^{(1)}(x_{j,n})|>ct_n\right\}\cap  \mathcal{A}_{n,\eta} \right) =o(1).
\end{equation}

\smallskip
\noindent\textsc{Step 5.} Next, we move to analyzing random variables $\epsilon_{i,n}^{(2)}$ defined in \eqref{eq:centered_eps_remainder}. Using their definition, we have that
\begin{align*}
    \epsilon_{i,n}^{(2)} &= \epsilon_i - \epsilon_i\Ind\{|\epsilon_i|\leq\tau_n\} + \E\left[ \epsilon_i\Ind\{|\epsilon_i|\leq\tau_n\} \mid X_i \right] \nonumber\\
    &=  \epsilon_i\Ind\{|\epsilon_i|>\tau_n\} - \E\left[ \epsilon_i\Ind\{|\epsilon_i|>\tau_n\} \mid X_i \right],
\end{align*}
where in the last equality we used that by Assumption~\ref{assumption:regression_errors_L2}, $\E[\epsilon_i\mid X_i]=0$. We hence have that 
\begin{equation}
\label{eq:remainder_conditional_mean}
\E\left[|\epsilon_{i,n}^{(2)}| \mid X_i \right] \leq 2\E\left[ |\epsilon_i|\Ind\{|\epsilon_i|>\tau_n\} \mid X_i \right].
\end{equation}
We now can bound the maximum of the second summand $\check{h}_{n,\lambda}^{(2)}(x)$ in the decomposition of $\check{h}_{n,\lambda}(x)$ over $\mathcal{C}_n$ as follows 
\begin{align*}
\max_{1\leq j\leq N_n}|\check{h}_{n,\lambda}^{(2)}(x_{j,n})| &=\max_{1\leq j\leq N_n} \left|\frac{1}{n}\sum_{i=1}^n a_{i,n}(x_{j,n})\epsilon_{i,n}^{(2)}\right|\\
&\leq \frac{1}{n}\sum_{i=1}^n \max_{1\leq j\leq N_n}\left|  a_{i,n}(x_{j,n})\right|\left|\epsilon_{i,n}^{(2)}\right| \leq \frac{\zeta_{K_n}^2}{\lambda_{K_n,n}} \frac{1}{n}\sum_{i=1}^n \left| \epsilon_{i,n}^{(2)} \right| ,
\end{align*}
where the first equality follows from the definition \eqref{eq:h_tilde_l} with $\ell=2$, while in the last inequality we used that, uniformly over $x\in\mathcal{X}$ and $1\leq i\leq n$, the bound in \eqref{eq:a_abs} holds. Finally, by Markov's inequality, for any $c>0$,
\begin{align}
\label{eq:second_summand}
    &\Prob\left( \max_{1\leq j\leq N_n} \left|\check{h}_{n,\lambda}^{(2)}(x_{j,n})\right|>ct_n\right) 
        \leq \frac{1}{n}\frac{\zeta_{K_n}^2}{c\lambda_{K_n,n}t_n} \sum_{i=1}^n \E\left[\left| \epsilon_{i,n}^{(2)} \right|\right]\nonumber\\
    &\qquad\leq \frac{2}{c}\frac{\zeta_{K_n}^2}{\lambda_{K_n,n}t_n} \E\big[ |\epsilon_i|\Ind\{|\epsilon_i|>\tau_n\} \big]  
        \leq \frac{2}{c}\frac{\zeta_{K_n}^2}{\lambda_{K_n,n}t_n} \E\left[ \tau_n^{-(1+\delta)} |\epsilon_i|^{2+\delta} \Ind\{|\epsilon_i|>\tau_n\}\right]\nonumber\\
    &\qquad = \frac{2}{c}\frac{\zeta_{K_n}}{\lambda_{K_n,n}\overline{\psi}_{K_n,n}}\sqrt{\frac{n}{\log(n)}}\frac{1}{\tau_n^{1+\delta}}\E\left[  |\epsilon_i|^{2+\delta} \Ind\{|\epsilon_i|>\tau_n\}\right]\nonumber\\
    &\qquad = \frac{2}{c}\E\left[  |\epsilon_i|^{2+\delta} \Ind\{|\epsilon_i|>\tau_n\}\right]=o(1),
\end{align}
where the second inequality follows from \eqref{eq:remainder_conditional_mean} and the tower property, in the third inequality under the expectation operator we used the higher-order moment bound on a truncated random variable, the first equality follows after using the definition of $t_n$ in \eqref{eq:def_tn}, and the second one is a result of substituting the choice of truncation $\tau_n$ from \eqref{eq:def_tau}, while the last one follows from the fact that  $\tau_n\to \infty$ by construction,  by Assumption~\ref{assumption:higher_moment} and by dominated convergence. 

\smallskip
\noindent\textsc{Step 6.}
Choose $c>0$ sufficiently large so that \eqref{eq:first_summand} holds. By the definition $\check{h}_{n,\lambda}=\check{h}_{n,\lambda}^{(1)}+\check{h}_{n,\lambda}^{(2)}$, the triangle inequality gives
\begin{align}
\label{eq:net_truncation_decomposition}
\mathbb{P}\left(\left\{\max_{1\leq j\leq N_n}|\check{h}_{n,\lambda}(x_{j,n})|>2c t_n\right\}\cap \mathcal{A}_{n,\eta}\right)&\leq\mathbb{P}\left(\left\{\max_{1\leq j\leq N_n}|\check h_{n,\lambda}^{(1)}(x_{j,n})|>ct_n\right\}\cap \mathcal{A}_{n,\eta}\right)\nonumber\\
&\qquad+\mathbb{P}\left(\max_{1\leq j\leq N_n}|\check{h}_{n,\lambda}^{(2)}(x_{j,n})|> ct_n\right)=o(1),
\end{align}
where the last equality follows from \eqref{eq:first_summand} and \eqref{eq:second_summand}.

Recall that $\mathcal{B}_{n,\eta}\subseteq\mathcal{A}_{n,\eta}$ and that, on $\mathcal{B}_{n,\eta}$, \eqref{eq:h_check_inf} gives
\begin{equation*} 
\|\check{h}_{n,\lambda}\|_\infty \leq \max_{x_{j,n}\in\mathcal{C}_n} |\check{h}_{n,\lambda}(x_{j,n})| + \frac{t_n}{4}. 
\end{equation*} 
Hence we can write that
\begin{align*}
\mathbb{P}\left(\|\check{h}_{n,\lambda}\|_\infty> \left(2c +\frac{1}{4}\right)t_n  \right)&\leq \mathbb{P}\left(\left\{\|\check{h}_{n,\lambda}\|_\infty> \left(2c +\frac{1}{4}\right)t_n \right\}\cap \mathcal{B}_{n,\eta} \right) + \mathbb{P}(\mathcal{B}_{n,\eta}^{\complement})\nonumber\\
&\leq \mathbb{P}\left(\left\{ \|\check{h}_{n,\lambda}\|_\infty > \left(2c +\frac{1}{4}\right)t_n \right\}\cap \mathcal{B}_{n,\eta}\right)+\eta\nonumber\\ &\leq \mathbb{P}\left(\left\{\max_{1\leq j\leq N_n} |\check{h}_{n,\lambda}(x_{j,n})|> 2ct_n\right\}  \cap \mathcal{B}_{n,\eta}\right)+\eta \nonumber\\ &\leq \mathbb{P}\left(\left\{\max_{1\leq j\leq N_n} |\check{h}_{n,\lambda}(x_{j,n})|> 2ct_n\right\}  \cap \mathcal{A}_{n,\eta}\right)+\eta \nonumber\\
&= o(1) +\eta,
\end{align*}
where the last equality follows from \eqref{eq:net_truncation_decomposition}.

Since the choice of $\eta>0$ is arbitrary, it follows that
\begin{equation*}
    \|\check{h}_{n,\lambda}\|_\infty 
    = O_P(t_n) 
    = O_P\left( \zeta_{K_n}\overline\psi_{K_n,n}\sqrt{\frac{\log(n)}{n}} \right),
\end{equation*}
which proves the first claim.
We conclude by noticing that the first condition in Assumption~\ref{assumption:growth_variance} gives $t_n=o(1)$ and hence it also holds that
\begin{equation*}
    \|\check{h}_{n,\lambda}\|_\infty=o_P(1),
\end{equation*}
which proves the second claim, as required.
\end{proof}

\begin{proof}[Proof of Lemma~\ref{lemma:RWNN_regular}]
    We consider the two cases separately. Suppose, first, that condition~(i) holds. Since $\mathcal{X}$ and $\Pi$ are compact and $\sigma$ is Lipschitz, the constants $M_{\sigma}:=\sup_{\substack{x\in\mathcal{X}\\(A,B)\in\Pi}}\left|{\sigma}(A^\top x+B)\right|<\infty$ and $C_{A,\Pi}:=\sup_{(A,B)\in\Pi}\|A\|<\infty$ are well defined.
    For every $K\in \mathbb{N}^+$ and $x\in \mathcal{X}$, 
    \begin{equation*}
        \sup_{x\in\mathcal{X}} \|g_K(x)\|^2
        = \sup_{x\in\mathcal{X}} \sum_{k=1}^K\left|{\sigma}(A_k^\top x+B_k)\right|^2 
        \leq K M_{\sigma}^2.
    \end{equation*}
    Consequently, $\sup_{x\in\mathcal{X}}\|g_K(x)\|\leq M_{\sigma} \sqrt{K}$.

    Moreover, for any $x_1,x_2\in \mathcal{X}$, by the Lipschitz continuity of $\sigma$ and the Cauchy-Schwarz inequality, one has
    \begin{equation*}
        \|g_K(x_1)-g_K(x_2)\|^2 \leq \sum_{k=1}^K \left|{\sigma}(A_k^\top x_1+B_k) - {\sigma}(A_k^\top x_2+B_k) \right|^2 \leq K L_{\sigma}^2 C_{A,\Pi} ^2 \|x_1-x_2\|^2.
    \end{equation*}
    
Hence, in case~(i) one may take $M_g=\max\{1,M_{\sigma}\}$ and $L_g=\max\{1,L_{\sigma}C_{A,\Pi}\}$.

    Assume now that condition~(ii) holds. Since $\mathcal{X}=[-M,M]^d$ and $\Pi=\left\{A\in \mathbb{R}^d \mid \|A\|_1\leq C_A\right\}\times [-C_B, C_B]$, then, for every $x\in \mathcal{X}$ and $(A,B)\in \Pi$,  
    \begin{equation*}
    \left|A^\top x +B\right|\leq \|A\|_1 \|x\|_\infty + |B|\leq M C_A+C_B. 
    \end{equation*}
    Set $R:=M C_A+C_B$ and notice that $|\sigma_m(u)|\leq R^m$, $u\in [-R,R]$. Hence,   $\sup_{x\in\mathcal{X}}\|g_K(x)\|\leq R^m \sqrt{K}$. Moreover, $\sigma_m$ is Lipschitz on $[-R,R]$ with Lipschitz constant $mR^{m-1}$ and hence, for   any $x_1,x_2\in \mathcal{X}$,
    \begin{align*}
        \|g_K(x_1)-g_K(x_2)\|^2&\leq \sum_{k=1}^K \left|{\sigma_m}(A_k^\top x_1+B_k) -{\sigma_m}(A_k^\top x_2+B_k) \right|^2\\
        &\leq m^2 R^{2m-2} \sum_{k=1}^K |A_k^\top(x_1-x_2)|^2\leq  m^2 R^{2m-2} \sum_{k=1}^K \|A_k\|^2 \|x_1-x_2\|^2\\
        &\leq m^2 R^{2m-2} C_A^2 K \|x_1-x_2\|^2.
    \end{align*}
Therefore, in case~(ii), one may take $M_g=\max\{1,R^m\}$ and $L_g=\max\{1,mR^{m-1}C_A\}$.

    In both cases, since the sampled parameters belong to $\Pi$ almost surely, these bounds hold almost surely for every $K$. Consequently, Assumption~\ref{assumption:uniform_sieve_regularity}~(i)-(ii) holds with $\zeta_K=M_g \sqrt{K}$ and $L_K=L_g\sqrt{K}$ as required. 
    
    Finally, it remains to check the statement in Assumption~\ref{assumption:uniform_sieve_regularity}~(iii), namely \eqref{eq:pol_growth}. Indeed, 
    \begin{equation*}
        \frac{L_{K_n}\sqrt{m_n}}{\zeta_{K_n}\sqrt{\log(n)}} 
        = \frac{L_g \sqrt{K_n}}{M_g \sqrt{K_n}} \, \frac{\sqrt{m_n}}{\sqrt{\log(n)}}
        \leq \frac{L_g}{M_g} \sqrt{\frac{n}{\log(n)}}
        \leq C n^{1/2}
    \end{equation*}
for all sufficiently large $n$ and some constant $C>0$. Hence \eqref{eq:pol_growth} is satisfied, for example with $q=1/2$, which completes the proof.
\end{proof}
\begin{proof}[Proof of Corollary~\ref{cor:RWNN_variance}]
    By Lemma~\ref{lemma:RWNN_regular}, Assumption~\ref{assumption:uniform_sieve_regularity} holds and hence the claim follows from Lemma~\ref{lemma:uniform_bound_variance}.
\end{proof}

\subsection{Ridge pseudo-projection and spectral stability}
\label{app:supnorm_spectral_proofs}

\begin{proof}[Proof of Lemma~\ref{lemma:pseudo-proj_sup_bound}]
Let $f:\mathcal{X}\to\mathbb{R}$ be bounded. Then, for every $x\in\mathcal{X}$, we write 
\begin{align}
    \left|\bigl(\overline{P}_{K_n,n,\lambda}f\bigr)(x)\right|
    &\leq\|g_{K_n}(x)\|\left\|\widehat{\Gamma}_{K_n,n,\lambda}^{-1}\left( n\inv G_{K_n,n}^\top F\right)\right\|\nonumber\\ 
&=\|g_{K_n}(x)\|\left\|\frac{1}{\sqrt{n}}V_{m_n,n}\operatorname{diag}\left(\left(\frac{s_{k,n}}{s_{k,n}^2+\lambda_{K_n,n}}\right)_{k=1}^{m_n}\right)U_{m_n,n}^\top F\right\|\nonumber\\
    &\leq \|g_{K_n}(x)\| \frac{\psi_{K_n,n,\lambda}}{\sqrt{n}}\|U_{m_n,n}^\top F\| \leq \|g_{K_n}(x)\| \frac{\psi_{K_n,n,\lambda}}{\sqrt{n}}\| F\|, %
\label{eq:pseudo_projection_pointwise_bound}
\end{align}
where in the first equality we applied the singular value decomposition in \eqref{eq:uniform_design_svd}, in the second inequality we used the definition of $\psi_{K_n,n,\lambda}$ in \eqref{eq:psi_def}, while in the same and the subsequent inequality we also used that $U_{m_n,n}$ and $V_{K_n,n}$ are matrices with orthonormal columns. 

Next, notice that 
\begin{equation*}
\frac{1}{\sqrt{n}}\|F\|=\left(\frac{1}{n}\sum_{i=1}^nf(X_i)^2\right)^{1/2}\leq\|f\|_\infty.
\end{equation*}
Hence, taking in \eqref{eq:pseudo_projection_pointwise_bound} the supremum over $x\in\mathcal{X}$ and then over all bounded $f$ satisfying $\|f\|_\infty\leq1$ together with using Assumption~\ref{assumption:uniform_sieve_regularity}~(i), yields
\begin{align*}
\| \overline{P}_{K_n,n,\lambda}\|_{\infty} \leq \zeta_{K_n}\psi_{K_n,n,\lambda},
\end{align*}
as required.
\end{proof}

\begin{proof}[Proof of Lemma~\ref{lemma:uniform_bias_bound}]
We start by writing that 
\begin{align}
\left\|\widetilde{h}_{n,\lambda}-h\right\|_\infty&=\left\|\overline{P}_{K_n,n,\lambda}\left(h-h_{K_n}^{\mathrm{IS}}\right)+\left(\overline{P}_{K_n,n,\lambda}h_{K_n}^{\mathrm{IS}}-h_{K_n}^{\mathrm{IS}}\right)+\left(h_{K_n}^{\mathrm{IS}}-h\right)\right\|_\infty\nonumber \\
& \leq\left(1+\left\|\overline{P}_{K_n,n,\lambda}\right\|_\infty\right)\left\|h-h_{K_n}^{\mathrm{IS}}\right\|_\infty +\left\|\overline{P}_{K_n,n,\lambda}h_{K_n}^{\mathrm{IS}}-h_{K_n}^{\mathrm{IS}}\right\|_\infty.
\label{eq:uniform_bias_decomposition_bound}
\end{align}
where we used the definition of the pseudo-projection operator in the first equality and the definition of $\|\overline{P}_{K_n,n,\lambda}\|_\infty$ in the subsequent inequality.

It remains to control the second term. Let $H_{K_n,n}^{\mathrm{IS}}:=\left(h_{K_n}^{\mathrm{IS}}(X_1),\ldots,h_{K_n}^{\mathrm{IS}}(X_n)\right)^\top=G_{K_n,n}W_{K_n}$. Then
\begin{align*}
    \left(\overline{P}_{K_n,n,\lambda}h_{K_n}^{\mathrm{IS}}\right)(x)
    &=g_{K_n}(x)^\top\widehat{\Gamma}_{K_n,n,\lambda}^{-1}\left( n\inv G_{K_n,n}^\top H_{K_n,n}^{\mathrm{IS}}\right)\nonumber\\
    &=g_{K_n}(x)^\top\widehat{\Gamma}_{K_n,n,\lambda}^{-1}\left( n\inv G_{K_n,n}^\top G_{K_n,n}\right)W_{K_n}\nonumber\\
    &=g_{K_n}(x)^\top\widehat{\Gamma}_{K_n,n,\lambda}^{-1}\widehat{\Gamma}_{K_n,n}W_{K_n}\nonumber\\
    &=g_{K_n}(x)^\top\widehat{\Gamma}_{K_n,n,\lambda}^{-1}\left(\widehat{\Gamma}_{K_n,n,\lambda}-\lambda_{K_n,n}I_{K_n}\right)W_{K_n}\nonumber\\
    &=g_{K_n}(x)^\top W_{K_n}-\lambda_{K_n,n}g_{K_n}(x)^\top\widehat{\Gamma}_{K_n,n,\lambda}^{-1}W_{K_n}.
\end{align*}
Using this expression, one immediately gets that
\begin{align}
\label{eq:pseudo_projection_IS}
\sup_{x\in\mathcal{X}}\left|\left(\overline{P}_{K_n,n,\lambda}h_{K_n}^{\mathrm{IS}}-h_{K_n}^{\mathrm{IS}}\right)(x)\right|&=\sup_{x\in\mathcal{X}}\left|-\lambda_{K_n,n}g_{K_n}(x)^\top\widehat{\Gamma}_{K_n,n,\lambda}^{-1}W_{K_n}\right|\nonumber\\
&\leq \lambda_{K_n,n}\left\|\widehat{\Gamma}_{K_n,n,\lambda}^{-1}W_{K_n}\right\|\sup_{x\in\mathcal{X}}\|g_{K_n}(x)\| \nonumber\\
&\leq\lambda_{K_n,n}\zeta_{K_n}\left\|\widehat{\Gamma}_{K_n,n,\lambda}^{-1}W_{K_n}\right\|,
\end{align}
where we used the Cauchy-Schwarz inequality and Assumption~\ref{assumption:uniform_sieve_regularity}(i). Substituting this bound into \eqref{eq:uniform_bias_decomposition_bound} proves
\eqref{eq:uniform_bias_bound}.

We next bound $R_{K_n,n,\lambda}$ as follows:
\begin{align*}
    R_{K_n,n,\lambda} &\leq\lambda_{K_n,n} \zeta_{K_n} \left\|\widehat{\Gamma}_{K_n,n,\lambda}^{-1} \right\| \|W_{K_n}\| =\zeta_{K_n}\frac{\lambda_{K_n,n}}{s_{K_n,n}^2+\lambda_{K_n,n}}\|W_{K_n}\|\nonumber\\
    &\leq \sqrt{6}M_f\zeta_{K_n}K_n^{-1/2}\frac{\lambda_{K_n,n}}{s_{K_n,n}^2+\lambda_{K_n,n}},
\end{align*}
where in the first equality we used again the singular value decomposition in \eqref{eq:uniform_design_svd} and applied in the second inequality Lemma~\ref{lem:stratified_weight_norms} for the bound of the importance sampling weights constructed in Theorem~\ref{theorem:sharp_is_linf_unified}. This proves \eqref{eq:ridge_shrinkage_spectral}.

Finally, Lemma~\ref{lemma:pseudo-proj_sup_bound} gives $\left\| \overline{P}_{K_n,n,\lambda} \right\|_\infty\leq\zeta_{K_n}\psi_{K_n,n,\lambda}$ and combining it with \eqref{eq:uniform_bias_bound} and \eqref{eq:pseudo_projection_IS} yields \eqref{eq:uniform_bias_bound_crude}, as required.
\end{proof}

\begin{proof}[Proof of Theorem~\ref{theorem:uniform_estimation_bound}]
We recall decomposition in \eqref{eq:sup_error_split} which follows from the application of the triangle inequality, namely,
\begin{equation*}
    \|\widehat{h}_{n,\lambda}-h \|_\infty
    \leq \|\widehat{h}_{n,\lambda}-\widetilde{h}_{n,\lambda}\|_\infty
    + \|\widetilde{h}_{n,\lambda}-h\|_\infty.
\end{equation*}
Applying \eqref{eq:uniform_bias_bound_crude} from Lemma~\ref{lemma:uniform_bias_bound} to the second term proves \eqref{eq:uniform_estimation_bound}.

Finally, we apply Lemma~\ref{lemma:uniform_bound_variance} to the first term and notice that by \eqref{eq:sup_norm_unified} in Theorem~\ref{theorem:sharp_is_linf_unified} it holds, via Markov's inequality, that
\begin{equation*}
    \| h-h_{K_n}^{\mathrm{IS}} \|_\infty = O_P\left( \big( \sqrt d+\sqrt{\log(K_n)} \,\big) K_n^{-1/2-\alpha} \right).
\end{equation*}
Combining these bounds in \eqref{eq:uniform_estimation_bound} and using that by the first statement of Assumption~\ref{assumption:growth_variance}, $\psi_{K_n,n,\lambda}=O_P\left(\, \overline{\psi}_{K_n,n} \right)$, proves
\eqref{eq:uniform_estimation_rate}, as required.
\end{proof}
\begin{proof}[Proof of Corollary~\ref{corollary:uniform_estimation_RWNN}]
    The proof is an immediate application of Lemma~\ref{lemma:RWNN_regular} in the statement of  Theorem~\ref{theorem:uniform_estimation_bound}.
\end{proof}

\subsection{Prediction space and leverage based stability}
\label{app:supnorm_leverage_proofs}

\begin{proof}[Proof of Lemma~\ref{lemma:ridge_smoothing_representation}]
Let $F:=(f(X_1),\ldots,f(X_n))^\top$. By \eqref{eq:ridge_pseudo_projection}, one has that
\begin{equation*}
    \bigl(\overline{P}_{K_n,n,\lambda}f\bigr)(x)
    =
    g_{K_n}(x)^\top\widehat{\Gamma}_{K_n,n,\lambda}^{-1}\left( n\inv G_{K_n,n}^\top F\right)=\sum_{i=1}^np_{i,n,\lambda}(x)f(X_i),
\end{equation*}
which proves \eqref{eq:ridge_smoother_representation}. Hence,
\begin{equation*}
\left|\bigl(\overline{P}_{K_n,n,\lambda}f\bigr)(x)\right|\leq\|f\|_\infty\sum_{i=1}^n|p_{i,n,\lambda}(x)|,
\end{equation*}
and taking the supremum over $x\in\mathcal{X}$ and $\|f\|_\infty\leq1$ together with the first definition in \eqref{eq:def_Lambda_nu} proves \eqref{eq:ridge_lebesgue_bound}.

Next, we write that
\begin{align*}
    \sum_{i=1}^np_{i,n,\lambda}(x)^2
    &=\frac{1}{n^2} \left( g_{K_n}(x)^\top\widehat{\Gamma}_{K_n,n,\lambda}^{-1}G_{K_n,n}^\top G_{K_n,n}\widehat{\Gamma}_{K_n,n,\lambda}^{-1}g_{K_n}(x) \right) \nonumber\\
    &=\frac{1}{n} \left( g_{K_n}(x)^\top\widehat{\Gamma}_{K_n,n,\lambda}^{-1}\widehat{\Gamma}_{K_n,n}\widehat{\Gamma}_{K_n,n,\lambda}^{-1}g_{K_n}(x) \right) ,
\end{align*}
where the first equality follows from definition \eqref{eq:ridge_smoother_weights} and in the second one we used $n^{-1}G_{K_n,n}^\top G_{K_n,n}=\widehat{\Gamma}_{K_n,n}$. Moreover, $\widehat{\Gamma}_{K_n,n}\preceq\widehat{\Gamma}_{K_n,n,\lambda}$ implies
\begin{equation*}
\widehat{\Gamma}_{K_n,n,\lambda}^{-1}\widehat{\Gamma}_{K_n,n}\widehat{\Gamma}_{K_n,n,\lambda}^{-1}\preceq\widehat{\Gamma}_{K_n,n,\lambda}^{-1},
\end{equation*}
which proves the first inequality in \eqref{eq:ridge_variance_leverage_bound}. 

Finally, the Cauchy-Schwarz inequality for the inner product induced by $\widehat{\Gamma}_{K_n,n,\lambda}^{-1}$ together with the definition of $\ell_{n,\lambda}$ in \eqref{eq:def_l_m} gives
\begin{equation*}
|p_{i,n,\lambda}(x)|\leq\frac{1}{n}\left(g_{K_n}(x)^\top\widehat{\Gamma}_{K_n,n,\lambda}^{-1}g_{K_n}(x)\right)^{1/2}\left(g_{K_n}(X_i)^\top\widehat{\Gamma}_{K_n,n,\lambda}^{-1}g_{K_n}(X_i)\right)^{1/2}\leq\frac{\ell_{n,\lambda}}{n}.
\end{equation*}
Taking the supremum over $x\in\mathcal{X}$ and $i=1,\ldots,n$ proves the second inequality in \eqref{eq:ridge_variance_leverage_bound}.
\end{proof}

\begin{proof}[Proof of Lemma~\ref{lemma:uniform_bound_variance_stable}]
Set $\check{h}_{n,\lambda}:=\widehat{h}_{n,\lambda}-\widetilde{h}_{n,\lambda}$ and $e:=(\epsilon_1,\ldots,\epsilon_n)^\top$. By \eqref{eq:ridge_smoother_representation}, 
\begin{equation*} 
    \check{h}_{n,\lambda}(x)=\sum_{i=1}^np_{i,n,\lambda}(x)\epsilon_i. 
\end{equation*} 
We follow again the steps used in the proof of Lemma~\ref{lemma:uniform_bound_variance}.
Indeed, the same argument used to obtain \eqref{eq:mean_value_h_check} gives 
\begin{equation} 
\label{eq:stable_variance_lipschitz} 
    |\check{h}_{n,\lambda}(x_1)-\check{h}_{n,\lambda}(x_2)|\leq L_{K_n}\psi_{K_n,n,\lambda}\widetilde U_n\|x_1-x_2\|, 
\end{equation} 
where $\widetilde U_n:=\|n^{-1/2}U_{m_n,n}^\top e\|$. 
As in \eqref{eq:norm_gamma_G}, conditioning on $X_{1:n}$ and $\theta_{1:K_n}$ gives
\begin{equation*}
\E[\widetilde U_n^2\mid X_{1:n},\theta_{1:K_n}]\leq\overline\sigma^2\frac{m_n}{n},
\end{equation*}
and hence $\widetilde U_n=O_P(\sqrt{m_n/n})$. 
Fix $\eta>0$. By the hypothesis $\ell_{n,\lambda}=O_P(K_n)$, the condition $\psi_{K_n,n,\lambda}=O_P(\overline{\psi}_{K_n,n})$, and the preceding bound on $\widetilde U_n$, there exists $M_\eta>0$ such that, for all sufficiently large $n$, the events
\begin{align*}
\mathcal{A}_{n,\eta}&:=\left\{\psi_{K_n,n,\lambda}\leq M_\eta\overline{\psi}_{K_n,n},\ \ell_{n,\lambda}\leq M_\eta K_n\right\},\\
\mathcal{B}_{n,\eta}&:=\mathcal{A}_{n,\eta}\cap\left\{\widetilde U_n\leq M_\eta\sqrt{\frac{m_n}{n}}\right\}
\end{align*}
satisfy $\Prob(\mathcal{B}_{n,\eta}^{\complement})\leq\eta$. Moreover, $\mathcal{A}_{n,\eta}$ is measurable with respect to $\sigma(X_{1:n},\theta_{1:K_n})$. On $\mathcal{A}_{n,\eta}$, Lemma~\ref{lemma:ridge_smoothing_representation} implies 
\begin{equation} 
\label{eq:stable_variance_weight_bounds} 
\nu_{n,\lambda}\leq M_\eta\frac{K_n}{n}\quad\text{and}\quad \mathfrak m_{n,\lambda}\leq M_\eta\frac{K_n}{n}. 
\end{equation} 

Let $\mathcal{C}_n$ be a minimal $\rho_n$-net of $\mathcal{X}$, where 
\begin{equation*} 
    \rho_n:=\frac{\sqrt{K_n\log(n)}}{4M_\eta^2L_{K_n}\overline{\psi}_{K_n,n}\sqrt{m_n}}. 
\end{equation*} 

Let 
\begin{equation*} 
t_n:=\sqrt{\frac{K_n\log(n)}{n}}. 
\end{equation*} 
For each $x\in\mathcal{X}$, let $x_{j(x),n}$ be a closest element of $\mathcal{C}_n$. It follows from \eqref{eq:stable_variance_lipschitz} that, on $\mathcal{B}_{n,\eta}$,
\begin{equation}
\label{eq:stable_variance_net_oscillation}
\sup_{x\in\mathcal{X}}|\check{h}_{n,\lambda}(x)-\check{h}_{n,\lambda}(x_{j(x),n})|\leq\frac{t_n}{4}.
\end{equation}

Since $\mathcal{X}=[-M,M]^d$, the volumetric covering bound (see, for example, Corollary 4.2.11 in~\cite{vershynin2026high})and \eqref{eq:stable_variance_covering_growth} imply 
\begin{equation} 
\label{eq:stable_variance_net_cardinality} 
\log(|\mathcal{C}_n|)=O(\log(n)). 
\end{equation} 

Define 
\begin{equation*} 
    \tau_n:=\left(\frac{K_n n}{\log(n)}\right)^{\frac{1}{2(1+\delta)}} 
\end{equation*} 
and decompose $\epsilon_i = \epsilon_{i,n}^{(1)} + \epsilon_{i,n}^{(2)}$, where 
\begin{equation*} 
    \epsilon_{i,n}^{(1)} := \epsilon_i\Ind\{|\epsilon_i|\leq\tau_n\}-\E[\epsilon_i\Ind\{|\epsilon_i|\leq\tau_n\}\mid X_i]
    \quad\text{and}\quad 
    \epsilon_{i,n}^{(2)} := \epsilon_i - \epsilon_{i,n}^{(1)} . 
\end{equation*} 

Conditionally on $X_{1:n}$ and $\theta_{1:K_n}$, the random variables $p_{i,n,\lambda}(x_{j,n})\epsilon_{i,n}^{(1)}$, $i=1,\ldots,n$, are independent and centered. On $\mathcal{A}_{n,\eta}$, their conditional variance and maximal absolute value satisfy
\begin{equation*}
    \sum_{i=1}^n \E\left[p_{i,n,\lambda}(x_{j,n})^2(\epsilon_{i,n}^{(1)})^2\mid X_{1:n},\theta_{1:K_n}\right]\leq\overline\sigma^2\nu_{n,\lambda}\leq M_\eta\overline\sigma^2\frac{K_n}{n}
\end{equation*}
and
\begin{equation*}
    \max_{1\leq i\leq n}\left|p_{i,n,\lambda}(x_{j,n})\epsilon_{i,n}^{(1)}\right|\leq 2\tau_n\mathfrak m_{n,\lambda}\leq 2M_\eta\tau_n\frac{K_n}{n}.
\end{equation*}

Next, 
\begin{equation*} 
    (\tau_nt_n)^{2(1+\delta)} = \left(\frac{K_n^{1+2/\delta}\log(n)}{n}\right)^\delta=O(1) 
\end{equation*} 
by \eqref{eq:stable_variance_moment_growth}. 
Therefore, the conditional Bernstein inequality gives, for every sufficiently large constant $C_\eta>0$,
\begin{equation*}
\Prob\left(\left|\sum_{i=1}^np_{i,n,\lambda}(x_{j,n})\epsilon_{i,n}^{(1)}\right|>C_\eta t_n \:\middle\vert\: X_{1:n},\theta_{1:K_n}\right)\Ind_{\mathcal A_{n,\eta}}\leq 2n^{-c_\eta C_\eta}
\end{equation*}
for some constant $c_\eta>0$ independent of $C_\eta$. Combining this inequality with \eqref{eq:stable_variance_net_cardinality} and applying the union bound yields
\begin{equation}
\label{eq:stable_variance_truncated_bound}
\Prob\left(\left\{\max_{x_{j,n}\in\mathcal C_n}\left|\sum_{i=1}^np_{i,n,\lambda}(x_{j,n})\epsilon_{i,n}^{(1)}\right|>C_\eta t_n\right\}\cap\mathcal A_{n,\eta}\right)=o(1).
\end{equation}

It remains to control the remainder. Since $\E[\epsilon_i\mid X_i]=0$,
\begin{equation*}
\epsilon_{i,n}^{(2)}=\epsilon_i\Ind\{|\epsilon_i|>\tau_n\}-\E[\epsilon_i\Ind\{|\epsilon_i|>\tau_n\}\mid X_i].
\end{equation*}
On $\mathcal A_{n,\eta}$, \eqref{eq:stable_variance_weight_bounds} gives
\begin{equation*}
\sup_{x\in\mathcal{X}}\left|\sum_{i=1}^np_{i,n,\lambda}(x)\epsilon_{i,n}^{(2)}\right|\leq\mathfrak m_{n,\lambda}\sum_{i=1}^n|\epsilon_{i,n}^{(2)}|\leq M_\eta\frac{K_n}{n}\sum_{i=1}^n|\epsilon_{i,n}^{(2)}|.
\end{equation*}
Consequently, for every $c>0$,
\begin{equation}
\label{eq:stable_variance_remainder_bound}
\Prob\left(\left\{\sup_{x\in\mathcal{X}}\left|\sum_{i=1}^np_{i,n,\lambda}(x)\epsilon_{i,n}^{(2)}\right|>ct_n\right\}\cap\mathcal A_{n,\eta}\right)\leq\frac{2M_\eta K_n}{ct_n\tau_n^{1+\delta}}\E[|\epsilon_i|^{2+\delta}\Ind\{|\epsilon_i|>\tau_n\}]=o(1).
\end{equation}
By the definitions of $t_n$ and $\tau_n$, 
\begin{equation*} 
t_n\tau_n^{1+\delta}=K_n. 
\end{equation*} 
The preceding probability is therefore $o(1)$ by Assumption~\ref{assumption:higher_moment} and dominated convergence. 
Combining \eqref{eq:stable_variance_net_oscillation} with the decomposition $\epsilon_i=\epsilon_{i,n}^{(1)}+\epsilon_{i,n}^{(2)}$, we obtain, on $\mathcal B_{n,\eta}$,
\begin{equation*}
\|\check h_{n,\lambda}\|_\infty\leq\max_{x_{j,n}\in\mathcal C_n}\left|\sum_{i=1}^np_{i,n,\lambda}(x_{j,n})\epsilon_{i,n}^{(1)}\right|+\sup_{x\in\mathcal X}\left|\sum_{i=1}^np_{i,n,\lambda}(x)\epsilon_{i,n}^{(2)}\right|+\frac{t_n}{4}.
\end{equation*}
Consequently, for every $\eta>0$ and a sufficiently large constant $C_\eta>0$,
\begin{align*}
\Prob&\left(\|\check h_{n,\lambda}\|_\infty>(C_\eta+\frac{5}{4})t_n\right)\nonumber\\&\quad \leq\Prob(\mathcal B_{n,\eta}^{\complement})+\Prob\left(\left\{\max_{x_{j,n}\in\mathcal C_n}\left|\sum_{i=1}^np_{i,n,\lambda}(x_{j,n})\epsilon_{i,n}^{(1)}\right|>C_\eta t_n\right\}\cap\mathcal A_{n,\eta}\right)\\
&\quad \enspace\enspace +\Prob\left(\left\{\sup_{x\in\mathcal X}\left|\sum_{i=1}^np_{i,n,\lambda}(x)\epsilon_{i,n}^{(2)}\right|>t_n\right\}\cap\mathcal A_{n,\eta}\right)\leq \eta+o(1),
\end{align*}
where we used $\mathcal B_{n,\eta}\subseteq\mathcal A_{n,\eta}$, \eqref{eq:stable_variance_truncated_bound}, and the remainder estimate in \eqref{eq:stable_variance_remainder_bound}. Since $\eta>0$ is arbitrary, this establishes $\|\check h_{n,\lambda}\|_\infty=O_P(t_n)$ and proves \eqref{eq:uniform_variance_stable}.
\end{proof}
\begin{proof}[Proof of Lemma~\ref{lemma:ridge_rremainder_bound}]
Since $h_{K_n}^{\mathrm{IS}}(x)=g_{K_n}(x)^\top W_{K_n}$, the definition of $\overline{P}_{K_n,n,\lambda}$ gives 
\begin{equation*} 
\bigl(\overline{P}_{K_n,n,\lambda}h_{K_n}^{\mathrm{IS}}-h_{K_n}^{\mathrm{IS}}\bigr)(x)=-\lambda_{K_n,n}g_{K_n}(x)^\top\widehat{\Gamma}_{K_n,n,\lambda}^{-1}W_{K_n}. 
\end{equation*} 
The Cauchy-Schwarz inequality for the inner product induced by $\widehat{\Gamma}_{K_n,n,\lambda}^{-1}$ yields 
\begin{equation*} 
R_{K_n,n,\lambda}\leq\lambda_{K_n,n}\ell_{n,\lambda}^{1/2}\left(W_{K_n}^\top\widehat{\Gamma}_{K_n,n,\lambda}^{-1}W_{K_n}\right)^{1/2}\leq \lambda_{K_n,n}\ell_{n,\lambda}^{1/2}\left(\lambda_{K_n,n}^{-1}\|W_{K_n}\|^2\right)^{1/2},
\end{equation*} 
where in the second inequality we used that $\widehat{\Gamma}_{K_n,n,\lambda}\succeq\lambda_{K_n,n}I_{K_n}$, and  
which proves \eqref{eq:ridge_remainder_leverage_bound}. The bound \eqref{eq:ridge_remainder_IS_bound} follows from $\|W_{K_n}\|\leq\sqrt{6}M_fK_n^{-1/2}$. Finally, $\ell_{n,\lambda}=O_P(K_n)$ proves \eqref{eq:ridge_remainder_rate}. 
\end{proof}
\begin{proof}[Proof of Theorem~\ref{theorem:uniform_estimation_bound_stable}]
The decomposition in \eqref{eq:sup_error_split} and Lemma~\ref{lemma:uniform_bias_bound} give 
\begin{equation*} 
\|\widehat{h}_{n,\lambda}-h\|_\infty\leq\|\widehat{h}_{n,\lambda}-\widetilde{h}_{n,\lambda}\|_\infty+(1+\|\overline P_{K_n,n,\lambda}\|_\infty)\|h-h_{K_n}^{\mathrm{IS}}\|_\infty+R_{K_n,n,\lambda}. 
\end{equation*} 
Applying \eqref{eq:ridge_lebesgue_bound} and \eqref{eq:ridge_remainder_IS_bound} proves \eqref{eq:uniform_estimation_bound_stable}. The rate in \eqref{eq:uniform_estimation_rate_stable} follows from Lemma~\ref{lemma:uniform_bound_variance_stable}, Assumption~\ref{assumption:supnorm_leverage_stability}, Theorem~\ref{theorem:sharp_is_linf_unified}, Markov's inequality, and \eqref{eq:ridge_remainder_rate}. 
\end{proof}
\begin{proof}[Proof of Corollary~\ref{corollary:uniform_estimation_rate_stable}]
Under \eqref{eq:stable_optimal_Kn}, one has $\log(K_n)\asymp\log(n)$ and 
\begin{equation*} 
\sqrt{\frac{K_n\log(n)}{n}}\asymp\sqrt{\log(K_n)}K_n^{-1/2-\alpha}\asymp n^{-\frac{1+2\alpha}{4(1+\alpha)}}\sqrt{\log(n)}. 
\end{equation*} 
Moreover, \eqref{eq:stable_moment_condition} implies \eqref{eq:stable_variance_moment_growth}, while \eqref{eq:stable_lambda_condition} gives 
\begin{equation*} 
\sqrt{\lambda_{K_n,n}}=O\left(n^{-\frac{1+2\alpha}{4(1+\alpha)}}\sqrt{\log(n)}\right). 
\end{equation*} 
The conclusion follows from \eqref{eq:uniform_estimation_rate_stable}. 
\end{proof}

\clearpage

\section{Theory for generic sieve estimation and inference}
\label{app:generic_sieve_estimation}

This appendix develops complementary results for constrained RWNN sieve estimation under general loss functions. We first recall a baseline $L^2$ risk bound, then combine the approximation results of Section~\ref{sec:approximation_RWNNs} with generic sieve arguments to obtain convergence rates under suitable empirical-process and dependence conditions. Finally, we establish asymptotic normality for smooth functionals using a local perturbation argument. Unlike the ridge-normalized theory of Section~\ref{sec:Ridge_RWNN_least_squares}, these results concern constrained, unregularized sieve estimators, although the approximation bounds developed here also enter the ridge analysis.
\subsection{A baseline $L^2$ risk bound}%
Consider the regression problem \eqref{eq:regression} and the associated RWNN least squares estimator $h^{\theta,\widehat W}_{K,n}$ in \eqref{eq:sieve_RWNN_estimator}. As a baseline, we recall the $L^2$ risk bound of \citet{gononRandomFeatureNeural2023} for the $L$-clipped RWNN least-squares estimator. More explicitly, for a clipping level $L>0$, define the clipped RWNN least-squares estimator by
\begin{equation*}
    \widehat h_{K,n,L}(x):=\min\{\max\{\widehat h_{K,n}(x),-L\},L\},\qquad x\in\mathcal X.
\end{equation*}
Whenever this estimator is used, we assume that $h$ is bounded and $L\geq\|h\|_\infty$. The $L^2$ error guarantees are derived in \citet{gononRandomFeatureNeural2023} under the following assumptions.
\begin{assumption}
\label{assumption:h_conv}
    Function $h : \R^d \to \R$ is {convolutional}, meaning that a $\mathbb{R}^d$-valued random vector $V$ and a function $\Phi \in L^1(\R^d)$ exist such that
    \begin{equation*}
        h(x) = \E_V[ \Phi( x + V ) ] .
    \end{equation*}
\end{assumption}
\begin{assumption}\label{assumption:subgauss_charfun}
    There exists $C_V > 0$ such that 
    \begin{equation*}
        \lvert \E[e^{i \xi ^\top V}] \rvert \leq \exp(-C_V \lVert \xi \rVert^2_2) \quad \textnormal{for all } \xi \in \R^d .
    \end{equation*}
\end{assumption}
\begin{assumption}\label{assumption:pi_A_pi_B_unbounded}
    The weight density $\pi_A$ is a multivariate $t$-distribution with $\eta > 1$ degrees of freedom, $t_\eta(0, \mathbb{I}_d)$. The bias density $\pi_B$ satisfies a {polynomial} lower-tail bound; that is, there exists a polynomial function $p(z)$ such that
        $\inf_{z \in \R} p(z) \pi_B(z) \geq 1$.
\end{assumption}
 The following theorem (\citealt{gononRandomFeatureNeural2023}, Theorem 17) provides an $L^2$ error bound for the RWNN least squares $L$-clipped  estimator $\widehat{h}_{K,n,L}$. Let 	$(X,Y)$ have the same distribution as $(X_1,Y_1)$ but be independent from $\{(Y_i, X_i)\}_{i=1}^n$, $A$, and $B$.

\begin{theorem}[\citealt{gononRandomFeatureNeural2023}, Theorem 17]\label{theorem:L2_regress_rate_gonon}
    Let Assumptions~\ref{assumption:X_domain}, \ref{assumption:h_conv}, \ref{assumption:subgauss_charfun} and \ref{assumption:pi_A_pi_B_unbounded} hold with $C_V > (2^{3/2}\pi)^{-1}$. Let $L>0$ and assume  $\varsigma^2 := \sup_{x \in \mathbb{R}^d} \E\left[ (Y_1 - h(X_1))^2  \vert  X_1 = x \right] < \infty$ and $\sup_{x \in \mathbb{R}^d}|h(x)| \leq L$.
    Then, there exists $q \in \N$ and $C_{app}>0$ such that
    \begin{equation*}
        \E\big[ |  h(X) - \widehat{h}_{K,n,L}(X) |^2 \big]^{1/2}
        \leq C_{app}\left(
        \max(\varsigma, L) \sqrt{\frac{K (\log(n) + 1)}{n}}
        + \norm{\Phi}_{L^1(\R^d)} \frac{(\eta + d)^{q+3}}{\sqrt{K}}\right) .
    \end{equation*}
    The constant $q$ depends only on $\pi_B$ while $C_{app}$ depends on $C_V$, $\eta$, $\pi_B$, $M$, but does not depend on $d$, $n$, or $K$.
\end{theorem}
\begin{remark}
The bound in Theorem~\ref{theorem:L2_regress_rate_gonon} implies the $\mathcal{O}( \sqrt{{K (\log(n) + 1)}/{{n}}} + 1/\sqrt{K} )$ error. Balancing the two terms yields the choice
\begin{equation*}
K \asymp n^{1/2} (\log(n) + 1)^{-1/2},
\end{equation*}
which leads to the convergence rate
\begin{equation*}
    \mathcal{O}\big(n^{-1/4} (\log(n) + 1)^{1/4}\big).
\end{equation*}

Up to logarithmic factors, this rate coincides with the classical Barron approximation rate \citep{barronUniversalApproximationBounds1993,barron1994approximation}. However, it remains slower than the rates typically required for nonparametric regression estimators under smoothness assumptions. In particular, sieve estimators based on fully trainable shallow neural networks can achieve rates $o_P(n^{-1/4})$ under suitable~regularity conditions \citep{Chen1999}. This rate is the upper rate bound needed to obtain standard asymptotic CLT results for the CEF estimators under weak dependence (\cite{shenMethodsSievesPenalization1997}).

Consequently, improving upon the convergence rate in Theorem~\ref{theorem:L2_regress_rate_gonon} is important for obtaining sharper statistical guarantees for RWNN estimators.
\end{remark}

\subsection{Generic sieve estimation and $\beta$-H\"older RWNN rates}%
\label{subsec:L2_Chen_Shen}

We now combine the $L^2$ approximation result under the $\beta$-H\"older source condition with a sieve-estimation argument for RWNN estimators. To derive our asymptotic sieve results, we use the framework of~\citet{shenWong1994}, its extension to weakly dependent data in~\cite{chenSieveExtremumEstimates1998}, and its neural-network specialization in~\citet{Chen1999}. We view $h$ as a function-valued parameter belonging to a class $ \mathcal{H}$, and collect outcomes and regressors in the random variables $Z_i := (Y_i,X_i^\top)^\top \in \mathcal{Z}\subseteq \mathbb{R}^{d+1}$, $i=1,\ldots, n$, where we assume that $X_i\in\mathcal{X}\subset\mathbb{R}^d$ almost surely, with $\mathcal{X}$ compact.

\begin{assumption}[Data]
\label{assumption:c_stationarity}
    The sequence $\{Z_i\}_{i\in\mathbb{N}^+}$ is stationary, with common distribution $P_Z$, and satisfies one of the following conditions:
    \begin{enumerate}
        \item[(i)] $\{Z_i\}$ is i.i.d.\ or $\nu$-dependent for some fixed $\nu\in\mathbb{N}^+$;
        \item[(ii)] $\{Z_i\}$ is $\beta$-mixing with $\beta_{\mathrm{mix}}(j)\leq\beta_0j^{-\zeta}$, $j\geq1$, for some $\beta_0>0$ and $\zeta>2$.
    \end{enumerate}
\end{assumption}

Let $\mathcal{V}$ be a linear space of functions on $\mathcal{X}$ containing $ \mathcal{H}$ and let $\|\cdot\|_\circ$ be a seminorm on $ \mathcal{V}$. Recall from Section~\ref{sec:rwnn_regression} that $ \mathcal{H}_K$ denotes the RWNN class of width $K$. In this subsection, $K_n = K(n)$ may depend on the sample size. Conditional on the sampled $\theta_k=(A_k^\top,B_k)^\top$, $k=1,\ldots,K_n$, we work with the $\ell_2$-constrained RWNN sieve classes
\begin{equation}
\label{eq:constrained_RWNN}
 \mathcal{H}_{K_n}(M_n) := \left\{ x \mapsto W^\top g_{K_n}(x) = \sum_{k=1}^{K_n} W_k \sigma(A_k^\top x+B_k) \middle| \|W\| \le M_n \right\}\subseteq\mathcal{H},
\end{equation}
where
\begin{equation*}
	g_{K_n}(x) =\left(\sigma(A_1^\top x+B_1),\dots,\sigma(A_{K_n}^\top x+B_{K_n})\right)^\top.%
\end{equation*}
All entropy bounds below are understood conditionally on the hidden-layer parameters, while the final $O_P$ statements are with respect to the joint law of the data and the random features.

The generic sieve rate in Theorem~\ref{theorem:generic_pseudometric_rate} and the entropy bound in Lemma~\ref{lemma:rwnn_entropy} hold for any deterministic sequence $M_n>0$ under their respective assumptions. For the explicit $L^2$ regression rate, we take $M_n =C_\alpha M_f\sqrt{\frac{\log(K_n)}{K_n}}$ with $C_\alpha$ as in Lemma~\ref{lem:nn_l2_bound}. This choice is dictated by the constrained approximation result in Corollary~\ref{cor:constrained_approx}, since with this radius, the control-neighbors approximant belongs to $\mathcal{H}_{K_n}(M_n)$ on the high-probability event quantified in Lemma~\ref{lem:nn_l2_bound}. This choice is also used subsequently in Theorem~\ref{theorem:rwnn_generic_rate}, which delivers an explicit convergence rate.

We first state the identification, curvature and empirical-process conditions used in the generic sieve argument. The population identification condition (a) and the local separation condition~(b) are similar to the corresponding conditions~(a) and~(b) in Assumption 3.2 in~\cite{Chen1999}, with the local separation condition~(b) also analogous to  condition~C1 of \cite{shenWong1994}. The local variance~(c), entropy~(d), and concentration~(e) conditions are analogous to conditions C2$'$-C4$'$ for unbounded criteria, respectively. Our specific formulation of the local variance and concentration conditions follows more closely Assumption 3.4 of~\cite{Chen1999}. The entropy condition is stated explicitly, while in~\citet{Chen1999} the corresponding entropy bound is derived from the ANN sieve structure in the proof of their Theorem~3.1. All conditions are adapted to the loss-minimization setting and to uniformity over the realized random features.

\begin{assumption}[Generic loss conditions]
\label{assumption:c_generic_loss}
Let $\rho_\circ(g,h):=\|g-h\|_\circ$, $g,h\in \mathcal{H}$. There exists a constant $\eta_0>0$ such that the following conditions hold on
\begin{equation*}
\mathcal{H}_{K_n}(M_n;\eta_0):=\{g\in\mathcal{H}_{K_n}(M_n)\mid \rho_\circ(g,h) \leq\eta_0\}.
\end{equation*}
The constant $C_1>0$ and exponent $s\in(0,2)$ do not depend on $n$ or on the realization of $\theta_1,\ldots,\theta_{K_n}$. The variance and local concentration constants may depend on $n$ but not  on the realization of $\theta_1,\ldots,\theta_{K_n}$, and are uniformly bounded by a deterministic sequence $\Lambda_n\geq1$, with $\Lambda_n=O((\log(n))^m)$ for some fixed $m<\infty$.
\begin{itemize}
\item[(a)] [Identification] $\E_Z[\ell(Z_i,h)]<\infty$ and
\begin{equation*}
\E_Z[\ell(Z_i,h)]\leq\E_Z[\ell(Z_i,\widetilde{h})]
\end{equation*}
for all $\widetilde{h}\in\mathcal{H}$ with $\E_Z[\ell(Z_i,\widetilde{h})]<\infty$.

\item[(b)] [Local quadratic separation] For all $g\in\mathcal{H}_{K_n}(M_n;\eta_0)$, conditionally on $\theta_1,\ldots,\theta_{K_n}$,
\begin{equation*}
\E_Z[\ell(Z_i,g)-\ell(Z_i,h)]\geq 2C_1\rho_\circ(g,h)^2.
\end{equation*}

\item[(c)] [Local variance] There exists $C_{2,n}\leq\Lambda_n$ such that, conditionally on $\theta_1,\ldots,\theta_{K_n}$, for all sufficiently small $\varepsilon>0$,
\begin{equation*}
\sup_{\{g\in\mathcal{H}_{K_n}(M_n;\eta_0)\mid \rho_\circ(g,h)\leq\varepsilon\}}\mathrm{Var}_Z[\ell(Z_i,g)-\ell(Z_i,h)]\leq C_{2,n}\varepsilon^2.
\end{equation*}

\item[(d)] [Local bracketing entropy] For $\delta>0$, define
\begin{equation*}
    \mathcal{M}_n(\delta):=\{z\mapsto\ell(z,g)-\ell(z,h)\mid g\in\mathcal{H}_{K_n}(M_n;\eta_0),\ \rho_\circ(g,h)\leq\delta\}.
\end{equation*}
Let $\mathcal{N}_{[]}(\varepsilon,\mathcal{F},L^2(P_Z))$ denote the $L^2(P_Z)$ bracketing number of $\mathcal{F}$ (see, for example, Definition 2.1.6 in \cite{vandervaartWeakConvergenceEmpirical2023}).
There exists a deterministic sequence $\delta_n \to 0$ such that $\delta_n \geq 0$ is the smallest value for which it holds
\begin{equation}
\label{eq:delta_def}
    \delta^{-2}\int_{\delta^2}^{\delta}\sqrt{\log\mathcal{N}_{[]}(w,\mathcal{M}_n(\delta),L^2(P_Z))}\,dw\lesssim\sqrt{n},
\end{equation}
for all $\delta \in [\delta_n, \eta_0]$, uniformly over the realization of $\theta_1,\ldots,\theta_{K_n}$.

\item[(e)] [Local concentration] There exist constants $\kappa>0$, a deterministic constant $C_{3,n}\leq\Lambda_n$, and a measurable function $\psi_n:\mathcal{Z}\to[0,\infty)$ such that, conditionally on $\theta_1,\ldots,\theta_{K_n}$, for every $\Delta>0$,
\begin{equation*}
\sup_{\{g\in\mathcal{H}_{K_n}(M_n;\eta_0)\mid \rho_\circ(g,h)\leq\Delta\}}|\ell(Z_i,g)-\ell(Z_i,h)|\leq\Delta^s\psi_n(Z_i),
\end{equation*}
with
\begin{equation*}
\E_Z[|\psi_n(Z_i)|^\kappa]\leq C_{3,n},
\end{equation*}
where the moment exponent satisfies
\begin{equation*}
\begin{cases}
\kappa>2, & \text{under Assumption~\ref{assumption:c_stationarity}~{(i)},}\\
\kappa>4, & \text{with $\zeta=\kappa-2$ under Assumption~\ref{assumption:c_stationarity}~{(ii)}.}
\end{cases}
\end{equation*}
\end{itemize}
\end{assumption}

Define the empirical risk for the regression problem, for $\widetilde{h}\in\mathcal{H}$, by
\begin{equation*}
	L_n(\widetilde{h}) := \frac{1}{n}\sum_{i=1}^n \ell(Z_i,\widetilde{h}),
\end{equation*}
where $\ell: \mathcal{Z}\times \mathcal{H} \rightarrow [0,\infty)$ is a loss function. 
Additionally, for $g\in\mathcal{H}_{K_n}(M_n)$, define 
\begin{equation}
\label{eq:q_def}
    q(z,g,h):=\ell(z,g)-\ell(z,h)
\end{equation}
and the corresponding population and empirical risk differences as
\begin{equation}
\label{eq:Q_def}
    Q(g,h):=\E_Z[q(Z_i,g,h)] \enspace \text{and}\enspace Q_n(g,h):=\frac{1}{n}\sum_{i=1}^nq(Z_i,g,h)=L_n(g)-L_n(h),
\end{equation}
respectively.

As in the sieve-extremum formulations of \cite{shenWong1994}, \cite{chenSieveExtremumEstimates1998}, and the ANN specialization of \cite{Chen1999}, we allow for approximate empirical minimization.

\begin{assumption}[Near-minimizer]
\label{assumption:sieve_extremum_estimator}
There exists a measurable sequence $\widehat{h}_n\equiv h_{K_n}^{\theta,\widehat{W}}\in\mathcal{H}_{K_n}(M_n)$ such that
\begin{equation*}
    L_n(\widehat{h}_n)\leq\inf_{g\in\mathcal{H}_{K_n}(M_n)}L_n(g)+R_n, \quad \text{where} \quad R_n=O_P(\varepsilon_n^2) \quad \text{and} \quad \varepsilon_n\to0 \quad \text{as} \quad n\rightarrow \infty.
\end{equation*}
\end{assumption}

We now introduce a best sieve approximation of $h$ with respect to the pseudo-metric. Assume that the infimum is attained by a measurable selection by $\pi_n h\in\mathcal{H}_{K_n}(M_n) $, and define it by 
\begin{equation}
\label{eq:sieve_proj}
	\rho_\circ(\pi_n h,h):=\inf_{g\in\mathcal{H}_{K_n}(M_n)} \|h-g\|_\circ.
\end{equation}
Define the projection loss discrepancy by
\begin{equation*}
    \chi(h,\pi_n h):=\E_Z[q(Z_i,\pi_n h,h)]=\E_Z[\ell(Z_i,\pi_n h)-\ell(Z_i,h)].
\end{equation*}

Assumption~\ref{assumption:c_generic_loss}(b)-(e) is imposed only on the local class $\mathcal{H}_{K_n}(M_n;\eta_0)$. In the subsequent proof of our result in Theorem~\ref{theorem:generic_pseudometric_rate}, these bounds can be applied to $\widehat{h}_n$ provided that $\widehat{h}_n\in\mathcal{H}_{K_n}(M_n;\eta_0)$ with probability tending to one. 
The uniformity of the entropy integral in Assumption~\ref{assumption:c_generic_loss}(d) posed for $\delta_n$ allows us to derive the shell tail bound of Lemma~\ref{lem:chen_shen_maximal}(ii) below, and is stronger than imposing that \eqref{eq:delta_def} hold only for $\delta = \delta_n$. Comparable uniform conditions are also imposed, e.g., in \cite{shenWong1994}, where an explicit rate for the bracketing entropy is assumed (cf. their Assumption~C3'). In fact, when considering RWNNs, we show in Lemma~\ref{lemma:loss_entropy_transfer} that such uniformity is not restrictive since the entropy of local loss difference classes does not depend explicitly on the chosen radius $\delta$.
The following lemma provides the localization result under a uniform law of large numbers, global well-separatedness of the population criterion away from $h$, and under vanishing population excess risk of the sieve projection.

\begin{lemma}[Preliminary localization]
\label{lemma:preliminary_localization}
    Let Assumption~\ref{assumption:sieve_extremum_estimator} hold. Assume that the following uniform law of large numbers holds:
    \begin{equation}
    \label{eq:ULLN}
        \sup_{g\in\mathcal{H}_{K_n}(M_n)}|Q_n(g,h)-Q(g,h)|=o_P(1).
    \end{equation}
    Additionally, assume global well-separation condition: for every $\eta>0$ there exists $\gamma(\eta)>0$, independent of $n$ and of the realization of $\theta_1,\ldots,\theta_{K_n}$, such that
    \begin{equation}
    \label{eq:well_separation}
        \inf_{\{g\in\mathcal{H}_{K_n}(M_n)\mid \rho_\circ(g,h)\geq\eta\}}Q(g,h)\geq\gamma(\eta).
    \end{equation}
    Finally, assume that
    \begin{equation}
    \label{eq:approx_of_proj}
        Q(\pi_nh,h)=\chi(h,\pi_nh)=o_P(1).
    \end{equation}
    Then, for every fixed $\eta_0>0$,
    \begin{equation*}
        \mathbb{P}(\rho_\circ(\widehat{h}_n,h)>\eta_0)\to0.
    \end{equation*}
\end{lemma}

\begin{proof}[Proof of Lemma~\ref{lemma:preliminary_localization}]
By Assumption~\ref{assumption:sieve_extremum_estimator},
\begin{equation*}
    Q_n(\widehat{h}_n,h)\leq Q_n(\pi_nh,h)+R_n,
\end{equation*}
where $R_n=O_P(\varepsilon_n^2)=o_P(1)$. Since $\pi_nh\in\mathcal{H}_{K_n}(M_n)$, \eqref{eq:ULLN} gives
\begin{equation*}
    Q_n(\pi_nh,h)=Q(\pi_nh,h)+o_P(1)=o_P(1),
\end{equation*}
where in the last equality we used hypothesis \eqref{eq:approx_of_proj}.
Hence, for some $U_n=o_P(1)$,
\begin{equation*}
    Q_n(\widehat{h}_n,h)\leq U_n.
\end{equation*}
Since $\widehat{h}_n\in\mathcal{H}_{K_n}(M_n)$, the uniform law of large numbers \eqref{eq:ULLN} implies
\begin{equation*}
    Q(\widehat{h}_n,h)\leq U_n+\sup_{g\in\mathcal{H}_{K_n}(M_n)}|Q_n(g,h)-Q(g,h)|=:V_n,
\end{equation*}
where $V_n=o_P(1)$.

On the event $\{\rho_\circ(\widehat{h}_n,h)>\eta_0\}$, \eqref{eq:well_separation} gives $Q(\widehat{h}_n,h)\geq\gamma(\eta_0)$. Therefore
\begin{equation*}
\mathbb{P}(\rho_\circ(\widehat{h}_n,h)>\eta_0)\leq\mathbb{P}(V_n\geq\gamma(\eta_0))\to0,
\end{equation*}
since $\gamma(\eta_0)>0$ is fixed.
\end{proof}
The following lemma isolates the two empirical process estimates corresponding to those used in the proof of Theorem~1 of~\citet{chenSieveExtremumEstimates1998}, and hence also the sieve-rate argument in the proof of Theorem~3.1 of~\citet{Chen1999}. Part (i) controls the empirical fluctuation at the sieve approximation $\pi_n h$, corresponding to the single point term treated separately in that proof. Part (ii) is the shell-restricted, two-sided version of the large deviation inequality in Theorem~3 in~\citet{chenSieveExtremumEstimates1998}. All probabilities and stochastic order statements are understood conditionally on the realized hidden-layer parameters $\theta_1,\ldots,\theta_{K_n}$, with the corresponding bounds and constants uniform over their realization.

\begin{lemma}[Local modulus and shell tail maximal inequality]
\label{lem:chen_shen_maximal}
    Define
    \begin{equation*}
    \mu_n(g,h):=(Q_n-Q)(g,h)\quad \text{and} \quad a_n:=\max\{\delta_n,\rho_\circ(\pi_nh,h)\}.
    \end{equation*}
    Let Assumptions~\ref{assumption:c_stationarity} and~\ref{assumption:c_generic_loss}~{(c)--(e)} hold on $\mathcal{H}_{K_n}(M_n;\eta_0)$, and let $\Lambda_n\geq1$ be as in Assumption~\ref{assumption:c_generic_loss} and chosen sufficiently large to dominate $C_{2,n}$, $C_{3,n}$, and the fixed algebraic combinations of these constants appearing below. Let  $a_n\to0$,
    \begin{equation}
    \label{eq:CS_growth_condition}
    {na_n^2}/{\Lambda_n}\to\infty,\quad \text{and}\quad \Lambda_n n^{-\kappa/2}a_n^{-(2-s)\kappa}\to0.
    \end{equation}
    Then the following hold:
    \begin{enumerate}
    \item[(i)] %
    \begin{equation}
    \label{eq:CS_near_ball}
    |\mu_n(\pi_nh,h)|=O_P(a_n^2).
    \end{equation}
    
     \item[(ii)] %
    For every fixed $c>0$, there exist constants $A_0,A_1,A_2>0$, independent of the realization of $\theta_1,\ldots,\theta_{K_n}$, such that, for all $x\geq1$ and all sufficiently large $n$,
    \begin{align}
    \label{eq:CS_maximal_present}
    & \mathbb{P}\left(\sup_{\{g\in\mathcal{H}_{K_n}(M_n;\eta_0) \,\mid\, xa_n<\rho_\circ(g,h)\leq2xa_n\}}|\mu_n(g,h)|\geq c(xa_n)^2\right) \\
    & \qquad\qquad\qquad\qquad \leq \, A_0 e^{-A_1nx^{2-s}a_n^2/\Lambda_n} + A_2 \Lambda_n n^{-\kappa/2}a_n^{-(2-s)\kappa}x^{-(2-s)\kappa}. \nonumber
    \end{align}
    
    \item[(iii)] %
    For every sequence $b_n\geq a_n$ and every $C>0$, define
    \begin{align*}
    p_{j,n}(C):=&A_0\exp\left\{-A_1\frac{na_n^2}{\Lambda_n}\left(\frac{2^jCb_n}{a_n}\right)^{2-s}\right\}\\&+A_2\Lambda_n n^{-\kappa/2}a_n^{-(2-s)\kappa}\left(\frac{2^jCb_n}{a_n}\right)^{-(2-s)\kappa}.
    \end{align*}
    Then
    \begin{equation}
    \label{eq:CS_shell_summability}
    \lim_{C\to\infty}\limsup_{n\to\infty}\sum_{j\geq0}p_{j,n}(C)=0.
    \end{equation}
    \end{enumerate}
\end{lemma}

\begin{proof}[Proof of Lemma~\ref{lem:chen_shen_maximal}]
For part~{(i)}, we proceed as in Theorem~1 of \citet{chenSieveExtremumEstimates1998}. Since $a_n\to0$, and   $\rho_\circ(\pi_nh,h)\leq a_n$, Assumption~\ref{assumption:c_generic_loss}(e) yields that $|q(Z_i,\pi_nh,h)|\leq a_n^s\psi_n(Z_i)$. Consequently, $\E_Z\big[|q(Z_i,\pi_nh,h)|^\kappa\big]\leq a_n^{s\kappa}\E_Z\big[|\psi_n(Z_i)|^\kappa\big]\leq C_{3,n}a_n^{s\kappa}$. Recall that, by definition, $\mu_n(\pi_nh,h)=\frac{1}{n}\sum_{i=1}^n \left\{ q(Z_i,\pi_nh,h)-\E_Z[q(Z_i,\pi_nh,h)]\right\}$. Applying the same partial-sum moment inequality as in the bound for $P_2$ in the proof of Theorem~1 of \citet{chenSieveExtremumEstimates1998}, there exists a constant $C_{\mathrm{mom}}>0$, independent of $n$ and of the realization of the hidden-layer parameters, such that, for every $u>0$,
\begin{align*}
    \mathbb{P}\left(|\mu_n(\pi_nh,h)|>u a_n^2\right)&\leq 2^\kappa C_{\mathrm{mom}} u^{-\kappa}C_{3,n}n^{-\kappa/2}a_n^{-(2-s)\kappa}\\&\leq 2^\kappa C_{\mathrm{mom}} u^{-\kappa}\Lambda_n n^{-\kappa/2}a_n^{-(2-s)\kappa}\to0
\end{align*}
where we used the second condition in \eqref{eq:CS_growth_condition} and which proves \eqref{eq:CS_near_ball}. %

In part~{(ii)}, under Assumption~\ref{assumption:c_stationarity}~(i) we follow the i.i.d. argument of \citet{shenWong1994}, as used in the proof of Theorem~3.1 of \citet{Chen1999}, while  under Assumption~\ref{assumption:c_stationarity}~(ii), we follow the proof of Theorem~3 of \citet{chenSieveExtremumEstimates1998},  applied separately  to $q(z,g,h)=\ell(z,g)-\ell(z,h)$ and to $-q(z,g,h)$. Restricting the supremum to the shell and using the union bound gives 
\begin{align*}
    & \mathbb{P}\left(\sup_{\{g\in\mathcal{H}_{K_n}(M_n;\eta_0) \,\mid\, xa_n<\rho_\circ(g,h)\leq2xa_n\}}|\mu_n(g,h)|\geq c(xa_n)^2\right) \\
    & \qquad\qquad\qquad\qquad \leq \, 14 e^{-C_n nx^{2-s}a_n^2} + D_n  n^{-\kappa/2}a_n^{-(2-s)\kappa}x^{-(2-s)\kappa}, \nonumber
\end{align*}
where $C_n$ and $D_n$ are deterministic and uniform over the realization of the hidden-layer parameters, and $C_n^{-1}$ and $D_n$ depend on $C_{2,n}$ and $C_{3,n}$ in Assumption~\ref{assumption:c_generic_loss}(c) and (e), via their fixed powers. Hence, by the choice of $\Lambda_n$, there exist constants $A_1,A_2>0$, independent of $n$ and of the realization of the hidden-layer parameters, such that $C_n\geq \frac{A_1}{\Lambda_n}$ and $D_n\leq A_2\Lambda_n$. Substituting these in the expression above and taking $A_0=14
$ yields \eqref{eq:CS_maximal_present} as required.

For part~{(iii)}, the polynomial term is summable since $b_n\geq a_n$ gives $2^jCb_n/a_n\geq2^jC$ and $(2-s)\kappa>0$. Indeed,
\begin{align}
\label{eq:second_term}
\sum_{j\geq0}A_2\Lambda_n n^{-\kappa/2}a_n^{-(2-s)\kappa}\left(\frac{2^jCb_n}{a_n}\right)^{-(2-s)\kappa}&=A_2\Lambda_n n^{-\kappa/2}(Cb_n)^{-(2-s)\kappa}\sum_{j\geq0}2^{-j(2-s)\kappa}\nonumber \\
&\leq A_3 A_2\Lambda_n C^{-(2-s)\kappa} n^{-\kappa/2}a_n^{-(2-s)\kappa},
\end{align}
with $A_3:=\sum_{j\geq0}2^{-j(2-s)\kappa}<\infty$, where the last inequality follows from  $b_n\geq a_n$. The right hand side of this inequality  tends to zero under \eqref{eq:CS_growth_condition}. For the exponential term,
\begin{align*}
\sum_{j\geq0}A_0\exp\left\{-A_1\frac{na_n^2}{\Lambda_n}\left(\frac{2^jCb_n}{a_n}\right)^{2-s}\right\}&\leq A_0\sum_{j\geq0}\exp\left\{-A_1\frac{na_n^2}{\Lambda_n}C^{2-s}2^{j(2-s)}\right\}\\
&\leq A_0 \frac{\exp\{-A_1C^{2-s}na_n^2/\Lambda_n\}} {1-\exp\left\{-A_1C^{2-s}(2^{2-s}-1)na_n^2/\Lambda_n\right\}},
\end{align*}
where in the last inequality we used $2-s>0$ and $2^{j(2-s)}\geq 1+j(2^{2-s}-1)$. This expression  tends to zero for every fixed $C>0$ by $na_n^2/\Lambda_n\to\infty$, which  together with \eqref{eq:second_term} proves \eqref{eq:CS_shell_summability}.
\end{proof}

The preliminary localization result in Lemma~\ref{lemma:preliminary_localization} reduces the rate analysis to the local neighborhood  $\mathcal{H}_{K_n}(M_n;\eta_0)$, while Lemma~\ref{lem:chen_shen_maximal} provides the result which controls the empirical deviations on dyadic shells within this neighborhood. These results yield a version of the generic sieve argument of \citet{chenSieveExtremumEstimates1998}, as extended to approximate sieve extremum estimators by \cite{Chen1999}. Here we keep the approximation error and the corresponding population criterion discrepancy separate, which will be useful for the RWNN sieves considered below.

\begin{theorem}[Generic sieve convergence rate]
\label{theorem:generic_pseudometric_rate}
    Suppose Assumption~\ref{assumption:c_generic_loss}{(b)-(e)}, Assumption~\ref{assumption:sieve_extremum_estimator}, and the conditions of Lemmas~\ref{lemma:preliminary_localization} and~\ref{lem:chen_shen_maximal} hold. Then
    \begin{equation}
    \label{eq:norm_sieve_projection}
    \|h-\widehat{h}_n\|_\circ=O_P\left(\max\left\{\delta_n,\varepsilon_n,\|h-\pi_nh\|_\circ,[\chi(h,\pi_nh)]^{1/2}\right\}\right).
    \end{equation}
\end{theorem}

\begin{proof}[Proof of Theorem~\ref{theorem:generic_pseudometric_rate}]
Fix a realization of $\theta_1,\ldots,\theta_{K_n}$. Conditional on this realization, the sieve $\mathcal{H}_{K_n}(M_n)$ is deterministic. The constants and deterministic sequences in the assumptions are uniform over the realization of $\theta_1,\ldots,\theta_{K_n}$. Hence the conditional bound obtained below also holds under the joint law of the data and random features.

Define
\begin{equation}
\label{eq:appendix_rn_def}
r_n:=\max\{\delta_n,\varepsilon_n,\rho_\circ(\pi_nh,h),[\chi(h,\pi_nh)]^{1/2}\},
\end{equation}
where $\rho_\circ(g,h):=\|g-h\|_\circ$ as in Assumption~\ref{assumption:c_generic_loss} and recall the definitions introduced in \eqref{eq:q_def} and \eqref{eq:Q_def}.
Then
\begin{equation}
\label{eq:proj_bound}
\rho_\circ(\pi_nh,h)\leq r_n,\qquad Q(\pi_nh,h)=\chi(h,\pi_nh)\leq r_n^2.
\end{equation}
It suffices to prove
\begin{equation*}
\lim_{C\to\infty}\limsup_{n\to\infty}\mathbb{P}(\rho_\circ(\widehat{h}_n,h)>Cr_n)=0.
\end{equation*}

By Assumption~\ref{assumption:sieve_extremum_estimator},
\begin{equation}
\label{eq:near_min_q}
    Q_n(\widehat{h}_n,h)\leq Q_n(\pi_nh,h)+R_n,
\end{equation}
where $R_n=O_P(\varepsilon_n^2)=O_P(r_n^2)$. Set
\begin{equation*}
    a_n:=\max\{\delta_n,\rho_\circ(\pi_nh,h)\}.
\end{equation*}
Then $a_n\leq r_n$. 
Since $\pi_nh\in\mathcal{H}_{K_n}(M_n)$ and $\rho_\circ(\pi_nh,h)\leq a_n\to0$, we have $\pi_nh\in\mathcal{H}_{K_n}(M_n;\eta_0)$ for all sufficiently large $n$. Hence, \eqref{eq:CS_near_ball} together with $a_n\leq r_n$ gives
\begin{equation*}
    |Q_n(\pi_nh,h)-Q(\pi_nh,h)|=O_P(a_n^2)=O_P(r_n^2).
\end{equation*}
Combining this with $R_n=O_P(r_n^2)$, for every $\eta>0$ there exists $C_\eta>0$ such that, for all sufficiently large $n$,
\begin{equation}
\label{eq:good_event_prob}
\mathbb{P}(A_{n,C_\eta}^c)\leq\eta,
\end{equation}
where
\begin{equation*}
A_{n,C}:=\left\{R_n+|Q_n(\pi_nh,h)-Q(\pi_nh,h)|\leq\frac{C_1}{4}C^2r_n^2\right\}.
\end{equation*}
Since $A_{n,C}$ is increasing in $C$, the same bound holds for any $C\geq C_\eta$.

Fix $C>0$ and define
\begin{equation*}
    \mathcal{S}_{j,n}^{\eta_0}:=\{g\in\mathcal{H}_{K_n}(M_n; \eta_0)\mid 2^jCr_n<\rho_\circ(g,h)\leq2^{j+1}Cr_n\}, \quad j\geq 0,
\end{equation*}
where we recall that $\mathcal{H}_{K_n}(M_n; \eta_0)=\{g\in \mathcal{H}_{K_n}(M_n)\mid \rho_\circ(g,h) \leq \eta_0\}$.
\begin{equation*}
    J_n(C):=\max\{j\geq0\mid 2^{j}Cr_n < \eta_0\} %
\end{equation*}
so that
\begin{equation*}
    \{C r_n < \rho_\circ(\widehat{h}_n, h)\leq \eta_0\}\subseteq \bigcup_{j=0}^{J_n(C)} \{\widehat{h}_n \in \mathcal{S}_{j,n}^{\eta_0}  \} ,
\end{equation*}
with the convention that the union over $j=0,\ldots, J_n(C)$ is empty if the indexing set is empty.
If the set defining $J_n(C)$ is empty, then $C r_n \geq \eta_0$ and $\{C r_n < \rho_\circ(\widehat{h}_n, h)\leq \eta_0\} = \varnothing$, meaning that the corresponding probability is zero.

Next, $0\leq j\leq J_n(C)$, on $\{\widehat{h}_n\in\mathcal{S}_{j,n}^{\eta_0}\}$, Assumption~\ref{assumption:c_generic_loss}(b) gives
\begin{equation}
\label{eq:lower_bound_Q}
Q(\widehat{h}_n,h)\geq2C_1(2^jCr_n)^2.
\end{equation}
On the other hand, by \eqref{eq:near_min_q},
\begin{equation}
\label{eq:Q_n_deriv}
Q(\widehat{h}_n,h)\leq Q(\pi_nh,h)+|Q_n(\pi_nh,h)-Q(\pi_nh,h)|+R_n+|Q_n(\widehat{h}_n,h)-Q(\widehat{h}_n,h)|.
\end{equation}
Using \eqref{eq:proj_bound}, \eqref{eq:lower_bound_Q}, and \eqref{eq:Q_n_deriv}, on $A_{n,C}\cap\{\widehat{h}_n\in\mathcal{S}_{j,n}^{\eta_0}\}$ we have
\begin{equation}
\label{eq:C_bound}
2C_1(2^jCr_n)^2\leq r_n^2+\frac{C_1}{4}C^2r_n^2+|Q_n(\widehat{h}_n,h)-Q(\widehat{h}_n,h)|.
\end{equation}
Choose $C \geq \max\left\{1,2/\sqrt{C_1}\right\}$. With this choice 
\begin{equation*}
    r_n^2+\frac{C_1}{4}C^2r_n^2\le \frac{C_1}{2}C^2r_n^2 \le \frac{C_1}{2}(2^jCr_n)^2.
\end{equation*}
Hence, on $A_{n,C}\cap\{\widehat{h}_n\in\mathcal{S}_{j,n}^{\eta_0}\}$, using this bound in \eqref{eq:C_bound} and rearranging the terms after substitution, one obtains the following expression:
\begin{equation*}
    |Q_n(\widehat{h}_n,h)-Q(\widehat{h}_n,h)| \geq \frac{3C_1}{2}(2^jCr_n)^2.
\end{equation*}
In particular, taking any fixed constant $c_0 \in (0, 3C_1/2]$ (for example, $c_0=C_1$) gives
\begin{equation*}
    A_{n,C}\cap\{\widehat{h}_n\in\mathcal{S}_{j,n}^{\eta_0}\}\subseteq\left\{\sup_{g\in\mathcal{S}_{j,n}^{\eta_0}}|(Q_n-Q)(g,h)|\geq c_0(2^jCr_n)^2\right\}.
\end{equation*}

Set
\begin{equation*}
x_{j,n}:=\frac{2^jCr_n}{a_n}\geq1,
\end{equation*}
so that, by the definition of $\mathcal{S}_{j,n}^{\eta_0}$,
\begin{equation*}
\mathcal{S}_{j,n}^{\eta_0}=\{g\in\mathcal{H}_{K_n}(M_n;\eta_0)\mid x_{j,n}a_n<\rho_\circ(g,h)\leq2x_{j,n}a_n\}
\end{equation*}
is exactly the shell appearing in \eqref{eq:CS_maximal_present} at $x=x_{j,n}$. Applying the shell tail bound \eqref{eq:CS_maximal_present} with $x=x_{j,n}$ and $c=c_0$ gives, for $0\leq j\leq J_n(C)$,
\begin{equation*}
    \mathbb{P}(A_{n,C}\cap\{\widehat{h}_n\in\mathcal{S}_{j,n}^{\eta_0}\})\leq p_{j,n}(C),
\end{equation*}
where $p_{j,n}(C)$ is as in Lemma~\ref{lem:chen_shen_maximal}~{(iii)} with $b_n=r_n$. Therefore, by \eqref{eq:CS_shell_summability},
\begin{equation*}
    \lim_{C\to\infty}\limsup_{n\to\infty}\sum_{j=0}^{J_n(C)}p_{j,n}(C)=0.
\end{equation*}
Hence
\begin{align*}
\mathbb{P}(Cr_n<\rho_\circ(\widehat{h}_n,h)\leq\eta_0)&\leq\mathbb{P}\left(\bigcup_{j=0}^{J_n(C)}\{\widehat{h}_n\in\mathcal{S}_{j,n}^{\eta_0}\}\right)\nonumber\\
&\leq\mathbb{P}(A_{n,C}^c)+\sum_{j=0}^{J_n(C)}\mathbb{P}(A_{n,C}\cap\{\widehat{h}_n\in\mathcal{S}_{j,n}^{\eta_0}\})\leq\mathbb{P}(A_{n,C}^c)+\sum_{j=0}^{J_n(C)}p_{j,n}(C).
\end{align*}
Taking $\limsup_{n\to\infty}$, then $C\to\infty$, and finally using \eqref{eq:good_event_prob} with $\eta\downarrow0$, gives
\begin{equation}
\label{eq:local_shell_rate}
\lim_{C\to\infty}\limsup_{n\to\infty}\mathbb{P}(Cr_n<\rho_\circ(\widehat{h}_n,h)\leq\eta_0)=0.
\end{equation}
By Lemma~\ref{lemma:preliminary_localization},
\begin{equation}
\label{eq:prelim_localization_use}
\mathbb{P}(\rho_\circ(\widehat{h}_n,h)>\eta_0)\to0.
\end{equation}
Combining \eqref{eq:local_shell_rate} and \eqref{eq:prelim_localization_use}, we get
\begin{equation*}
\lim_{C\to\infty}\limsup_{n\to\infty}\mathbb{P}(\rho_\circ(\widehat{h}_n,h)>Cr_n)=0.
\end{equation*}
Therefore $\rho_\circ(\widehat{h}_n,h)=O_P(r_n)$. Since $\rho_\circ(g,h)=\|g-h\|_\circ$, the definition of $r_n$ in \eqref{eq:appendix_rn_def} gives \eqref{eq:norm_sieve_projection}, which completes the proof.
\end{proof}

We note that the result in Theorem~\ref{theorem:generic_pseudometric_rate} does not require $K_n<n$. Any restriction on the relative growth of $K_n$ with respect to $n$ enters through the verification of its assumptions for the particular RWNN sieve considered in our setting, in particular through the entropy and stochastic equicontinuity conditions.

We now verify the RWNN-specific ingredients entering Theorem~\ref{theorem:generic_pseudometric_rate}. More specifically, in Lemma~\ref{lemma:rwnn_entropy}, we provide the bracketing entropy of $\mathcal{H}_{K_n}(M_n)$, which determines $\delta_n$, Corollary~\ref{cor:constrained_approx} quantifies the sieve approximation error $\|h-\pi_n h\|_\circ$, and we bound the projection loss discrepancy $\chi(h,\pi_n h)$.

\begin{lemma}[Entropy of RWNN sieve with $\ell_2$-bounded readout weights]
\label{lemma:rwnn_entropy}
    Let Assumption~\ref{assumption:X_domain} hold, assume that $\Pi\subset\mathbb{R}^{d+1}$ is compact and define $M_\sigma := \sup_{x\in\mathcal{X}} \sup_{\theta=(A^\top,B)^\top\in\Pi} |\sigma(A^\top x+B)|<\infty$. For $M_n>0$, let the constrained RWNN class $\mathcal{H}_{K_n}(M_n)$ be as in \eqref{eq:constrained_RWNN}. Let  $\theta_k=(A_k^\top,B_k)^\top\in\Pi$, $k=1,\dots,K_n$. Then, for every realization of $\theta_1,\dots,\theta_{K_n}$ and every $r>0$,
    \begin{equation}
    \log \mathcal{N}_{[]} \left(r, \mathcal{H}_{K_n}(M_n),\|\cdot\|_{L^2(X)}\right)\le K_n \log \left(\frac{4 M_n M_\sigma \sqrt{K_n}}{r}+1\right).
    \label{eq:rwnn_entropy_bound}
    \end{equation}
    Consequently, for every $\delta\in(0,1)$,
    \begin{equation}
    \label{eq:rwnn_entropy_integral_bound}
    \delta^{-2} \int_{\delta^2}^{\delta} \sqrt{ \log \mathcal{N}_{[]} \left( r,\mathcal{H}_{K_n}(M_n),\|\cdot\|_{L^2(X)} \right) } dr \lesssim \delta^{-1} \sqrt{ K_n \log\left( \frac{1+4M_nM_\sigma\sqrt{K_n}}{\delta^2} \right)}.
    \end{equation}
    Let
    \begin{equation}
    \label{eq:delta_l2_def}
    \delta_{n,2} := \inf\left\{\delta>0\mid \delta^{-2} \int_{\delta^2}^{\delta} \sqrt{ \log \mathcal{N}_{[]} \left( r,\mathcal{H}_{K_n}(M_n),\|\cdot\|_{L^2(X)} \right) }\,dr\lesssim\sqrt{n}\right\}.
    \end{equation}
    If
    \begin{equation*}
    M_n\asymp \sqrt{\frac{\log(K_n)}{K_n}},\qquad\log(K_n)\asymp \log(n),\qquad K_n\log(K_n)=o(n),
    \end{equation*}
    then
    \begin{equation*}
    \delta_{n,2}
    \lesssim
    \sqrt{\frac{K_n\log(K_n)}{n}}.
    \end{equation*}
\end{lemma}

\begin{proof}[Proof of Lemma~\ref{lemma:rwnn_entropy}]
Fix a realization of $\theta_1,\dots,\theta_{K_n}\in\Pi$. Conditional on this realization, $\mathcal{H}_{K_n}(M_n)$ is a $K_n$-dimensional parametrized  class indexed by $W\in B_2(M_n):=\{W\in\mathbb{R}^{K_n}:\|W\|\le M_n\}$.

For $f_W(x):=W^\top g_{K_n}(x)$ and any $x\in\mathcal{X}$, the Cauchy-Schwarz inequality and $|\sigma(A_k^\top x+B_k)|\le M_\sigma$ give
\begin{equation*}
    |f_W(x)| \le \|W\| \|g_{K_n}(x)\| \le  \|W\| \sqrt{K_n} M_\sigma.
\end{equation*}
Hence, for any $W,\widetilde{W}\in B_2(M_n)$,
\begin{equation}
    \sup_{x\in\mathcal{X}}|f_W(x)-f_{\widetilde{W}}(x)|\le \|W-\widetilde{W}\|\sqrt{K_n}\,M_\sigma.
\label{eq:rwnn_supnorm_lipschitz}
\end{equation}
Let $\eta>0$, and let $\mathcal{W}_\eta$ be an $\eta$-net of $B_2(M_n)$ under $\|\cdot\|$. By the standard volumetric bound (see, for example, Corollary 4.2.11 in~\cite{vershynin2026high}),
\begin{equation*}
    |\mathcal{W}_\eta|\le\left(\frac{2M_n}{\eta}+1\right)^{K_n}.
\end{equation*}
For every $W\in B_2(M_n)$ there exists $\bar W\in\mathcal{W}_\eta$ such that $\|W-\bar W\|\le \eta$, so by 
\eqref{eq:rwnn_supnorm_lipschitz}, 
$\sup_{x\in\mathcal{X}}|f_W(x)-f_{\bar W}(x)|\le\eta\sqrt{K_n}M_\sigma$. Choosing $\eta=r/(2\sqrt{K_n}M_\sigma)$ gives $\sup_{x\in\mathcal{X}}|f_W(x)-f_{\bar W}(x)| \le \frac{r}{2}$. It follows that, for each $x \in \mathcal{X}$,
\begin{equation*}
    f_{\bar W}(x)-\frac{r}{2} \le f_W(x)\le f_{\bar W}(x)+\frac{r}{2}.
\end{equation*}
The brackets $\left[f_{\bar W}-\frac{r}{2},\ f_{\bar W}+\frac{r}{2}\right]$, $\bar W\in\mathcal{W}_\eta$, cover $ \mathcal{H}_{K_n}(M_n)$ pointwise, and each has $L^2(X)$-diameter equal to $r$. Hence they form an $r$-bracketing cover of $ \mathcal{H}_{K_n}(M_n)$ with respect to $\|\cdot\|_{L^2(X)}$. 

Hence,
\begin{equation*}
	\mathcal{N}_{[]} \left(r, \mathcal{H}_{K_n}(M_n),\|\cdot\|_{L^2(X)}\right)\le|\mathcal{W}_\eta|\le\left(\frac{4 M_n M_\sigma \sqrt{K_n}}{r}+1\right)^{K_n},
\end{equation*}
which proves \eqref{eq:rwnn_entropy_bound} after taking logarithms.

We next bound the entropy integral. Set $A_n:=4M_nM_\sigma\sqrt{K_n}$. By \eqref{eq:rwnn_entropy_bound},
\begin{equation*}
	\delta^{-2}\int_{\delta^2}^{\delta}\sqrt{\log \mathcal{N}_{[]}\left(r,\mathcal{H}_{K_n}(M_n),\|\cdot\|_{L^2(X)}\right)}dr\leq\delta^{-2}\int_{\delta^2}^{\delta}\sqrt{K_n\log\left(\frac{A_n}{r}+1\right)}dr .
\end{equation*}
For $r\in[\delta^2,\delta]$ and $\delta\in(0,1)$, $\log\left({A_n}/{r}+1\right)\leq\log\left({(1+A_n)}/{r}\right)\leq\log\left({(1+A_n)}/{\delta^2}\right)$. Hence, 
\begin{equation*}
    \delta^{-2}\int_{\delta^2}^{\delta}\sqrt{K_n\log\left(\frac{A_n}{r}+1\right)}dr\leq\delta^{-1}\sqrt{K_n\log\left(\frac{1+A_n}{\delta^2}\right)},
\end{equation*}
which proves \eqref{eq:rwnn_entropy_integral_bound}. 

Next, take $\bar\delta_n :=C\sqrt{({K_n\log(K_n)})/{n}}$, with $C>0$ sufficiently large. Since $M_n\asymp\sqrt{\log(K_n)/K_n}$, one has that $A_n=4M_nM_\sigma\sqrt{K_n}\lesssim\sqrt{\log(K_n)}$.  Moreover, $\log(K_n)\asymp\log(n)$ and $K_n\log(K_n)=o(n)$ imply $\bar\delta_n\to0$ and $
|\log\bar\delta_n|\lesssim \log(n)\asymp \log(K_n)$. Hence, it holds that $\log\left({(1+A_n)}/{\bar\delta_n^2}\right)\lesssim\log(K_n)$. Substituting $\delta=\bar\delta_n$ in \eqref{eq:rwnn_entropy_integral_bound} gives
\begin{equation*}
    \bar\delta_n^{-2}\int_{\bar\delta_n^2}^{\bar\delta_n}\sqrt{\log \mathcal{N}_{[]}\left(r,\mathcal{H}_{K_n}(M_n),\|\cdot\|_{L^2(X)}\right)}dr\lesssim \sqrt{n}.
\end{equation*}
Thus the entropy condition defining $\delta_{n,2}$ in \eqref{eq:delta_l2_def} is satisfied for $\delta=\bar\delta_n$, which yields
\begin{equation*}
	\delta_{n,2}\lesssim\sqrt{\frac{K_n\log(K_n)}{n}},
\end{equation*}
as required.
\end{proof}

The entropy condition in Theorem~\ref{theorem:generic_pseudometric_rate} is stated for the localized class of loss differences while Lemma~\ref{lemma:rwnn_entropy} provides the entropy of the RWNN class itself. The following lemma allows us to connect the latter bound to the loss class under a Lipschitz condition on the loss. More explicitly, we show that a Lipschitz constant growing polylogarithmically does not change the resulting loss class critical radius order.

\begin{lemma}[Loss-entropy transfer]
\label{lemma:loss_entropy_transfer}
    Assume that on the prediction range of $\mathcal{H}_{K_n}(M_n;\eta_0)\cup\{h\}$, $\ell$ is Lipschitz in its second argument with the constant $L_{\ell,n}>0$.
    Let
    \begin{equation*}
        \mathcal{M}_n(\delta):=\{z\mapsto \ell(z,g)-\ell(z,h)\mid g\in\mathcal{H}_{K_n}(M_n;\eta_0),\ \rho_\circ(g,h)\leq\delta\}.
    \end{equation*}
    Then, for every $w>0$ and every $\delta>0$,
    \begin{equation}
    \label{eq:loss_entropy_transfer}
        \log\mathcal{N}_{[]}\bigl(w,\mathcal{M}_n(\delta),L^2(P_Z)\bigr)\leq \log\mathcal{N}_{[]}\left(\frac{w}{2L_{\ell,n}},\mathcal{H}_{K_n}(M_n;\eta_0),L^2(X)\right).%
    \end{equation}
    Consequently, if $L_{\ell,n}\leq \Lambda_n$, $\Lambda_n=O((\log(n))^m)$, $M_n\asymp\sqrt{\log(K_n)/K_n}$, $\log(K_n)\asymp\log(n)$, and $K_n\log(K_n)=o(n)$, then the loss-class critical radius satisfies
    \begin{equation*}
        \delta_n\lesssim \sqrt{\frac{K_n\log(K_n)}{n}}.
    \end{equation*}
\end{lemma}

\begin{proof}[Proof of Lemma~\ref{lemma:loss_entropy_transfer}]
Fix $w>0$ and set $\eta:=w/(2L_{\ell,n})$. Let $\{[g_j^L,g_j^U]\}_{j=1}^N$ be an $\eta$-bracketing cover of $\mathcal{H}_{K_n}(M_n;\eta_0)$ in $L^2(X)$. Thus, for every $g\in\mathcal{H}_{K_n}(M_n;\eta_0)$, there is $j$ such that $g_j^L(x)\leq g(x)\leq g_j^U(x)$ for $P_X$-a.e. $x$ and $\|g_j^U-g_j^L\|_{L^2(X)}\leq\eta$. Discard empty brackets and for each $j$ choose $g_j\in\mathcal{H}_{K_n}(M_n;\eta_0)\cap[g_j^L,g_j^U]$. If $g\in[g_j^L,g_j^U]$, then  $|g(x)-g_j(x)|\leq g_j^U(x)-g_j^L(x)$ for $P_X$-a.e. $x$. Define $q_j^L(z):=\ell(z,g_j)-\ell(z,h)-L_{\ell,n}\bigl(g_j^U(x)-g_j^L(x)\bigr)$, $q_j^U(z):=\ell(z,g_j)-\ell(z,h)+L_{\ell,n}\bigl(g_j^U(x)-g_j^L(x)\bigr)$, with $z=(y,x)$.

By the Lipschitz assumption in the hypotheses of the lemma,
\begin{equation*}
    q_j^L(z)\leq \ell(z,g)-\ell(z,h)\leq q_j^U(z)\quad\text{for }P_Z\text{-a.e. }z.
\end{equation*}
Hence the brackets $\{[q_j^L,q_j^U]\}_{j=1}^N$ cover $\mathcal{M}_n(\delta)$. Their widths satisfy
\begin{equation*}
    \|q_j^U-q_j^L\|_{L^2(Z)}\leq 2L_{\ell,n}\|g_j^U-g_j^L\|_{L^2(X)}\leq w.
\end{equation*}
Thus
\begin{equation*}
    \mathcal{N}_{[]}\bigl(w,\mathcal{M}_n(\delta),L^2(Z)\bigr)\leq \mathcal{N}_{[]}\left(\frac{w}{2L_{\ell,n}},\mathcal{H}_{K_n}(M_n;\eta_0),L^2(X)\right),
\end{equation*}
which proves \eqref{eq:loss_entropy_transfer}.

Combining \eqref{eq:loss_entropy_transfer} with Lemma~\ref{lemma:rwnn_entropy}, with $A_n:=4M_nM_\sigma\sqrt{K_n}$, gives
\begin{equation*}
    \log\mathcal{N}_{[]}\bigl(w,\mathcal{M}_n(\delta),L^2(Z)\bigr)\leq K_n\log\left(\frac{2A_nL_{\ell,n}}{w}+1\right).
\end{equation*}
Therefore, for $\delta\in(0,1)$,
\begin{equation*}
    \delta^{-2}\int_{\delta^2}^{\delta}\sqrt{\log\mathcal{N}_{[]}\bigl(w,\mathcal{M}_n(\delta),L^2(Z)\bigr)}\,dw\lesssim \delta^{-1}\sqrt{K_n\log\left(\frac{1+2A_nL_{\ell,n}}{\delta^2}\right)}.
\end{equation*}
Under $M_n\asymp\sqrt{\log(K_n)/K_n}$, one has $A_n\lesssim\sqrt{\log(K_n)}$. If $L_{\ell,n}\leq\Lambda_n$ with $\Lambda_n=O((\log(n))^m)$ and $\log(K_n)\asymp\log(n)$, then at $\bar\delta_n:=C\sqrt{K_n\log(K_n)/n}$, with $C>0$ sufficiently large, the logarithmic factor satisfies
\begin{equation*}
    \log\left(\frac{1+2A_nL_{\ell,n}}{\bar\delta_n^2}\right)\lesssim \log(K_n),
\end{equation*}
provided $K_n\log(K_n)=o(n)$. Substituting $\delta=\bar\delta_n$ gives
\begin{equation*}
    \bar\delta_n^{-2}\int_{\bar\delta_n^2}^{\bar\delta_n}\sqrt{\log\mathcal{N}_{[]}\bigl(w,\mathcal{M}_n(\bar\delta_n),L^2(Z)\bigr)}\,dw\lesssim \sqrt{n}.
\end{equation*}
Hence, by the definition of the loss-class critical radius $\delta_n$,
\begin{equation*}
    \delta_n\lesssim \sqrt{\frac{K_n\log(K_n)}{n}},
\end{equation*}
as required.
\end{proof}

In the results below we will find it useful to impose the following domination condition.
\begin{assumption}[Pseudo-metric dominated by $L^2(X)$]
\label{assumption:pseudometric_dominated}
There exists a constant $C_\circ>0$, independent of $n$ and of the realized $\theta_1,\dots,\theta_{K_n}\in\Pi$, such that
\begin{equation*}
    \|g_1-g_2\|_\circ \leq C_\circ\|g_1-g_2\|_{L^2(X)}
\end{equation*}
for all $g_1,g_2 \in \operatorname{span}\bigl(\mathcal{H}_{K_n}(M_n)\cup\{h,\mathbf 1\}\bigr)$, where $\mathbf 1$ denotes the constant function on $\mathcal{X}$.
\end{assumption}
Assumption~\ref{assumption:pseudometric_dominated} ensures that the $L^2(X)$ approximation bounds derived for the RWNN approximation theorem also control the pseudo-metric error appearing in Theorem~\ref{theorem:generic_pseudometric_rate}. In particular,
\begin{equation*}
    \|h-\pi_n h\|_\circ=\inf_{g\in\mathcal{H}_{K_n}(M_n)}\|h-g\|_\circ \leq C_\circ \inf_{g\in\mathcal{H}_{K_n}(M_n)}\|h-g\|_{L^2(X)}.
\end{equation*}

The approximation result of Theorem~\ref{thm:prob-approx} is constructed without imposing the readout constraint defining the statistical sieve. Lemma~\ref{lem:nn_l2_bound} shows that the corresponding control-neighbors weights satisfy the required $\ell_2$ bound with probability tending to one. Hence the same approximation rate is attained within the constrained RWNN class.
\begin{corollary}[Constrained RWNN approximation]
\label{cor:constrained_approx}
    Under the assumptions of Theorem~\ref{thm:prob-approx} and Lemma~\ref{lem:nn_l2_bound}, define
    \begin{equation*}
    	M_n := C_\alpha M_f\sqrt{\frac{\log(K_n)}{K_n}}
    \end{equation*}
    for $\alpha>0$ and $C_\alpha>0$ as in Lemma~\ref{lem:nn_l2_bound}. Then
    \begin{equation*}
    	\inf_{g \in \mathcal{H}_{K_n}(M_n)} \|h - g\|_{L^2(X)}=O_P \left(K_n^{-1/2 - \beta/(d+1)}\right),
    \end{equation*}
    where the probability is with respect to  $\theta_1,\ldots,\theta_{K_n}$.
\end{corollary}

\begin{proof}[Proof of Corollary~\ref{cor:constrained_approx}]
Let $W^*\in\mathbb{R}^{K_n}$ be the $\sigma(\theta_1,\dots,\theta_{K_n})$-measurable control-neighbors weight vector produced by Theorem~\ref{thm:prob-approx}, with components $W^*_k = \omega_{k,K_n}f(A_k,B_k)$. By Theorem~\ref{thm:prob-approx},
\begin{equation}
\label{eq:approx_rate_prob}
    \|h-h_{K_n}^{\theta,W^*}\|_{L^2(X)} = O_P\left(K_n^{-1/2-\beta/(d+1)}\right).
\end{equation}
Moreover, by Lemma~\ref{lem:nn_l2_bound}, for the chosen $\alpha>0$ there exist constants $C_\alpha,C_\alpha'>0$ such that
\begin{equation}
\label{eq:weight_bound_prob}
    \mathbb{P}_{A,B}\left( \|W^*\| \leq C_\alpha M_f\sqrt{\frac{\log(K_n)}{K_n}} \right) \geq 1-C_\alpha'K_n^{-\alpha}.
\end{equation}
By the definition of $M_n$, the event in \eqref{eq:weight_bound_prob} is $\{\|W^*\|\leq M_n\}$. On this event, $h_{K_n}^{\theta,W^*}\in\mathcal{H}_{K_n}(M_n)$, and therefore
\begin{equation*}
    \inf_{g\in\mathcal{H}_{K_n}(M_n)} \|h-g\|_{L^2(X)} \leq \|h-h_{K_n}^{\theta,W^*}\|_{L^2(X)}.
\end{equation*}
Since $\mathbb{P}_{A,B}(\|W^*\|\leq M_n)\to1$, combining this inequality with \eqref{eq:approx_rate_prob} yields
\begin{equation*}
    \inf_{g\in\mathcal{H}_{K_n}(M_n)} \|h-g\|_{L^2(X)} = O_P\left(K_n^{-1/2-\beta/(d+1)}\right),
\end{equation*}
as required.
\end{proof}

Combining Assumption~\ref{assumption:pseudometric_dominated} with Corollary~\ref{cor:constrained_approx} allows us to write the constrained RWNN approximation rate in terms of the pseudo-metric $\|\cdot\|_\circ$, appropriate for Theorem~\ref{theorem:generic_pseudometric_rate} to be applied. 
\begin{corollary}[RWNN approximation in pseudo-metric]
\label{corollary:generic_pseudometric_rwnn}
    Under Assumption~\ref{assumption:pseudometric_dominated} and the assumptions of Corollary~\ref{cor:constrained_approx},
    \begin{equation*}
        \inf_{g\in \mathcal{H}_{K_n}(M_n)}\|h-g\|_\circ = O_P \left(K_n^{-1/2-\beta/(d+1)}\right).
    \end{equation*}
    Hence, if additionally the  assumptions of Theorem~\ref{theorem:generic_pseudometric_rate} hold,
    \begin{equation*}
        \|h-\widehat{h}_n\|_\circ = O_P \left( \max\Big\{ \delta_n, \varepsilon_n, K_n^{-1/2-\beta/(d+1)},  [\chi(h,\pi_n h)]^{1/2} \Big\} \right).
    \end{equation*}
\end{corollary}

\begin{proof}[Proof of Corollary~\ref{corollary:generic_pseudometric_rwnn}]
By definition of the sieve projection in \eqref{eq:sieve_proj},
\begin{equation*}
    \|h-\pi_n h\|_\circ = \inf_{g\in \mathcal{H}_{K_n}(M_n)} \|h-g\|_\circ.
\end{equation*}
By Assumption~\ref{assumption:pseudometric_dominated},
\begin{equation*}
    \|h-\pi_n h\|_\circ \le C_\circ \inf_{g\in \mathcal{H}_{K_n}(M_n)} \|h-g\|_{L^2(X)}.
\end{equation*}
Applying Corollary~\ref{cor:constrained_approx} yields
\begin{equation*}
    \inf_{g\in \mathcal{H}_{K_n}(M_n)} \|h-g\|_{L^2(X)} = O_P \left(K_n^{-1/2-\beta/(d+1)}\right),
\end{equation*}
and therefore
\begin{equation*}
    \|h-\pi_n h\|_\circ = O_P \left(K_n^{-1/2-\beta/(d+1)}\right).
\end{equation*}
The second claim follows by substitution into Theorem~\ref{theorem:generic_pseudometric_rate}.
\end{proof}

Similarly to Subsection~\ref{subsec:L2_Chen_Shen}, we consider an asymptotically underparametrized RWNN sieve case. The choice of $K_n$ satisfies $K_n=o(n)$ and therefore $K_n<n$ for all sufficiently large $n$.
\begin{theorem}[RWNN convergence rate in the pseudo-metric]
\label{theorem:rwnn_generic_rate}
   Suppose that the conditions of Corollary~\ref{corollary:generic_pseudometric_rwnn} and Lemmas~\ref{lemma:rwnn_entropy} and~\ref{lemma:loss_entropy_transfer} hold. 
    Suppose that
    \begin{equation*}
        M_n\asymp \sqrt{\frac{\log(K_n)}{K_n}}
    \end{equation*}
    and that $K_n$ is chosen so that
    \begin{equation*}
        K_n^{2(1+\beta/(d+1))}\log(K_n) \asymp n ,
    \end{equation*}
    meaning $K_n \asymp (n / \log(n))^{\frac{d+1}{2(\beta + d+1)}}$.
    Then
    \begin{equation*}
        \|h-\widehat{h}_n\|_\circ=O_P \left(\max\left\{(n/\log(n))^{-\frac{1+2\beta/(d+1)}{4(1+\beta/(d+1))}},\varepsilon_n,[\chi(h,\pi_n h)]^{1/2}\right\}\right).
    \end{equation*}
\end{theorem}

\begin{proof}[Proof of Theorem~\ref{theorem:rwnn_generic_rate}]
By Corollary~\ref{corollary:generic_pseudometric_rwnn},
\begin{equation}
\label{eq:K_n_rate}
    \|h-\widehat{h}_n\|_\circ = O_P \left(\max\left\{ \delta_n, \varepsilon_n, K_n^{-1/2-\beta/(d+1)}, [\chi(h,\pi_n h)]^{1/2}\right\}\right) ,
\end{equation}
and, by Lemma~\ref{lemma:rwnn_entropy} and Lemma~\ref{lemma:loss_entropy_transfer},
\begin{equation*}
    \delta_n\lesssim \sqrt{\frac{K_n\log(K_n)}{n}}.
\end{equation*}
Using this bound in \eqref{eq:K_n_rate} then gives
\begin{equation}
\label{eq:K_n_rate_final}
    \|h-\widehat{h}_n\|_\circ=O_P\left(\max\left\{\sqrt{\frac{K_n\log(K_n)}{n}},\varepsilon_n,K_n^{-1/2-\beta/(d+1)},[\chi(h,\pi_nh)]^{1/2}\right\}\right).
\end{equation}

The choice of $K_n$ is made by balancing the stochastic and approximation terms in \eqref{eq:K_n_rate_final}. More precisely,
after taking squares, we impose
\begin{equation*}
    \frac{K_n\log(K_n)}{n}\asymp K_n^{-1-2\beta/(d+1)},
\end{equation*}
or, equivalently,
\begin{equation*}
    K_n^{2(1+\beta/(d+1))}\log(K_n)\asymp n.
\end{equation*}
This gives
\begin{equation*}
    K_n\asymp \left(\frac{n}{\log(n)}\right)^{\frac{1}{2(1+\beta/(d+1))}}.
\end{equation*}
Indeed, for this choice one has that $ \log(K_n) \asymp \log(n)$ and hence
\begin{equation*}
    K_n^{2(1+\beta/(d+1))} \log(K_n) \asymp \frac{n}{\log(n)} \log(n) \asymp n.
\end{equation*}
Finally, substituting this choice into the approximation term yields
\begin{equation*}
    K_n^{-1/2-\beta/(d+1)}\asymp (n/\log(n))^{-\frac{1/2+\beta/(d+1)}{2(1+\beta/(d+1))}}\asymp (n/\log(n))^{-\frac{1+2\beta/(d+1)}{4(1+\beta/(d+1))}}.
\end{equation*}
By the balancing relation, the stochastic term is of the same order
\begin{equation*}
    \sqrt{\frac{K_n\log(K_n)}{n}} \asymp (n/\log(n))^{-\frac{1+2\beta/(d+1)}{4(1+\beta/(d+1))}}.
\end{equation*}
Substituting this into \eqref{eq:K_n_rate_final} yields the required rate.
\end{proof}

\begin{remark}[Choice of sieve radius]
\label{rem:sieve_radius}
    The rate $\delta_n\lesssim\sqrt{K_n\log(K_n)/n}$ and the balancing condition $K_n^{2(1+\beta/(d+1))}\log(K_n)\asymp n$ in Theorem~\ref{theorem:rwnn_generic_rate} hold for any $M_n$ satisfying $\sqrt{\log(K_n)/K_n}\lesssim M_n\lesssim K_n^C$ for some fixed $C>0$. The lower bound ensures that the control-neighbors approximant with $W^*$ belongs to $\mathcal{H}_{K_n}(M_n)$ with high probability by Lemma~\ref{lem:nn_l2_bound}, while the upper bound implies $\log(K_nM_n)\asymp\log(K_n)$, so that Lemmas~\ref{lemma:rwnn_entropy} and~\ref{lemma:loss_entropy_transfer} yield $\delta_n\lesssim\sqrt{K_n\log(K_n)/n}$. The choice $M_n\asymp\sqrt{\log(K_n)/K_n}$ is the smallest admissible order.
\end{remark}

We finally specialize the generic framework to squared-loss regression and provide other examples of practically useful loss functions that fit in this setting.

\subsubsection{Squared-loss regression and other examples}

\begin{assumption}[Squared-loss local empirical-process condition]
    \label{assumption:squared_loss_cs}
    For the squared-loss criterion with $\|g-h\|_\circ=\|g-h\|_{L^2(X)}$ and the corresponding local class
    \begin{equation*}
            \mathcal{M}_n(\delta):=\{z\mapsto \ell(z,g)-\ell(z,h)\mid g\in\mathcal{H}_{K_n}(M_n;\eta_0),\ \|g-h\|_{L^2(X)}\leq\delta\},
    \end{equation*}
    the conclusions of Lemma~\ref{lem:chen_shen_maximal} hold with $\Lambda_n$ growing at most polylogarithmically in $n$.
\end{assumption}

\begin{proposition}[Squared-loss regression with conditional noise moments]
\label{prop:squared_loss_specialization}
    Consider the regression model
    \begin{equation*}
        Y_i=h(X_i)+\epsilon_i
    \end{equation*}
    with squared loss
    \begin{equation*}
        \ell(Z_i,g)=\frac{1}{2}\bigl(Y_i-g(X_i)\bigr)^2
    \end{equation*}
    and pseudo-metric $\|g-h\|_\circ=\|g-h\|_{L^2(X)}$. Suppose Assumption~\ref{assumption:c_stationarity} holds, together with the assumptions of Theorem~\ref{theorem:sharp_l2}, Lemma~\ref{lem:nn_l2_bound}, 
    Assumption~\ref{assumption:sieve_extremum_estimator},
    and Assumption~\ref{assumption:squared_loss_cs}. Assume also that
    \begin{enumerate}
    \item[(i)] $\E[\epsilon_i\mid X_i]=0$ and $\E[\epsilon_i^2\mid X_i]\leq \bar\sigma^2<\infty$ almost surely;
    \item[(ii)] $\E[|\epsilon_i|^\kappa\mid X_i]\leq C_\epsilon<\infty$ for the moment exponent $\kappa$ in Assumption~\ref{assumption:c_generic_loss}{(e)}.
    \end{enumerate}
    Then Assumption~\ref{assumption:c_generic_loss}{(a)--(d)} holds with $C_1=1/4$ and with variance and entropy constants growing at most polylogarithmically. Moreover,
    \begin{equation*}
        \chi(h,\pi_nh)=\frac{1}{2}\|h-\pi_nh\|_{L^2(X)}^2.
    \end{equation*}
    Consequently, if
    \begin{equation*}
        K_n^{2(1+\beta/(d+1))}\log(K_n)\asymp n
    \end{equation*}
    and
    \begin{equation*}
        \varepsilon_n=O\left((n/\log(n))^{-\frac{1+2\beta/(d+1)}{4(1+\beta/(d+1))}}\right),
    \end{equation*}
    then
    \begin{equation}
    \label{eq:sq_rate}
        \|h-\widehat{h}_n\|_{L^2(X)}=O_P\left((n/\log(n))^{-\frac{1+2\beta/(d+1)}{4(1+\beta/(d+1))}}\right).
    \end{equation}
\end{proposition}

\begin{proof}[Proof of Proposition~\ref{prop:squared_loss_specialization}]
Let $v:=g-h$. Since $Y_i=h(X_i)+\epsilon_i$,
\begin{equation*}
    q(Z_i,g,h):=\ell(Z_i,g)-\ell(Z_i,h)=-\epsilon_i v(X_i)+\frac{1}{2}v(X_i)^2.
\end{equation*}
Using $\E[\epsilon_i\mid X_i]=0$, we get
\begin{equation*}
    Q(g,h)=\E_Z[q(Z_i,g,h)]=\frac{1}{2}\E[v(X_i)^2]-\E[\epsilon_i v(X_i)]=\frac{1}{2}\|g-h\|_{L^2(X)}^2.
\end{equation*}
Hence $h$ minimizes the population risk, so Assumption~\ref{assumption:c_generic_loss}{(a)} holds. Moreover,
\begin{equation*}
    Q(g,h)=\frac{1}{2}\|g-h\|_{L^2(X)}^2\geq 2C_1\|g-h\|_{L^2(X)}^2
\end{equation*}
for any $C_1\leq1/4$, proving Assumption~\ref{assumption:c_generic_loss}{(b)}.

For $g\in\mathcal{H}_{K_n}(M_n)$, the sieve sup-bound gives
\begin{equation*}
    \|g\|_\infty\leq M_n\sqrt{K_n}M_\sigma=:B_n.
\end{equation*}
Under $M_n\asymp\sqrt{\log(K_n)/K_n}$, $B_n\asymp\sqrt{\log(K_n)}$. Since $h$ is bounded under the source condition, write $\|h\|_\infty\leq H$. Then $\|v\|_\infty\leq B_n+H$. Using $q(Z_i,g,h)=-\epsilon_i v(X_i)+\frac{1}{2}v(X_i)^2$,
\begin{equation*}
    \mathrm{Var}_Z(q(Z_i,g,h))\leq \E[q(Z_i,g,h)^2]\leq 2\E[\epsilon_i^2v(X_i)^2]+\frac{1}{2}\E[v(X_i)^4].
\end{equation*}
By $\E[\epsilon_i^2\mid X_i]\leq\bar\sigma^2$ and $\|v\|_\infty\leq B_n+H$,
\begin{equation*}
    \mathrm{Var}_Z(q(Z_i,g,h))\leq \left(2\bar\sigma^2+\frac{1}{2}(B_n+H)^2\right)\|g-h\|_{L^2(X)}^2.
\end{equation*}
Thus Assumption~\ref{assumption:c_generic_loss}{(c)} holds with $C_{2,n}\lesssim 1+B_n^2$, which is polylogarithmic in $n$ under the stated calibration.

For the entropy condition, squared loss satisfies on the sieve prediction range
\begin{equation*}
    |\ell((y,x),u)-\ell((y,x),v)|\leq (|\epsilon|+B_n+H)|u-v|.
\end{equation*}
Together with $\E[|\epsilon_i|^\kappa\mid X_i]\leq C_\epsilon$, this gives the conditional-$L^2$ loss-entropy transfer bound
\begin{equation*}
    \log\mathcal{N}_{[]}\bigl(w,\mathcal{M}_n(\delta),L^2(P_Z)\bigr)\leq K_n\log\left(\frac{\widetilde{L}_n A_n}{w}+1\right),
\end{equation*}
where $A_n=4M_nM_\sigma\sqrt{K_n}$ and $\widetilde{L}_n^2\lesssim \bar\sigma^2+(B_n+H)^2$. Since $\widetilde{L}_nA_n$ is polylogarithmic, the same calculation as in Lemma~\ref{lemma:loss_entropy_transfer} gives
\begin{equation*}
    \delta_n\lesssim\sqrt{\frac{K_n\log(K_n)}{n}},
\end{equation*}
so Assumption~\ref{assumption:c_generic_loss}{(d)} holds.

The local empirical-process condition required for Lemma~\ref{lem:chen_shen_maximal} is imposed in Assumption~\ref{assumption:squared_loss_cs}. This condition replaces the pointwise local loss concentration verification (Assumption~\ref{assumption:c_generic_loss}{(e)}), which is not implied by $L^2(X)$-localization alone.

Taking $g=\pi_nh$ in the identity for $Q(g,h)$ yields
\begin{equation*}
    \chi(h,\pi_nh)=Q(\pi_nh,h)=\frac{1}{2}\|\pi_nh-h\|_{L^2(X)}^2.
\end{equation*}
By Corollary~\ref{cor:constrained_approx},
\begin{equation*}
    [\chi(h,\pi_nh)]^{1/2}=O_P\left(K_n^{-1/2-\beta/(d+1)}\right).
\end{equation*}
Under $K_n^{2(1+\beta/(d+1))}\log(K_n)\asymp n$,
\begin{equation*}
    K_n^{-1/2-\beta/(d+1)}\asymp (n/\log(n))^{-\frac{1+2\beta/(d+1)}{4(1+\beta/(d+1))}}.
\end{equation*}

We now check that $\Prob( \norm{ \widehat{h}_n - h }_{L^2(X)} > \eta_0 ) \to 0$ as needed to apply Theorem~\ref{theorem:generic_pseudometric_rate}. 
By Assumption~\ref{assumption:sieve_extremum_estimator}, $Q_n(\widehat{h}_n, h) \leq Q_n(\pi_n h, h) + o_P(1)$, cf. \eqref{eq:near_min_q}; by \eqref{eq:CS_near_ball}, we also have $Q_n(\pi_n h, h) = o_P(1)$.
We proceed by contradiction: assume that 
$\norm{ \widehat{h}_n - h }_{L^2(X)} > \eta_0$ while $\norm{ \pi_n h - h }_{L^2(X)} < \eta_0 / 2$. By convexity of the squared loss, there exists an element $\Breve{h} \in \mathcal{H}_{K_n}(M_n)$ on the segment with endpoints $\widehat{h}_n$ and $\pi_n h$ such that $\norm{ \Breve{h} - h }_{L^2(X)} = \eta_0 / 2$, population excess loss $Q(\Breve{h}, h) = \eta_0^2 / 8$, and $Q_n(\Breve{h}, h) = o_P(1)$. 
We apply the local shell bound from Lemma~\ref{lem:chen_shen_maximal}(ii) with $x = \eta_0 / (4 a_n)$ and $c = 1/4$, and find
\begin{align}
\label{eq:prob_shell_breve_h}
    & \Prob\left( \sup_{\{g\in\mathcal{H}_{K_n}(M_n) \,\mid\, \eta_0/4 < \norm{g - h}_{L^2(X)} \leq \eta_0/2 \}} | (Q_n - Q)(g, h) | \geq \frac{\eta_0^2}{64} \right) \notag \\
    & \qquad\qquad \leq A_0 \exp\left( - A_1 \frac{n a_n^2}{\Lambda_n} \left( \frac{\eta_0}{4 a_n}^{2-s} \right) \right) + A_2 (\eta_0 / 4)^{-(2-s) \kappa} \Lambda_n n^{-\kappa/2} = o(1) .
\end{align}
However, with probability tending to one,
\begin{equation*}
    \abs{ (Q_n - Q)(\Breve{h}, h) } \geq Q(\Breve{h}, h) -  Q_n(\Breve{h}, h) > \frac{\eta_0^2}{16}
\end{equation*}
which contradicts \eqref{eq:prob_shell_breve_h} since $\Breve{h} \in \{g\in\mathcal{H}_{K_n}(M_n) \,\mid\, \eta_0/4 < \norm{g - h}_{L^2(X)} \leq \eta_0 \}$.

Combining the preceding bounds with the rate argument in the proof of  Theorem~\ref{theorem:rwnn_generic_rate} and using the assumed bound on $\varepsilon_n$ gives \eqref{eq:sq_rate}.
\end{proof}

The following result shows that post-estimation clipping of the estimator preserves our established $L^2(X)$ convergence rate provided that the clipping threshold exceeds $\|h\|_\infty$.
\begin{corollary}[Post-estimation clipping]
\label{cor:post_estimation_clipping}
    Let $T_{\tau}$ denote the clipping operator
    \begin{equation*}
        T_\tau u:=\max\{-\tau,\min\{u,\tau\}\}.
    \end{equation*}
    Let the assumptions of Proposition~\ref{prop:squared_loss_specialization} hold and define the clipped estimator $T_{\tau_n}\widehat{h}_n$ by applying $T_{\tau_n}$ pointwise after estimation for a sequence $\tau_n > 0$ of clipping thresholds. 
    Then
    \begin{equation*}
        \|T_{\tau_n}\widehat{h}_n-h\|_{L^2(X)}\leq \|\widehat{h}_n-h\|_{L^2(X)}+\|T_{\tau_n}h-h\|_{L^2(X)}.
    \end{equation*}
    In particular, if $\|h\|_\infty\leq \tau_n$ for all sufficiently large $n$, then
    \begin{equation*}
        \|T_{\tau_n}\widehat{h}_n-h\|_{L^2(X)}\leq \|\widehat{h}_n-h\|_{L^2(X)}.
    \end{equation*}
    Consequently, under the conditions of Proposition~\ref{prop:squared_loss_specialization},
    \begin{equation*}
        \|T_{\tau_n}\widehat{h}_n-h\|_{L^2(X)}=O_P\left((n/\log(n))^{-\frac{1+2\beta/(d+1)}{4(1+\beta/(d+1))}}\right)
    \end{equation*}
    whenever $\|h\|_\infty\leq\tau_n$ for all sufficiently large $n$.
\end{corollary}

\begin{proof}[Proof of Corollary~\ref{cor:post_estimation_clipping}]
The clipping map $T_\tau:\mathbb{R}\to[-\tau,\tau]$ is nonexpansive, that is
\begin{equation*}
    |T_\tau a-T_\tau b|\leq |a-b|
\end{equation*}
for all $a,b\in\mathbb{R}$. Therefore,
\begin{align*}
    |T_{\tau_n}\widehat{h}_n(X)-h(X)|&\leq |T_{\tau_n}\widehat{h}_n(X)-T_{\tau_n}h(X)|+|T_{\tau_n}h(X)-h(X)|\\
    &\leq |\widehat{h}_n(X)-h(X)|+|T_{\tau_n}h(X)-h(X)|.
\end{align*}
Taking $L^2(X)$ norms gives the first claim. If $\|h\|_\infty\leq\tau_n$, then $T_{\tau_n}h=h$, and the contraction bound gives the second claim. The rate follows immediately from Proposition~\ref{prop:squared_loss_specialization}.
\end{proof}

\begin{remark}[Other losses]
The generic sieve convergence theorem, Theorem~\ref{theorem:generic_pseudometric_rate}, can also be applied to other losses once the corresponding local separation, variance, entropy, and Chen-Shen empirical-process conditions are verified. Losses that are Lipschitz in the second argument are especially convenient, since Lemma~\ref{lemma:loss_entropy_transfer} transfers the RWNN prediction-class entropy to the loss-difference class. Examples include:
\begin{itemize}
\item[(i)] absolute loss $\ell((y,x),u)=|y-u|$, which is globally $1$-Lipschitz in $u$;
\item[(ii)] Huber loss with fixed threshold, which is globally Lipschitz in $u$;
\item[(iii)] quantile or pinball loss, which is globally Lipschitz in $u$;
\item[(iv)] logistic loss $\ell((y,x),u)=\log(1+e^u)-yu$ for $y\in\{0,1\}$, whose derivative with respect to $u$ is uniformly bounded;
\item[(v)] hinge loss $\ell((y,x),u)=(1-yu)_+$ for $y\in\{-1,1\}$, which is globally Lipschitz in $u$.
\end{itemize}
For smooth likelihood-type losses, the local quadratic separation condition follows from local positive curvature of the population risk. For nonsmooth losses, such as absolute or quantile loss, separation typically requires an additional condition on the conditional distribution of $Y$ given $X$, for instance a density bounded away from zero near the conditional median or quantile.
\end{remark}
\begin{remark}[Comparison with the sieve rate in \citet{Chen1999}]
For certain integral Sobolev classes, \citet{Chen1999} obtained the consistency rate
\begin{equation*}
    O_P\left(\left({n}/{\log(n)}\right)^{-\frac{1+2\alpha/(d+1)}{4(1+\alpha/(d+1))}}\right),
\end{equation*}
where $\alpha\in(0,1]$ is related to the regularity of the activation function used to construct fully trainable shallow neural networks. The rate in Theorem~\ref{theorem:rwnn_generic_rate} has the same algebraic form, with the source regularity parameter $\beta$ replacing $\alpha$. In particular, when $\beta=\alpha$, the RWNN estimator achieves the same statistical rate as \citet{Chen1999}, although the underlying approximation classes and network constructions are different.
\end{remark}

\subsection{Asymptotic normality for smooth functionals}
\label{subsec:rwnn_functional_clt}

We now consider inference on smooth functionals of the target function. Let $\Psi:\mathcal{H}\to\mathbb{R}$ be a known functional and suppose that the object of interest is $\Psi(h)$. We follow the local-perturbation method of \citet{shenMethodsSievesPenalization1997} and \citet{chenSieveExtremumEstimates1998}, and provide the proof in our notation.

We now define all the ingredients that we need in order to prove asymptotic normality of smooth functionals of RWNN estimators. For any function $u\in \mathcal{H}$ considered below, let $\pi_n u$ denote a measurable sieve projection of $u$ onto $\mathcal{H}_{K_n}(M_n)$, that is $\pi_n u\in\arg\min_{g\in\mathcal{H}_{K_n}(M_n)}\|u-g\|_\circ$. %
For a direction $v$ such that $h+v\in\mathcal{H}$, define the loss increment 
\begin{equation}
\label{eq:loss_increment}
    q_h(v,Z_i):=\ell(Z_i,h+v)-\ell(Z_i,h).
\end{equation}
Hence, for $g\in\mathcal{H}$, we can write the expression of its population excess risk and its empirical analogue, with respect to $h$ , using the notation introduced in \eqref{eq:Q_def}-\eqref{eq:q_def}, as
\begin{equation*}
    Q(g,h)=\E_Z[q(Z_i,g,h)]=\E_Z[q_h(g-h,Z_i)]
\end{equation*}
and 
\begin{equation*}
    Q_n(g,h)=\frac{1}{n}\sum_{i=1}^n q(Z_i,g,h)=L_n (g)-L_n(h) = \frac{1}{n}\sum_{i=1}^n q_h(g-h,Z_i),
\end{equation*}
respectively.

For an integrable measurable function $f$, define the corresponding centered empirical process as
\begin{equation*}
    \nu_n(f):=\frac{1}{\sqrt{n}}\sum_{i=1}^n\left(f(Z_i)-\E_Z[f(Z_i)]\right).
\end{equation*}
and, finally, for $g_1,g_2\in\mathcal{H}$, write, using the notation introduced in Lemma~\ref{lem:chen_shen_maximal},
\begin{equation}
\label{eq:mu_n_def}
    \mu_n(h, g_1,g_2)=\mu_n(g_1,h)-\mu_n( g_2,h)=(Q_n(g_1,h)-Q(g_1,h))-(Q_n(g_2,h)-Q(g_2,h)).
\end{equation}

\begin{assumption}[Smooth functional and Riesz representer]
\label{assumption:functional_smooth_riesz}
Let $\mathcal{V}:=\overline{\operatorname{span}(\mathcal{H}-\{h\})}^{\|\cdot\|_\circ}$. The norm $\|\cdot\|_\circ$ is induced by an inner product $\langle\cdot,\cdot\rangle_\circ$ on $\mathcal{V}$. The functional $\Psi$ is differentiable at $h$ in the sense that there exists a continuous linear functional $\Psi'_h:\mathcal{V}\to\mathbb{R}$ such that, as $\|\widetilde{h}-h\|_\circ\to0$,
\begin{equation}
\label{eq:func_expansion}
    \Psi(\widetilde{h})-\Psi(h)=\Psi'_h(\widetilde{h}-h)+O(\|\widetilde{h}-h\|_\circ^\omega)
\end{equation}
for some $\omega>0$. Let $\psi\in\mathcal{V}$ denote the Riesz representer of $\Psi'_h$, so that
\begin{equation}
\label{eq:Riesz_representer}
    \Psi'_h(v)=\langle v,\psi\rangle_\circ,\qquad v\in\mathcal{V}.
\end{equation}
\end{assumption}

\begin{assumption}[Local criterion linearization]
\label{assumption:criterion_linearization}
For every $g$ in a neighborhood of $h$, write $v:=g-h$ and suppose that there exists a path $t\mapsto h_t(v)$ in $\mathcal{H}$, $t\in[0,1]$, such that $h_0(v) = h$ and $h_1(v)=g$. Define the pathwise derivative of the loss difference at $h$ in the direction $v$ by 
\begin{equation*}
    q'_h(v,Z_i):=\lim_{t \,\downarrow\, 0} \frac{\ell(Z_i,h_t(v))-\ell(Z_i,h)}{t},
\end{equation*}
whenever the limit exists. The loss difference \eqref{eq:loss_increment} admits the expansion
\begin{equation}
\label{eq:def_remainder}
    q_h(v,Z_i)=q'_h(v,Z_i)+r_h(v,Z_i),
\end{equation}
where $r_h(v,Z_i)$ is the corresponding remainder. 
We assume that $q'_h(v,Z_i)-\E_Z[q'_h(v,Z_i)]$ is linear in $v$. The empirical contribution of the remainder is controlled by Assumption~\ref{assumption:shen_local_conditions}.
\end{assumption}

Use $\mathfrak{R}_n$ to denote the rate obtained via Theorem~\ref{theorem:rwnn_generic_rate}, as in
\citet[p.~2567]{shenMethodsSievesPenalization1997}, so that $\|\widehat{h}_n-h\|_\circ = { O_P(\mathfrak{R}_n) } ,$ and let $\mathfrak{R}_n = o(r_n)$ so that
\begin{equation*}
    \|\widehat{h}_n-h\|_\circ = o_P(r_n) .
\end{equation*}
Further, let 
$t_n>0$ be a deterministic sequence satisfying
\begin{equation*}
t_n=o(n^{-1/2}).
\end{equation*}
For $s\in\{-1,1\}$, define the local perturbation of $g$ in the Riesz direction $s\psi$ as
\begin{equation*}
g_{n,s}^*(g):=(1-t_n)g+t_n(h+s\psi).
\end{equation*}
Since $g_{n,s}^*(g)$ need not belong to $\mathcal{H}_{K_n}(M_n)$, let
\begin{equation*}
\widetilde{g}_{n,s}(g):=\pi_n g_{n,s}^*(g)\in \mathcal{H}_{K_n}(M_n)
\end{equation*}
be its sieve projection, where $\pi_n$ is the sieve projection introduced in \eqref{eq:sieve_proj}.
The following assumption is the analogue of Conditions (A)-(D) in \cite{shenMethodsSievesPenalization1997}, presented in our notation and in our loss-minimization setting.
\begin{assumption}
\label{assumption:shen_local_conditions}
The following conditions hold conditionally on realized random parameters $\theta_1, \theta_2, \ldots, \theta_{K_n}$.
\begin{enumerate}
\item[(a)] [Stochastic equicontinuity of the remainder] For $r_h$ defined in \eqref{eq:def_remainder},
\begin{equation*}
\sup_{\substack{\{g\in \mathcal{H}_{K_n}(M_n)\mid \|g-h\|_\circ\le r_n\}\\ s\in\{-1,1\}}} n^{-1/2} \left|\nu_n\left(
r_h(g-h,Z_i)-r_h(\widetilde{g}_{n,s}(g)-h,Z_i) \right)\right| =O_P(t_n^2).
\end{equation*}
\item[(b)] [Expectation of the criterion difference] 
\begin{equation*}
\sup_{\substack{\{g\in \mathcal{H}_{K_n}(M_n)\mid 0<\|g-h\|_\circ\le r_n\}\\ s\in\{-1,1\}}} \left| Q(\widetilde{g}_{n,s}(g),h)-Q(g,h)-\frac{1}{2}\left(\|\widetilde{g}_{n,s}(g)-h\|_\circ^2-\|g-h\|_\circ^2\right)\right|=O(t_n^2).
\end{equation*}
\item[(c)] [Projection error of the local perturbation] 
\begin{equation*}
\sup_{\substack{\{g\in \mathcal{H}_{K_n}(M_n)\mid 0<\|g-h\|_\circ\le r_n\}\\ s\in\{-1,1\}}}\|g_{n,s}^*(g)-\widetilde{g}_{n,s}(g)\|_\circ=O(r_n^{-1}t_n^2),
\end{equation*}
and
\begin{equation*}
\sup_{\substack{\{g\in \mathcal{H}_{K_n}(M_n)\mid \|g-h\|_\circ\le r_n\}\\ s\in\{-1,1\}}} n^{-1/2} \left| \nu_n \left(q'_h(g_{n,s}^*(g)-\widetilde{g}_{n,s}(g),Z_i)\right)\right|=O_P(t_n^2).
\end{equation*}
\item[(d)] [Local derivative bound] 
\begin{equation*}
\sup_{\substack{\{g\in \mathcal{H}_{K_n}(M_n)\mid \|g-h\|_\circ\le r_n\}}} n^{-1/2} \left|\nu_n \left(q'_h(g-h,Z_i)\right)\right|=O_P(t_n).
\end{equation*}
\end{enumerate}
\end{assumption}

\begin{assumption}[Optimization error and local rate]
\label{assumption:functional_near_minimization}
    The estimator satisfies
    \begin{equation}
    \label{eq:q_ineq}
    Q_n(\widehat{h}_n,h)\le \inf_{g\in \mathcal{H}_{K_n}(M_n)}Q_n(g,h)+\rho_n,
    \end{equation}
    where $\rho_n=o_P(t_n n^{-1/2})$. Moreover, \eqref{eq:func_expansion} holds with some $\omega \ge 2$ and the rate $r_n$ satisfies  
    \begin{equation*}
    \sqrt{n} r_n^2 \to 0 
    \quad \textnormal{and} \quad
    n^{-1/2} = o(r_n) .
    \end{equation*}
\end{assumption}

For comparison's sake, given the rate obtained in Theorem~\ref{theorem:rwnn_generic_rate}, from Proposition~\ref{prop:squared_loss_specialization} we can see that a valid choice of $r_n$ with the squared loss is
\begin{equation*}
    r_n = \log(n) (n / \log(n))^{-\gamma}, 
    \quad \textnormal{with} \ \gamma = \frac{1+2\beta/(d+1)}{4(1+\beta/(d+1))} \in \left(\frac{1}{4}, \frac{1}{2}\right) .
\end{equation*}
It is easy to show that this choice fulfills both conditions of Assumption~\ref{assumption:functional_near_minimization}.

\begin{assumption}[Finite variance and scalar CLT in the Riesz direction]
\label{assumption:functional_score_clt}
    The linearized loss increment in the Riesz direction has finite positive asymptotic variance
    \begin{equation*}
    0<\sigma_\Psi^2:=\lim_{n\to\infty}\frac{1}{n}\text{Var}\left[\sum_{i=1}^n q'_h(\psi,Z_i)\right]<\infty.
    \end{equation*}
    Moreover,
    \begin{equation}
    \label{eq:normal_assump}
    \nu_n(q'_h(\psi,Z_i))
    =\frac{1}{\sqrt n}\sum_{i=1}^n\left(q'_h(\psi,Z_i)-\E_Z[q'_h(\psi,Z_i)]\right)
    \todist \mathcal{N}(0,\sigma_\Psi^2).
\end{equation}
\end{assumption}

In the i.i.d. case of Assumption~\ref{assumption:c_stationarity}, \eqref{eq:normal_assump} follows from the usual scalar central limit theorem under the finite-variance condition. In the dependent cases of Assumption~\ref{assumption:c_stationarity}, it is the analogue of Condition B.5 in \citet{chenSieveExtremumEstimates1998}. 

\begin{theorem}[Asymptotic normality of smooth RWNN functionals]
\label{theorem:rwnn_functional_clt}
    Let Assumptions~\ref{assumption:functional_smooth_riesz}-\ref{assumption:functional_score_clt} hold. Then
    \begin{equation*}
    \sqrt{n}\{\Psi(\widehat{h}_n)-\Psi(h)\}\todist \mathcal{N}(0,\sigma_\Psi^2).
    \end{equation*}
\end{theorem}

\begin{proof}[Proof of Theorem~\ref{theorem:rwnn_functional_clt}]
We start by defining
\begin{equation*}
v_n:=\widehat{h}_n-h.
\end{equation*}
By Assumption~\ref{assumption:functional_smooth_riesz}, it holds that
\begin{equation*}
\Psi(\widehat{h}_n)-\Psi(h)=\Psi'_h(v_n)+O_P(\|v_n\|_\circ^\omega)=\langle v_n,\psi\rangle_\circ+O_P(\|v_n\|_\circ^\omega),
\end{equation*}
where in the first equality we used \eqref{eq:func_expansion}, while in the second one used the Riesz representation \eqref{eq:Riesz_representer}. 
Next, since $\| v_n \|_\circ \equiv \|\widehat{h}_n-h\|_\circ = o_P(r_n)$, we observe that the event
\begin{equation}
    \mathcal{A}_n := \left\{ \|\widehat{h}_n-h\|_\circ > r_n \right\}
\end{equation}
fulfills $\Prob_Z( \mathcal{A}_n ) = o(1)$ as $n \to \infty$, and hence is asymptotically negligible. Thus, in the following we can work conditionally on the event $ \mathcal{A}_n^\complement = \{  \|\widehat{h}_n-h\|_\circ \leq r_n \}$.
Then, as $\omega\geq 2$ and $\sqrt{n}\,r_n^\omega\to0$, by Assumption~\ref{assumption:functional_near_minimization}, we have
\begin{equation}
\label{eq:functional_linearization}
\sqrt{n}\{\Psi(\widehat{h}_n)-\Psi(h)\}=\sqrt{n}\langle v_n,\psi\rangle_\circ+o_P(1).
\end{equation}
Hence, it remains to quantify the leading term $\sqrt{n}\langle v_n,\psi\rangle_\circ$. 

Fix $s\in\{-1,1\}$ and define
\begin{equation*}
\widetilde{h}_{n,s}:=\widetilde{g}_{n,s}(\widehat{h}_n)=\pi_n g_{n,s}^*(\widehat{h}_n)\in\mathcal{H}_{K_n}(M_n).
\end{equation*}
We now notice that \eqref{eq:q_ineq} in Assumption~\ref{assumption:functional_near_minimization} gives
\begin{equation}
\label{eq:minimization_ineq_functional}
Q_n(\widetilde{h}_{n,s},h)-Q_n(\widehat{h}_n,h)\geq -\rho_n.
\end{equation}
We now decompose the left-hand side into its population and empirical-process parts:
\begin{equation}
\label{eq:emp_pop_decomp_functional}
Q_n(\widetilde{h}_{n,s},h)-Q_n(\widehat{h}_n,h)=Q(\widetilde{h}_{n,s},h)-Q(\widehat{h}_n,h)+\mu_n(h, \widetilde{h}_{n,s},\widehat{h}_n),
\end{equation}
where, by definition \eqref{eq:mu_n_def},
\begin{equation*}
\mu_n(h,\widetilde{h}_{n,s},\widehat{h}_n)=(Q_n(\widetilde{h}_{n,s},h)-Q(\widetilde{h}_{n,s},h))-(Q_n(\widehat{h}_n,h)-Q(\widehat{h}_n,h)).
\end{equation*}
We analyze first the empirical-process term in \eqref{eq:emp_pop_decomp_functional}. More specifically, by Assumption~\ref{assumption:criterion_linearization}, 
\begin{align}
\label{eq:mu_expand}
\mu_n(h,\widetilde{h}_{n,s},\widehat{h}_n)&=n^{-1/2}\nu_n\left(q'_h(\widetilde{h}_{n,s}-h,Z_i)-q'_h(v_n,Z_i)\right)\nonumber\\
&\quad+n^{-1/2}\nu_n\left(r_h(\widetilde{h}_{n,s}-h,Z_i)-r_h(v_n,Z_i)\right),
\end{align}
where we applied \eqref{eq:def_remainder} to each of the brackets in the previous expression.

 Moreover, one can write down the following equality:
\begin{equation*}
\widetilde{h}_{n,s}-h=(\widetilde{h}_{n,s}-g_{n,s}^*(\widehat{h}_n))+(g_{n,s}^*(\widehat{h}_n)-\widehat{h}_n)+(\widehat{h}_n-h).
\end{equation*}
Since by Assumption~\ref{assumption:criterion_linearization}, the centered derivative is linear in its direction, we obtain that the first term in \eqref{eq:mu_expand} can be expressed as
\begin{align*}
&n^{-1/2}\nu_n\left(q'_h(\widetilde{h}_{n,s}-h,Z_i)-q'_h(v_n,Z_i)\right)\\
&=n^{-1/2}\nu_n\left(q'_h(\widetilde{h}_{n,s}-g_{n,s}^*(\widehat{h}_n),Z_i)\right)+n^{-1/2}\nu_n\left(q'_h(g_{n,s}^*(\widehat{h}_n)-\widehat{h}_n,Z_i)\right).
\end{align*}
By Assumption~\ref{assumption:shen_local_conditions}~{(c)}, the first term on the right-hand side is $O_P(t_n^2)$. Next, since
\begin{equation*}
g_{n,s}^*(\widehat{h}_n)-\widehat{h}_n=t_n(s\psi-v_n),
\end{equation*}
linearity of the centered derivative gives
\begin{align}
\label{eq:term_1}
n^{-1/2}\nu_n \left(q'_h(g_{n,s}^*(\widehat{h}_n)-\widehat{h}_n,Z_i)\right) &=t_n n^{-1/2}s\nu_n\left(q'_h(\psi,Z_i)\right)-t_n n^{-1/2}\nu_n\left(q'_h(v_n,Z_i)\right).
\end{align}
By Assumption~\ref{assumption:shen_local_conditions}~{(d)},
\begin{equation*}
n^{-1/2}\nu_n\left(q'_h(v_n,Z_i)\right)=O_P(t_n),
\end{equation*}
and hence
\begin{equation}
\label{eq:term_2}
t_n n^{-1/2}\nu_n\left(q'_h(v_n,Z_i)\right)=O_P(t_n^2).
\end{equation}
Using the fact that, by Assumption~\ref{assumption:shen_local_conditions}~{(a)}, the remainder contribution in \eqref{eq:mu_expand} is $O_P(t_n^2)$ and combining expressions \eqref{eq:term_2} and  \eqref{eq:term_1} allows to write \eqref{eq:mu_expand} as
\begin{equation}
\label{eq:empirical_expansion_functional}
\mu_n(h,\widetilde{h}_{n,s},\widehat{h}_n)=t_n n^{-1/2}s\nu_n\left(q'_h(\psi,Z_i)\right)+O_P(t_n^2).
\end{equation}
We next analyze the population term in \eqref{eq:emp_pop_decomp_functional}. By Assumption~\ref{assumption:shen_local_conditions}~{(b)},
\begin{equation*}
    Q(\widetilde{h}_{n,s},h)-Q(\widehat{h}_n,h)=\frac{1}{2}\left(\|\widetilde{h}_{n,s}-h\|_\circ^2-\|v_n\|_\circ^2\right)+O_P(t_n^2).
\end{equation*}
By Assumption~\ref{assumption:shen_local_conditions}~{(c)}, replacing $\widetilde{h}_{n,s}$ by $g_{n,s}^*(\widehat{h}_n)$ in the squared norm introduces only an $O_P(t_n^2)$ error. Indeed, by the reverse triangle inequality,
\begin{align*}
\left|\|\widetilde{h}_{n,s}-h\|_\circ^2-\|g_{n,s}^*(\widehat{h}_n)-h\|_\circ^2\right| &\le \|\widetilde{h}_{n,s}-g_{n,s}^*(\widehat{h}_n)\|_\circ \left(\|\widetilde{h}_{n,s}-h\|_\circ+\|g_{n,s}^*(\widehat{h}_n)-h\|_\circ\right).
\end{align*}
By Assumption~\ref{assumption:shen_local_conditions}~{(c)}, the first factor is $O_P(r_n^{-1}t_n^2)$. Since $\|\widehat{h}_n-h\|_\circ=O_P(r_n)$, 
$t_n = o(n^{-1/2}) = o(r_n)$, and $\psi\in\mathcal{V}$ is fixed, $\|g_{n,s}^*(\widehat{h}_n)-h\|_\circ \leq (1 - t_n) \norm{ v_n }_\circ + t_n \norm{ \psi }_\circ =O_P(r_n)$; as $t_n^2 / r_n = o(r_n)$, the same holds for $\|\widetilde{h}_{n,s}-h\|_\circ$. 
Hence the displayed difference is $O_P(t_n^2)$.

Hence, 
\begin{equation}
\label{eq:Q_equality}
Q(\widetilde{h}_{n,s},h)-Q(\widehat{h}_n,h)=\frac{1}{2}\left(\|g_{n,s}^*(\widehat{h}_n)-h\|_\circ^2-\|v_n\|_\circ^2\right)+O_P(t_n^2).
\end{equation}
Since
\begin{equation*}
g_{n,s}^*(\widehat{h}_n)-h=(1-t_n)v_n+t_n s\psi=v_n+t_n(s\psi-v_n),
\end{equation*}
we have
\begin{align*}
\frac{1}{2}\left(\|g_{n,s}^*(\widehat{h}_n)-h\|_\circ^2-\|v_n\|_\circ^2\right)&=t_n\langle v_n,s\psi-v_n\rangle_\circ+\frac{1}{2}t_n^2\|s\psi-v_n\|_\circ^2\\
&=t_ns\langle v_n,\psi\rangle_\circ-t_n\|v_n\|_\circ^2+\frac{1}{2}t_n^2\|s\psi-v_n\|_\circ^2.
\end{align*}
Now as $\|v_n\|_\circ=O_P(r_n)$ and $\sqrt{n} r_n^2 \to 0$, one gets
\begin{equation*}
t_n\|v_n\|_\circ^2=o_P(t_n n^{-1/2}).
\end{equation*}
Also, since $t_n=o(n^{-1/2})$ and $\psi$ is fixed in $\mathcal{V}$,
\begin{equation*}
t_n^2\|s\psi-v_n\|_\circ^2=o_P(t_n n^{-1/2}).
\end{equation*}
The $O_P(t_n^2)$ term is also $o_P(t_n n^{-1/2})$ since $t_n=o(n^{-1/2})$. Hence in \eqref{eq:Q_equality} one gets
\begin{equation}
\label{eq:population_expansion_functional}
Q(\widetilde{h}_{n,s},h)-Q(\widehat{h}_n,h)=t_ns\langle v_n,\psi\rangle_\circ+o_P(t_n n^{-1/2}).
\end{equation}
Combining \eqref{eq:minimization_ineq_functional}, \eqref{eq:emp_pop_decomp_functional}, \eqref{eq:empirical_expansion_functional}, and \eqref{eq:population_expansion_functional}, we obtain
\begin{equation}
\label{eq:risk_dif_rhs}
0\leq t_ns\langle v_n,\psi\rangle_\circ+t_n n^{-1/2}s\nu_n\left(q'_h(\psi,Z_i)\right)+O_P(t_n^2)+\rho_n.
\end{equation}
Since $t_n=o(n^{-1/2})$, we have $t_n^2=o(t_n n^{-1/2})$, and, by Assumption~\ref{assumption:functional_near_minimization},
\begin{equation*}
\rho_n=o_P(t_n n^{-1/2}).
\end{equation*}
Hence, we rewrite \eqref{eq:risk_dif_rhs} as
\begin{equation*}
0\leq t_ns\langle v_n,\psi\rangle_\circ+t_n n^{-1/2}s\nu_n\left(q'_h(\psi,Z_i)\right)+o_P(t_n n^{-1/2}).
\end{equation*}
Dividing by $t_n$ and multiplying by $\sqrt{n}$ gives
\begin{equation}
\label{eq:two_sided_ineq_functional}
s\sqrt{n}\langle v_n,\psi\rangle_\circ+s\nu_n\left(q'_h(\psi,Z_i)\right)\geq o_P(1).
\end{equation}
Finally, $s=1$ in \eqref{eq:two_sided_ineq_functional} gives
\begin{equation*}
\sqrt{n}\langle v_n,\psi\rangle_\circ\geq-\nu_n\left(q'_h(\psi,Z_i)\right)+o_P(1),
\end{equation*}
while taking $s=-1$ results in
\begin{equation*}
\sqrt{n}\langle v_n,\psi\rangle_\circ\leq-\nu_n\left(q'_h(\psi,Z_i)\right)+o_P(1),
\end{equation*}
which yield
\begin{equation}
\label{eq:asym_linear_functional}
\sqrt{n}\langle \widehat{h}_n-h,\psi\rangle_\circ=-\nu_n\left(q'_h(\psi,Z_i)\right)+o_P(1).
\end{equation}
Combining \eqref{eq:functional_linearization} and \eqref{eq:asym_linear_functional}, we obtain
\begin{equation*}
\sqrt{n}\{\Psi(\widehat{h}_n)-\Psi(h)\}=-\nu_n\left(q'_h(\psi,Z_i)\right)+o_P(1).
\end{equation*}
By Assumption~\ref{assumption:functional_score_clt} and Slutsky's theorem, we obtained the required result
\begin{equation*}
\sqrt{n}\{\Psi(\widehat{h}_n)-\Psi(h)\}\todist \mathcal{N}(0,\sigma_\Psi^2),
\end{equation*}
since multiplication by $-1$ does not change the limiting variance.
\end{proof}

\begin{remark}[Relation to Shen's local perturbation argument]
    The proof does not rely on an empirical score equation at $\widehat{h}_n$. Instead, it compares the empirical loss contrast at $\widehat{h}_n$ with that at the projected local perturbations $\widetilde{h}_{n,s}$, $s\in\{-1,1\}$. This is the device that makes the argument compatible with sieve and constrained estimators, where $\widehat{h}_n$ may lie on the boundary of the approximating space.
\end{remark}

\clearpage
\section{Auxiliary results}
\label{app:auxiliary_results}

\subsection{Geometric and sampling inequalities}
\label{app:geometric_sampling_inequalities}
\begin{lemma}[Kabatianski, Zeger]
\label{lemma:KabatianskiZeger}
	Given a point set $P \subset \mathbb{R}^d$ and $x \in P$, the maximum number of points in $P$ that can have $x$ as nearest neighbor is bounded by the $d$-dimensional kissing number $\psi_d$, where $\psi_d \leq 2^{0.401 d(1+\mathrm{o}(1))}$ as $d \rightarrow \infty$.
\end{lemma}

\subsection{Matrix concentration and perturbation inequalities}
\label{app:matrix_inequalities}
The following matrix-valued law of large numbers is needed for the $L^2$ consistency result in Section~\ref{subsec:L2_estimation}.

\begin{lemma}[Rudelson's Matrix LLN, \citealp{Belloni2015}, Lemma 6.2]\label{lemma:matrix_lln_iid}
    Let \( Q_1, \ldots, Q_n \) be a sequence of independent symmetric non-negative \( K \times K \)-matrix valued random variables with \( K \geq 2 \) such that \( Q = \mathbb{E}[Q_i] \) and \( \|Q_i\| \leq M \) a.s., then for \( \widehat{Q} = \mathbb{E}_n[Q_i] \)
    \[
        \mathbb{E} \|\widehat{Q} - Q\| \lesssim \frac{M \log(K)}{n} + \sqrt{\frac{M \|Q\| \log(K)}{n}}.
    \]
    In particular, if \( Q_i = p_i p_i^{\top} \), with \( \|p_i\| \leq \xi_K \) a.s., then
    \[
        \mathbb{E} \|\widehat{Q} - Q\| \lesssim \frac{\xi_K^2 \log(K)}{n} + \sqrt{\frac{\xi_K^2 \|Q\| \log(K)}{n}}.
    \]
\end{lemma}

\subsection{Entropy and empirical-process inequalities}
\label{app:empirical_process_inequalities}

The following statement of Bernstein's inequality, together with its conditional statement in Lemma~\ref{lemma:conditional_bernstein} are used in Subsection~\ref{subsubsec:variance_uniform_bound}.

\begin{theorem}[Bernstein's inequality, \citealp{Boucheron2013}]
\label{theorem:bernstein_boucheron}
    Let $X_1,\ldots,X_n$ be independent real-valued random variables. Assume that there exist positive numbers $v$ and $c$ such that $\sum_{i=1}^n \mathbb{E}[X_i^2]\leq v$ and
    \begin{equation*}
    \sum_{i=1}^n \mathbb{E}\bigl[(X_i)_+^q\bigr]\leq\frac{q!}{2}\,v\,c^{q-2}\qquad\text{for all integers }q\geq 3,
    \end{equation*}
    where $x_+:=\max(x,0)$.
    
    If $S:=\sum_{i=1}^n\bigl(X_i-\mathbb{E}X_i\bigr)$, then, for every $\lambda\in(0,1/c)$ and every $t>0$,
    \begin{equation*}
    \psi_S(\lambda)\leq\frac{v\lambda^2}{2(1-c\lambda)}
    \end{equation*}
    and
    \begin{equation*}
    \psi_S^*(t)\geq\frac{v}{c^2}h_1\left(\frac{ct}{v}\right),
    \end{equation*}
    where $h_1(u)=1+u-\sqrt{1+2u}$ for $u>0$. In particular, for every $t>0$,
    \begin{equation*}
    \mathbb{P}\left(S\geq\sqrt{2vt}+ct \right) \leq e^{-t}.
    \end{equation*}
\end{theorem}
The following conditional formulation is a consequence of Theorem~\ref{theorem:bernstein_boucheron} applied under the regular conditional distribution given $\mathcal{G}$, and then symmetrically applied to both $\xi_i$ and $-\xi_i$.

\begin{lemma}[Conditional Bernstein inequality]
\label{lemma:conditional_bernstein}
    Let $\mathcal{G}$ be a sub-$\sigma$-field, and let $\xi_1,\ldots,\xi_n$ be random variables that are conditionally independent
    given $\mathcal{G}$. Suppose that, almost surely, $ \E[\xi_i\mid\mathcal{G}]=0$ and $|\xi_i|\leq b$ 
    for every $i=1,\ldots,n$. 
    Define the conditional variance proxy
    \begin{equation*}
        v := \sum_{i=1}^n \E\big[ \xi_i^2 \mid \mathcal{G} \big].
    \end{equation*}
    Then, for every $t>0$,
    \begin{equation*}
        \mathbb{P}\left( \left|\sum_{i=1}^n\xi_i \right| \geq t \, \middle| \, \mathcal{G} \right) 
        \leq 
        2\exp\left( -\frac{t^2}{2\left(v+bt/3\right)} \right)
    \end{equation*}
    almost surely.
\end{lemma}
\begin{lemma}
\label{lemma:entropy_integral_bound}
    For all $a \geq 1$, it holds
    \begin{equation*}
        \int_0^1 \sqrt{\log\left( \frac{a}{x} \right)} \,dx \leq \sqrt{\log(a)} + 1 .
    \end{equation*}
\end{lemma}

\begin{proof}[Proof of Lemma~\ref{lemma:entropy_integral_bound}]
    The definite integral, given that $a$ is strictly positive, is equal to
    \begin{equation*}
        \int_0^1 \sqrt{\log\left( \frac{a}{x} \right)} \,dx
        =
        \sqrt{\log(a)} + a \frac{\sqrt{\pi}}{2}\textnormal{erfc}\left( \sqrt{\log(a)} \right) ,
    \end{equation*}
    where $\textnormal{erfc}(\cdot)$ is the complementary error function. Applying, for all $t > 0$, the bound
    \begin{equation*}
        \textnormal{erfc}(t)  
         \leq
         \frac{2}{\sqrt{\pi}} \frac{e^{-t^2}}{t + \sqrt{t^2 + \frac{4}{\pi}}} ,
    \end{equation*}
    (see, e.g., \textit{NIST Handbook of Mathematical Functions}, \cite{NISTHandbookMathematical2010}, Section~7.8.2.) the latter term in the integral can be immediately controlled by
    \begin{equation*}
        a \frac{\sqrt{\pi}}{2}\textnormal{erfc}\left( \sqrt{\log(a)} \right) 
        \leq 
        \frac{1}{\sqrt{\log(a)} + \sqrt{\log(a) + \frac{4}{\pi}}}
        \leq 
        1 
    \end{equation*}
    since $a \geq 1$ and $\frac{4}{\pi} > 1$.
\end{proof}

\clearpage

\section{Additional numerical experiments and results}
\label{appendix:additional_simulations}

\subsection{Residual, wild and score bootstrap inference for evaluation functional}
\label{appendix:sims_resid_wild_score_bootstrap}

We now consider only the pointwise evaluation functional $\Psi_x(h) = h(x)$, which is treated as an irregular sieve functional using the sieve-dependent studentization described below.
We evaluate the RWNN estimate and construct pointwise confidence intervals on a uniformly spaced grid of $31$ points over the interval $[-0.9, 0.9]$. Restricting the evaluation grid to the interior of $[-1,1]$ avoids boundary estimation issues and allows us to analyze the local coverage properties of the inference procedures.
We discuss two RWNN specifications: we consider both ReLU and tanh activations with the same weight distributions and specifications used in the simulations of Section~\ref{subsec:sims_statistical_inference}. %
We also apply the same Monte Carlo design as in that setting, redrawing RWNN weights across different data draws.
We construct pointwise confidence intervals over the grid by using three bootstrapping approaches: residual (RB), wild (WB), and score-based (SB) bootstrap.

We generate bootstrap response variables as described in Section~\ref{subsec:sims_bootstrap} for the residual bootstrap.
As before, in the RB the innovations are sampled independently from the centered empirical residual distribution, meaning $\epsilon_i^* := \widehat{\epsilon}_{i}^* - \overline{\widehat{\epsilon}}$, with
$\widehat{\epsilon}_i^* \overset{\mathrm{i.i.d.}}{\sim} \Prob_{\widehat{\epsilon},n}$ and $\overline{\widehat{\epsilon}}:= n\inv \sum_{i=1}^n \widehat{\epsilon}_i$. 
For WB, we instead set $\epsilon_i^* := (\widehat{\epsilon}_{i} - \overline{\widehat{\epsilon}}\,) \xi_i$, where  $\{\xi_i\}_{i=1}^{n}$ are i.i.d. multipliers, independent of data and satisfying $\E[{\xi}_i] = 0$, $\text{Var}({\xi}_i) = 1$, and having a finite absolute third moment. In the simulations, we take ${\xi}_i \sim \mathcal{N}(0,1)$.\footnote{Alternative choices include, for example, Rademacher and recentered exponential multipliers.}

For bootstrap schemes $b\in\{\mathrm{RB},\mathrm{WB}\}$, let $\widehat{h}_{n,\lambda}^{*,b}$ denote the RWNN estimator obtained by refitting the model to the corresponding bootstrap sample, using the same realized random parameters, dimension $K_n$, and the same $\lambda_{K_n,n}$ as in the original sample. For each $x$ in the evaluation grid, let $z^{*,b}_{\alpha}(x)$ denote the conditional $\alpha$-quantile of the bootstrap distribution of $\widehat{h}_{n,\lambda}^{*,b}(x)$, given the original sample. The corresponding percentile-bootstrap confidence interval is
\begin{equation*}
    \text{CI}_{1-\alpha,b}(\Psi;x) := \left[ z^{*,b}_{\alpha/2}(x), \; z^{*,b}_{1-\alpha/2}(x) \right], \quad b\in\{\mathrm{RB},\mathrm{WB}\}.
\end{equation*}
Both RB and WB schemes avoid direct estimation of the asymptotic sieve-dependent variance of the pointwise evaluation functional.

The score bootstrap procedure we implement is adapted from the scheme proposed by \cite{Chen2018} for uniform inference in nonparametric instrumental-variable models. 
For the evaluation functional $\Psi_x(h)=h(x)$, it is easy to show that $\Psi'_{\widehat{h}_{n,\lambda}}(v) = v(x)$ for every sieve direction $v$. Hence,  $\Psi'_{\widehat{h}_{n,\lambda}}(g_{K_n}) = g_{K_n}(x)$. Similarly to \eqref{eq:def_ridge-adjusted_representer_psi2}, the empirical ridge-adjusted Riesz representer of $\Psi_x$ is
\begin{equation*}
    \widehat{\psi}_{\Psi_x,\lambda}(z):=g_{K_n}(z)^\top \widehat{\Gamma}_{K_n,n,\lambda}^{-1} g_{K_n}(x), \quad z\in \mathcal{X},
\end{equation*}
Consequently, the sieve variance estimator for pointwise evaluation admits the equivalent representation
\begin{equation*}
     \widehat{\sigma}^2_{\Psi_x}(x) = g_{K_n}(x)^\top \widehat{\Omega}_{K_n, \lambda} g_{K_n}(x), \quad x\in \mathcal{X}.
\end{equation*}
Let $\{\xi_i\}_{i=1}^n$ be i.i.d.\ standard Gaussian multipliers, independent of the data. The pointwise score-bootstrap $t$-statistic is defined by
\begin{align*}
    \frak{z}_{n}^*(x)
    & := \frac{1}{\widehat{\sigma}_{\Psi_x}(x)\sqrt n} \sum_{i=1}^n \widehat{\epsilon}_i \widehat\psi_{\Psi_x,\lambda}(X_i) \xi_i\: =
    \sqrt{\frac{n}{\widehat{\sigma}^2_{\Psi_x}(x)}} \sum_{i=1}^n p_{i,n,\lambda}(x) \, \widehat{\epsilon}_i \, \xi_i
    ,
\end{align*}
where $p_{i,n,\lambda}(x)$ denotes the ridge sieve smoothing weight defined in \eqref{eq:ridge_smoother_weights}, and where the second equality follows from $\widehat{\psi}_{\Psi_x,\lambda}(X_i) = g_{K_n}(X_i)^\top \widehat{\Gamma}_{K_n,n,\lambda}^{-1} g_{K_n}(x) = n\,p_{i,n,\lambda}(x)$.\footnote{Note that, by construction, conditional on the data, the score bootstrap distribution is, in fact, Gaussian at each fixed $x$: \cite{Chen2018} take the supremum over $x \in \mathcal{X}$ to obtain uniform critical values for $\widehat{h}_{n,\lambda}$, while we skip this step.}

Let $z^\circ_{\alpha}(x)$ denote the conditional $\alpha$-quantile, given the original sample, of $\abs{ \frak{z}_{n}^*(x) }$. The corresponding $100(1-\alpha)\%$ pointwise score bootstrap CI is
\begin{equation*}
    \text{CI}_{1-\alpha, \textsc{sb}}(\Psi_x; x)
    :=
    \left[ \widehat{h}_{n,\lambda}(x) - \frac{\widehat{\sigma}_{\Psi_x}(x)}{\sqrt{n}} z^\circ_{1-\alpha}(x), \: \widehat{h}_{n,\lambda}(x) + \frac{\widehat{\sigma}_{\Psi_x}(x)}{\sqrt{n}} z^\circ_{1-\alpha}(x) \right]
\end{equation*}
We use $\text{MC} = 1000$ Monte Carlo replications and $\frak{B} = 1000$ bootstrap resamplings. To match the pipeline of real-world applications of RWNN, we again independently resample the RWNN weights $\{ A_k, B_k \}_{k=1}^{K_n}$ at each Monte Carlo data draw. The reported Monte Carlo coverage probabilities thus average over both the sampling variation in the data and the randomness in the hidden-layer parameters, although each bootstrap confidence interval is conditional on the realized random features. The exact model specifications used are summarized in Table~\ref{tab:rwnn_specs}.
 
We report the results for both coverage and bias in the plots of Appendix~\ref{appendix:additional_figures}. 
We find that all bootstrap schemes perform comparatively well for both tanh and ReLU RWNN specifications in the simpler specification involving $g_3$. More pronounced differences arise for $g_1$ and $g_2$. For $g_2$, the ReLU-based confidence intervals tend to systematically undercover in regions where the regression function is steeper, while the tanh CIs are substantially better calibrated. In these regions, the RB procedure generally attains the highest coverage among the three bootstrap methods.
For $g_1$, the ReLU RWNN confidence intervals perform particularly poorly near the origin, where the regression function has a sharp peak, with observed coverage at times falling below $50\%$ on the interval $[-0.25, 0.25]$. The confidence intervals based on the $\tanh$ RWNN perform better over the entire evaluation region, and the RB procedure again generally yields the highest coverage probabilities.

\newpage

\subsection{Additional tables}
\label{appendix:additional_tables}

~\\[4em]

\begin{table}[h!]
\caption{Asymptotic inference results for $\Psi_{2}(\widehat{h}_{n, \lambda})$ with $\lambda = 10^{-2}$.}
\label{tab:coverage_ridge_lambda_1e-2}
\centering
\setlength\tabcolsep{4pt}
\setlength\extrarowheight{1pt}
\setlength\extracolsep{0.1pt}
\linespread{1.}\selectfont\centering
\small
\begin{tabular}{lllccccccccc}
\toprule
 &  &  & \multicolumn{3}{c}{Gaussian--Gaussian} & \multicolumn{3}{c}{Gaussian--Uniform} & \multicolumn{3}{c}{Uniform--Uniform} \\
    \cmidrule(lr){4-6} \cmidrule(lr){7-9} \cmidrule(lr){10-12}
 $n$ & $K_n$ & Method & Bias & Coverage & Length & Bias & Coverage & Length & Bias & Coverage & Length \\
\midrule
\multicolumn{12}{l}{$h = g_1$ ~\raisedrule[0.2em]{0.2pt}} \\
\multirow[c]{2}{*}{100} & \multirow[c]{2}{*}{15} & RSZ & -1.565 & 0.582 & 0.356 & -1.474 & 0.611 & 0.360 & -1.780 & 0.575 & 0.395 \\
 &  & SNW & -1.565 & 0.570 & 0.348 & -1.474 & 0.605 & 0.355 & -1.780 & 0.559 & 0.385 \\
\multirow[c]{2}{*}{300} & \multirow[c]{2}{*}{20} & RSZ & -1.294 & 0.318 & 0.195 & -1.231 & 0.370 & 0.199 & -1.495 & 0.297 & 0.218 \\
 &  & SNW & -1.294 & 0.293 & 0.189 & -1.231 & 0.342 & 0.194 & -1.495 & 0.273 & 0.210 \\
\multirow[c]{2}{*}{1000} & \multirow[c]{2}{*}{30} & RSZ & -0.914 & 0.090 & 0.099 & -0.849 & 0.132 & 0.100 & -1.076 & 0.078 & 0.110 \\
 &  & SNW & -0.914 & 0.081 & 0.097 & -0.849 & 0.124 & 0.098 & -1.076 & 0.069 & 0.106 \\
\multirow[c]{2}{*}{3000} & \multirow[c]{2}{*}{40} & RSZ & -0.746 & 0.003 & 0.055 & -0.706 & 0.016 & 0.055 & -0.890 & 0.002 & 0.060 \\
 &  & SNW & -0.746 & 0.003 & 0.053 & -0.706 & 0.013 & 0.054 & -0.890 & 0.002 & 0.058 \\
\multicolumn{12}{l}{$h = g_2$ ~\raisedrule[0.2em]{0.2pt}} \\
\multirow[c]{2}{*}{100} & \multirow[c]{2}{*}{15} & RSZ & -1.760 & 0.100 & 0.176 & -1.561 & 0.151 & 0.173 & -0.931 & 0.518 & 0.163 \\
 &  & SNW & -1.760 & 0.079 & 0.152 & -1.561 & 0.115 & 0.152 & -0.931 & 0.487 & 0.155 \\
\multirow[c]{2}{*}{300} & \multirow[c]{2}{*}{20} & RSZ & -1.403 & 0.015 & 0.101 & -1.214 & 0.040 & 0.098 & -0.632 & 0.391 & 0.093 \\
 &  & SNW & -1.403 & 0.012 & 0.090 & -1.214 & 0.033 & 0.088 & -0.632 & 0.373 & 0.090 \\
\multirow[c]{2}{*}{1000} & \multirow[c]{2}{*}{30} & RSZ & -1.036 & 0.001 & 0.054 & -0.846 & 0.005 & 0.053 & -0.391 & 0.278 & 0.051 \\
 &  & SNW & -1.036 & 0.001 & 0.049 & -0.846 & 0.002 & 0.049 & -0.391 & 0.263 & 0.049 \\
\multirow[c]{2}{*}{3000} & \multirow[c]{2}{*}{40} & RSZ & -0.808 & 0.000 & 0.031 & -0.655 & 0.000 & 0.030 & -0.284 & 0.111 & 0.029 \\
 &  & SNW & -0.808 & 0.000 & 0.029 & -0.655 & 0.000 & 0.028 & -0.284 & 0.110 & 0.029 \\
\multicolumn{12}{l}{$h = g_3$ ~\raisedrule[0.2em]{0.2pt}} \\
\multirow[c]{2}{*}{100} & \multirow[c]{2}{*}{15} & RSZ & -0.190 & 0.892 & 0.168 & -0.143 & 0.901 & 0.166 & -0.070 & 0.909 & 0.162 \\
 &  & SNW & -0.190 & 0.899 & 0.170 & -0.143 & 0.913 & 0.168 & -0.070 & 0.922 & 0.165 \\
\multirow[c]{2}{*}{300} & \multirow[c]{2}{*}{20} & RSZ & -0.133 & 0.887 & 0.097 & -0.100 & 0.907 & 0.097 & -0.050 & 0.927 & 0.096 \\
 &  & SNW & -0.133 & 0.886 & 0.097 & -0.100 & 0.911 & 0.097 & -0.050 & 0.924 & 0.096 \\
\multirow[c]{2}{*}{1000} & \multirow[c]{2}{*}{30} & RSZ & -0.081 & 0.905 & 0.053 & -0.059 & 0.928 & 0.053 & -0.024 & 0.945 & 0.053 \\
 &  & SNW & -0.081 & 0.903 & 0.053 & -0.059 & 0.928 & 0.053 & -0.024 & 0.942 & 0.053 \\
\multirow[c]{2}{*}{3000} & \multirow[c]{2}{*}{40} & RSZ & -0.070 & 0.846 & 0.031 & -0.053 & 0.878 & 0.031 & -0.027 & 0.929 & 0.031 \\
 &  & SNW & -0.070 & 0.841 & 0.031 & -0.053 & 0.877 & 0.031 & -0.027 & 0.929 & 0.031 \\
\bottomrule
\end{tabular}
\\[0.7em]
\justifying
\noindent
{\footnotesize
    \textit{Notes:} 
    Bias, coverage rate, and interval length of asymptotic confidence intervals for $\Psi_2(h)$ estimated using $\text{MC} = 1000$ replications.
    The method column reports the type of functional and variance estimators (sandwich-type or Riesz-type) used to compute asymptotic confidence intervals. 
    The Gaussian-Gaussian, Gaussian-Uniform, and Uniform-Uniform column groupings indicate random sampling laws of $A_k$ and $B_k$, respectively.
    Bias is scaled by a factor of 10 to improve legibility. 
}
\end{table}

\clearpage

\begin{table}[p!]
\caption{Asymptotic inference results for $\Psi_{2}(\widehat{h}_{n, \lambda})$ with $\lambda = 10^{-4}$.}
\label{tab:coverage_ridge_lambda_1e-4}
\centering
\setlength\tabcolsep{4pt}
\setlength\extrarowheight{1pt}
\setlength\extracolsep{0.1pt}
\linespread{1.}\selectfont\centering
\small
\begin{tabular}{lllccccccccc}
\toprule
 &  &  & \multicolumn{3}{c}{Gaussian--Gaussian} & \multicolumn{3}{c}{Gaussian--Uniform} & \multicolumn{3}{c}{Uniform--Uniform} \\
    \cmidrule(lr){4-6} \cmidrule(lr){7-9} \cmidrule(lr){10-12}
 $n$ & $K_n$ & Method & Bias & Coverage & Length & Bias & Coverage & Length & Bias & Coverage & Length \\
\midrule
\multicolumn{12}{l}{$h = g_1$ ~\raisedrule[0.2em]{0.2pt}} \\
\multirow[c]{2}{*}{100} & \multirow[c]{2}{*}{15} & RSZ & -0.115 & 0.900 & 0.259 & -0.120 & 0.896 & 0.265 & -0.335 & 0.865 & 0.298 \\
 &  & SNW & -0.115 & 0.925 & 0.269 & -0.120 & 0.925 & 0.277 & -0.335 & 0.888 & 0.311 \\
\multirow[c]{2}{*}{300} & \multirow[c]{2}{*}{20} & RSZ & -0.095 & 0.917 & 0.150 & -0.118 & 0.910 & 0.152 & -0.229 & 0.871 & 0.162 \\
 &  & SNW & -0.095 & 0.922 & 0.153 & -0.118 & 0.921 & 0.155 & -0.229 & 0.880 & 0.165 \\
\multirow[c]{2}{*}{1000} & \multirow[c]{2}{*}{30} & RSZ & -0.032 & 0.936 & 0.082 & -0.035 & 0.939 & 0.082 & -0.084 & 0.897 & 0.084 \\
 &  & SNW & -0.032 & 0.934 & 0.083 & -0.035 & 0.941 & 0.083 & -0.084 & 0.901 & 0.085 \\
\multirow[c]{2}{*}{3000} & \multirow[c]{2}{*}{40} & RSZ & -0.035 & 0.939 & 0.047 & -0.038 & 0.925 & 0.048 & -0.063 & 0.894 & 0.048 \\
 &  & SNW & -0.035 & 0.940 & 0.048 & -0.038 & 0.930 & 0.048 & -0.063 & 0.895 & 0.048 \\
\multicolumn{12}{l}{$h = g_2$ ~\raisedrule[0.2em]{0.2pt}} \\
\multirow[c]{2}{*}{100} & \multirow[c]{2}{*}{15} & RSZ & -0.085 & 0.875 & 0.157 & -0.034 & 0.884 & 0.156 & -0.027 & 0.886 & 0.156 \\
 &  & SNW & -0.085 & 0.890 & 0.162 & -0.034 & 0.901 & 0.161 & -0.027 & 0.907 & 0.163 \\
\multirow[c]{2}{*}{300} & \multirow[c]{2}{*}{20} & RSZ & -0.050 & 0.906 & 0.093 & -0.025 & 0.925 & 0.092 & -0.011 & 0.928 & 0.092 \\
 &  & SNW & -0.050 & 0.907 & 0.094 & -0.025 & 0.930 & 0.093 & -0.011 & 0.930 & 0.093 \\
\multirow[c]{2}{*}{1000} & \multirow[c]{2}{*}{30} & RSZ & -0.033 & 0.930 & 0.051 & -0.020 & 0.932 & 0.051 & -0.004 & 0.937 & 0.051 \\
 &  & SNW & -0.033 & 0.926 & 0.051 & -0.020 & 0.934 & 0.051 & -0.004 & 0.941 & 0.051 \\
\multirow[c]{2}{*}{3000} & \multirow[c]{2}{*}{40} & RSZ & -0.026 & 0.926 & 0.030 & -0.017 & 0.941 & 0.030 & -0.005 & 0.949 & 0.029 \\
 &  & SNW & -0.026 & 0.926 & 0.030 & -0.017 & 0.943 & 0.030 & -0.005 & 0.948 & 0.030 \\
\multicolumn{12}{l}{$h = g_3$ ~\raisedrule[0.2em]{0.2pt}} \\
\multirow[c]{2}{*}{100} & \multirow[c]{2}{*}{15} & RSZ & 0.037 & 0.916 & 0.161 & 0.050 & 0.916 & 0.161 & 0.065 & 0.907 & 0.159 \\
 &  & SNW & 0.037 & 0.926 & 0.167 & 0.050 & 0.923 & 0.166 & 0.065 & 0.923 & 0.165 \\
\multirow[c]{2}{*}{300} & \multirow[c]{2}{*}{20} & RSZ & 0.018 & 0.936 & 0.096 & 0.021 & 0.933 & 0.096 & 0.026 & 0.927 & 0.095 \\
 &  & SNW & 0.018 & 0.939 & 0.097 & 0.021 & 0.938 & 0.097 & 0.026 & 0.937 & 0.097 \\
\multirow[c]{2}{*}{1000} & \multirow[c]{2}{*}{30} & RSZ & 0.014 & 0.953 & 0.053 & 0.015 & 0.953 & 0.053 & 0.018 & 0.949 & 0.053 \\
 &  & SNW & 0.014 & 0.949 & 0.053 & 0.015 & 0.950 & 0.053 & 0.018 & 0.947 & 0.053 \\
\multirow[c]{2}{*}{3000} & \multirow[c]{2}{*}{40} & RSZ & -0.001 & 0.944 & 0.031 & -0.001 & 0.944 & 0.031 & 0.001 & 0.943 & 0.031 \\
 &  & SNW & -0.001 & 0.941 & 0.031 & -0.001 & 0.944 & 0.031 & 0.001 & 0.942 & 0.031 \\
\bottomrule
\end{tabular}
\\[0.7em]
\justifying
\noindent
{\footnotesize
    \textit{Notes:} 
    Bias, coverage rate, and interval length of asymptotic confidence intervals for $\Psi_2(h)$ estimated using $\text{MC} = 1000$ replications.
    The method column reports the type of functional and variance estimators (sandwich-type or Riesz-type) used to compute asymptotic confidence intervals. 
    The Gaussian-Gaussian, Gaussian-Uniform, and Uniform-Uniform column groupings indicate random sampling laws of $A_k$ and $B_k$, respectively.
    Bias is scaled by a factor of 10 to improve legibility. 
}
\end{table}

\clearpage

\begin{table}
\caption{Asymptotic inference results for $\Psi_{2}(\widehat{h}_{n, \lambda})$ with $\lambda = 10^{-8}$.}
\label{tab:coverage_ridge_lambda_1e-08}
\centering
\setlength\tabcolsep{4pt}
\setlength\extrarowheight{1pt}
\setlength\extracolsep{0.1pt}
\linespread{1.}\selectfont\centering
\small
\begin{tabular}{lllccccccccc}
\toprule
 &  &  & \multicolumn{3}{c}{Gaussian--Gaussian} & \multicolumn{3}{c}{Gaussian--Uniform} & \multicolumn{3}{c}{Uniform--Uniform} \\
    \cmidrule(lr){4-6} \cmidrule(lr){7-9} \cmidrule(lr){10-12}
 $n$ & $K_n$ & Method & Bias & Coverage & Length & Bias & Coverage & Length & Bias & Coverage & Length \\
\midrule
\multicolumn{12}{l}{$h = g_1$ ~\raisedrule[0.2em]{0.2pt}} \\
\multirow[c]{2}{*}{100} & \multirow[c]{2}{*}{15} & RSZ & 0.139 & 0.889 & 0.237 & 0.129 & 0.878 & 0.239 & 0.086 & 0.871 & 0.260 \\
 &  & SNW & 0.139 & 0.898 & 0.251 & 0.129 & 0.902 & 0.253 & 0.086 & 0.902 & 0.285 \\
\multirow[c]{2}{*}{300} & \multirow[c]{2}{*}{20} & RSZ & 0.021 & 0.932 & 0.144 & 0.022 & 0.928 & 0.145 & -0.000 & 0.930 & 0.148 \\
 &  & SNW & 0.021 & 0.937 & 0.149 & 0.022 & 0.935 & 0.149 & -0.000 & 0.936 & 0.152 \\
\multirow[c]{2}{*}{1000} & \multirow[c]{2}{*}{30} & RSZ & 0.021 & 0.936 & 0.081 & 0.021 & 0.937 & 0.081 & 0.023 & 0.936 & 0.081 \\
 &  & SNW & 0.021 & 0.938 & 0.082 & 0.021 & 0.939 & 0.082 & 0.023 & 0.937 & 0.082 \\
\multirow[c]{2}{*}{3000} & \multirow[c]{2}{*}{40} & RSZ & -0.006 & 0.948 & 0.047 & -0.006 & 0.951 & 0.047 & -0.005 & 0.950 & 0.047 \\
 &  & SNW & -0.006 & 0.953 & 0.047 & -0.006 & 0.954 & 0.047 & -0.005 & 0.952 & 0.047 \\
\multicolumn{12}{l}{$h = g_2$ ~\raisedrule[0.2em]{0.2pt}} \\
\multirow[c]{2}{*}{100} & \multirow[c]{2}{*}{15} & RSZ & 0.129 & 0.885 & 0.150 & 0.138 & 0.875 & 0.150 & 0.131 & 0.866 & 0.150 \\
 &  & SNW & 0.129 & 0.914 & 0.159 & 0.138 & 0.909 & 0.159 & 0.131 & 0.908 & 0.162 \\
\multirow[c]{2}{*}{300} & \multirow[c]{2}{*}{20} & RSZ & 0.052 & 0.932 & 0.090 & 0.052 & 0.936 & 0.090 & 0.056 & 0.926 & 0.090 \\
 &  & SNW & 0.052 & 0.935 & 0.093 & 0.052 & 0.937 & 0.093 & 0.056 & 0.932 & 0.093 \\
\multirow[c]{2}{*}{1000} & \multirow[c]{2}{*}{30} & RSZ & 0.016 & 0.937 & 0.051 & 0.017 & 0.936 & 0.051 & 0.020 & 0.934 & 0.050 \\
 &  & SNW & 0.016 & 0.943 & 0.051 & 0.017 & 0.937 & 0.051 & 0.020 & 0.937 & 0.051 \\
\multirow[c]{2}{*}{3000} & \multirow[c]{2}{*}{40} & RSZ & 0.006 & 0.949 & 0.029 & 0.006 & 0.949 & 0.029 & 0.007 & 0.948 & 0.029 \\
 &  & SNW & 0.006 & 0.952 & 0.029 & 0.006 & 0.951 & 0.029 & 0.007 & 0.951 & 0.029 \\
\multicolumn{12}{l}{$h = g_3$ ~\raisedrule[0.2em]{0.2pt}} \\
\multirow[c]{2}{*}{100} & \multirow[c]{2}{*}{15} & RSZ & 0.114 & 0.894 & 0.158 & 0.117 & 0.889 & 0.157 & 0.134 & 0.882 & 0.156 \\
 &  & SNW & 0.114 & 0.927 & 0.166 & 0.117 & 0.913 & 0.167 & 0.134 & 0.902 & 0.167 \\
\multirow[c]{2}{*}{300} & \multirow[c]{2}{*}{20} & RSZ & 0.042 & 0.923 & 0.095 & 0.042 & 0.916 & 0.095 & 0.046 & 0.918 & 0.094 \\
 &  & SNW & 0.042 & 0.933 & 0.097 & 0.042 & 0.936 & 0.097 & 0.046 & 0.930 & 0.097 \\
\multirow[c]{2}{*}{1000} & \multirow[c]{2}{*}{30} & RSZ & 0.023 & 0.948 & 0.053 & 0.023 & 0.948 & 0.053 & 0.026 & 0.947 & 0.053 \\
 &  & SNW & 0.023 & 0.948 & 0.053 & 0.023 & 0.949 & 0.053 & 0.026 & 0.945 & 0.053 \\
\multirow[c]{2}{*}{3000} & \multirow[c]{2}{*}{40} & RSZ & 0.002 & 0.943 & 0.031 & 0.003 & 0.944 & 0.031 & 0.004 & 0.941 & 0.031 \\
 &  & SNW & 0.002 & 0.943 & 0.031 & 0.003 & 0.944 & 0.031 & 0.004 & 0.945 & 0.031 \\
\bottomrule
\end{tabular}
\\[0.7em]
\justifying
\noindent
{\footnotesize
    \textit{Notes:} 
    Bias, coverage rate, and interval length of asymptotic confidence intervals for $\Psi_2(h)$ estimated using $\text{MC} = 1000$ replications.
    The method column reports the type of functional and variance estimators (sandwich-type or Riesz-type) used to compute asymptotic confidence intervals. 
    The Gaussian-Gaussian, Gaussian-Uniform, and Uniform-Uniform groupings indicate random sampling laws of $A_k$ and $B_k$, respectively.
    Bias is scaled by a factor of 10 to improve legibility. 
}
\end{table}

\clearpage

\begin{table}
\centering
\setlength\tabcolsep{10pt}
\setlength\extrarowheight{1pt}
\setlength\extracolsep{0.1pt}
\linespread{1.1}\selectfont\centering
\caption{RWNN model specifications for bootstrap inference}
\begin{tabular}{ccccccc}
    \toprule
     & \multicolumn{2}{c}{Weight Sampling} & \multicolumn{2}{c}{Scaling} & \\
     \cmidrule(lr){2-3} \cmidrule(lr){4-5}
     Activation & $A_k$ & $B_k$ & $A_k$ & $B_k$ & $\lambda$ \\
    \midrule
    Tanh & Gaussian & Gaussian & 5.0 & 1.0 & $10^{-6}$ \\
    Tanh & Gaussian & Uniform  & 5.0 & 2.0  & $10^{-6}$ \\
    Tanh & Uniform  & Uniform  & 10.0 & 5.0 & $10^{-6}$ \\[5pt]
    ReLU & Gaussian & Gaussian & 5.0 & 2.0 & $10^{-6}$ \\
    ReLU & Gaussian & Uniform  & 5.0 & 2.0 & $10^{-6}$ \\
    ReLU & Uniform  & Uniform  & 1.5 & 0.7 & $10^{-6}$ \\
    \bottomrule
\end{tabular}
\label{tab:rwnn_specs}
\\[0.7em]
\justifying
\noindent
{\footnotesize
}
\end{table}

\clearpage
\newpage

\subsection{Additional figures}
\label{appendix:additional_figures}

\begin{figure}[h!]
    \centering
    \includegraphics[width=0.95\linewidth]{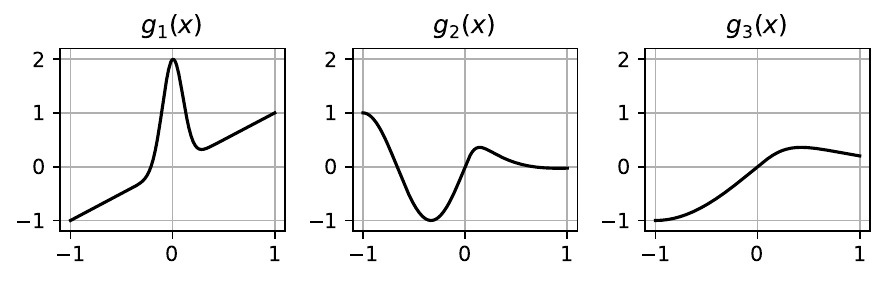}
    \caption{Regression functions from \cite{Hall2013}.}
    \label{fig:hh13_functions}
\end{figure}

In Figure~\ref{fig:hh13_functions} we provide a plot of the three functions---$g_1$, $g_2$ and $g_3$---from \cite{Hall2013} that we use in our simulations.

Figures~\ref{fig:inference_rayleigh_bias_cover_gaussian_uniform} and \ref{fig:inference_rayleigh_bias_cover_uniform_uniform} present the results obtained when running the simulations of Section~\ref{subsec:sims_bootstrap} with different sampling laws for the RWNN weights.

Figures~\ref{fig:inference_boot} to \ref{fig:inference_boot__UU} plot the pointwise coverage probabilities associated with the evaluation functional over a fixed grid for the residual, wild and score bootstrap CIs. Figures~\ref{fig:inference_bias__NU} to \ref{fig:inference_bias__UU} plot the pointwise bias of the RWNN ridge estimator over the same evaluation grid.

\vspace{10em}

\afterpage{%
\begin{figure}[p!]
    \centering
    \begin{subfigure}{0.98\textwidth}
    \centering
        \includegraphics[width=\textwidth]{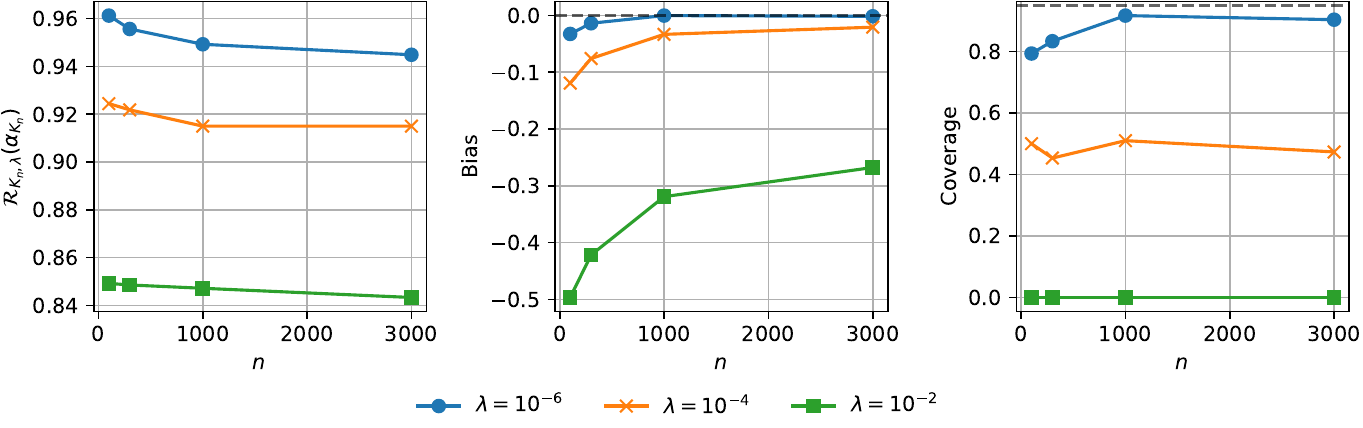}
        \caption{$\Psi_0(g_1) = g_1(0)$}
    \end{subfigure}
    \\[1em]
    \vspace{1em}
    \begin{subfigure}{0.98\textwidth}
    \centering
        \includegraphics[width=\textwidth]{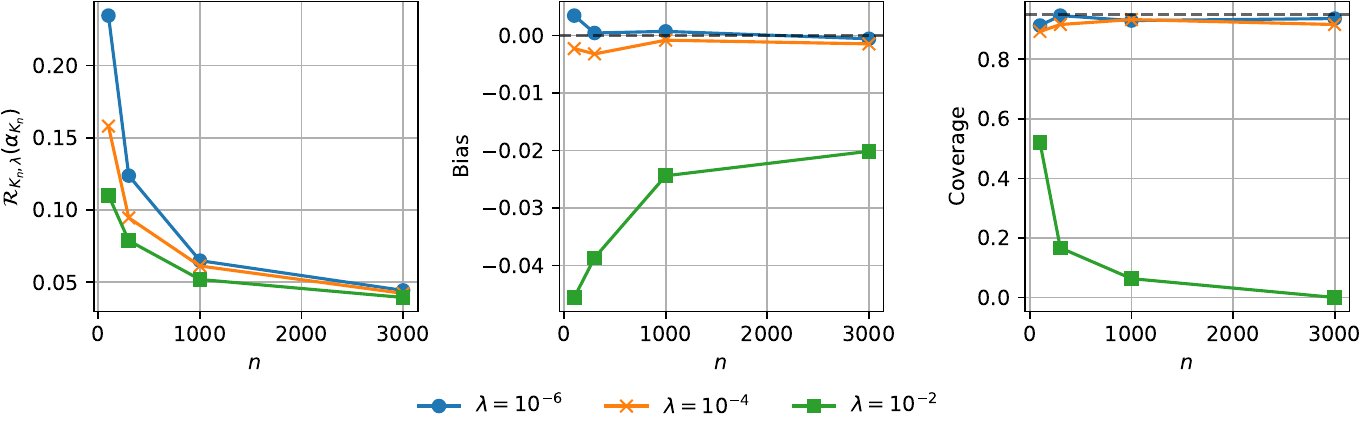}
        \caption{$\Psi_1(g_1) = \E[g_1(X) (1 + X)]$}
    \end{subfigure}
    \\[0.5em]
    \caption{Bootstrap inference results for functionals of $g_1$ with $X_i \overset{\text{i.i.d.}}{\sim} \mathcal{U}([-1,1])$, $\epsilon_i \sim \mathcal{N}(0, \sigma_\epsilon^2)$, $\sigma_\epsilon^2 = 0.04$. RWNNs use $\tanh$ activation, $A_k \overset{\text{i.i.d.}}{\sim} \mathcal{N}(0, \sigma_A^2)$ and $B_k \overset{\text{i.i.d.}}{\sim} \mathcal{U}([-\sigma_B, \sigma_B])$, with $\sigma_A = 5$ and $\sigma_B = 2$. Ridge penalty is set as $\lambda \in \{ 10^{-6}, 10^{-4}, 10^{-2} \}$. Lines indicate averages across Monte Carlo replications.}
    \label{fig:inference_rayleigh_bias_cover_gaussian_uniform}
\end{figure}
\clearpage}

\afterpage{%
\begin{figure}[p!]
    \centering
    \begin{subfigure}{0.98\textwidth}
    \centering
        \includegraphics[width=\textwidth]{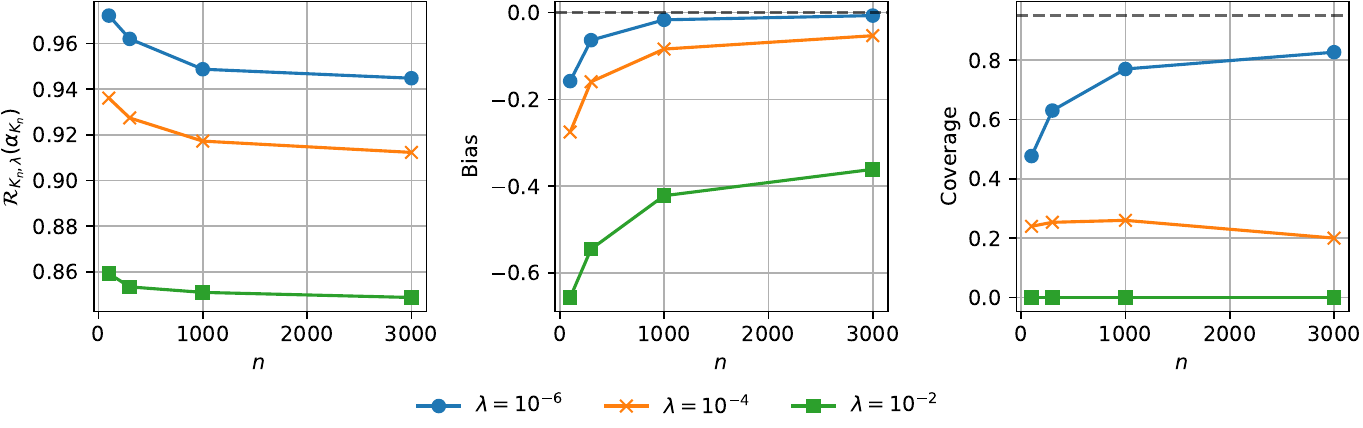}
        \caption{$\Psi_0(g_1) = g_1(0)$}
    \end{subfigure}
    \\[1em]
    \vspace{1em}
    \begin{subfigure}{0.98\textwidth}
    \centering
        \includegraphics[width=\textwidth]{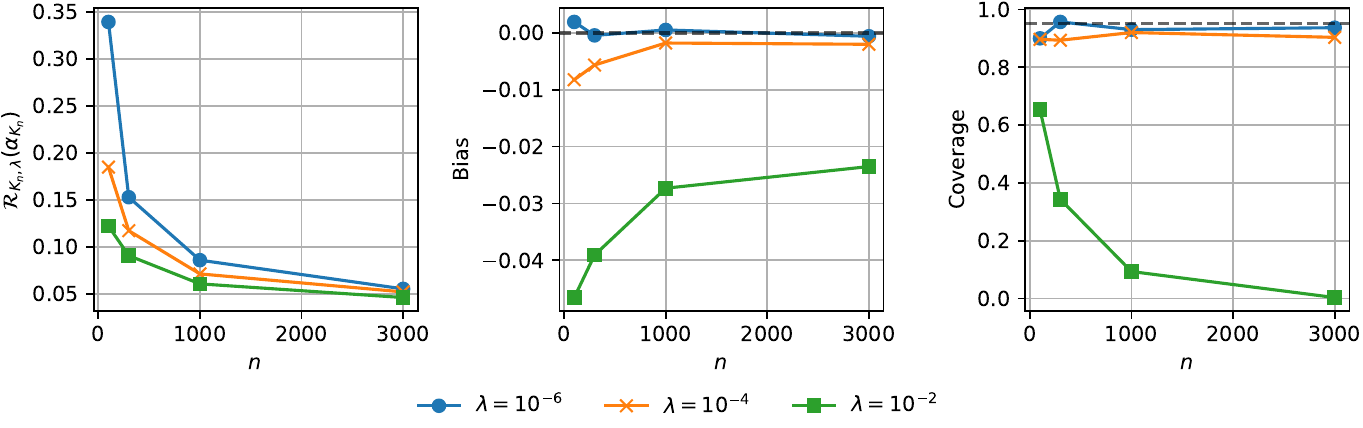}
        \caption{$\Psi_1(g_1) = \E[g_1(X) (1 + X)]$}
    \end{subfigure}
    \\[0.5em]
    \caption{Bootstrap inference results for functionals of $g_1$ with $X_i \overset{\text{i.i.d.}}{\sim} \mathcal{U}([-1,1])$, $\epsilon_i \sim \mathcal{N}(0, \sigma_\epsilon^2)$, $\sigma_\epsilon^2 = 0.04$. RWNNs use $\tanh$ activation, $A_k \overset{\text{i.i.d.}}{\sim} \mathcal{U}([-\sigma_A, \sigma_A])$ and $B_k \overset{\text{i.i.d.}}{\sim} \mathcal{U}([-\sigma_B, \sigma_B])$, with $\sigma_A = 10$ and $\sigma_B = 5$. Ridge penalty is set as $\lambda \in \{ 10^{-6}, 10^{-4}, 10^{-2} \}$. Lines indicate averages across Monte Carlo replications.}
    \label{fig:inference_rayleigh_bias_cover_uniform_uniform}
\end{figure}
\clearpage}

\afterpage{%
\begin{figure}[p]
    \centering
    \begin{subfigure}{0.48\textwidth}
    \centering
        \hspace{2em}
        $\sigma = \tanh$\\
        \includegraphics[width=\textwidth]{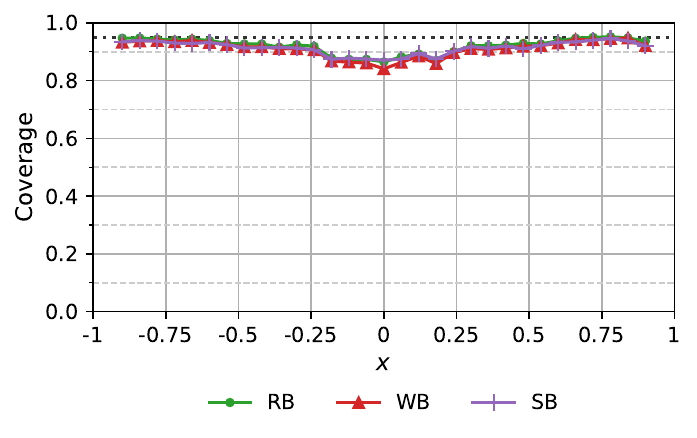}
    \end{subfigure}
    \hfill
    \begin{subfigure}{0.48\textwidth}
    \centering
        \hspace{2em}
        $\sigma = \text{ReLU}$\\
        \includegraphics[width=\textwidth]{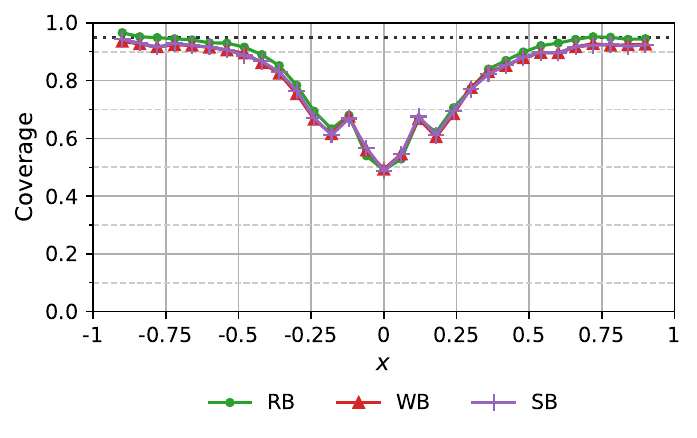}
    \end{subfigure}
    (a)~~{\small$%
        g_1(x)
    $}
    \\[20pt]
    \begin{subfigure}{0.48\textwidth}
    \centering
        \hspace{2em}
        $\sigma = \tanh$\\
        \includegraphics[width=\textwidth]{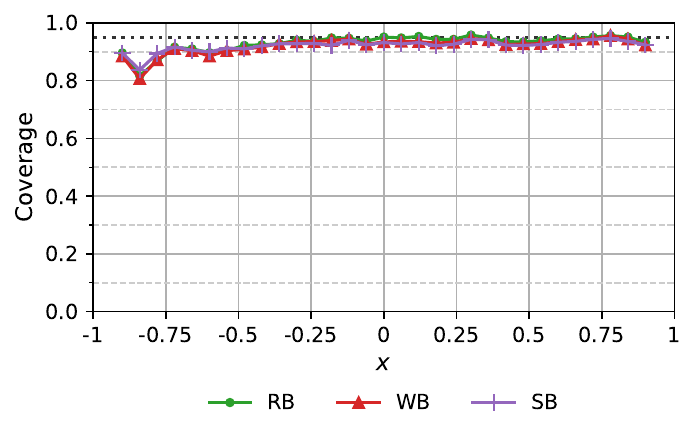}
    \end{subfigure}
    \hfill
    \begin{subfigure}{0.48\textwidth}
    \centering
        \hspace{2em}
        $\sigma = \text{ReLU}$\\
        \includegraphics[width=\textwidth]{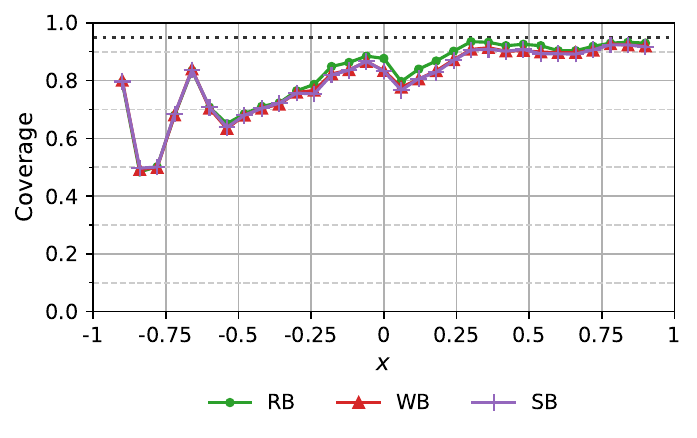}
    \end{subfigure}
    (b)~~{\small$%
        g_2(x)
    $}
    \\[20pt]
    \begin{subfigure}{0.48\textwidth}
    \centering
        \hspace{2em}
        $\sigma = \tanh$\\
        \includegraphics[width=\textwidth]{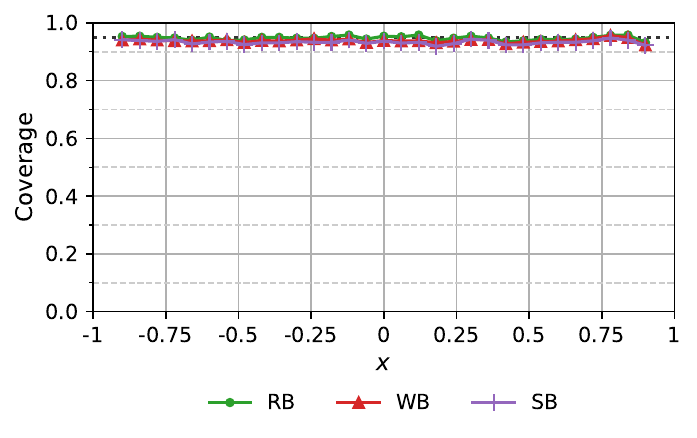}
    \end{subfigure}
    \hfill
    \begin{subfigure}{0.48\textwidth}
    \centering
        \hspace{2em}
        $\sigma = \text{ReLU}$\\
        \includegraphics[width=\textwidth]{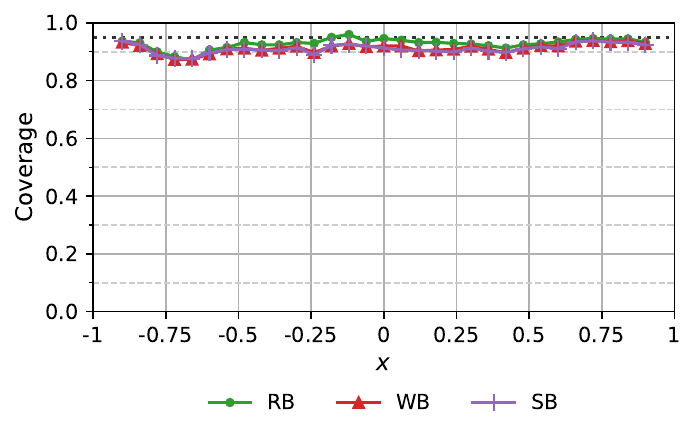}
    \end{subfigure}
    (c)~~{\small$%
        g_3(x)
    $}
    \vspace{1em}
    \caption{Pointwise coverage probabilities of residual bootstrap (RB), wild bootstrap (WB), and score bootstrap (SB) confidence intervals for  $\Psi_x(g_j)=g_j(x)$, $j \in \{1,2,3\}$ computed for $X_i \sim \mathcal{U}([-1,1])$ and RWNNs with $A_k \overset{\text{i.i.d.}}{\sim} \mathcal{N}$, $B_k \overset{\text{i.i.d.}}{\sim} \mathcal{U}$ and $\sigma \in \{\text{tanh}, \text{ReLU}\}$. $\text{MC} = 1000$ Monte Carlo replications and $\frak{B} = 1000$ bootstrap resamples.}
    \label{fig:inference_boot}
\end{figure}
\clearpage} %

\afterpage{
\begin{figure}[p]
    \centering
    \begin{subfigure}{0.48\textwidth}
    \centering
        \hspace{2em}
        $\sigma = \tanh$\\
        \includegraphics[width=\textwidth]{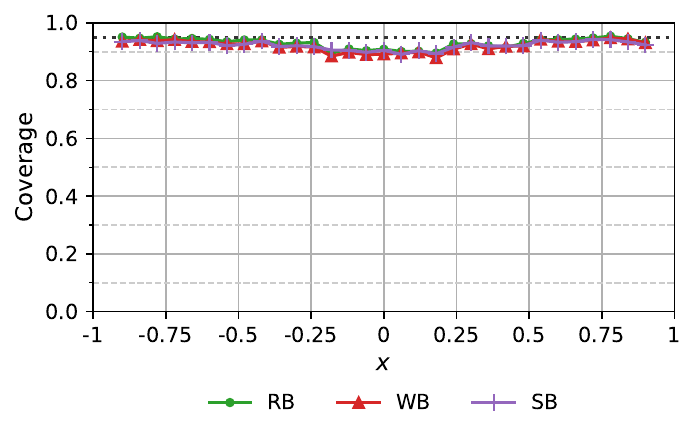}
    \end{subfigure}
    \hfill
    \begin{subfigure}{0.48\textwidth}
    \centering
        \hspace{2em}
        $\sigma = \text{ReLU}$\\
        \includegraphics[width=\textwidth]{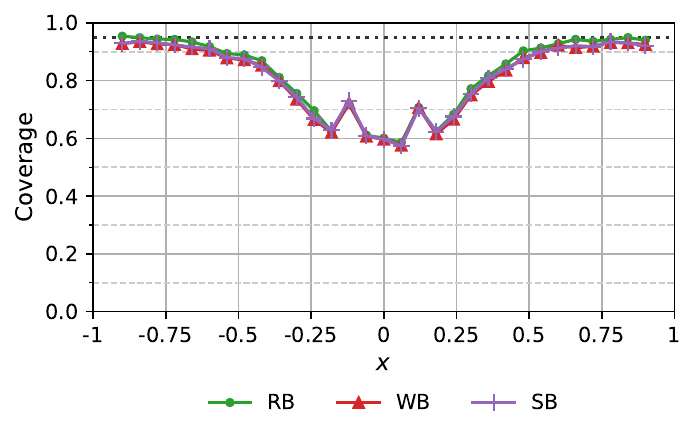}
    \end{subfigure}
    (a)~~{\small$%
        g_1(x)
    $}
    \\[20pt]
    \begin{subfigure}{0.48\textwidth}
    \centering
        \hspace{2em}
        $\sigma = \tanh$\\
        \includegraphics[width=\textwidth]{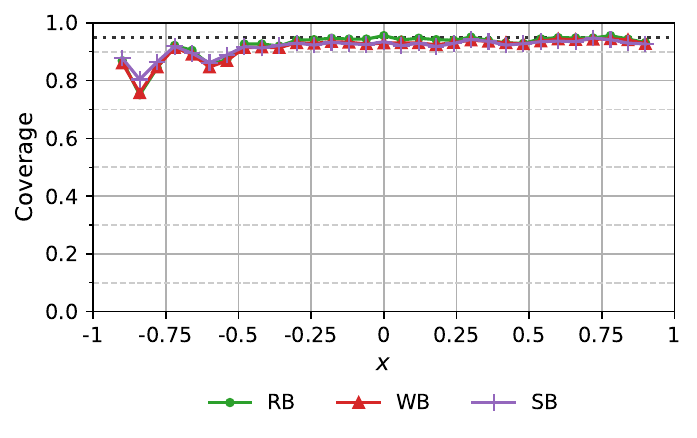}
    \end{subfigure}
    \hfill
    \begin{subfigure}{0.48\textwidth}
    \centering
        \hspace{2em}
        $\sigma = \text{ReLU}$\\
        \includegraphics[width=\textwidth]{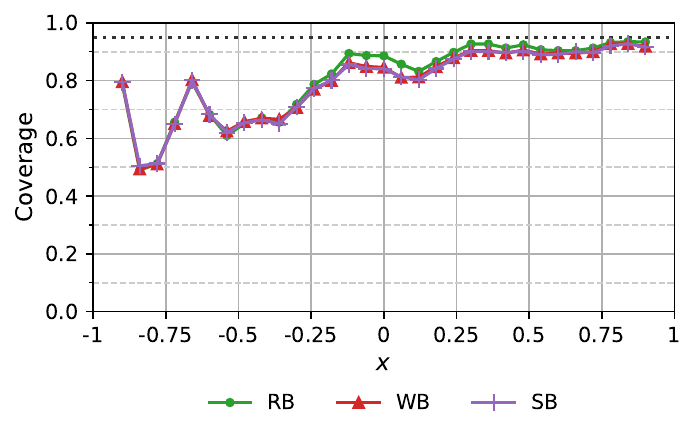}
    \end{subfigure}
    (b)~~{\small$%
        g_2(x)
    $}
    \\[20pt]
    \begin{subfigure}{0.48\textwidth}
    \centering
        \hspace{2em}
        $\sigma = \tanh$\\
        \includegraphics[width=\textwidth]{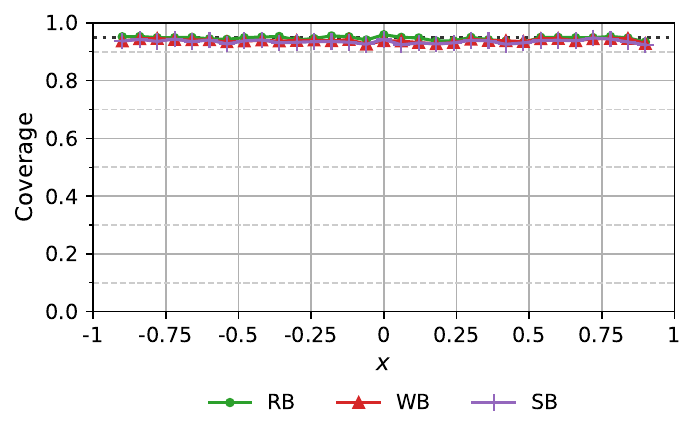}
    \end{subfigure}
    \hfill
    \begin{subfigure}{0.48\textwidth}
    \centering
        \hspace{2em}
        $\sigma = \text{ReLU}$\\
        \includegraphics[width=\textwidth]{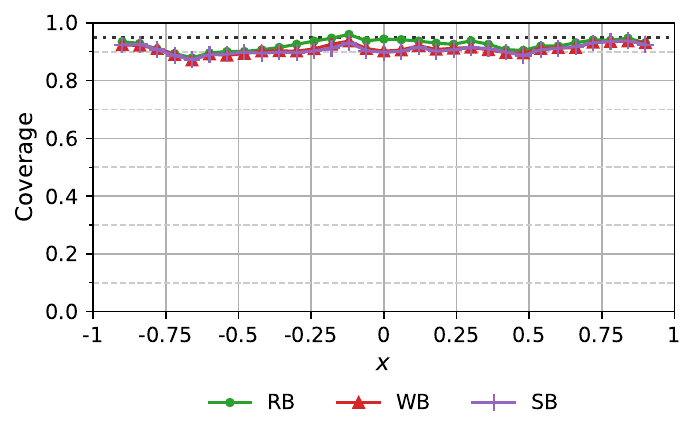}
    \end{subfigure}
    (c)~~{\small$%
        g_3(x)
    $}
    \vspace{1em}
    \caption{Pointwise coverage probabilities of residual bootstrap (RB), wild bootstrap (WB), and score bootstrap (SB) confidence intervals for  $\Psi_x(g_j)=g_j(x)$, $j \in \{1,2,3\}$ computed for $X_i \sim \mathcal{U}([-1,1])$ and RWNNs with $A_k \overset{\text{i.i.d.}}{\sim} \mathcal{N}$, $B_k \overset{\text{i.i.d.}}{\sim} \mathcal{N}$, and $\sigma \in \{\text{tanh}, \text{ReLU}\}$. $\text{MC} = 1000$ Monte Carlo replications and $\frak{B} = 1000$ bootstrap resamples.}
    \label{fig:inference_boot__NN}
\end{figure}
\clearpage} %

\afterpage{%
\begin{figure}[p]
    \centering
    \begin{subfigure}{0.48\textwidth}
    \centering
        \hspace{2em}
        $\sigma = \tanh$\\
        \includegraphics[width=\textwidth]{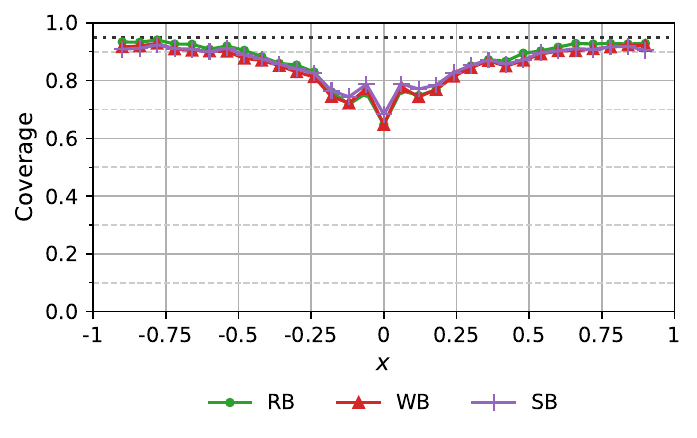}
    \end{subfigure}
    \hfill
    \begin{subfigure}{0.48\textwidth}
    \centering
        \hspace{2em}
        $\sigma = \text{ReLU}$\\
        \includegraphics[width=\textwidth]{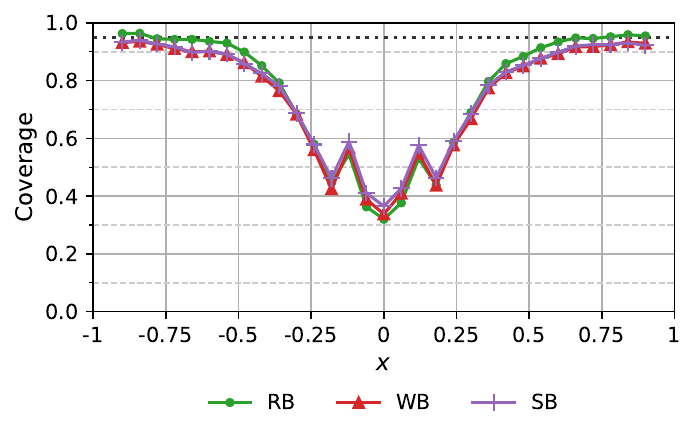}
    \end{subfigure}
    (a)~~{\small$%
        g_1(x)
    $}
    \\[20pt]
    \begin{subfigure}{0.48\textwidth}
    \centering
        \hspace{2em}
        $\sigma = \tanh$\\
        \includegraphics[width=\textwidth]{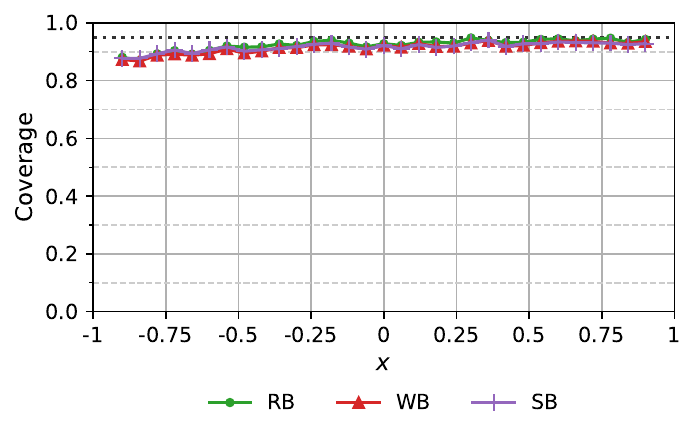}
    \end{subfigure}
    \hfill
    \begin{subfigure}{0.48\textwidth}
    \centering
        \hspace{2em}
        $\sigma = \text{ReLU}$\\
        \includegraphics[width=\textwidth]{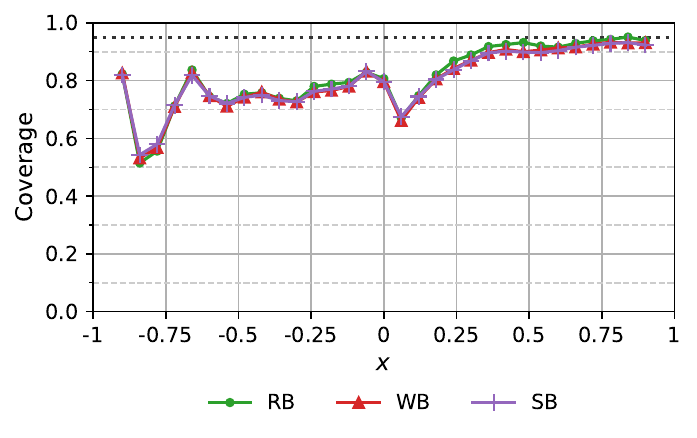}
    \end{subfigure}
    (b)~~{\small$%
        g_2(x)
    $}
    \\[20pt]
    \begin{subfigure}{0.48\textwidth}
    \centering
        \hspace{2em}
        $\sigma = \tanh$\\
        \includegraphics[width=\textwidth]{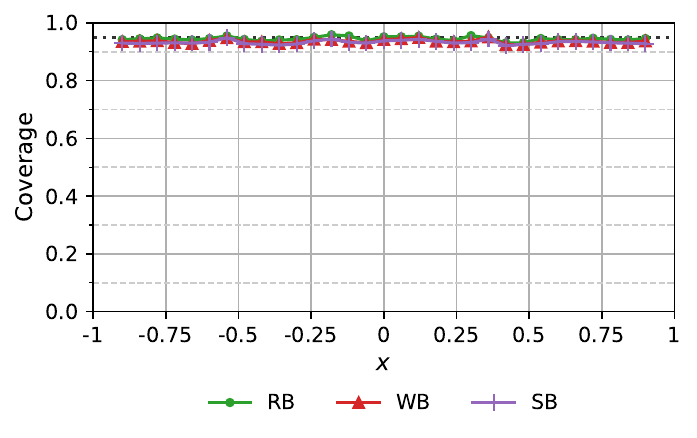}
    \end{subfigure}
    \hfill
    \begin{subfigure}{0.48\textwidth}
    \centering
        \hspace{2em}
        $\sigma = \text{ReLU}$\\
        \includegraphics[width=\textwidth]{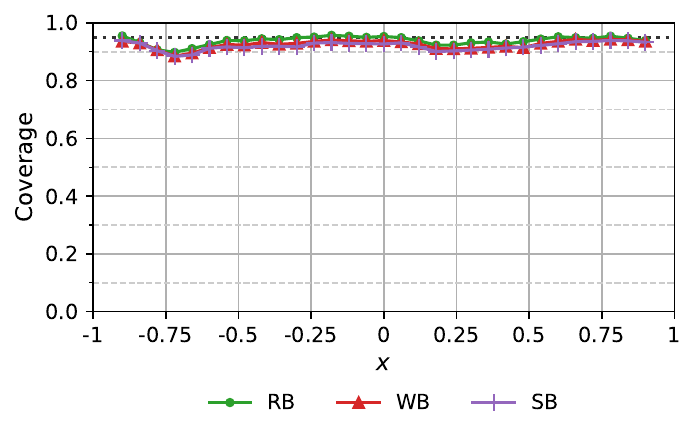}
    \end{subfigure}
    (c)~~{\small$%
        g_3(x)
    $}
    \vspace{1em}
    \caption{Pointwise coverage probabilities of residual bootstrap (RB), wild bootstrap (WB), and score bootstrap (SB) confidence intervals for  $\Psi_x(g_j)=g_j(x)$, $j \in \{1,2,3\}$ computed for $X_i \sim \mathcal{U}([-1,1])$ and RWNNs with $A_k \overset{\text{i.i.d.}}{\sim} \mathcal{U}$, $B_k \overset{\text{i.i.d.}}{\sim} \mathcal{U}$ and $\sigma \in \{\text{tanh}, \text{ReLU}\}$. $\text{MC} = 1000$ Monte Carlo replications and $\frak{B} = 1000$ bootstrap resamples.}
    \label{fig:inference_boot__UU}
\end{figure}
\clearpage} %

\afterpage{%
\begin{figure}[p]
    \centering
    \begin{subfigure}{0.48\textwidth}
    \centering
        \hspace{2em}
        $\sigma = \tanh$\\
        \includegraphics[width=\textwidth]{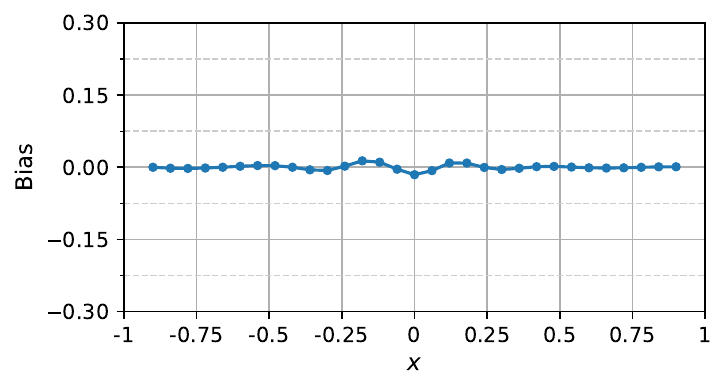}
    \end{subfigure}
    \hfill
    \begin{subfigure}{0.48\textwidth}
    \centering
        \hspace{2em}
        $\sigma = \text{ReLU}$\\
        \includegraphics[width=\textwidth]{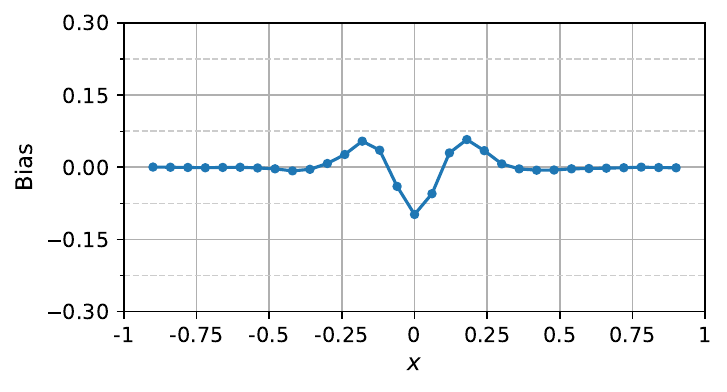}
    \end{subfigure}
    (a)~~{\small$%
        g_1(x)
    $}
    \\[20pt]
    \begin{subfigure}{0.48\textwidth}
    \centering
        \hspace{2em}
        $\sigma = \tanh$\\
        \includegraphics[width=\textwidth]{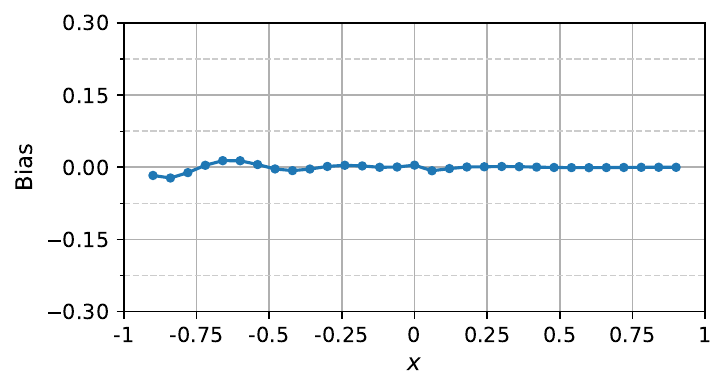}
    \end{subfigure}
    \hfill
    \begin{subfigure}{0.48\textwidth}
    \centering
        \hspace{2em}
        $\sigma = \text{ReLU}$\\
        \includegraphics[width=\textwidth]{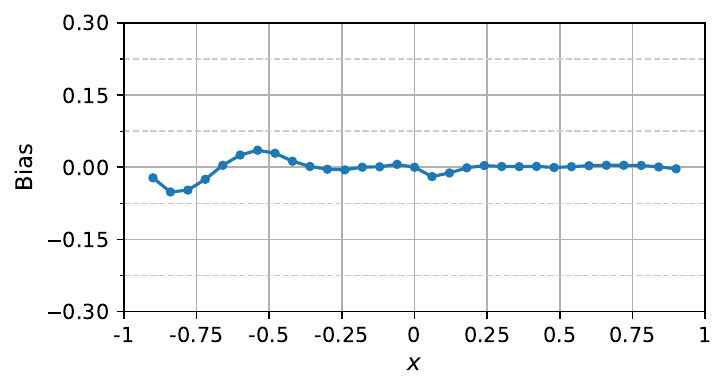}
    \end{subfigure}
    (b)~~{\small$%
        g_2(x)
    $}
    \\[20pt]
    \begin{subfigure}{0.48\textwidth}
    \centering
        \hspace{2em}
        $\sigma = \tanh$\\
        \includegraphics[width=\textwidth]{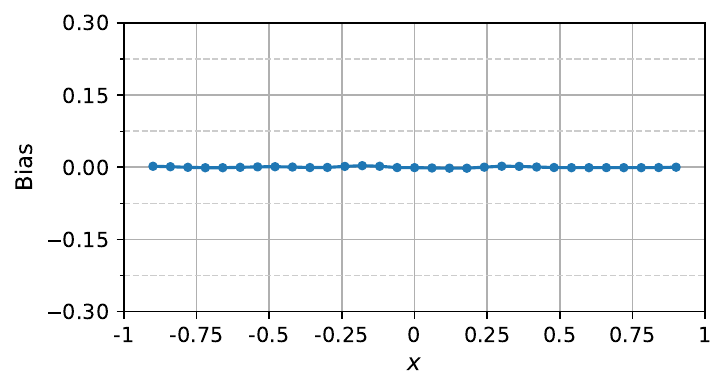}
    \end{subfigure}
    \hfill
    \begin{subfigure}{0.48\textwidth}
    \centering
        \hspace{2em}
        $\sigma = \text{ReLU}$\\
        \includegraphics[width=\textwidth]{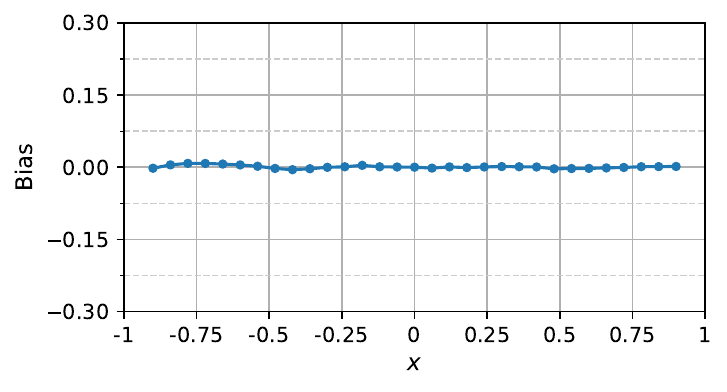}
    \end{subfigure}
    (c)~~{\small$%
        g_3(x)
    $}
    \vspace{1em}
    \caption{Pointwise bias for $\Psi_x(g_j)=g_j(x)$, $j \in \{1,2,3\}$ computed for $X_i \sim \mathcal{U}([-1,1])$ and RWNNs with $A_k \overset{\text{i.i.d.}}{\sim} \mathcal{N}$, $B_k \overset{\text{i.i.d.}}{\sim} \mathcal{U}$ and $\sigma \in \{\text{tanh}, \text{ReLU}\}$. $\text{MC} = 1000$ Monte Carlo replications.}
    \label{fig:inference_bias__NU}
\end{figure}
\clearpage} %

\afterpage{%
\begin{figure}[p]
    \centering
    \begin{subfigure}{0.48\textwidth}
    \centering
        \hspace{2em}
        $\sigma = \tanh$\\
        \includegraphics[width=\textwidth]{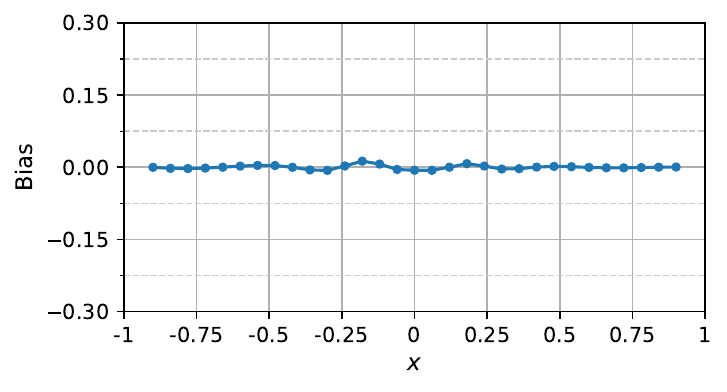}
    \end{subfigure}
    \hfill
    \begin{subfigure}{0.48\textwidth}
    \centering
        \hspace{2em}
        $\sigma = \text{ReLU}$\\
        \includegraphics[width=\textwidth]{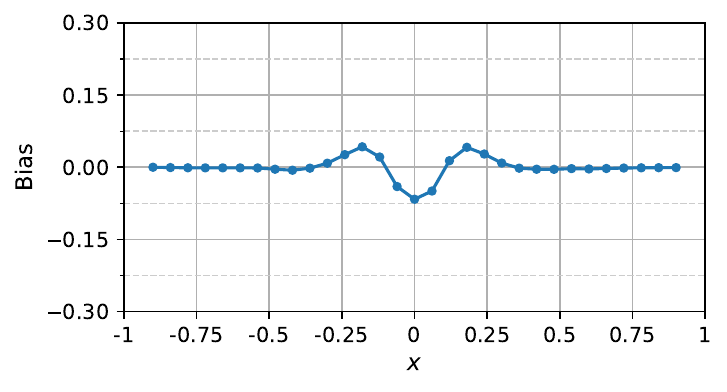}
    \end{subfigure}
    (a)~~{\small$%
        g_1(x)
    $}
    \\[20pt]
    \begin{subfigure}{0.48\textwidth}
    \centering
        \hspace{2em}
        $\sigma = \tanh$\\
        \includegraphics[width=\textwidth]{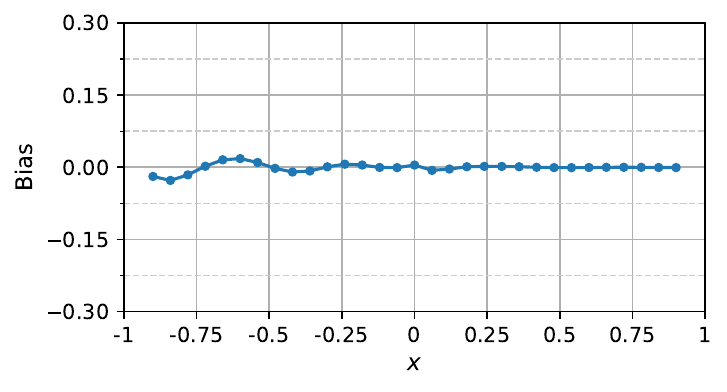}
    \end{subfigure}
    \hfill
    \begin{subfigure}{0.48\textwidth}
    \centering
        \hspace{2em}
        $\sigma = \text{ReLU}$\\
        \includegraphics[width=\textwidth]{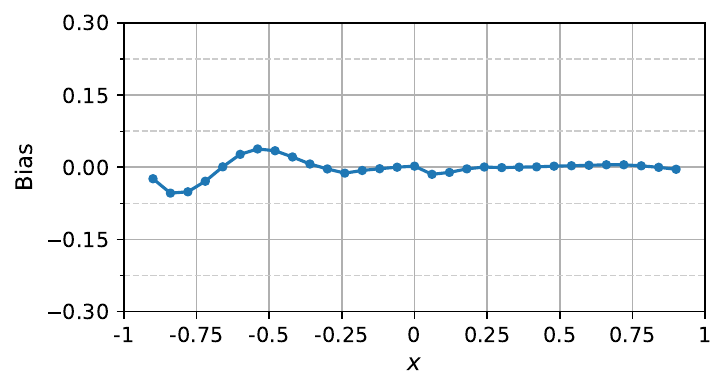}
    \end{subfigure}
    (b)~~{\small$%
        g_2(x)
    $}
    \\[20pt]
    \begin{subfigure}{0.48\textwidth}
    \centering
        \hspace{2em}
        $\sigma = \tanh$\\
        \includegraphics[width=\textwidth]{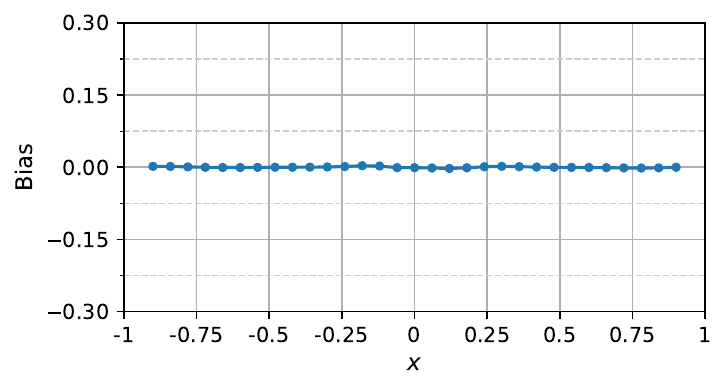}
    \end{subfigure}
    \hfill
    \begin{subfigure}{0.48\textwidth}
    \centering
        \hspace{2em}
        $\sigma = \text{ReLU}$\\
        \includegraphics[width=\textwidth]{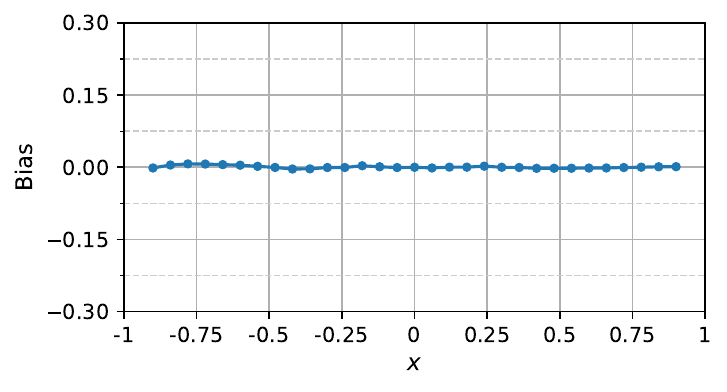}
    \end{subfigure}
    (c)~~{\small$%
        g_3(x)
    $}
    \vspace{1em}
    \caption{Pointwise bias for  $\Psi_x(g_j)=g_j(x)$, $j \in \{1,2,3\}$ computed for $X_i \sim \mathcal{U}([-1,1])$ and RWNNs with $A_k \overset{\text{i.i.d.}}{\sim} \mathcal{N}$, $B_k \overset{\text{i.i.d.}}{\sim} \mathcal{N}$ and $\sigma \in \{\text{tanh}, \text{ReLU}\}$. $\text{MC} = 1000$ Monte Carlo replications.}
    \label{fig:inference_bias__NN}
\end{figure}
\clearpage} %

\afterpage{%
\begin{figure}[p]
    \centering
    \begin{subfigure}{0.48\textwidth}
    \centering
        \hspace{2em}
        $\sigma = \tanh$\\
        \includegraphics[width=\textwidth]{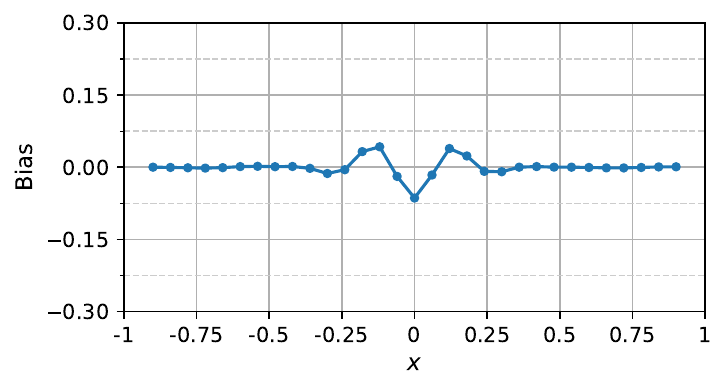}
    \end{subfigure}
    \hfill
    \begin{subfigure}{0.48\textwidth}
    \centering
        \hspace{2em}
        $\sigma = \text{ReLU}$\\
        \includegraphics[width=\textwidth]{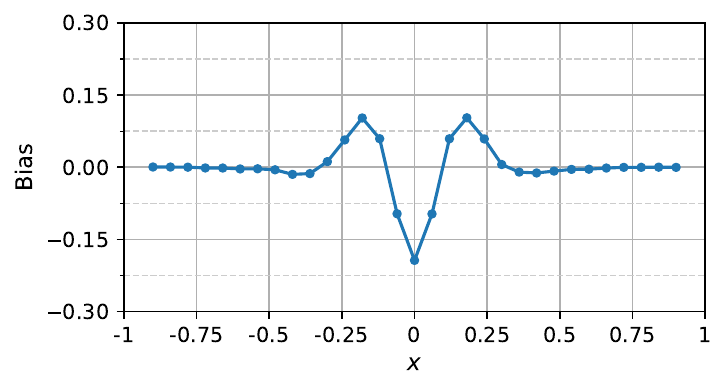}
    \end{subfigure}
    (a)~~{\small$%
        g_1(x)
    $}
    \\[20pt]
    \begin{subfigure}{0.48\textwidth}
    \centering
        \hspace{2em}
        $\sigma = \tanh$\\
        \includegraphics[width=\textwidth]{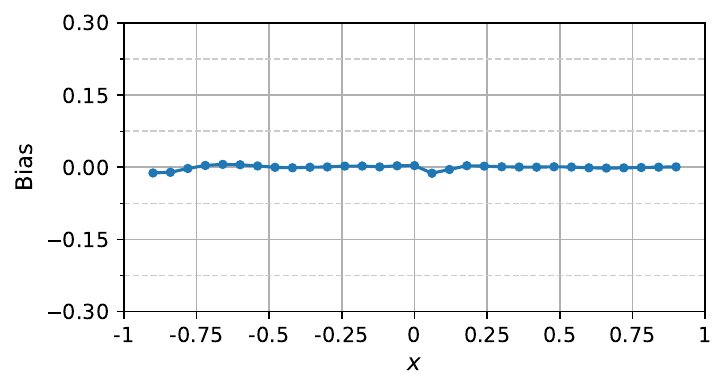}
    \end{subfigure}
    \hfill
    \begin{subfigure}{0.48\textwidth}
    \centering
        \hspace{2em}
        $\sigma = \text{ReLU}$\\
        \includegraphics[width=\textwidth]{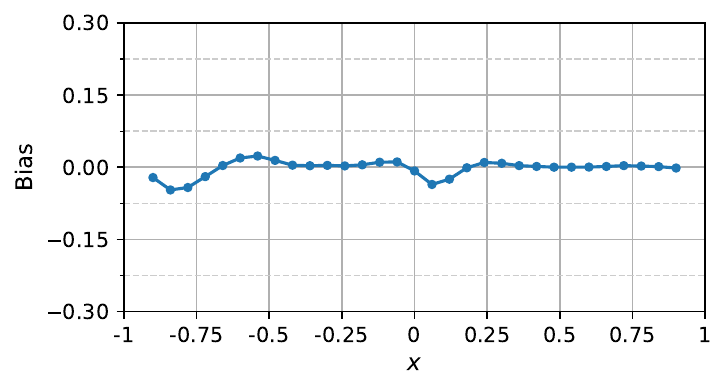}
    \end{subfigure}
    (b)~~{\small$%
        g_2(x)
    $}
    \\[20pt]
    \begin{subfigure}{0.48\textwidth}
    \centering
        \hspace{2em}
        $\sigma = \tanh$\\
        \includegraphics[width=\textwidth]{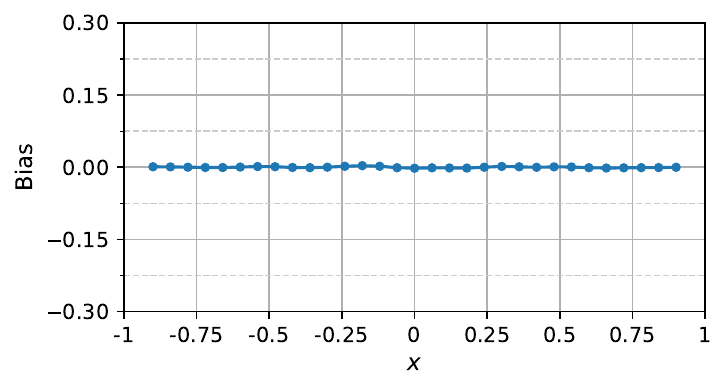}
    \end{subfigure}
    \hfill
    \begin{subfigure}{0.48\textwidth}
    \centering
        \hspace{2em}
        $\sigma = \text{ReLU}$\\
        \includegraphics[width=\textwidth]{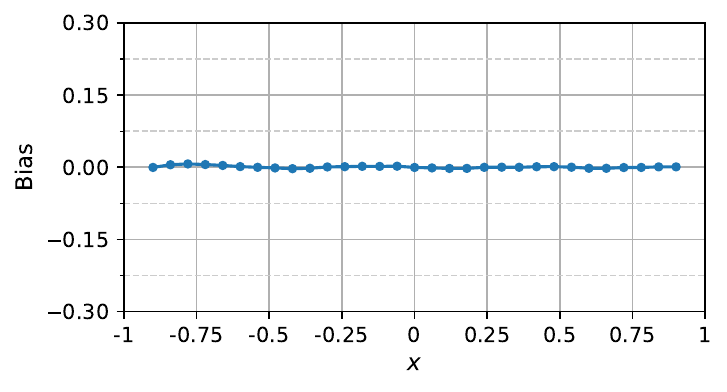}
    \end{subfigure}
    (c)~~{\small$%
        g_3(x)
    $}
    \vspace{1em}
    \caption{Pointwise bias for  $\Psi_x(g_j)=g_j(x)$, $j \in \{1,2,3\}$ computed for $X_i \sim \mathcal{U}([-1,1])$ and RWNNs with $A_k \overset{\text{i.i.d.}}{\sim} \mathcal{U}$, $B_k \overset{\text{i.i.d.}}{\sim} \mathcal{U}$ and $\sigma \in \{\text{tanh}, \text{ReLU}\}$. $\text{MC} = 1000$ Monte Carlo replications.}
    \label{fig:inference_bias__UU}
\end{figure}
\clearpage} %

\end{document}